\documentclass[aps,rmp,reprint,amsmath,amssymb,graphicx,longbibliography]{revtex4-1} 

\usepackage{bm,upgreek,bm,appendix,soul,color,graphicx}
\usepackage[normalem]{ulem}

\usepackage[T1]{fontenc} 

\def\kb{{k_{\rm B}}}
\def\kt{{k_{\rm B}T}}
\def\bk{{\bf k}}

\def\br{{\bf r}}
\def\bx{{\bf x}}
\def\bq{{\bf q}}

\def\m{{\bf m}}
\def\a{{\alpha}}

\def\bu{{\bf u}}
\def\be{{\bf e}}
\def\bR{{\bf R}}
\def\bRp{{\bf R}_p}
\def\btau{{\bm \tau}}
\def\D{\partial}
\def\d{\delta}
\def\w{\omega}
\def\ba{{\bf a}}

\def\bE{{\bf E}}

\def\bb{{\bf b}}
\def\bv{{\bf v}}
\def\bG{{\bf G}}
\def\bT{{\bf T}}

\def\bJ{{\bf F}}
\def\bD{{\bf D}}
\def\<{\langle}
\def\>{\rangle}
\def\k{\kappa}
\def\ve{\varepsilon}
\def\e{\epsilon}
\def\ef{\varepsilon_{\rm F}}
\def\kf{k_{\rm F}}
\def\wf{{\rm w}}

\def\ha{{\hat{a}}}
\def\hc{{\hat{c}}}
\def\hn{{\hat{n}}}
\def\hne{{\hat{n}_{\rm e}}}
\def\hnn{{\hat{n}_{\rm n}}}
\def\hp{{\hat{\psi}}}
\def\hpd{{\hat{\psi}^\dagger}}

\def\hT{{\hat{T}}}
\def\hU{{\hat{U}}}

\def\hP{{\hat{P}}}
\def\hH{{\hat{H}}}

\begin{document}

\raggedbottom 

\title{Electron-phonon interactions from first principles}

\author{Feliciano Giustino}
\affiliation{Department of Materials, University of Oxford, Parks Road, Oxford OX1 3PH, United Kingdom}

\date{\today{}}

\begin{abstract}
This article reviews the theory of electron-phonon interactions in solids from the point 
of view of {\it ab~initio} calculations. While the electron-phonon interaction has been 
studied for almost a century, predictive non-empirical calculations have become feasible 
only during the past two decades. Today it is possible to calculate from first principles 
many materials properties related to the electron-phonon interaction, including the critical 
temperature of conventional superconductors, the carrier mobility in semiconductors, the 
temperature dependence of optical spectra in direct and indirect-gap semiconductors, the 
relaxation rates of photoexcited carriers, the electron mass renormalization in angle-resolved 
photoelectron spectra, and the non-adiabatic corrections to phonon dispersion relations.
Here we review the theoretical and computational framework underlying modern electron-phonon 
calculations from first principles, as well as landmark investigations of the electron-phonon 
interaction in real materials. In the first part of the article we summarize the elementary 
theory of electron-phonon interactions and their calculations based on density-functional 
theory. In the second part we discuss a general field-theoretic formulation of the 
electron-phonon problem, and establish the connection with practical first-principles 
calculations. In the third part we review a number of recent investigations of electron-phonon 
interactions in the areas of vibrational spectroscopy, photoelectron spectroscopy, optical 
spectroscopy, transport, and superconductivity.
\end{abstract}


\maketitle

\tableofcontents{}

\section{Introduction}\label{sec.intro}

The interaction between fermions and bosons is one of the cornerstones of many-particle
physics. It is therefore unsurprising that, despite being one of the most thoroughly studied 
chapters of solid state physics, the interaction between electrons and phonons in solids 
continues to attract unrelenting attention.

Electron-phonon interactions (EPIs) are ubiquitous in condensed matter and materials physics.
For example, they underpin the temperature dependence of the electrical resistivity in metals 
and the carrier mobility in semiconductors, they give rise to conventional superconductivity, 
and contribute to optical absorption in indirect-gap semiconductors. In addition, EPIs enable 
the thermalization of hot carriers, determine the temperature dependence of electron energy 
bands in solids, and distort band structures and phonon dispersion relations of metals, leading 
to characteristic kinks and Kohn anomalies in photoemission and Raman/neutron spectra, respectively. 
EPIs also play a role in the areas of spintronics and quantum information, for example by coupling 
lattice and spin degrees of freedom in electromagnons, or by modulating the lifetimes of electron 
spins in color centers.

Given the fundamental and practical importance of electron-phonon interactions, it is perhaps 
surprising that the majority of theoretical studies in this area still rely on semi-empirical 
model Hamiltonians, especially in times when {\it ab~initio} calculations have become pervasive 
in every area of condensed matter and materials physics. The reason for this lag can be found 
in the complexity of electron-phonon calculations: while density functional theory (DFT) 
calculations of total energies and structural properties were already well established in the 
early 1980s \cite{Martin2004}, systematic {\it ab~initio} calculations of EPIs had to wait 
for the development of density functional perturbation theory (DFPT) for lattice dynamics 
between the late 1980s and the mid 1990s \cite{Baroni1987,Gonze1992,Savrasov1992}.

Despite this delayed start, the past two decades have witnessed tremendous progress in this 
area, and new exciting applications are becoming accessible as first-principles 
techniques for studying EPIs catch up with more established DFT methods. These advances
are driving the evolution from {\it qualitative} and {\it descriptive} theories of 
electron-phonon effects in model solids to {\it quantitative} and {\it predictive} 
theories of real materials. As the methodology for calculating EPIs from first principles 
is rapidly reaching maturity, it appears that the time is ripe for reviewing this vast, 
complex and fascinating landscape. 

One of the most authoritative reviews on the theory of EPIs is the 
classic book by \textcite{Grimvall1981}. This monumental work represents an 
unmissable reference for the specialist. However, as this book pre-dates the rise of 
{\it ab~initio} computational methods based on DFT, it inevitably misses the 
most recent developments in this area. The present article constitutes an attempt at
filling this gap by reflecting on what DFT calculations can contribute to the study
of electron-phonon physics. In addition, this article is also an opportunity
to establish a unified conceptual and mathematical framework in this incredibly diverse 
landscape, shed light on the key approximations, and identify some of the challenges 
and opportunities ahead.

As emphasized by the title `Electron-phonon interactions
from first principles', the aim of this article is to review the {\it ab~initio} 
theory of EPIs and to survey modern advances in {\it ab~initio} calculations of EPIs.
The reader interested in the fundamentals of electron-phonon physics or in theoretical
developments relating to model Hamiltonians is referred to the outstanding monographs by 
\textcite{Ziman1960}, \textcite{Grimvall1981}, \textcite{Schrieffer1983},
\textcite{Mahan1993}, and \textcite{Alexandrov2010}.

Among significant recent advances that are covered in this review we mention the 
zero-point renormalization and the temperature dependence of electronic band structures; 
the calculation of phonon-assisted optical absorption spectra; the electron mass 
renormalization and the kinks in angle-resolved photoemission spectra; the thermalization 
of hot carriers in semiconductors; the calculation of phonon-limited mobility;
the development of efficient computational techniques for calculating EPIs; 
and efforts to improve the predictive power of EPI calculations by going beyond 
standard density functional theory.

The review is organized as follows: Sec.~\ref{sec.history} provides an historical
perspective on the development of theories of the EPI, from early semi-empirical
approaches to modern first-principles calculations. In Sec.~\ref{sec.formal} we
examine the various components of DFT calculations of EPIs in solids, and set 
the formalism which will be used throughout this article. 
Section~\ref{sec.green} provides a synthesis of the most advanced field-theoretic
approaches employed to study EPIs, and Sec.~\ref{sec.green-recsp} makes the link
between the most general formalism and DFT calculations for real materials. In this
section the reader will find a number of expressions which are useful for practical
implementations. Section~\ref{sec.wannier} reviews advanced computational techniques 
for performing calculations of EPIs efficiently and accurately, such as Wannier 
interpolation and Fermi surface harmonics. Here we also discuss recent progress in 
the study of electron-phonon couplings in polar semiconductors. 
In Sec.~\ref{sec.nonadiab} we discuss recent calculations of phonons beyond the
adiabatic Born-Oppenheimer approximation. Section~\ref{sec.kinks} reviews calculations
of EPIs in the context of photoelectron spectroscopy. 
Section~\ref{sec.semicond} focuses
on the optical properties of semiconductors and insulators, in particular the
temperature dependence of the band structure and phonon-assisted optical processes.
In Sec.~\ref{sec.transport} we review calculations on the effects of EPIs on
carrier dynamics and transport, including carrier thermalization rates and mobilities.
Section~\ref{sec.supercond} discusses EPI calculations 
in the area of phonon-mediated superconductivity. 
Attempts at improving the accuracy and predictive power of {\it ab~initio} EPI calculations
by using more sophisticated electronic structure methods are discussed in 
Sec.~\ref{sec.beyonddft}. 
Finally in Sec.~\ref{sec.conclusions} we highlight the most pressing challenges in the
study of EPIs from first principles, and we present our conclusions. We leave to the
appendices some notational remarks and more technical discussions.

\section{Historical development}\label{sec.history}

The notion of `electron-phonon interactions' is as old as the quantum theory of solids.
In fact in the very same work where \textcite{Bloch1928} discussed the formal solutions 
of the Schr\"odinger equation in periodic potentials, Sec.~V begins with the all-telling 
title: ``The interaction of the electrons and the elastic waves of the lattice''. 
In this work the first quantum theory of the temperature-dependent electrical resistivity 
of metals was developed. It took only a few years for Bloch's `elastic waves' to be replaced 
by the brand-name `phonon' by \textcite{Frenkel1932}, thus establishing a tradition that 
continues unaltered almost a century later \cite{Walker1970}.

In order to discuss the early approaches to the electron-phonon problem, it is useful
to state right from the start the standard form of the Hamiltonian describing a coupled 
electron-phonon system:
  \begin{eqnarray}\label{eq.epi-hamilt}
  \hH &=& \sum_{n\bk} \ve_{n\bk} \hc_{n\bk}^\dagger \hc_{n\bk} +
      \sum_{\bq\nu} \hbar\w_{\bq\nu} (\ha_{\bq\nu}^\dagger \ha_{\bq\nu}+1/2) \nonumber \\
      &+& \!N_p^{-\frac{1}{2}}\!\! \sum_{\substack{\bk,\bq \\ m n \nu }} \!g_{mn\nu}(\bk,\bq) \,
      \hc_{m\bk+\bq}^\dagger \hc_{n\bk}(\ha_{\bq\nu}+\ha_{-\bq\nu}^\dagger) \nonumber \\
      \Bigg[\!\!&+&\!  N_p^{-1}\!\!\!\!
     \sum_{\substack{\bk,\bq,\bq'\\m n \nu\nu'}} \!g^{\rm DW}_{mn\nu\nu'}(\bk,\bq,\bq')\,
     \hc_{m\bk+\bq+\bq'}^\dagger \hc_{n\bk} \nonumber \\
    &\times&
   (\ha_{\bq\nu}+\ha^\dagger_{-\bq\nu}) (\ha_{\bq'\nu'}+\ha^\dagger_{-\bq'\nu'}) \Bigg]. \label{eq.ham-2-dw}
    \end{eqnarray}
In this expression the first line describes the separate electron and phonon subsystems
using the usual second-quantized formalism, while the second line specifies the mutual 
coupling between electrons and phonons to first order in the atomic displacements \cite{Mahan1993}. 
Here $\ve_{n\bk}$ is the single-particle eigenvalue 
of an electron with crystal momentum $\bk$ in the band $n$, $\w_{\bq\nu}$ is the frequency 
of a lattice vibration with crystal momentum $\bq$ in the branch $\nu$, and $\hc^\dagger_{n\bk}/\hc_{n\bk}$
($\ha^\dagger_{\bq\nu}/\ha_{\bq\nu}$) are the associated fermionic (bosonic) creation/destruction operators.
$N_p$~is the number of unit cells in the Born-von K\'arm\'an supercell (see Appendix~\ref{sec.bvk}).
The third and fourth lines of Eq.~(\ref{eq.epi-hamilt}) describe the electron-phonon coupling
Hamiltonian to second order in the atomic displacements. This contribution is rarely found in the
early literature (hence the square brackets), but it plays an important role in the theory of
temperature-dependent band structures (Sec.~\ref{sec.elec-SE}). 
The matrix elements $g_{mn\nu}(\bk,\bq)$ and $g^{\rm DW}_{mn\nu\nu'}(\bk,\bq,\bq')$ 
measure the strength of the coupling between the electron and the phonon subsystems, 
and have physical dimensions of an energy. 
Here the superscript `DW' stands for Debye-Waller, and relates
to the Debye-Waller self-energy to be discussed in Sec.~\ref{eq.explic-elec}.
Complete details as well as a derivation 
of Eq.~(\ref{eq.epi-hamilt}) will be provided in Sec.~\ref{sec.formal}.

The formal simplicity of Eq.~(\ref{eq.epi-hamilt}) conceals some important difficulties
that one faces when attempting to use this equation for predictive calculations.
For example, the electronic Hamiltonian relies on the assumption that the system under consideration
can be described in terms of well-defined quasi-particle excitations. Similarly,
the phonon term is meaningful only within the harmonic and the adiabatic approximations.
More importantly, Eq.~(\ref{eq.epi-hamilt}) does not provide us with any prescription for determining
the numerical parameters $\ve_{n\bk}$, $\w_{\bq\nu}$, $g_{mn\nu}(\bk,\bq)$, and
$g^{\rm DW}_{mn\nu\nu'}(\bk,\bq,\bq')$. 

In a sense the history of the study of electron-phonon interactions is really the history 
of how to calculate the parameters entering Eq.~(\ref{eq.epi-hamilt}) using procedures that
can be at once rigorous, reliable, and practical. As it will become clear in Sec.~\ref{sec.green}, 
despite enormous progress in this area, some conceptual difficulties still remain.

\subsection{Early approaches to the electron-phonon interaction}\label{sec.history-early}

\subsubsection{Metals}\label{sec.metals}

A clear account of the theory of EPIs until the late 1950s 
is given by \textcite{Ziman1960}. In the following we highlight only those aspects
that are relevant to the subsequent
discussion in this article. 

Early studies of electron-phonon interactions 
in solids were motivated by the quest for a quantum theory of the electrical resistivity 
in metals \cite{Hoddeson1980}.
The common denominator of most early approaches is that the electronic excitations
in Eq.~(\ref{eq.epi-hamilt}) were described using the free electron 
gas model, $\ve_{n\bk} = \hbar^2\bk^2/2m_{\rm e} -\ve_{\rm F}$, $m_{\rm e}$ being the electron 
mass and $\ve_{\rm F}$ the Fermi energy; the lattice vibrations were described 
as acoustic waves using the Debye model, $\w_{\bq\nu} = v_{\rm s} |\bq|$, $v_{\rm s}$ being
the speed of sound in the solid. Both approximations 
were reasonable given that the systems of interest included almost exclusively
elemental metals, and primarily monovalent alkali and noble metals~\cite{Mott1936}.
While these approximations were fairly
straightforward, it was considerably more challenging to determine the EPI matrix elements 
$g_{mn\nu}(\bk,\bq)$ using realistic approximations.

The very first expression of the electron-phonon matrix element was derived by \textcite{Bloch1928};
using contemporary notation it can be written as: 
  \begin{equation}\label{eq.matel-noumkl-2}
    g_{mn\nu}(\bk,\bq) = -i \left(\frac{\hbar}{2 N_p M_\k\w_{\bq\nu}}\right)^{\!\!1/2}\!\!\!
       \bq\cdot\be_{\kappa\nu}(\bq) \,V_0.
  \end{equation}
Here $M_\k$ is the mass of the $\k$-th nucleus, and $\be_{\kappa\nu}(\bq)$ is the polarization of the acoustic
wave corresponding to the wavevector $\bq$ and mode $\nu$.
The term $V_0$ represents a unit-cell average of the `effective' potential experienced by the electrons
in the crystal. 
Equation~(\ref{eq.matel-noumkl-2}) was meant to describe the scattering from an initial
electronic state with wavevector $\bk$ to a final state with wavevector $\bk+\bq$, via
an acoustic phonon of wavevector $\bq$ and frequency $\w_{\bq\nu}$. The formula was developed for
free electron metals, and neglects so-called `umklapp' (folding) processes, i.e.\ scattering
events whereby $\bk$ goes into $\bk+\bq+\bG$ with $\bG$ being a reciprocal lattice vector.
A derivation of Eq.~(\ref{eq.matel-noumkl-2}) is provided in Sec.~\ref{sec.matel-approx}.
In order to determine $V_0$ \textcite{Bloch1928} argued that the crystal may be described
as a continuous deformable medium. Starting from this assumption he reached the conclusion
that the average potential can be approximated as $V_0 = \hbar^2/(16m_e a_0^2)$ 
($a_0$ is the Bohr radius).
Even though Bloch's matrix element is no longer in use, this model provides helpful
insight into the nature of EPIs in monovalent metals. For example the 
so-called `polarization factor' in Eq.~(\ref{eq.matel-noumkl-2}), $\bq\cdot\be_{\kappa\nu}(\bq)$,
shows that (in the absence of umklapp processes) only longitudinal sound waves scatter 
electrons. 

\textcite{Nordheim1931} proposed a refinement of Bloch's model whereby the 
average potential $V_0$ in Eq.~(\ref{eq.matel-noumkl-2}) is replaced by the Fourier component 
$V_\k(\bq)$ of the ionic Coulomb potential (see Sec.~\ref{sec.matel-approx}). The key assumption
underlying this model is that the effective potential experienced by the electrons is simply
the sum of the individual bare ionic potentials of each nucleus. When a nucleus
is displaced from its equilibrium position, the corresponding potential also shifts rigidly.
This is the so-called the `rigid-ion' approximation.

The main difficulty that arises with the rigid-ion model is that the Fourier transform of the
Coulomb potential diverges as $q^{-2}$ for $q=|\bq|\rightarrow 0$; this leads to unrealistically strong
EPIs. In order to circumvent this difficulty
\textcite{Mott1936} proposed to truncate the ionic potential at the boundary of the
Wigner-Seitz unit cell of the crystal. This choice represents the first attempt at including 
the {\it electronic screening} of the nuclear potential in a rudimentary form. In practice \textcite{Mott1936} 
calculated the Fourier transform of $V_\k(\br)$ by restricting the integration over a Wigner-Seitz cell;
the resulting potential 
is no longer singular at long wavelengths.
A detailed discussion of this model can be found in \cite{Ziman1960}.

Despite some initial successes in the study of the electrical conductivity of metals,
the descriptive power of these early models was undermined by the complete neglect 
of the electronic response to the ionic displacements.
The first attempt at describing the effect of the electronic screening
was made by \textcite{Bardeen1937}. In his model the average potential $V_0$ in Eq.~(\ref{eq.matel-noumkl-2})
is replaced by: 
  \begin{equation}\label{eq.bardeen-history}
     V_0 \rightarrow V_\k(\bq)/\e(q),
  \end{equation}
where $\e(q)$ is the Lindhard function \cite{Mahan1993}:
  \begin{equation}\label{eq.lindhard-history}
  \e(q) = 1+ (k_{\rm TF}/q)^2 F(q/2\kf).
  \end{equation}
Here $k_{\rm TF}$ and $\kf$ are the Thomas-Fermi screening wavevector and the Fermi wavevector,
respectively, and
 $ F(x) = 1/2 + (4x)^{-1}(1-x^2)\log|1+x|/|1-x|$.
A derivation of Bardeen's model is provided in Sec.~\ref{sec.matel-approx}.
Since $\e(q) \rightarrow (k_{\rm TF}/q)^2$ for $q\rightarrow 0$,
the sigularity of the electron-nuclei potential is removed in Bardeen's matrix element.
The work of \textcite{Bardeen1937} can be considered as a precursor
of modern {\it ab~initio} approaches, insofar the calculation of the matrix element
was carried out using a self-consistent field method within the linearized Hartree theory.
This strategy is similar in spirit to modern DFPT calculations.

The key qualitative difference between the approach of \textcite{Bardeen1937} and modern techniques
lies in the neglect of exchange and correlation effects in the screening. A possible route
to overcome this limitation was proposed by \textcite{Bardeen1955}. In this work the
authors considered the role of a screened exchange interaction in the electron-phonon
problem (see Appendix~B of their work), however the mathematical complexity of the
formalism prevented further progress along this direction. Similar efforts were undertaken 
by \textcite{Hone1960}, and a more detailed account of the early approximations to exchange
and correlation can be found in \cite{Grimvall1981}.

The most interesting aspect of the work by \textcite{Bardeen1955}, as well as previous work 
along the same lines by \textcite{Nakajima1954}, is that for the first time the electron-phonon
problem was addressed using a {\it field-theoretic} approach. 

One interesting feature in the theory of \citeauthor{Bardeen1955} 
is that their field-theoretic formulation naturally leads to a {\it retarded}  electron-phonon vertex: the
effective potential experienced by electrons upon the displacement of nuclei depends
on how fast this displacement takes place. In this approach the effective potential $V_0$
in Eq.~(\ref{eq.matel-noumkl-2}) is replaced by the {\it dynamically} screened potential:
    \begin{equation}\label{eq.bardeen-history-1955}
     V_0 \rightarrow  V_\k(\bq)/\e(q,\w_{\bq\nu}).
  \end{equation}
Here $\e(q,\w)$ is the frequency-dependent Lindhard function \cite{Mahan1993},
and the effect of electronic screening is evaluated at the phonon frequency, $\w=\w_{\bq\nu}$. 
Somewhat surprisingly, this development was not followed up in the literature on {\it ab~initio} calculations
of EPIs.

\subsubsection{Semiconductors}\label{sec.semiconductors}

While the investigation of electron-phonon effects was initially restricted to monovalent
metals, the formal developments were soon extended to the case of more complex systems
such as semiconductors. Carriers in semiconductors are typically confined within
a narrow energy range near the band extrema; consequently it is expected 
that the dominant electron-phonon scattering mechanisms will involve long-wavelength 
phonons ($q\rightarrow 0$). This concept was formalized by \textcite{Shockley1950,Bardeen1950},
laying the foundations of the `deformation-potential' method.

In the deformation potential approach it is assumed that the atomic displacements can be
described by long-wavelength acoustic waves, and these can be related in turn to the elastic strain 
of the crystal. Using concepts from the effective mass theory, \citeauthor{Bardeen1950} showed that
in this approximation the potential $V_0$ in Eq.~(\ref{eq.matel-noumkl-2}) can be replaced by:
   \begin{equation}\label{eq.defpot}
   V_0\rightarrow E_{1,n\bk} = \Omega\, \D\ve_{n\bk}/\D\Omega,
   \end{equation}
where $\Omega$ represents the volume of the unit cell, and the electron eigenvalues correspond
to the valence or conduction band extrema. The derivation of this result can be found in
Appendix~B of \cite{Bardeen1950}. The deformation potentials $E_1$ were
obtained empirically; for example \citeauthor{Bardeen1950} determined these values 
for the band extrema of silicon by fitting mobility data.
More complex scenarios such as anisotropic constant-energy surfaces in semiconductors
were subsequently addressed by considering the effects of shear deformations \cite{Dumke1956}.
While the concept of deformation potentials has become a classic in semiconductor physics,
this method relies on a semi-empirical approach and lacks predictive power.

\subsubsection{Ionic crystals}

A class of materials that played an important role in the development 
of the theory of EPIs is that of ionic crystals. The qualitative difference between ionic solids
and the systems discussed in Secs.~\ref{sec.metals}-\ref{sec.semiconductors} is that
the atomic displacements can generate long-ranged electric fields; these fields provide 
a new scattering channel for electrons and holes.

The theory of polar electron-phonon coupling started with the investigation of the
electron mean free path in ionic crystals, in search for a theoretical model of
dielectric breakdown in insulators \cite{Froehlich1937,Froehlich1939}.
The central idea of these models is that in insulators the density of free
carriers is very low, therefore it is sensible to consider a 
single electron interacting with the polarization field of the ionic lattice.

The Fr\"ohlich model is similar in spirit to the contemporary work of \textcite{Bardeen1937} 
for metals. The main difference is that Fr\"ohlich considered the screening arising 
from the dielectric polarization of an insulating crystal, while Bardeen considered
the screening arising from the response of the Fermi sea.

\textcite{Froehlich1950b} showed that in the 
case of isotropic ionic crystals the effective potential $V_0$ appearing in Eq.~(\ref{eq.matel-noumkl-2}) 
must be replaced by:
  \begin{equation}\label{eq.V0-polar}
    V_0 \rightarrow - \left[\frac{e^2 M_\k \w_{\bq\nu}^2}{\epsilon_0 \,\Omega}
        \left(\frac{1}{\e^\infty}-\frac{1}{\e^0}\right)\right]^{\!\frac{1}{2}} \!\!\!\frac{1}{\,\,|\bq|^2}.
  \end{equation}
In this expression $e$ is the electron charge, $\epsilon_0$ is the dielectric permittivity of vacuum, 
$\e^0$ and $\e^\infty$ are the static and the high-frequency relative permittivities, respectively.
This result is derived in Sec.~\ref{sec.polar}. 
Using Eqs.~(\ref{eq.V0-polar}) and (\ref{eq.matel-noumkl-2}) we see that when
$\e_0 > \e_\infty$ the matrix element $g_{mn\nu}(\bk,\bq)$ diverges as $|\bq|^{-1}$ 
at long wavelengths. This singular behavior can lead to very strong EPIs,
and provides the physical basis for the phenomenon of electron self-trapping in polarons 
(\citeauthor{Pekar1946}, \citeyear{Pekar1946}; \citeauthor{Emin2013}, \citeyear{Emin2013}). 
The initial studies in this area were rapidly followed by more refined approaches based on
field-theoretic methods \cite{Lee1953}.
A comprehensive discussion of the various models can be found in the original review article 
by \textcite{Froehlich1954}.

\subsection{The pseudopotential method}\label{sec.emp-pseudo}

The approximations underpinning the models discussed in Sec.~\ref{sec.history-early} 
become inadequate when one tries 
to study EPIs for elements across the periodic table. This and other limitations stimulated 
the development of the {\it pseudopotential} 
method, starting in the late 1950s with the work of \textcite{Phillips1959}. The theory of pseudopotentials 
is too vast to be summarized in a few lines, and the reader is referred to Chapter~11 of \cite{Martin2004} 
for a thorough discussion. Here we only highlight the aspects that are relevant to the calculation
of EPIs.

The genesis of the pseudopotential method is linked with the question on how the valence electrons
of metals could be described using the electron gas model, even though the orthogonality
to the core states imposes rapid fluctuations of the valence wavefunctions near the atomic cores. 
In order to address this question, it is useful to go through the key steps of the orthogonalized planewaves 
method \cite{Herring1940}. In this method one considers planewaves $|\bk+\bG\>$ for 
the wavevector $\bk+\bG$, and projects out the component belonging the Hilbert subspace spanned 
by core electrons. This is done by defining $|\bk+\bG\>_{\rm OPW} = |\bk+\bG\> - {\sum}_c |\phi_c\> 
\<\phi_c|\bk+\bG\>$, where the $|\phi_c\>$ represent the core states
of all atoms in the system. The functions $|\bk+\bG\>_{\rm OPW}$ are by construction
orthogonal to core states, therefore they can be used to expand the valence electron wavefunctions
$|\psi_{n\bk}\>$ using only a few basis elements: $|\psi_{n\bk}\> = \sum_\bG c_\bk(\bG) |\bk+\bG\>_{\rm OPW}$.
In the language of pseudopotential theory $|\psi_{n\bk}\>$ is referred to as the `all-electron'
wavefunction, while the function $|\tilde{\psi}_{n\bk}\> = \sum_\bG c_\bk(\bG) |\bk+\bG\>$ is
referred to as the `pseudo' wavefunction. The all-electron and the pseudo wavefunctions are
simply related as follows:
  \begin{equation}\label{eq.OPW}
  |\psi_{n\bk}\> = \hat{\mathcal{T}}\,|\tilde{\psi}_{n\bk}\>, \mbox{ with } 
  \hat{\mathcal{T}} = 1-{\sum}_c |\phi_c\> \<\phi_c|.
  \end{equation}
Here we used a modern notation borrowed from the projector-augmented wave (PAW) method 
of \textcite{Bloechl1994}. By construction, the pseudo-wavefunction $|\tilde{\psi}_{n\bk}\>$
does not exhibit rapid fluctuations near the atomic cores.
The projector operator $\hat{\mathcal{T}}$ is now used to
rewrite the single-particle Schr\"odinger equation for the all-electron 
wavefunction (e.g.~the Kohn-Sham equations)
in terms of the pseudo-wavefunctions. Using $\hH |\psi_{n\bk}\> = \ve_{n\bk} |\psi_{n\bk}\>$
and Eq.~(\ref{eq.OPW}) we have:
  \begin{equation}\label{eq.paw}
  \hat{\mathcal{T}}^\dagger\,\hH\, \hat{\mathcal{T}}\,\,|\tilde{\psi}_{n\bk}\> = \ve_{n\bk} \,
         \hat{\mathcal{T}}^\dagger \hat{\mathcal{T}}\,\,|\tilde{\psi}_{n\bk}\>,
  \end{equation}
which is a generalized eigenvalue problem.
By replacing the definition of $\hat{\mathcal{T}}$ given above one finds \cite{Phillips1959}:
  \begin{equation}
  ( \hH + \hat{V}^{\rm rep}) |\tilde{\psi}_{n\bk}\> = \ve_{n\bk}
         |\tilde{\psi}_{n\bk}\>,
  \end{equation}
with $\hat{V}^{\rm rep} = {\sum}_c(\ve_{n\bk}-\ve_c)|\phi_c\> \<\phi_c|$ and
$\ve_c$ being the eigenvalue of a core electron. 
Clearly the additional potential $\hat{V}_{\rm rep}$ is strongly repulsive and is localized near the
atomic cores. \textcite{CohenMH1961} showed that this extra potential largely cancels
the attractive potential of the nuclei. This is the reason why valence electrons in metals
behave almost like free electrons.

The practical consequence of these developments is that it is possible to define smooth effective 
`pseudo-potentials' for systematic band structure calculations, whose form factors include
only a few Fourier components 
(\citeauthor{Phillips1959b}, \citeyear{Phillips1959b}; \citeauthor{Heine1964}, \citeyear{Heine1964};
\citeauthor{Animalu1965}, \citeyear{Animalu1965}; \citeauthor{Cohen1966}, \citeyear{Cohen1966}).

The use of pseudopotentials in electron-phonon calculations started with the works of \textcite{Sham1961,Sham1963}.
\textcite{Sham1961} showed that, if the pseudopotential can be described by a local function, then 
the electron-phonon matrix element $g_{mn\nu}(\bk,\bq)$ can be calculated
by replacing the all-electron potentials 
and wavefunctions by the corresponding pseudo-potentials and pseudo-wavefunctions.
In this approach the pseudo-potentials move around rigidly with the ionic cores, therefore
we are dealing effectively with an improved version of the rigid-ion approximation discussed 
in Sec.~\ref{sec.history-early}.

The pseudopotential method was employed by \textcite{Shuey1965} in order to calculate the
electron-phonon matrix elements in germanium.
Shortly afterwards many calculations of electron-phonon interactions based on the pseudopotential
method appeared in the literature, including work on the resistivity of metals 
\cite{Carbotte1967,Dynes1968,Hayman1971,Kaveh1972}, the electron mass-enhancement in metals
(\citeauthor{Ashcroft1965}, \citeyear{Ashcroft1965};
\citeauthor{Grimvall1969}, \citeyear{Grimvall1969};
\citeauthor{Allen1970}, \citeyear{Allen1970}; 
\citeauthor{Allen1972b}, \citeyear{Allen1972b};
\citeauthor{Allen1972}, \citeyear{Allen1972})
the superconducting 
transition temperatures within the McMillan formalism 
(\citeauthor{Allen1968}, \citeyear{Allen1968}; \citeauthor{Allen1969}, \citeyear{Allen1969}),
the mobility of semiconductors \cite{Ralph1970}, and the temperature dependence of semiconductor
band structures \cite{Allen1981,Allen1983}.
These calculations were mostly based on phonon dispersion relations extracted from neutron scattering
data, and the results were in reasonable agreement with experiment. It seems fair to say 
that the pseudopotential method enabled the evolution from {\it qualitative} to {\it quantitative} 
calculations of electron-phonon interactions.

Before proceeding we note that, although Eqs.~(\ref{eq.OPW}) and (\ref{eq.paw}) 
were introduced starting from the method of orthogonalized planewaves, there exists considerable
freedom in the choice of the operator $\hat{\mathcal{T}}$. 
In practice $\hat{\mathcal{T}}$ can be chosen so as to make $\tilde{\psi}_{n\bk}$ as smooth
as possible, while retaining information on the all-electron wavefunctions near the ionic cores. 
This was achieved by the PAW method of \textcite{Bloechl1994}. Broadly speaking it is also
possible to re-interpret the historical development of the pseudopotential method as the evolution
of the projector $\hat{\mathcal{T}}$. In fact \citeauthor{Bloechl1994} showed how the
most popular pseudopotential methods 
(\citeauthor{Hamann1979}, \citeyear{Hamann1979}; 
\citeauthor{Bachelet1982}, \citeyear{Bachelet1982}; 
\citeauthor{Troullier1991}, \citeyear{Troullier1991};
\citeauthor{Vanderbilt1990}, \citeyear{Vanderbilt1990})
can be derived from the PAW method under specific approximations.

\subsection{\textit{Ab initio} self-consistent field calculations}

Predictive calculations of EPIs became possible with the development of {\it ab~initio} 
DFT techniques. The key advantage of DFT methods is the
possibility of calculating electron band structures, phonon dispersion relations, and electron-phonon
matrix elements entirely from first principles. Historically, DFT started with
the works of \textcite{Hohenberg1964,Kohn1965}. However, its widespread use had to wait for 
the development of accurate parametrizations of the exchange and correlation energy of the electron gas
(\citeauthor{Hedin1971}, \citeyear{Hedin1971};
\citeauthor{vonBarth1972}, \citeyear{vonBarth1972};
\citeauthor{Gunnarrsson1974}, \citeyear{Gunnarrsson1974};
\citeauthor{Ceperley1980}, \citeyear{Ceperley1980};
\citeauthor{Perdew1981}, \citeyear{Perdew1981}).
An introduction to DFT 
techniques can be found in the books by \citeauthor{Parr1994} (\citeyear{Parr1994}, advanced), 
\citeauthor{Martin2004} (\citeyear{Martin2004}, intermediate), and \citeauthor{Giustino2014} 
(\citeyear{Giustino2014}, elementary).

The first calculation of electron-phonon interactions using DFT was carried out by \textcite{Dacarogna1985b}
using a `frozen-phonon' approach (see Sec.~\ref{sec.matel-dft}).
In this work the authors computed electron bands, phonon dispersions, and electron-phonon matrix elements
of Al entirely from first principles. 
 Quoting from the original manuscript: 
 ``This calculation is {\it ab~initio} since only information about the Al atom, i.e.~the atomic 
 number and atomic mass, is used as input''.
\citeauthor{Dacarogna1985b} calculated the so-called electron-phonon coupling strength $\lambda_{\bq\nu}$ for several
phonon branches $\nu$ and momenta $\bq$ throughout the Brillouin zone, as well as the phonon linewidths arising from the
EPI (see Secs.~\ref{sec.nonadiab} and \ref{sec.sc-standard}). The average coupling strength was found to be in good 
agreement with that extracted from the superconducting transition temperature.
In the approach of \textcite{Chang1985,Dacarogna1985a,Dacarogna1985b,Lam1986}
the electron-phonon matrix element was calculated using:
  \begin{equation}\label{eq.lam}
  g_{mn\nu}(\bk,\bq) = \< u_{m\bk+\bq} | \Delta_{\bq\nu} v^{\rm KS} | u_{n\bk} \>_{\rm uc},
  \end{equation}
with $u_{n\bk}$ and $u_{m\bk+\bq}$ being the Bloch-periodic components of the Kohn-Sham electron wavefunctions,
$\Delta_{\bq\nu} v^{\rm KS}$ being the phonon-induced variation of the {\it self-consistent} potential 
experienced by the electrons, and the integral extending over one unit cell. Equation~(\ref{eq.lam}) 
will be discussed in Sec.~\ref{sec.general-matel}.
The scattering potential $\Delta_{\bq\nu} v^{\rm KS}$ was calculated by explicitly taking into account
the re-arrangement of the electronic charge following a small displacement of the nuclei.
The inclusion of the self-consistent response of the electrons constitutes a considerable step forward
beyond the rigid-ion approximation of Sec.~\ref{sec.emp-pseudo}.

The next and most recent step in the evolution of electron-phonon calculations
came with the development of DFPT for lattice dynamics
\cite{Baroni1987,Gonze1992,Savrasov1992}. 
In contrast to frozen-phonon calculations, which may require large supercells, DFPT enables
the calculations of vibrational frequencies and eigenmodes at arbitrary wavevectors in the Brillouin zone.
This innovation was critical in the context of electron-phonon physics, since the calculation 
of many physical quantities requires the evaluation of nontrivial integrals over the Brillouin zone.
The first calculations of EPIs using DPFT were reported by 
\textcite{Savrasov1994}, \textcite{Liu1996}, \textcite{Mauri1996}, and \textcite{Bauer1998}.
They calculated the electrical resistivity, thermal conductivity, mass enhancement
and superconducting critical temperature of a number of elemental metals 
(e.g.~Al, Au, Cu, Mo, Nb, Pb, Pd, Ta, V, and Te), and reported good agreement with experiment.

By the late 1990s most of the basic ingredients required for the {\it ab~initio} calculation of EPIs were available;
subsequent studies focused on using these techniques for calculating a variety of materials
properties, and on improving the efficiency and accuracy of the methodology. The most recent advances
will be reviewed in Secs.~\ref{sec.wannier}-\ref{sec.beyonddft}.

\section{Electron-phonon interaction in density-functional theory}\label{sec.formal}

In this section we review the basic formalism underlying the calculation of EPIs
using DFT, and we establish the link with the Hamiltonian
in Eq.~(\ref{eq.epi-hamilt}). We start by introducing the standard formalism for lattice
vibrations, and the electron-phonon coupling Hamiltonian. 
Then we briefly summarize established methods of DFPT for calculating electron-phonon matrix elements.
For the time being we describe electrons and phonons as separate subsystems;
a rigorous theoretical framework for addressing the coupled electron-phonon system
will be discussed in Sec.~\ref{sec.green}.

\subsection{Lattice vibrations in crystals}\label{sec.vibr-theory}

The formalism for studying lattice dynamics in crystals is covered in many excellent textbooks 
such as 
(\citeauthor{Born1954}, \citeyear{Born1954}; \citeauthor{Ziman1960}, \citeyear{Ziman1960};
\citeauthor{Kittel1963}, \citeyear{Kittel1963}; \citeauthor{Ashcroft1976}, \citeyear{Ashcroft1976};
\citeauthor{Kittel1976}, \citeyear{Kittel1976}). Here we introduce the
notation and summarize those aspects which will be useful for subsequent discussions 
in this section and in Secs.~\ref{sec.green} and~\ref{sec.green-recsp}.

We consider $M$ nuclei or ions in the unit cell. The position vector and Cartesian coordinates
of the nucleus $\kappa$ in the primitive unit cell are denoted by $\btau_\kappa$ and 
$\tau_{\kappa\a}$, respectively.
We describe the infinitely extended solid using Born-von K\'arm\'an (BvK) boundary conditions.
In this approach, periodic boundary conditions are applied to a large supercell which contains $N_p$
unit cells, identified by the direct lattice vectors $\bRp$, with $p = 1, \dots, N_p$.
The position of the nucleus $\k$ belonging to 
the unit cell $p$ is indicated by $\btau_{\kappa p} = \bRp + \btau_\kappa$.
The Bloch wavevectors~$\bq$ are taken to define a uniform grid of $N_p$ points in one unit cell 
of the reciprocal lattice, and the vectors of the reciprocal lattice are indicated by $\bG$.
In Appendix~\ref{sec.bvk} we provide additional details on the notation, and we state
the Fourier transforms between direct and reciprocal lattice.

Using standard DFT techniques it is possible to calculate the
total potential energy of electrons and nuclei in the BvK supercell. This quantity
is denoted as $U(\{ \btau_{\kappa p} \})$, where the braces are a short-hand notation for the
coordinates of all the ions. The total potential energy refers to electrons in their
ground state, with the nuclei being represented as {\it classical} particles 
{\it clamped} at the coordinates $\btau_{\kappa p}$. Every DFT software
package available today provides the quantity $U$ as a standard output.

In order to study lattice vibrations, one begins by making the {\it harmonic} approximation.
Accordingly, the total potential energy is expanded to second order in the displacements 
$\Delta \tau_{\kappa\a p}$ of the ions in the BvK supercell away from their 
equilibrium positions $\btau_{\kappa p}^0$: 
  \begin{equation}
  U = U_0 + \frac{1}{2}\!\! \sum_{\substack{\kappa\a p\\\kappa'\a' p'}}
   \frac{\D^2 U}{\D \tau_{\kappa\a p}\D \tau_{\kappa'\a' p'}}\Delta \tau_{\kappa\a p}
    \Delta \tau_{\kappa'\a' p'}, \label{eq.harm}
  \end{equation}
where $U_0$ denotes the total energy calculated for the ions in their equilibrium positions,
and the derivatives are evaluated for the equilibrium structure.
The second derivatives of the total energy with respect to the
nuclear coordinates define the matrix of `interatomic force constants':
  \begin{equation}\label{eq.ifc}
   C_{\kappa\a p,\kappa'\a' p'} = \D^2 U/\D \tau_{\kappa\a p}\D \tau_{\kappa'\a' p'}.
  \end{equation} 
The Fourier transform of the interatomic force constants yields the `dynamical matrix' \cite{Maradudin1968}: 
  \begin{equation}\label{eq.dynmat}
  D^{\rm dm}_{\kappa\a,\kappa'\a'}(\bq) = (M_\kappa M_{\kappa'})^{-\frac{1}{2}} {\sum}_p
     C_{\kappa\a 0,\kappa'\a' p} \exp(i\bq\cdot\bRp),
  \end{equation}
where $M_\kappa$ is the mass of the $\kappa$-th ion. The superscript `dm' is there to distinguish
this quantity from the many-body phonon propagators $D(12)$ and $D_{\k\a p,\k'\a'p'}$ that
will be introduced in Sec.~\ref{sec.GWph}.
The dynamical matrix is Hermitian and therefore admits real eigenvalues, which we
denote as $\w_{\bq\nu}^2$:
  \begin{equation}\label{eq.dynmat2}
  {\sum}_{\kappa'\a'}
  D^{\rm dm}_{\kappa\a,\kappa'\a'}(\bq)\, e_{\kappa'\a',\nu}(\bq) = \w_{\bq\nu}^2\, e_{\kappa\a,\nu}(\bq).
  \end{equation}
In classical mechanics, each $\w_{\bq\nu}$ corresponds to the vibrational frequency of an independent
harmonic oscillator. The hermiticity of the dynamical matrix allows us to choose the eigenvectors 
$e_{\kappa\a,\nu}(\bq)$ to be orthonormal for each $\bq$:
  \begin{eqnarray}
  {\sum}_{\nu} e_{\kappa'\a',\nu}^*(\bq) e_{\kappa\a,\nu}(\bq) &=& \delta_{\kappa\kappa'}
  \delta_{\a\a'}, \label{eq.ortho1}\\
  {\sum}_{\kappa\a} e_{\kappa\a,\nu}^*(\bq) e_{\kappa\a,\nu'}(\bq) &=& \delta_{\nu\nu'}.  
  \label{eq.ortho2}
  \end{eqnarray}
Here the index $\nu$ runs from 1 to $3M$.
The column vectors $e_{\kappa\a,\nu}(\bq)$ for a given 
$\nu$ are called the `normal modes of vibration' or the `polarization' of the vibration wave.
The following relations can be derived from Eq.~(\ref{eq.dynmat}):
  \begin{equation}\label{eq.e-cc}
  \w_{-\bq\nu}^2 = \w_{\bq\nu}^2; \qquad e_{\kappa\a,\nu}(-\bq) = e_{\kappa\a,\nu}^*(\bq). 
  \end{equation}
These relations between normal modes carry a degree of arbitrariness in the choice of phases;
here we have chosen to follow the same phase convention as \textcite{Maradudin1968}.

Using Eqs.~(\ref{eq.harm}) and (\ref{eq.ifc}) the Hamiltonian for nuclei considered as
quantum particles can be written as:
  \begin{equation}\label{eq.H-harm-adiab}
  \hH_{\rm p} = 
  \frac{1}{2}\! \sum_{\substack{\kappa\a p\\\kappa'\a' p'}}
   \!\!C_{\kappa\a p,\kappa'\a' p'} \Delta \tau_{\kappa\a p} \Delta \tau_{\kappa'\a' p'}
  -\sum_{\kappa\a p}\!\frac{\hbar^2}{2M_\kappa}\frac{\D^2}{\D \tau_{\kappa\a p}^2},
  \end{equation}
where the ground-state energy $U_0$ has been omitted and the last term is the kinetic energy operator. 
The Hamiltonian in the above expression corresponds to the energy of an entire BvK supercell.
Equation~(\ref{eq.H-harm-adiab}) relies on two approximations:
(i) the harmonic approximation, which coincides with the truncation of Eq.~(\ref{eq.harm}) to 
second order in the displacements; and (ii) the Born-Oppenheimer {\it adiabatic} approximation. This latter approximation
is made when one calculates the interatomic force constants with the electrons in their ground state.
The scope and validity of the adiabatic approximation will be discussed in detail in Sec.~\ref{sec.phonons-bo}.
We note incidentally that, strictly speaking, the Born-Oppenheimer approximation does not need
to be invoked were one to use the generalization of DFT to multicomponent systems introduced 
by \textcite{Kreibich2001,Kreibich2008}.

For practical purposes it is convenient to rewrite Eq.~(\ref{eq.H-harm-adiab}) by introducing
the quanta of lattice vibrations. This is accomplished by defining the standard creation ($\ha^\dagger_{\bq\nu}$)
and destruction ($\ha_{\bq\nu}$) operators for each phonon of energy $\hbar\w_{\bq\nu}$ and
polarization $e_{\kappa\a,\nu}(\bq)$. This operation is not entirely trivial and is described
in detail in Appendix~\ref{sec.normalcoord}. The formal definition of the ladder operators is
given in Eqs.~(\ref{eq.complex-ladder2})-(\ref{eq.complex-ladder}). These 
operators obey the commutation relations
  $[\ha_{\bq\nu},\ha_{\bq'\nu'}^\dagger] = \delta_{\nu\nu'}\delta_{\bq\bq'}$ and
  $[\ha_{\bq\nu},\ha_{\bq'\nu'}] = [\ha^\dagger_{\bq\nu},\ha^\dagger_{\bq'\nu'}] = 0$,
where $\delta$ is the Kronecker symbol.
From these relations we know that the quanta of the harmonic oscillations in crystals obey
Bose-Einstein statistics.
In Appendix~\ref{sec.normalcoord} it is shown that the atomic displacements can be expressed in
terms of the ladder operators as follows:
  \begin{equation}\label{eq.tau-from-x}
  \Delta\tau_{\kappa\a p}  = \left(\frac{M_0}{N_p M_\k}\right)^{\!\!\frac{1}{2}}
  \sum_{\bq\nu} e^{i\bq\cdot\bRp} e_{\kappa\a,\nu}(\bq)\, l_{\bq\nu} \,(\ha_{\bq\nu}+\ha^\dagger_{-\bq\nu}),
  \end{equation}
with $l_{\bq\nu}$ being the `zero-point' displacement amplitude:\label{page.3modes}
  \begin{equation}\label{eq.zeropdisp2}
  l_{\bq\nu} = [\hbar/(2M_0\w_{\bq\nu})]^{1/2}.
  \end{equation}
Here $M_0$ is an arbitrary reference mass which is introduced to ensure that $l_{\bq\nu}$ has the dimensions
of a length and is similar in magnitude to $\Delta\tau_{\kappa\a p}$. Typically $M_0$
is chosen to be the proton mass.

Using Eqs.~(\ref{eq.ifc})-(\ref{eq.zeropdisp2}) the 
nuclear Hamiltonian can be written in terms of $3MN_p$ independent harmonic oscillators
as follows:
  \begin{equation}\label{eq.herm-compl}
  \hH_{\rm p} = {\sum}_{\bq\nu} \hbar\w_{\bq\nu} \left( \ha_{\bq\nu}^\dagger \ha_{\bq\nu} + 1/2 \right),
  \end{equation}
where the sum is over all wavevectors. The ground-state wavefunction of this Hamiltonian is a product
of Gaussians, and all other states can be generated by acting on the ground state with the
operators $\ha_{\bq\nu}^\dagger$.
In the case of $|\bq|=0$ there are three normal modes for which $\w_{\bq\nu}\!=\!0$.
For these modes, which correspond to global translations of the crystal,
the zero-point displacement $l_{\bq\nu}$ is not defined.
Throughout this article it is assumed that these modes are skipped in summations containing
zero-point amplitudes. A detailed derivation of Eq.~(\ref{eq.herm-compl}) and a discussion
of the eigenstates of $\hH_{\rm p}$ are provided
in Appendix~\ref{sec.normalcoord}.

\subsection{Electron-phonon coupling Hamiltonian}\label{sec.epham-linear}

Having outlined the standard formalism for addressing lattice vibrations in crystals,
we now proceed to make the connection between DFT calculations and
the remaining terms of Eq.~(\ref{eq.epi-hamilt}). The electronic band structure $\ve_{n\bk}$
and electron-phonon matrix elements $g_{mn\nu}(\bk,\bq)$ are almost invariably
calculated by using the Kohn-Sham (KS) Hamiltonian \cite{Hohenberg1964,Kohn1965}.
A justification for these choices will be provided in Sec.~\ref{sec.green-recsp}; for now we
limit ourselves to outline the key elements of practical calculations.

\subsubsection{Kohn-Sham Hamiltonian}

Let us denote the Kohn-Sham eigenfunctions by $\psi_{n\bk}(\br)$, and use $\bk$ to indicate both
the wavevector and spin. We shall restrict ourselves to systems with collinear spins.
The KS eigenfunctions satisfy the equation $\hH^{\rm KS} \psi_{n\bk}(\br) = \ve_{n\bk} 
\psi_{n\bk}(\br)$, with the Hamiltonian given by:
  \begin{equation}\label{eq.ham-ks}
  \hH^{\rm KS} = -\frac{\hbar^2}{2m_e}\nabla^2 + V^{\rm KS}(\br;\{\tau_{\k\a p}\}).
  \end{equation}
Here the potential $V^{\rm KS}$ is the sum of the nuclear (or ionic) contribution $V^{\rm en}$, the
Hartree electronic screening $V^{\rm H}$, and the exchange and correlation potential $V^{xc}$
\cite{Martin2004}:
  \begin{equation}\label{eq.pot-ks}
  V^{\rm KS} = V^{\rm en} + V^{\rm H} + V^{xc}.
  \end{equation}
The potentials appearing in Eq.~(\ref{eq.pot-ks}) are defined as follows.
The electron-nuclei potential energy is given by:
  \begin{equation}\label{eq.pot-en}
  V^{\rm en}(\br;\{\tau_{\k\a p}\}) = {\sum}_{\kappa p,\bT} V_\k(\br-\btau_{\kappa p} - \bT),
  \end{equation}
where $V_\k(\br)$ is the interaction between an electron and the nucleus $\k$
located at the
center of the reference frame, and $\bT$ denotes a lattice vector of the BvK supercell.
In the case of all-electron DFT calculations, $V_\k(\br)$ 
is the Coulomb interaction:
  \begin{equation}\label{eq.Vk}
  V_\k(\br) = -\frac{e^2}{4\pi\epsilon_0} \frac{Z_\k}{|\br|},
  \end{equation}
where $Z_\k$ is the atomic number of the nucleus $\k$.
In the case of pseudopotential implementations $V_\k$ is a function that
goes as in Eq.~(\ref{eq.Vk}) at large $|\br|$, but remains 
finite at $|\br|\!=\!0$. Furthermore the nuclear charge is replaced by the ionic charge,
that is the difference between the nuclear charge and the number of core electrons
described by the pseudopotential.
In all modern pseudopotential implementations $V_\k(\br)$ is nonlocal
due to the separation of the angular momentum channels \cite{Martin2004}. However, since
this nonlocality is short-ranged and is inconsequential in the following discussion, 
it will be ignored here in order to maintain a light notation.
The Hartree term is obtained from the electron density, $n(\br';\{\tau_{\k\a p}\})$:
  \begin{equation}\label{eq.hartree}
  V^{\rm H}(\br;\{\tau_{\k\a p}\}) =  \frac{e^2}{4\pi\e_0}{\sum}_{\bT}\int_{\rm sc} 
    \frac{n(\br';\{\tau_{\k\a p}\})}
     {|\br-\br'-\bT|}\,d\br',
  \end{equation}
where the integral extends over the supercell. The
exchange and correlation potential is the functional derivative of the exchange
and correlation energy with respect to the electron density \cite{Kohn1965}:
  \begin{equation}\label{eq.Vxc}
  V^{xc}(\br;\{\tau_{\k\a p}\}) =  \delta E^{xc}[n] / \delta n \big|_{n(\br;\{\tau_{\k\a p}\})}.
  \end{equation}
The eigenfunctions $\psi_{n\bk}$ of $\hH^{\rm KS}$ can be expressed in the Bloch form:
  \begin{equation}\label{eq.utilde}
  \psi_{n\bk}(\br) = N_p^{-\frac{1}{2}} u_{n\bk}(\br) e^{i\bk\cdot\br},
  \end{equation}
with $u_{n\bk}$ a lattice-periodic function. The wavefunction $\psi_{n\bk}$ is taken to be normalized to one 
in the supercell, while the periodic part $u_{n\bk}(\br)$ is normalized to one 
in the crystal unit cell. 
The electron density is $n(\br)=\sum_{v\bk}|\psi_{v\bk}(\br)|^2$, where
$v$ indicates occupied states.
In order to determine $\psi_{n\bk}$ and $\ve_{n\bk}$ the Kohn-Sham equations are solved
{\it self-consistently}. This requires one to start from a reasonable guess for the electron density
(for example a superposition of atomic electron densities),
calculate the potentials in Eq.~(\ref{eq.pot-ks}), and determine the solutions of the 
KS Hamiltonian in Eq.~(\ref{eq.ham-ks}). The electron density is re-calculated using these
solutions, and the cycle is repeated until convergence.

In order to establish the link with Eq.~(\ref{eq.epi-hamilt}), we can regard the KS Hamiltonian
as an effective one-body operator, and make the transition to a second-quantized formalism
by using the standard prescription \cite{Merzbacher1998}:
  \begin{equation}\label{eq.ham-elec-2ndq}
  \hH_{\rm e} = 
               \sum_{n\bk,n'\bk'} \< \psi_{n\bk} | \hH^{\rm KS} | \psi_{n'\bk'} \>
             \hc_{n\bk}^\dagger \hc_{n'\bk'} 
  = \sum_{n\bk} \ve_{n\bk} \, \hc_{n\bk}^\dagger \hc_{n\bk}.
  \end{equation}
This expression is useful for performing formal manipulations
in the study of coupled electron-phonon systems. 
However, Eq.~(\ref{eq.ham-elec-2ndq}) implicitly introduces the drastic approximation that the electronic
system can be described in terms of sharp quasiparticle excitations.
A field-theoretic approach that does not rely on any such approximation is discussed in Sec.~\ref{sec.green}.

\subsubsection{Electron-phonon coupling Hamiltonian
to first- and second-order in the atomic displacements}\label{sec.general-matel}

Within the DFT Kohn-Sham formalism, the coupling Hamiltonian appearing in the second line 
of Eq.~(\ref{eq.epi-hamilt}) is obtained by expanding the Kohn-Sham effective potential
in terms of the nuclear displacements $\Delta \btau_{\k p}$ from their equilibrium positions 
$\btau_{\k p}^0$. The potential to first order in the displacements is:
  \begin{equation}\label{eq.ks-taylor-1}
  V^{\rm KS}(\{ \btau_{\k p} \}) = V^{\rm KS}(\{ \btau^0_{\k p} \}) + {\sum}_{\k\a p} \,
      \frac{\D V^{\rm KS}}{\D \tau_{\k\a p}} \Delta \tau_{\k\a p}.
  \end{equation}
This expression can be rewritten into normal mode coordinates using Eq.~(\ref{eq.tau-from-x}):
  \begin{equation}\label{eq.dV-KS}
  V^{\rm KS} \!=\! V^{\rm KS}(\{ \btau^0_{\k p} \}\!) + \,
   N_p^{-\frac{1}{2}}\!\sum_{\bq\nu} \Delta_{\bq\nu} V^{\rm KS} (\ha_{\bq\nu}+\ha^\dagger_{-\bq\nu}),
  \end{equation}
having defined:
  \begin{eqnarray} 
  \Delta_{\bq\nu} V^{\rm KS} & = & e^{i\bq\cdot\br} \Delta_{\bq\nu} v^{\rm KS}, \label{eq.dV-all-1}\\
  \Delta_{\bq\nu} v^{\rm KS} & = & l_{\bq\nu} 
   {\sum}_{\k\a} (M_0/M_\kappa)^\frac{1}{2}
    e_{\kappa\a,\nu}(\bq)\, \partial_{\k\a,\bq}v^{\rm KS}, \hspace{0.3cm} \label{eq.deltavq} \\
  \label{eq.dV-sq}
  \partial_{\k\a,\bq}v^{\rm KS} & = & 
   {\sum}_{p}e^{-i\bq\cdot(\br-\bRp)} \left.\frac{\D V^{\rm KS}}
  {\D \tau_{\k\a}}\right|_{\br-\bR_p}\!\!\!. \label{eq.dV-all-3}
  \end{eqnarray}
From the last expression we see that $\partial_{\k\a,\bq}v^{\rm KS}$ and $\Delta_{\bq\nu} v^{\rm KS}$ are 
lattice-periodic functions. The transition to second quantization is performed as in 
Eq.~(\ref{eq.ham-elec-2ndq}) \cite{Merzbacher1998}:
  \begin{equation}
  \hH_{\rm ep} = \!\!\!\!\sum_{n\bk,n'\bk'}\!\!\! \< \psi_{n\bk}| 
   V^{\rm KS}(\{ \btau_{\k p} \})-V^{\rm KS}(\{ \btau^0_{\k p} \})  
  |\psi_{n'\bk'}\> \hc_{n\bk}^\dagger \hc_{n'\bk'},
  \end{equation}
where the brakets indicate an integral over the supercell. 
After using Eqs.~(\ref{eq.utilde}), (\ref{eq.dV-KS})-(\ref{eq.dV-sq}), and (\ref{eq.psum}) we have:
  \begin{equation}\label{eq.ep-2nd}
  \hH_{\rm ep} = N_p^{-\frac{1}{2}}\! \sum_{\substack{\bk,\bq \\ m n \nu }} g_{mn\nu}(\bk,\bq)\,
  \hc_{m\bk+\bq}^\dagger \hc_{n\bk}(\ha_{\bq\nu}+\ha_{-\bq\nu}^\dagger),
  \end{equation}
where the electron-phonon matrix element is given by:
  \begin{equation}\label{eq.matel}
  g_{mn\nu}(\bk,\bq) = \< u_{m\bk+\bq} | \Delta_{\bq\nu} v^{\rm KS} | u_{n\bk} \>_{\rm uc}.
  \end{equation}
Here the subscript `uc' indicates that the integral is carried out within one unit cell of the crystal.
The coupling Hamiltonian in Eq.~(\ref{eq.ep-2nd}) yields the energy of an entire supercell.
In the case of the three translational modes at $|\bq|\!=\!0$ we set the matrix elements $g_{mn\nu}(\bk,\bq)$
to zero, as a consequence of the {\it acoustic sum rule} (see discussion in Sec.{\ref{sec.temper}}).

Taken together, Eqs.~(\ref{eq.herm-compl}), (\ref{eq.ham-elec-2ndq}), and (\ref{eq.ep-2nd})
constitute the starting point of most first-principles calculations 
of electron-phonon interactions. It remains to be seen how one calculates the electron-phonon
matrix elements $g_{mn\nu}(\bk,\bq)$; the most common procedures are described in Sec.~\ref{sec.matel-dft}.

Before proceeding, we discuss briefly the second-order coupling Hamiltonian which appears
in the third and fourth lines of Eq.~(\ref{eq.epi-hamilt}). The rationale for incorporating
this extra term is that the expansion of the Kohn-Sham potential
to first order in the atomic displacements, Eq.~(\ref{eq.ks-taylor-1}), is somewhat inconsistent
with the choice of expanding the total potential energy in Eq.~(\ref{eq.harm}) to second
order in the atomic displacements. This aspect was discussed by \textcite{Allen1976} and \textcite{Allen1978}.
In order to obtain an electron-phonon coupling Hamiltonian including terms of second-order in the
displacements, we must include the second derivatives of the Kohn-Sham potential in 
Eq.~(\ref{eq.ks-taylor-1}), and follow the same steps which led to Eq.~(\ref{eq.ep-2nd}).
By calling the extra term $\hH_{\rm ep}^{(2)}$ we have:
 \begin{eqnarray}
 \hH_{\rm ep}^{(2)} &=& N_p^{-1}\!\!\!
  \sum_{\substack{\bk,\bq,\bq'\\m n \nu\nu'}} g^{\rm DW}_{mn\nu\nu'}(\bk,\bq,\bq')
  \hc_{m\bk+\bq+\bq'}^\dagger \hc_{n\bk} \nonumber \\
  &\times&
 (\ha_{\bq\nu}+\ha^\dagger_{-\bq\nu}) (\ha_{\bq'\nu'}+\ha^\dagger_{-\bq'\nu'}), \label{eq.ham-2-dw}
 \end{eqnarray}
where 
 \begin{equation}\label{eq.matel-dw-dft}
  g^{\rm DW}_{mn\nu\nu'}(\bk,\bq,\bq') =
  \frac{1}{2} \< u_{m\bk+\bq+\bq'}| \Delta_{\bq\nu} \Delta_{\bq'\nu'} v^{\rm KS} |u_{n\bk}\>_{\rm uc}.
 \end{equation}
The variations $\Delta_{\bq\nu}$ are the same as in Eqs.~(\ref{eq.dV-all-1})-(\ref{eq.dV-sq}).

The second-order coupling Hamiltonian in Eq.~(\ref{eq.ham-2-dw}) is considerably more
involved than its first-order counterpart; the increased complexity partly explains why in the literature 
this term has largely been ignored. So far the Hamiltonian $\hH_{\rm ep}^{(2)}$ 
has only been described using an approximation based on first-order perturbation theory \cite{Allen1976}.
In this special case, the only terms in Eq.~(\ref{eq.ham-2-dw}) 
that can modify the electron excitation spectrum are those with $\bq'\!=\!-\bq$. The
corresponding energy shift is $\Delta \ve_{n\bk} \!=\! N_p^{-1}\sum_{\bq\nu} 
  g^{\rm DW}_{nn\nu\nu}(\bk,\bq,-\bq) (2n_{\bq\nu}\!+\!1)$,
with $n_{\bq\nu}$ being the number of phonons in each mode.
We will come back to this point in Sec.~\ref{sec.temper}.

\subsubsection{Calculation of electron-phonon matrix elements using density-functional
perturbation theory}
\label{sec.matel-dft}

In this section we review how the scattering potential $\Delta_{\bq\nu} v^{\rm KS}$ appearing 
in Eq.~(\ref{eq.matel}) is calculated in first-principles approaches. The most intuitive approach
is to evaluate the derivatives appearing in Eq.~(\ref{eq.dV-sq}) by using finite atomic displacements 
in a supercell:
  \begin{equation} 
  \left.\frac{\D V^{\rm KS}} {\D \tau_{\k\a p}}\right|_{\btau_{\k p}^0}    
  \!\simeq \left[ V^{\rm KS}(\br;\tau_{\k\a p}^0+b) - V^{\rm KS}(\br;\tau_{\k\a p}^0) \right]\!/b.
  \end{equation}
In this expression $b$ is a small displacement of the order of the zero-point amplitude 
($\sim$~0.1~\AA), and the atom~$\k$ 
in the unit cell~$p$ is displaced along the direction~$\a$. 
The first calculations of electron-phonon interactions within DFT employed a variant
of this `supercell approach' 
whereby all atoms are displaced according to a chosen vibrational eigenmode 
\cite{Chang1985,Dacarogna1985a,Dacarogna1985b,Lam1986}; this strategy is usually referred to
as the `frozen-phonon' method.

One disadvantage of the frozen-phonon method is that the supercell may become impractically large
when evaluating matrix elements corresponding to long-wavelength phonons.
This difficulty can be circumvented by using DFPT
\cite{Baroni1987,Gonze1992,Savrasov1992}. The main strength of DFPT
is that the scattering potential $\Delta_{\bq\nu} v^{\rm KS}$ in Eq.~(\ref{eq.matel})
is obtained by performing calculations within a single unit cell. Since
the computational workload of standard (non linear-scaling) 
DFT calculations scales as the cube of the number of electrons,
the saving afforded by DFPT over the frozen-phonon method is proportional to $N_p^2$, 
and typically corresponds to a factor $>10^3$.

In the DFPT approach of \textcite{Baroni2001}
one calculates the lattice-periodic scattering potential $\partial_{\k\a,\bq}v^{\rm KS}$ defined by
Eq.~(\ref{eq.dV-sq}). 
By differentiating Eq.~(\ref{eq.pot-ks}) via Eq.~(\ref{eq.dV-sq}) this potential is written as:
  \begin{eqnarray}\label{eq.potentials}
  \partial_{\k\a,\bq}v^{\rm KS} &=& \partial_{\k\a,\bq}v^{\rm en} +
    \partial_{\k\a,\bq}v^{\rm H} + \partial_{\k\a,\bq}v^{xc}.
  \end{eqnarray}
The variation of the ionic potential is obtained from Eqs.~(\ref{eq.pot-en}) and (\ref{eq.dV-sq}). The
result is conveniently expressed in reciprocal space:
  \begin{equation}\label{eq.dv-G}
  \partial_{\k\a,\bq}v^{\rm en}(\bG)  = -i\, 
   (\bq+\bG)_\a V_\k(\bq+\bG) e^{-i(\bq+\bG)\cdot\btau_\k},
  \end{equation}
where the convention for the Fourier transform is 
$f(\bG) = \Omega^{-1}\!\!\int_{\rm uc} d\br\,e^{-i\bG\cdot\br} f(\br)$, and $\Omega$ is the
volume of the unit cell.
In order to keep the presentation as general as possible we avoid indicating 
explicitly the non-locality of $V_\kappa$ which arises in pseudopotential
implementations. The adaptation of this equation and the following ones to the case of
nonlocal pseudopotentials, ultrasoft pseudopotentials, and the projector-augmented
wave method can be found in \cite{Giannozzi1991}, \cite{DalCorso1997}, and \cite{Audoze2006},
respectively.
The variation of the Hartree and exchange-correlation contributions to the
Kohn-Sham potential is obtained from the self-consistent charge
density response to the perturbation in Eq.~(\ref{eq.dv-G}).
After a few manipulations using Eqs.~(\ref{eq.hartree}) and (\ref{eq.dV-sq}) one obtains:
  \begin{equation}\label{eq.dvh-q-rec}
  \partial_{\k\a,\bq}v^{\rm H}(\bG) = \Omega\,v^{\rm C}(\bq+\bG) \,\partial_{\k\a,\bq} n(\bG),
  \end{equation}
where $v^{\rm C}(\bq) = \Omega^{-1}\!\int d\br \, e^{-i\bq\cdot\br} e^2/4\pi\e_0|\br|$ is the
Fourier transform of the Coulomb potential. For the exchange and correlation potential
we use Eq.~(\ref{eq.Vxc}) and the Taylor expansion of a functional to find:
  \begin{equation}\label{eq.dvxc-q}
  \partial_{\k\a,\bq}v^{xc}(\bG) = \Omega\,{\sum}_{\bG'} f^{xc}(\bq+\bG,\bq+\bG')\, \partial_{\k\a,\bq}n(\bG'),
  \end{equation}
  where 
$f^{xc}$ indicates the standard exchange and correlation kernel, which is the second-order functional
derivative of the exchange and correlation energy $E^{xc}$ with respect to the electron 
density~\cite{Hohenberg1964}:
  \begin{equation}\label{eq.fxc}
  f^{xc}(\br,\br') = \left.\frac{\delta^2 E^{xc}[n]}{\delta n(\br)\delta n(\br')}
   \right|_{n(\br;\{\btau_{\k p}^0\})}.
  \end{equation}
In the case of the local density approximation (LDA) to DFT the exchange and correlation kernel
reduces to a local function \cite{Parr1994}, 
and Eq.~(\ref{eq.dvxc-q}) is more conveniently evaluated in real space. 
Today DFPT calculations can be performed using one of several exchange and correlation kernels. The effect of the
kernel on the calculation of lattice-dynamical properties of solids has been analyzed 
in several works, see for example \textcite{DalCorso2013,He2014}.
The formal structure of the DFPT equations discussed in this section remains unchanged if
we replace the DFT kernel in Eq.~(\ref{eq.fxc}) by more sophisticated versions. For example
both DFPT calculations based on Hubbard-corrected DFT \cite{Floris2011}
and DFPT coupled with dynamical mean-field theory \cite{Savrasov2003} have been demonstrated.

It should be noted that in Eqs.~(\ref{eq.dvxc-q})-(\ref{eq.fxc}) we are implicitly
assuming a spin-unpolarized system. The adaptation of these equations as well as
the other DFPT equations to the most general case of non-collinear spin systems
can be found in \cite{DalCorso2007,Verstraete2008,DalCorso2008}.

From Eqs.~(\ref{eq.dvh-q-rec}) and (\ref{eq.dvxc-q}) we see that the evaluation of $g_{mn\nu}(\bk,\bq)$ 
goes through the calculation of the variation of the electron density induced by the
perturbation $\partial_{\k\a,\bq}v^{\rm KS} (\br)\, e^{i\bq\cdot\br}$.
Within DFPT such a variation is obtained by evaluating the
change of the Kohn-Sham wavefunctions to first order in perturbation theory. After inspection
of the perturbed Hamiltonian it becomes evident that the
wavefunction change must be of the form $\partial u_{n\bk,\bq}\,e^{i\bq\cdot\br}$,
with $\partial u_{n\bk,\bq}$ a lattice-periodic function. Using this observation
the first-order variation of the Kohn-Sham equations can be written as a Sternheimer 
equation \cite{Sternheimer1954}:
  \begin{equation}\label{eq.dfpt.1}
  \left(\hH^{\rm KS}_{\bk+\bq} - \ve_{v\bk}\right)\partial u_{v\bk,\bq} =  -\partial_{\k\a,\bq}v^{\rm KS} u_{v\bk},
  \end{equation}
with $\hH^{\rm KS}_{\bk+\bq} = e^{-i(\bk+\bq)\cdot\br}\, \hH^{\rm KS}\, e^{i(\bk+\bq)\cdot\br}$. 
In this equation the index $v$ indicates an occupied state.
For $|\bq|\!=\!0$ one needs also to consider a shift of the energy eigenvalues which
introduces an additional term $\< u_{v\bk} | \partial_{\k\a,0}v | u_{v\bk} \>_{\rm uc}\, u_{v\bk}$
on the right-hand side of Eq.~(\ref{eq.dfpt.1}). In practice this term is canceled by the use 
of the projectors described in Eq.~(\ref{eq.dfpt-projs}) below, unless one is dealing with metallic systems. 
This aspect is discussed in detail by \textcite{deGironcoli1995} and \textcite{Baroni2001}.
The principal advantage of Eq.~(\ref{eq.dfpt.1}) over standard perturbation theory is that 
it does not involve unoccupied electronic states.

A practical problem arises when attempting to solve Eq.~(\ref{eq.dfpt.1}): the linear system
on the left-hand side is ill-conditioned owing to small eigenvalues corresponding
to $\ve_{v\bk}\simeq\ve_{v'\bk+\bq}$; furthermore in the case of accidental degeneracies,
$\ve_{v\bk}=\ve_{v'\bk+\bq}$, the system becomes singular.
In order to make the system non-singular 
\textcite{Giannozzi1991} noted that the variation of the electron density only involves the
component of $\partial u_{v\bk,\bq}$ belonging to the unoccupied manifold of Kohn-Sham states.
As a consequence, what is really needed is only $\partial \tilde{u}_{v\bk,\bq} 
=(1-\hP^{\,\rm occ}_{\bk+\bq})\,\partial u_{v\bk,\bq}$,
having denoted by $\hP^{\,\rm occ}_{\bk+\bq} = \sum_v | u_{v\bk+\bq} \> \< u_{v\bk+\bq} |$
the projector over the occupied states with wavevector $\bk+\bq$. The equation for
this `trimmed' wavefunction variation is simply obtained by projecting both side
of Eq.~(\ref{eq.dfpt.1}) onto $(1-\hP^{\,\rm occ}_{\bk+\bq})$, 
and noting that $\hP^{\,\rm occ}_{\bk+\bq}$ and $\hH^{\rm KS}_{\bk+\bq}$
do commute:
  \begin{equation}\label{eq.dfpt-projs}
  \left(\hH^{\rm KS}_{\bk+\bq} - \ve_{v\bk}\right)\partial \tilde{u}_{v\bk,\bq} =  
    -(1-\hP^{\,\rm occ}_{\bk+\bq})\,\partial_{\k\a,\bq}v^{\rm KS} u_{v\bk}.
  \end{equation}
At this point it is possible to remove all small or null eigenvalues of the operator on the
left-hand side by adding a term $\alpha \hP^{\,\rm occ}_{\bk+\bq}$ to the Hamiltonian. This term has no
effect on the wavefunction variation, since $\hP^{\,\rm occ}_{\bk+\bq}\,\partial \tilde{u}_{v\bk,\bq}=0$
by construction. The operator is made non-singular by choosing the parameter $\alpha$ larger than 
the valence bandwidth \cite{Baroni2001}. 
From the wavefunction variation obtained by solving Eq.~(\ref{eq.dfpt-projs}), it is now
possible to construct the density response associated with the wavevector $\bq$:
  \begin{equation}\label{eq.dn-dfpt}
  \partial n_{\k\a,\bq}(\br) = 2\, N_p^{-1}{\sum}_{v\bk} u^*_{v\bk} \,\partial \tilde{u}_{v\bk,\bq}.
  \end{equation}
For simplicity a spin-degenerate system 
has been assumed (a factor of 2 is implicitly included in the sum over $\bk$), 
and time-reversal symmetry has been used in order 
to make the expression more compact (yielding the factor of 2 on the right-hand side). 

In practical DFPT calculations, Eq.~(\ref{eq.dfpt-projs}) is solved using an iterative procedure
which is similar to standard DFT total energy calculations. 
One sets the starting perturbation $\partial_{\k\a,\bq}v^{\rm KS}$ to be equal to the electron-nuclei 
potential in Eq.~(\ref{eq.dv-G}). By solving Eq.~(\ref{eq.dfpt-projs}) for each occupied 
state $v$ and each wavevector $\bk$ using standard linear algebra techniques, one obtains the
induced density in Eq.~(\ref{eq.dn-dfpt}). The new density is now used to construct the variations of the
Hartree and exchange-correlation potentials in Eqs.~(\ref{eq.dvh-q-rec}) and (\ref{eq.dvxc-q}).
These induced potentials are added to the electron-nuclei potential, yielding a `screened'
perturbation $\partial_{\k\a,\bq}v^{\rm KS}$ in Eq.~(\ref{eq.dfpt-projs}). The cycle is repeated until 
the change of $\partial n_{\k\a,\bq}$ between two successive cycles is smaller than a set 
tolerance.

It can be shown that the screened perturbation $\partial_{\k\a,\bq}v^{\rm KS}$ described in this section 
is also the key ingredient required for calculating the interatomic force constants
in Eq.~(\ref{eq.ifc}) \cite{Baroni2001}. As a practical consequence, every software implementation 
that supports DFPT calculations already contains all the information
necessary for evaluating the electron-phonon matrix elements $g_{mn\nu}(\bk,\bq)$.

All the quantities introduced in this section can equivalently be calculated using
an alternative, variational formulation of density-functional perturbation theory 
(\citeauthor{Gonze1992}, \citeyear{Gonze1992};
\citeauthor{Gonze1995}, \citeyear{Gonze1995},
\citeyear{Gonze1997};
\citeauthor{Gonze1997b}, \citeyear{Gonze1997b}). A thorough discussion of the connection 
between the Sternheimer approach and the variational approach to DFPT is provided by \textcite{Gonze1995b}.

The second-order matrix elements $g^{\rm DW}_{mn,\nu\nu'}(\bk,\bq,\bq')$ given by 
Eq.~(\ref{eq.matel-dw-dft}) involve the second derivative of the Kohn-Sham potential with respect
to the nuclear displacements. The evaluation of these quantities would require
the solution of second-order Sternheimer equations for the second variations of the Kohn-Sham
wavefunctions. The general structure of second-order Sternheimer equations can be found in Sec.~IV.H of
\cite{Gonze1995b}. Since these calculations are rather involved, most practical implementations
employ an approximation whereby the Debye-Waller matrix elements are expressed 
in terms of products of the standard matrix elements 
$g_{mn\nu}(\bk,\bq)$. Such an alternative formulation was developed by
\citeauthor{Allen1976} (\citeyear{Allen1976}) and \citeauthor{Allen1981} (\citeyear{Allen1981}), 
and will be discussed in Sec.~\ref{sec.temper}. All recent {\it ab~initio} 
calculations of electron-phonon interactions based on DFPT employed this latter 
approach.\footnote{ See for example \citeauthor{Marini2008} (\citeyear{Marini2008}); 
\citeauthor{Giustino2010} (\citeyear{Giustino2010});
\citeauthor{Gonze2011} (\citeyear{Gonze2011});
\citeauthor{Cannuccia2013} (\citeyear{Cannuccia2013});
\citeauthor{Ponce2014} (\citeyear{Ponce2014});
\citeauthor{Ponce2014b} (\citeyear{Ponce2014b});
\citeauthor{Antonius2014} (\citeyear{Antonius2014});
\citeauthor{Kaway2014} (\citeyear{Kaway2014});
\citeauthor{Ponce2015} (\citeyear{Ponce2015}).}

\subsubsection{The dielectric approach}\label{sec.dielec-mbody}

Besides the DFPT method described in the previous section, it is also
possible to calculate the screened perturbation $\partial_{\k\a,\bq}v^{\rm KS}$ using the so-called
`dielectric approach' \cite{Pick1970,Quong1992}. This latter approach did not find as widespread an application
as those of \textcite{Baroni1987,Gonze1992,Savrasov1992}, but it is useful to establish a link between 
DFT calculations of electron-phonon matrix elements and the field-theoretic
formulation to be discussed in Sec.~\ref{sec.green}.

For consistency with Sec.~\ref{sec.matel-dft}, we derive the key expressions of the dielectric approach
starting from DFPT. To this aim we expand the variation of the wavefunction $\partial \tilde{u}_{v\bk,\bq}$ 
using the complete set of states $u_{n\bk+\bq}$ (with $n$ referring to both occupied and empty Kohn-Sham states).
Then we replace this expansion inside Eq.~(\ref{eq.dfpt-projs}), project onto an arbitrary conduction state,
and insert the result in Eq.~(\ref{eq.dn-dfpt}). After taking into account time-reversal
symmetry, these steps lead to the following result:
  \begin{equation}\label{eq.chi0dv}
  \partial_{\k\a,\bq}n(\br) = \int_{\rm uc}\!\!\!d\br' 
  \chi_\bq^0(\br,\br') \, \partial_{\k\a,\bq}v^{\rm KS}(\br'),
  \end{equation}
having defined:
  \begin{eqnarray}\label{eq.chi0-q}
  \chi^0_\bq(\br,\br') & = & N_p^{-1}\,{\sum}_{mn\bk} \,
  \frac{f_{n\bk}-f_{m\bk+\bq}}{\ve_{n\bk}-\ve_{m\bk+\bq}} \nonumber \\ 
  & \times & u^*_{n\bk}(\br) u_{m\bk+\bq}(\br) u_{m\bk+\bq}^*(\br')u_{n\bk}(\br'). 
  \end{eqnarray}
In this expression $f_{n\bk}$ and $f_{m\bk+\bq}$ are the occupations of each state, 
and the indices run over all bands. A factor of 2 for the
spin degeneracy is implicitly included in the sum over $\bk$.
The quantity $\chi_\bq^0$ in Eq.~(\ref{eq.chi0-q}) is the lattice-periodic component 
for the wavevector~$\bq$ of the `independent-electron polarizability' 
(\citeauthor{Adler1962}, \citeyear{Adler1962}; \citeauthor{Wiser1963}, \citeyear{Wiser1963};
\citeauthor{Pick1970}, \citeyear{Pick1970}; \citeauthor{Quong1992}, \citeyear{Quong1992}).

For ease of notation we can write Eq.~(\ref{eq.chi0dv}) in symbolic form as 
$\partial n= \chi^0 \,\partial v^{\rm KS}$.
Using the same symbolic notation it is also possible to formally rewrite 
Eqs.~(\ref{eq.potentials}), (\ref{eq.dvh-q-rec}), 
and~(\ref{eq.dvxc-q}) as follows:
  \begin{equation}
  \partial v^{\rm KS} = 
             \partial v^{\rm en} + (v^{\rm C} + f^{xc}) \chi^0 \,\partial v^{\rm KS},
  \end{equation}
from which one obtains:
  \begin{equation}\label{eq.v-ven}
  \partial v^{\rm KS} = \left(\e^{{\rm H}xc}\right)^{-1}\partial v^{\rm en},
  \end{equation}
having defined the dielectric matrix:
  \begin{equation}\label{eq.eps-rpa+xc}
  \e^{{\rm H}xc} = 1-(v^{\rm C}+f^{xc})\,\chi^0.
  \end{equation}
The superscript `H$xc$' 
refers to the Hartree and exchange and correlation components of the screening.
In the language of many-body perturbation theory $\e^{{\rm H}xc}$ is referred to
as the `test electron' dielectric matrix, hinting at the fact that the electron density redistribution
in response to a perturbation arises both from classical electrostatics (the Hartree term $v^{\rm C}\,\chi^0$) 
and from quantum effects (the exchange and correlation term $f^{xc}\,\chi^0$).
If we neglect the kernel $f^{xc}$ in this expression, then we obtain the `test charge' 
dielectric matrix, which is most commonly known as the dielectric matrix in the 
random-phase approximation (RPA) \cite{Bohm1952}:
  \begin{equation}\label{eq.eps-rpa}
  \e^{\rm H} = 1-v^{\rm C}\chi_0.
  \end{equation}
The symbolic expressions outlined here remain almost unchanged when using a reciprocal-space
representation. As an example, Eq.~(\ref{eq.eps-rpa}) becomes simply:
  \begin{equation}\label{eq.eps-rpa-full}
   \e^{{\rm H}}_{\bG\bG'}(\bq) = \delta_{\bG\bG'} -\! 
   \Omega^2 \!\sum_{\bG''} \chi_{\bG''\bG'}^0(\bq) v^{\rm C}(\bq\!+\!\bG)\delta_{\bG\bG''}.\hspace{-0.3cm}
  \end{equation}
Taken together Eqs.~(\ref{eq.matel}) and (\ref{eq.v-ven}) show that the calculation of electron-phonon
matrix elements using DFPT is equivalent to screening the bare electron-nucleus interaction using
$\e^{{\rm H}xc}$; in this case we say that the screening is described at the `RPA+$xc$' level of approximation. 

At this point it is worth to point out that so far we only considered the screening of {\it static} 
perturbations: in fact $\partial v^{\rm en}$ was implicitly taken to be frequency-independent.
Physically this choice corresponds to describing phonons as quasi-static perturbations, so that
at each set of instantaneous atomic positions during a vibration cycle, 
the electrons have enough time to re-adjust and reach 
their ground state. This is a statement of the adiabatic approximation \cite{Born1927}.
The importance of {\it retardation} effects in the electron-phonon problem was already recognized
in the early work of \textcite{Bardeen1955}, but the first {\it ab~initio} calculations of
these effects appeared much later (see \citeauthor{Lazzeri2006}, \citeyear{Lazzeri2006}).
The formal framework required to incorporate retardation in the study of EPIs
will be presented in Sec.~\ref{sec.green}. 

\subsubsection{Connection with early formulations}\label{sec.matel-approx}

For completeness, we illustrate the link between electron-phonon matrix elements obtained
within DFPT (Sec.~\ref{sec.matel-dft}) and the 
early approaches of \textcite{Bloch1928} and \textcite{Bardeen1937} (Sec.~\ref{sec.history-early}).

The Bloch matrix element can be derived as follows. We assume that the scattering potential 
is unscreened and corresponds to the bare pseudopotentials $V_\k$ in Eq.~(\ref{eq.dv-G}); that
there is only one atom at the origin of the unit cell; and the Kohn-Sham wavefunctions can be
approximated by free electrons, $u_{n\bk}(\br) = \Omega^{-\frac{1}{2}}\exp(i\bG_n\cdot\br)$. 
In the last expression, the subscript in $\bG_n$ is used in order to stress the one-to-one correspondence between 
the reciprocal lattice vectors and the energy bands of the free electron gas in the reduced zone scheme.
Using these approximations in Eqs.~(\ref{eq.zeropdisp2}), (\ref{eq.deltavq}), (\ref{eq.matel}), and (\ref{eq.dv-G}), 
we find:
  \begin{eqnarray}\label{eq.matel-wumkl}
   g_{mn\nu}(\bk,\bq) &=& -i \left[\hbar/(2 N_p M_\k\w_{\bq\nu})\right]^{\frac{1}{2}}
       V_\k(\bq+\bG_m-\bG_n) \nonumber \\
     & & \;\; \times (\bq+\bG_m-\bG_n)\cdot\be_{\kappa,\nu}(\bq).
  \end{eqnarray}
By further neglecting umklapp processes ($\bG_m\!\ne\!\bG_n$) the previous result becomes
(\citeauthor{Grimvall1981}, \citeyear{Grimvall1981}, Sec.~3.4):
  \begin{equation}\label{eq.matel-noumkl}
    g_{mn\nu}(\bk,\bq) = -i \!\left[\hbar/\!(2 N_p M_\k\w_{\bq\nu})\right]^{\frac{1}{2}}
       \!\bq\cdot\be_{\kappa,\nu}(\bq)V_\k(\bq).
  \end{equation}
The expression obtained by \textcite{Bloch1928} and reproduced in Eq.~(\ref{eq.matel-noumkl-2})
is simply obtained by replacing $V_\k(\bq)$ with the effective potential $V_0$.

The Bardeen matrix element is more elaborate and can be derived as follows.
We describe the screening of the bare ionic potential within the RPA approximation,
and determine the dielectric matrix by replacing the Kohn-Sham wavefunctions by
free electrons. Using $u_{n\bk}(\br) = \Omega^{-\frac{1}{2}}\exp(i\bG_n\cdot\br)$ and
$\ve_{n\bk} = \hbar^2(\bk+\bG_n)^2/2m_{\rm e}-\ef$ in Eq.~(\ref{eq.chi0-q}), the polarizability 
reduces to: 
  \begin{equation}\label{eq.chi0-lindhard}
  \chi^0_{\bG\bG'}(\bq) =   
   -\frac{m_{\rm e}\kf}{\pi^2\hbar^2\Omega} F\left(|\bq+\bG|/2\kf\right)\delta_{\bG\bG'},
  \end{equation}
where $F$ is the function defined below Eq.~(\ref{eq.lindhard-history}). The derivation of 
this result requires making the 
transition from the first Brillouin zone to the extended zone scheme. 
If we use Eq.~(\ref{eq.chi0-lindhard}) inside Eq.~(\ref{eq.eps-rpa-full}), neglect the
exchange and correlation kernel, and use the Fourier transform of the Coulomb
potential, we find: 
  \begin{equation}\label{eq.lind-matrix}
  \e_{\bG\bG'}(\bq) = \delta_{\bG\bG'}
  \left[ 1\!+ \!\left(k_{\rm TF}^2/|\bq\!+\!\bG|^2\right)\!F(|\bq\!+\!\bG|/2\kf) \right],
  \end{equation}
where the Thomas-Fermi
screening length is given by $k_{\rm TF} = [4 m e^2 \kf / (4\pi \ve_0 \pi \hbar^2)]^{1/2}$.
Equation~(\ref{eq.lind-matrix}) is the well-known Lindhard dielectric matrix,
and the diagonal matrix elements are the same as in Eq.~(\ref{eq.lindhard-history})
(see \citeauthor{Mahan1993}, \citeyear{Mahan1993}, and
\citeauthor{Giuliani2005}, \citeyear{Giuliani2005},
for in-depth discussions of the Lindhard function).
By following the same steps that led to Eq.~(\ref{eq.matel-noumkl}), replacing the
bare ionic potential by its screened counterpart, and using Eq.~(\ref{eq.v-ven}) with $\e$
instead of $\e^{{\rm H}xc}$, we obtain:
  \begin{equation}\label{eq.ep-bardeen}
    g_{mn\nu}(\bk,\bq) = -i \left[\hbar/(2 N_p M_\k\w_{\bq\nu})\right]^{\frac{1}{2}}
      \bq\cdot\be_{\kappa,\nu}(\bq)\,
      V_\k(\bq)/\e(q).
  \end{equation}
Here we considered only one atom at the center of the unit cell, and we 
neglected umklapp processes.
This is essentially the result derived by \textcite{Bardeen1937} and reproduced in Eq.~(\ref{eq.bardeen-history}).

\section{Field-theoretic approach to the electron-phonon interaction}\label{sec.green}

In Sec.~\ref{sec.formal} we discussed how the materials parameters entering the electron-phonon
Hamiltonian in Eq.~(\ref{eq.epi-hamilt}), namely $\ve_{n\bk}$, $\w_{\bq\nu}$, and $g_{mn\nu}(\bk,\bq)$, 
can be calculated from first principles using DFT and DFPT.
Today the formalism and techniques described in Sec.~\ref{sec.formal} constitute
{\it de facto} the standard tool in quantitative studies of electron-phonon interactions in solids
(see Secs.~\ref{sec.nonadiab}-\ref{sec.beyonddft}).

However, it should be noted that the DFT approach to EPIs does not rest on strong theoretical foundations.
For one, the evaluation of the EPI matrix elements via Eq.~(\ref{eq.matel}) relies on the assumption
that the interaction between electrons and nuclei is governed by the effective Kohn-Sham potential;
therefore we can expect the matrix elements to be sensitive to the exchange and correlation
functional (see Sec.~\ref{sec.beyonddft}). Furthermore, the very definition of phonons
starting from Eq.~(\ref{eq.harm}) relies on the Born-Oppenheimer approximation, and one might
ask whether this choice is accurate enough in metals and narrow-gap semiconductors (see Sec.~\ref{sec.nonadiab}).
Finally, if one were to go beyond the Born-Oppenheimer approximation, then it would seem
sensible to also incorporate retardation effects in the calculation of the EPI matrix elements.

On top of these practical points, and at a more fundamental level, we expect that the electron-phonon
interaction will modify both the electronic structure and the lattice dynamics of a solid,
and these modifications will in turn affect the coupling between electrons and phonons. It is therefore
clear that a complete theory of interacting electrons and phonons must be 
{\it self-consistent}.
In order to address these issues it is necessary to formulate the
electron-phonon problem using a rigorous and general theory of interacting electrons
and phonons in solids. 

The most systematic and elegant approach is based
on quantum field theory \cite{Schwinger1951}, and is tightly connected with the development of
the $GW$ method \cite{Hedin1965}. 
The first attempts in this direction were from \textcite{Nakajima1954}, \textcite{Bardeen1955},
\textcite{Migdal1958}, and \textcite{Engelsberg1963}. However, from the point of view of the present article,
these works are of limited usefulness since they were mostly developed around the homogeneous electron gas.

A completely general formulation of the problem, which seamlessly applies to metals,
semiconductors, and insulators, was first provided by \textcite{Baym1961} and subsequently 
by \textcite{Hedin1969}.
The formalism developed in these articles constitutes today the most complete
theory of the electron-phonon problem. In fact, many aspects
of this formalism are yet to be explored within the context of {\it ab~initio} calculations.
After these seminal works several authors contributed to clarifying various aspects 
of the many-body theory of the coupled electron-phonon system, including \textcite{Keating1968}, 
\textcite{Gillis1970}, \textcite{Sjolander1965}, \textcite{Maksimov1976}, \textcite{Vogl1976}, 
and more recently \textcite{vanLeeuwen2004,Marini2015}. In particular, \citeauthor{vanLeeuwen2004}
focused on the issues of translational and rotational invariance of the resulting theory,
while \citeauthor{Marini2015} analyzed the connection between many-body perturbation theory approaches
and DFT calculations.

Since the mathematical notation of the original articles is obsolete and rather difficult to follow,
in Secs.~\ref{sec.sq-ham}-\ref{sec.hedin-baym} we cover the theory in some detail using contemporary notation.
The following derivations can be found across the works of \textcite{Kato1960}, \textcite{Baym1961},
\textcite{Hedin1969}, and \textcite{Maksimov1976}. Here we provide a synthesis of these contributions
using a unified notation, and we fill the gaps wherever it is necessary. 
The presentation requires some familiarity with field operators
(see for example \citeauthor{Merzbacher1998}, \citeyear{Merzbacher1998} for a succinct introduction).

\subsection{Operators and distinguishability}\label{sec.sq-ham}

The starting point for studying EPIs using a field-theoretic approach is to define the
Fock space and the field operators for electrons and nuclei. In the case of electrons 
the choice is unambiguous, since any many-body state can be represented as a linear combination
of Slater determinants constructed using a basis of single-particle wavefunctions.
In the case of nuclei the situation is slightly more ambiguous: in principle we might 
proceed in a very general way by choosing to focus on the nuclei as our quantum particles,
as opposed to their displacements from equilibrium. 
In practice this choice leads to a dead end for two reasons. Firstly, the quantum statistics
of nuclei would be dependent on their spin, therefore we would end up with an unwieldy mix of
fermions and bosons depending on the solid. Secondly, the notion of `indistinguishable'
particles, which is central to second quantization, does not apply to nuclei in solids
(at least in thermodynamic equilibrium and far from a solid-liquid phase transition).
In fact, in many cases we can directly label the nuclei, for example by means of experimental probes 
such as scanning tunneling microscopy and electron diffraction. 
In order to avoid these issues, it is best to study the electron-phonon problem by
considering (i) {\it indistinguishable} electrons, for which it is convenient to use second-quantized
operators; (ii) {\it distinguishable} nuclei, for which it is best to use first quantization
in the displacements; (iii) {\it indistinguishable} phonons, resulting from the quantization
of the nuclear displacements; in this latter case the distinction between first and second quantization
is irrelevant. These aspects are briefly mentioned by \textcite{Baym1961} and \textcite{Maksimov1976}.

With these choices, the dynamical variables of the problem are the electronic field operators
$\hp$ (discussed below) and the nuclear displacements from equilibrium $\Delta\hat{\btau}$ 
(discussed in Sec.~\ref{sec.pgf}). In this theory the equilibrium coordinates of the nuclei 
are regarded as {\it external} parameters, and are to be obtained for example from crystallography
or DFT calculations. Throughout this section, we limit ourselves to consider {\it equilibrium}
Green's functions at zero temperature. As a result, all expectation values
will be evaluated for the electron-nuclei ground state $|0\>$.
The extension of the main results to finite temperature is presented in Sec.~\ref{sec.green-recsp}.
We will not specify how to obtain the ground state, since the following discussion is independent 
on the precise shape of this state. In order to derive expressions that are useful for
first-principles calculations, at the very end the ground state will be approximated 
using standard DFT wavefunctions and phonons (see Sec.~\ref{sec.green-recsp}).

The electronic field creation/destruction operators are denoted by $\hpd(\bx)$/$\hp(\bx)$,
where the variable $\bx$ 
indicates both the position $\br$ and the spin label $\sigma$. 
These operators obey the anti-commutation relations \cite{Merzbacher1998}:
$\{\hp(\bx),\hp(\bx')\}\! =\! \{\hpd(\bx),\hpd(\bx')\}\!=\!0$, $\{\hp(\bx),\hpd(\bx')\} = \d(\bx\!-\!\bx')$.
The most general non-relativistic Hamiltonian for a system of coupled electrons and nuclei can be written as: 
  \begin{equation}\label{eq.H-mb}
  \hH = \hT_{\rm e} + \hT_{\rm n} + \hU_{\rm ee} + \hU_{\rm nn} + \hU_{\rm en},
  \end{equation}
where each term will be introduced hereafter. The electron kinetic energy is:
  \begin{equation}\label{eq.sq-kin}
  \hT_{\rm e} = -\frac{\hbar^2}{2m_{\rm e}}\int\!d\bx \, \hp^\dagger(\bx) \,\nabla^2 \, \hp(\bx),
  \end{equation}
with $m_{\rm e}$ being the electron mass, and the integrals $\int\!d\bx$ denoting the 
sum over spin and the integration over space, ${\sum}_\sigma\!\int d\br$. 
The electron-electron interaction is:
  \begin{equation}\label{eq.H-mb-e}
  \hU_{\rm ee} = \frac{1}{2}\!\int\!\!d\br\!\!\int \!\!d\br'\, \hne(\br)\left[\hne(\br') -
   \d(\br-\br')\right]v(\br,\br'),
  \end{equation}
where the electron particle density operator is given by $\hne(\br)={\sum}_\sigma\hpd(\bx)\hp(\bx)$,
and where $v(\br,\br')= e^2/(4\pi\ve_0|\br-\br'|)$ is the Coulomb interaction between two particles 
of charge~$e$. In Eqs.~(\ref{eq.sq-kin}) and (\ref{eq.H-mb-e}) the integrals are over the entire 
crystal. This corresponds to considering a supercell of infinite size (therefore the lattice
vectors $\bT$ of the supercell drop out) and a dense sampling of wavevectors
$\bq$ in the Brillouin zone. This choice is useful in order to maintain the formalism as light as possible.
Accordingly, all sums over $\bq$ are replaced using
$N_p^{-1}{\sum}_\bq \rightarrow \Omega_{\rm BZ}^{-1}\int\!d\bq$, where 
the integral is over the Brillouin zone of volume $\Omega_{\rm BZ}$. Similarly
the closure relations in Eq.~(\ref{eq.psum}) are replaced by:
$\int\!d\bq \exp(i\bq\cdot\bRp) = \Omega_{\rm BZ}\, \delta_{p0}$ and
$\sum_p \exp(i\bq\cdot\bRp) = \Omega_{\rm BZ}\,\d(\bq)$.
The nuclear kinetic energy operator is the same as the last term in Eq.~(\ref{eq.H-harm-adiab}).
Using the same notation as in Sec.~\ref{sec.formal} the nucleus-nucleus interaction energy is:
  \begin{equation}\label{eq.V-manybody-nucl}
  \hU_{\rm nn} = \frac{1}{2}\!\!\sum_{\,\,\k' p' \ne \k p}
  Z_\k Z_{\k'} v( \btau^0_{\kappa p}+\Delta\hat{\bm\tau}_{\kappa p}, 
  \btau^0_{\kappa' p'}+\Delta\hat{\bm\tau}_{\kappa' p'}).
  \end{equation}
Here $\btau^0_{\kappa p}$ denotes 
the classical equilibrium position of each nucleus, and the displacement operators 
$\Delta\hat{\bm\tau}_{\kappa p}$ will later be expressed in terms of the ladder operators from 
Appendix~\ref{sec.normalcoord}. The electron-nucleus interaction energy is:
  \begin{equation}\label{eq.H-manybody-en}
  \hU_{\rm en} = \!\int\!\!d\br\!\!\int \!\!d\br' \, \hne(\br) \hnn(\br') v(\br,\br'),
  \end{equation}
where the nuclear charge density operator is given by:
  \begin{equation}\label{eq.n-density-op}
  \hnn(\br) = - {\sum}_{\kappa p} Z_\k\, \d(\br-\btau^0_{\kappa p}-\Delta\hat{\bm\tau}_{\kappa p}).
  \end{equation}
Here the density operators are
expressed in units of the electron charge, so that the expectation value of the total charge density 
is $-e \<0| \hn(\br) |0\>$ with $\hn(\br) = \hne(\br) + \hnn(\br)$.

We underline the asymmetry between Eqs.~(\ref{eq.H-mb-e}) and (\ref{eq.V-manybody-nucl}):
in the case of electrons one considers the electrostatic energy of a continuous distribution of charge,
and the unphysical self-interaction is removed by the Dirac delta; whereas in the case of nuclei, the particles
are distinguishable therefore one has to take into account all pairwise interactions individually.

\subsection{Electron Green's function}\label{sec.GWph}

\subsubsection{Equation of motion and self-energy}\label{sec.GW-1}

In this section we focus on the electrons. By combining Eqs.~(\ref{eq.H-mb})-(\ref{eq.H-manybody-en})
and using the anti-commutation relations for the field operators one finds the standard expression:
  \begin{eqnarray}\label{eq.H-mb-3}
  \hH &=& \hT_{\rm n} + \hU_{\rm nn} +\int \!d\bx \, \hp^\dagger(\bx) \left[-\frac{\hbar^2}{2m_{\rm e}}\nabla^2 
   + \hat{V}_{\rm n}(\br)\right] \hp(\bx) \nonumber \\ &+& \frac{1}{2}\int\!\!d\bx\, d\bx' \, 
  v(\br,\br')\hpd(\bx)\hpd(\bx')\hp(\bx')\hp(\bx),
  \end{eqnarray}
where the nuclear potential $\hat{V}_{\rm n}$ is given by:
  \begin{equation}\label{eq.sq-npot}
  \hat{V}_{\rm n}(\br) = \int\! d\br' v(\br,\br') \hnn(\br').
  \end{equation}
In order to study the excitation spectrum of the many-body Hamiltonian $\hH$
at equilibrium we need to determine the time-ordered 
one-electron Green's function (\citeauthor{Kato1960}, \citeyear{Kato1960};
\citeauthor{Fetter2003}, \citeyear{Fetter2003}). 
At zero temperature this function is defined as:
  \begin{equation}\label{eq.green}
  G(\bx t, \bx' t') = -\frac{i}{\hbar}\< 0 |\, \hT\, \psi(\bx t) \psi^\dagger(\bx' t') | 0 \>,
  \end{equation}
where $\hT$ is Wick's time-ordering operator for fermions, and ensures that the times of the subsequent 
operators increase towards the left. The formal definition of the Wick operator is:
$\hT \psi(\bx t) \psi^\dagger(\bx' t')  = \theta(t\!-\!t') \psi(\bx t) \psi^\dagger(\bx' t')
- \theta(t'\!-\!t) \psi^\dagger(\bx' t')  \psi(\bx t)$, where $\theta$ is the Heaviside function.
Based on this definition we see that for $t>t'$ the Green's function in Eq.~(\ref{eq.green})
corresponds to the scalar product between the initial state $\psi^\dagger(\bx' t') | 0 \>$ and the
final state $\psi^\dagger(\bx t) | 0 \>$. This product is precisely the probability amplitude 
for finding an electron in the position $\bx$ at the time $t$, after having introduced an electron 
in $\bx'$ at an earlier time~$t'$. In the case $t<t'$ the situation is reversed and the 
Green's function describes the propagation of a hole created in the system at the time $t'$.

In order to determine $G(\bx t, \bx' t')$ we need to establish an equation of motion for
the field operators. This can be done by describing the time-dependence of the operators
within the Heisenberg picture:
  \begin{equation}\label{eq.heisen1}
  \hp(\bx t) = e^{i t\hH/\hbar} \,\hp(\bx) \,e^{-i t\hH/\hbar},
  \end{equation}
where $\hH$ was defined in Eq.~(\ref{eq.H-mb-3}). From this definition it follows
immediately:
  \begin{equation}\label{eq.heisen2}
  i\hbar\frac{\D}{\D t}\hp(\bx t) = [\hp(\bx t),\hH].
  \end{equation}
By combining Eqs.~(\ref{eq.H-mb-3}) and (\ref{eq.heisen2}) and using the anti-commutation relations
for the field operators 
one obtains:
  \begin{equation}\label{eq.heisen3}
  i\hbar\frac{\D}{\D t}\hp(\bx t) = \left[-\frac{\hbar^2}{2m_{\rm e}}\nabla^2
    + \int\!d\br' v(\br,\br')\,\hn(\br't) \right] \hp(\bx t), 
  \end{equation}
where the time-dependence in $\hn(\br' t)$ is to be understood
in the Heisenberg sense, as in Eq.~(\ref{eq.heisen1}).
This equation of motion allows us
to write the corresponding equation for the electron Green's function in Eq.~(\ref{eq.green}):
  \begin{eqnarray}\label{eq.green-tmp2}
   && \left[ i\hbar \frac{\D}{\D t}
     + \frac{\hbar^2}{2m_{\rm e}}\nabla^2 -\varphi(\br t) \right] G(\bx t, \bx' t') = \d(\bx t,\bx' t')
   \nonumber \\
   &&  -\frac{i}{\hbar} \int\!\!d\br''dt'' v(\br t,\br'' t'') \< \hT \, \hn(\br''t'')
   \psi(\bx t) \psi^\dagger(\bx' t')  \>.\hspace{0.2cm}  
  \end{eqnarray}
Here $v(\br t,\br'' t'') = v(\br,\br'') \d(t-t'')$, 
the brakets $\< \cdots \>$ are a short-hand notation for $\<0| \cdots |0\>$, and the additional
term $\varphi$ is discussed below.
In order to obtain Eq.~(\ref{eq.green-tmp2}) we used once again the anti-commutation relations, 
and we noted that the derivative of the Heaviside function is a Dirac delta.

The new term $\varphi(\br t)$ which appeared in Eq.~(\ref{eq.green-tmp2}) is a scalar 
electric potential which couples to both electronic and nuclear charges. This potential has
been introduced in order to perturb the system via the additional Hamiltonian
$\hH_1(t) = \int d\br\, \hn(\br t) \varphi(\br t)$.
The physical idea behind this choice is to use $\varphi(\br t)$ in order to induce
forced oscillations in the system. When the system resonates with the perturbation
we know that the resonant frequency must correspond to a free oscillation, that is a many-body eigenmode.
From a formal point of view, the potential $\varphi(\br t)$ is introduced 
in order to exploit Schwinger's functional derivative technique (Kato, \citeyear{Kato1960},
Appendix~II) and is set to zero at the end of the derivation.

One complication arising from the introduction of $\varphi(\br t)$ in Eq.~(\ref{eq.green-tmp2})
is that the time evolution in Eq.~(\ref{eq.heisen1}) is no longer valid, since
the perturbed Hamiltonian now depends on the time variable. The way around this complication is
to switch from the Heisenberg picture to the interaction picture. This change amounts to replacing
the exponentials in Eq.~(\ref{eq.heisen1}) by the time-ordered Dyson series
$\hU(t) = \hT \exp\left(-i \hbar^{-1}\int_{0}^{t}\hH(t') dt' \right)$ (Fetter and Walecka, \citeyear{Fetter2003}, 
p.~57). Since this would lead to an overlong derivation, we prefer to leave this
aspect aside and refer the reader to \textcite{Aryasetiawan1998} for a more comprehensive discussion.

In order to write Eq.~(\ref{eq.green-tmp2}) in a manageable form, we use the
identity \cite{Kato1960}:
  \begin{equation}\label{eq.centralequation} 
  \frac{\d \< \hT \hat{a}(t_1)\hat{b}(t_2)\> }{\d \varphi(\br'' t'')} \!= \!
    -\frac{i}{\hbar} \< \hT\!\left[ \hn(\br'' t'')\!-\!\< \hn(\br'' t'') \> \right] \hat{a}(t_1)\hat{b}(t_2)\>.
  \end{equation}
In this and the following expressions $\d /\d \varphi(\br'' t'')$ denotes the {\it functional} derivative
with respect to $\varphi(\br'' t'')$, and should not be confused with the Dirac delta functions 
$\d(\bx t,\bx' t')$.
Equation~(\ref{eq.centralequation}) is proven by \citeauthor{Kato1960} (\citeyear{Kato1960}, Appendix~II)
and by \citeauthor{Hedin1969} (\citeyear{Hedin1969}, Appendix~B.a).
After identifying $\hat{a}$ and $\hat{b}$ with $\hp(\bx t)$ and $\hpd(\bx' t')$, respectively,
Eq.~(\ref{eq.green-tmp2}) becomes:
  \begin{eqnarray}\label{eq.green-tmp3}
  && \,\,\left[ i\hbar \frac{\D}{\D t}
  + \frac{\hbar^2}{2m_{\rm e}}\nabla^2 -V_{\rm tot}(\br t)\! -\!i\hbar\!\!\int d\br''dt'' v(\br t+\eta,\br'' t'')
      \right. \nonumber \\
   &&\qquad\qquad \left. \times \frac{\d }{\d \varphi(\br'' t'')}   \right] G(\bx t, \bx' t') = \d(\bx t,\bx' t'), 
  \end{eqnarray}
where $\eta$ is a positive infinitesimal arising from the time-ordering, and 
$V_{\rm tot}(\br t)$ is the total potential acting on the electronic and nuclear charges,
averaged over the many-body quantum state $|0\>$:
  \begin{equation}
  V_{\rm tot}(\br t) = \int\!\!d\br' v(\br,\br') \< \hn(\br' t) \> + \varphi(\br t).
  \end{equation} 
Equation~(\ref{eq.green-tmp3}) was first derived by \textcite{Kato1960}.
In order to avoid a proliferation of variables, it is common practice to replace the letters
by integer numbers, using the convention $(\bx t)$ or $(\br t)\rightarrow 1$, 
$(\bx' t')$ or $(\br' t')\rightarrow 2$, $(\br  t\!+\!\eta)\rightarrow 1^+$, and so on.
Using this convention the last two equations become:
    \begin{eqnarray}\label{eq.green-tmp4}
  && \hspace{-0.7cm}\left[ i\hbar \frac{\D}{\D t_1}
  + \frac{\hbar^2}{2m_{\rm e}}\nabla^2(1) -V_{\rm tot}(1)\! \right.\nonumber \\ && 
   \hspace{0.5cm} \left. -i\hbar\!\!\int d3\, v(1^+ 3)
    \frac{\d }{\d \varphi(3)}   \right] G(1 2) = \d(12), 
  \end{eqnarray}
and
    \begin{equation}
   V_{\rm tot}(1) = \int\!\!d2\, v(12) \< \hn(2) \> + \varphi(1). \label{eq.vtot-den}
  \end{equation}
In these expressions the spin labels are implied for the Green's function and for the Dirac delta.

At this point, a set of self-consistent equations for coupled electrons and phonons can be generated by
eliminating the functional derivative in Eq.~(\ref{eq.green-tmp4}). For this purpose
one first relates the total screened electrostatic potential $V_{\rm tot}$
to the external potential $\varphi$ by introducing the inverse dielectric matrix $\e^{-1}$
as a functional derivative:
  \begin{equation}\label{eq.idm-mb}
   \e^{-1}(12) = \d V_{\rm tot}(1) \,/\, \d \varphi(2).
  \end{equation}
The function $\e^{-1}(12)$ is the many-body counterpart of the dielectric matrix discussed
in Sec.~\ref{sec.dielec-mbody}. The form given by Eq.~(\ref{eq.idm-mb}) is the most
general field-theoretic formulation for a system of interacting electrons and nuclei.

The next step is to rewrite $\d G / \d \varphi$ inside Eq.~(\ref{eq.green-tmp4}) in terms of the inverse Green's function,
$G^{-1}$. By using the fact that 
$\d \int d2\, G(12)\,G^{-1}(23) = 0$
and the rule for the functional derivative of a product \cite{Kadanoff1962} one obtains:
 \begin{equation}\label{eq.gphi}
 \frac{\d G(12)}{\d \varphi(3)}   = 
 -\!\int \!d(45) G(14) \frac{\d G^{-1}(45)}{\d \varphi(3)} G(52).
 \end{equation}
In order to eliminate any explicit reference to $\varphi$ we can express the functional derivative
on the right-hand side using the chain rule for functional differentiation \cite{Kadanoff1962}:
 \begin{equation}\label{eq.gm1}
 \frac{\d G^{-1}(45)}{\d \varphi(3)} = \int d6 \frac{\d G^{-1}(45)}{\d V_{\rm tot}(6)} 
 \frac{\d V_{\rm tot}(6)}{\d \varphi(3)}.
 \end{equation}
It is customary to call `vertex' the three-point quantity defined by:
 \begin{equation}\label{eq.vertex}
 \Gamma(123) = -\,\d G^{-1}(12) \,/\, \d V_{\rm tot}(3).
 \end{equation}
By combining Eqs.~(\ref{eq.green-tmp4}) and (\ref{eq.idm-mb})-(\ref{eq.vertex}) one finds:
  \begin{eqnarray}\label{eq.all-mb}
  && \hspace{-1cm}\left[ i\hbar \frac{\D}{\D t_1} + \frac{\hbar^2}{2m_{\rm e}}\nabla^2(1)
  -V_{\rm tot}(1) \right]G(12) \nonumber \\
  &&\hspace{2cm} - \int\!\!d3\,\Sigma(13) \,G(32)  = \d(12),
  \end{eqnarray}
having introduced the so-called `electron self-energy' $\Sigma$:
  \begin{equation}\label{eq.selfen-numbers}
  \Sigma(12) = i\hbar\int\!\! d(34) G(13) \Gamma(324) W(41^+),
  \end{equation}
which in turn contains the `screened Coulomb interaction' $W$, defined as:
  \begin{equation}\label{eq.W-numbers}
  W(12) =\! \int \!\!d3\,\e^{-1}(13) \,v(32) =\! \int \!\!d(3)\, v(13) \e^{-1}(23).\!\!
  \end{equation}
The last equality can be obtained by observing that
$\d \< \hn(1)\>/\d \varphi(2) = \d \< \hn(2)\>/\d \varphi(1)$ after Eq.~(\ref{eq.centralequation}),
therefore $W(12)=W(21)$.

Now, by inverting Eq.~(\ref{eq.all-mb}) and using Eq.~(\ref{eq.vertex}), we can express the 
vertex $\Gamma$ in terms of $\Sigma$ and $G$:
  \begin{equation}\label{eq.vertex3}
  \Gamma(123) \!=\! \d(12)\d(13) +\!\int \!\!d(4567) \frac{\d\Sigma(12)}{\d G(45)} 
       G(46) G(75) \Gamma(673).
  \end{equation}
The derivation of this result is rather lengthy: it requires the use of the chain rule,
in symbols $\d/\d V_{\rm tot} =(\d G/\d V_{\rm tot}) \d / \d G$, as well as
Eq.~(\ref{eq.gphi}) with $\Sigma$ and $V_{\rm tot}$ instead of $G$ and $\varphi$, respectively.

Equations~(\ref{eq.all-mb})-(\ref{eq.vertex3}) form a nonlinear system of equations
for the electron Green's function, $G$, the electron self-energy, $\Sigma$, the {\it total}
screened Coulomb interaction, $W$, and the vertex, $\Gamma$. In order to close the loop
it remains to specify the relation between $W$ and the other quantities.
The next section is devoted to this aspect.

\subsubsection{The screened Coulomb interaction}\label{sec.sci-J}

An equation for the screened Coulomb interaction can be found by combining Eqs.~(\ref{eq.vtot-den}), 
(\ref{eq.idm-mb}), and (\ref{eq.W-numbers}):
  \begin{equation}\label{eq.W-dn}
  W(12) = v(12) + \!\int \!d(34)\, v(13) \,\frac{\d \< \hn(3) \> }{\d V_{\rm tot}(4)}\, W(42).
  \end{equation}
By defining the `polarization propagator' as:
  \begin{equation}\label{eq.irrpol}
  P(12) = \frac{\d \< \hn(1) \> }{\d V_{\rm tot}(2)},
  \end{equation}
the previous expression takes the usual form \cite{Hedin1965}:
  \begin{equation}\label{eq.W-irrpol}
  W(12) = v(12) + \!\int \!d(34)\, v(13) P(34) W(42).
  \end{equation}
This result can be combined with Eq.~(\ref{eq.W-numbers}) in order to express the
dielectric matrix in terms of the polarization:
  \begin{equation}\label{eq.eps-hedin}
  \e(12) = \d(12) - \!\int \!\!d(3) v(13) P(32).
  \end{equation}
We now consider the special case whereby the nuclei are regarded as
classical point charges {\it clamped} to their equilibrium positions. In this
situation, the variation of the charge density $\d \< \hn \>$ in Eq.~(\ref{eq.irrpol})
corresponds to the re-distribution of the electronic charge in response to the
perturbation $\d V_{\rm tot}$. In order to describe this special case it is
convenient to introduce a new polarization propagator, $P_{\rm e}$, associated 
with the electronic response only:
  \begin{equation}\label{eq.P-el}
  P_{\rm e}(12) = \!\frac{\d \< \hn_{\rm e}(1) \> }{\d V_{\rm tot}(2)} =\!
    -i\hbar\! \sum_{\sigma_1} \!\!\int \!d(34) G(13) G(41^+) \Gamma(342).
  \end{equation}
The last equality in this expression is obtained by 
using Eq.~(\ref{eq.gphi}) with $V_{\rm tot}$ instead of $\varphi$, together with Eq.~(\ref{eq.vertex}),
and by considering that the electron density is related to the Green's function via the relation:
  \begin{equation}\label{eq.elec-density}
  \< \hn_{\rm e}(1) \> = -i\hbar {\sum}_{\sigma_1} G(11^+).
  \end{equation}
In conjunction with $P_{\rm e}$ it is natural to define the
Coulomb interaction screened by the electronic polarization only:
  \begin{equation}\label{eq.W-irrpol-e}
  W_{\rm e}(12) = v(12) + \!\int \!d(34)\, v(13) P_{\rm e}(34) W_{\rm e}(42),
  \end{equation}
as well as the associated dielectric matrix, in analogy with Eq.~(\ref{eq.eps-hedin}):
  \begin{equation}\label{eq.eps-hedin-el}
  \e_{\rm e}(12) = \d(12) - \!\int \!d3\, v(13) P_{\rm e}(32).
  \end{equation}
Taken together, Eqs.~(\ref{eq.all-mb})-(\ref{eq.vertex3}) with $W$ replaced by $W_{\rm e}$ 
constitute the well-known {\it Hedin's equations} for a system of interacting electrons 
when the nuclei are clamped to their equilibrium positions \cite{Hedin1965}.

In order to go back to the most general case whereby the nuclei are {\it not} clamped to their
equilibrium positions, one has to describe the re-adjustment of both electronic and 
nuclear charge. 
To this aim we combine Eq.~(\ref{eq.idm-mb}), (\ref{eq.W-numbers}), (\ref{eq.W-dn}),  (\ref{eq.P-el}),
(\ref{eq.W-irrpol-e}), and (\ref{eq.eps-hedin-el}). The result is:
  \begin{equation}\label{eq.W-D-tmp1}
   W(12)= W_{\rm e}(12)+\!\int \!d(34)\, W_{\rm e}(13) \frac{\d \< \hnn(3) \> }{\d \varphi(4)}  v(42).
  \end{equation}
An explicit expression for $\d \< \hnn \>/\d \varphi$ can be obtained using the following
reasoning. We go into the details since this is a delicate passage.
Equation~(\ref{eq.centralequation}) provides a recipe for evaluating the variation
of any operator with respect to a potential $\varphi(\br t)$ which couples to the
total charge density operator $\hn(\br t)$ via $\hH_1(t) = \int d\br\, \hn(\br t) \varphi(\br t)$.
Therefore we can replace $\hat{a}\,\hat{b}$ in Eq.~(\ref{eq.centralequation})
by $\hnn$ to obtain:
  \begin{equation}\label{eq.centralequation-n1}
  \frac{\d \< \hnn(1)\> }{\d \varphi(2)} =
    -\frac{i}{\hbar} \< \hT\!\left[ \hn(2)\!-\!\< \hn(2) \> \right] \![\hnn(1)-\!\<\hnn(1)\>]\>.\!\!\!
  \end{equation}
In addition, 
if we introduce a second perturbation, $\hH_2(t) = \int d\br\, \hnn(\br t) J(\br t)$,
which couples only to the nuclear charges, we can repeat the same reasoning 
as in Eq.~(\ref{eq.centralequation-n1}) after replacing $\varphi$ by $J$ and
$\hn$ by $\hnn$:
  \begin{equation}\label{eq.centralequation-n2}
  \frac{\d \< \hn(1)\> }{\d J(2)} \!= \!
    -\frac{i}{\hbar} \< \hT\!\left[ \hnn(2)\!-\!\< \hnn(2) \> \right] [\hn(1)-\<\hn(1)\>]\>.
  \end{equation}
The comparison between Eqs.~(\ref{eq.centralequation-n1}) and (\ref{eq.centralequation-n2}) yields:
  \begin{equation}\label{eq.W-D-tmp2}
  \frac{\d \< \hnn(1)\> }{\d \varphi(2)} =  \frac{\d \< \hn(2)\> }{\d J(1)}.
  \end{equation}
This can be restated by using the chain rule, $\d \< \hne\> /\d J =$ 
$\d \< \hne\> /\d V_{\rm tot} \times \d V_{\rm tot}/\d \<n\> \times \d \<n\>/ \d J$:
  \begin{equation}\label{eq.W-D-tmp3}
  \frac{\d \< \hn(1)\> }{\d J(2)} = \int\!\! d3\, \e_{\rm e}^{-1}(13)\, \frac{\d \< \hnn(3)\> }{\d J(2)}.
  \end{equation}
The variation $\d \< \hnn\>/\d J$ on the right-hand side can be expressed 
as in Eq.~(\ref{eq.centralequation-n2}):
  \begin{equation}\label{eq.den-den0}
  \frac{\d \< \hnn(1)\> }{\d J(2)} \!= \!
    -\frac{i}{\hbar} \< \hT\!\left[ \hnn(2)\!-\!\< \hnn(2) \> \right] \hnn(1)\>,
  \end{equation}
and since $\< \hnn-\< \hnn \> \>=0$ this can also be rewritten as:
  \begin{equation}
 \d \< \hnn(1)\> / \d J(2) \!= \! D(21),
  \end{equation}
having defined:
  \begin{equation}\label{eq.den-den}
  D(12) =  -\frac{i}{\hbar} \< \hT\!\left[ \hnn(1)\!-\!\< \hnn(1) \> \right] 
  \left[ \hnn(2)\!-\!\< \hnn(2) \> \right]\>.
  \end{equation}
This quantity is called the `density-density correlation function' for the nuclei.
Finally, we can combine Eqs.~(\ref{eq.W-D-tmp1}), (\ref{eq.W-D-tmp2}), (\ref{eq.W-D-tmp3}),
and (\ref{eq.den-den}) to obtain:
  \begin{equation}\label{eq.W=We+Wp}
  W(12)= W_{\rm e}(12)+ W_{\rm ph}(12),
  \end{equation}
where $W_{\rm ph}$ is the {\it nuclear} contribution to the screened Coulomb interaction,
and is given by:
  \begin{equation}\label{eq.WeDWe}
  W_{\rm ph}(12) = \int \!\!d(34)\, W_{\rm e}(13) D(34) W_{\rm e}(24).
  \end{equation}
This important result was derived first by \textcite{Hedin1969}.

\subsubsection{Nuclear contribution to the screened Coulomb interaction}\label{eq.e-self-phon}

In view of the forthcoming discussion, it is useful to derive a more explicit expression for
the screened interaction $W_{\rm ph}$ in Eq.~(\ref{eq.WeDWe}).
Here we follow \textcite{Baym1961,Maksimov1976}. 
The Taylor expansion of the Dirac delta to second order in the displacement $\bu$ reads:
  \begin{equation}\label{eq.delta-taylor}
  \d(\br-\bu) = \d(\br) - \bu\cdot \nabla \d(\br) + \frac{1}{2}\bu \!\cdot\! \nabla \nabla \d(\br)\! \cdot\! \bu,
  \end{equation}
where $\bu \cdot \nabla \nabla \cdot \bu$ is a short-hand notation for the second-order derivative
${\sum}_{\a\a'} u_\a u_{\a'} \nabla_{\a}\nabla_{\a'}$. The above expression derives from
the Fourier representation of the Dirac delta. 
Using Eq.~(\ref{eq.delta-taylor}) inside (\ref{eq.n-density-op}) we deduce:
  \begin{eqnarray}\label{eq.density-nuc-disp}
  \hnn(\br) &=& n_{\rm n}^0(\br) +
  {\sum}_{\k p} Z_\k\, \Delta\hat{\bm\tau}_{\k p}\cdot \nabla \d(\br-\bm\tau_{\k p}^0) \nonumber \\
  &-&\frac{1}{2}{\sum}_{\k p} Z_\k\, \Delta\hat{\bm\tau}_{\k p} \cdot \nabla \nabla
  \d(\br-\bm\tau_{\k p}^0) \cdot \Delta\hat{\bm\tau}_{\k p},\,\,\,\,\,\,\,
  \end{eqnarray}
where $n_{\rm n}^0(\br)$ is the density of nuclear point charges at the classical equilibrium
positions $\bm\tau_{\k p}^0$. After taking into account the time evolution in the Heisenberg
picture as in Eq.~(\ref{eq.heisen1}), we can replace this
expansion inside Eq.~(\ref{eq.den-den}) to obtain:
  \begin{equation} 
   D(12) =  \sum_{\substack{\k \a p \\ \k' \!\a' \!p'}} Z_\k \nabla_{1,\a} \d(\br_1-\bm\tau_{\k p}^0) 
    D_{\k \a p,\k' \a' p'}(t_1 t_2) \nonumber
  \end{equation} \vspace{-1.0cm}
  \begin{equation} \label{eq.den-den-tmp0}
    \hspace{0.5cm}\times \,Z_{\k'} \nabla_{2,\a'} \d(\br_2-\bm\tau_{\k' p' }^0).
  \end{equation}
On the right-hand side we introduced the
`displacement-displacement correlation function':
  \begin{equation}\label{eq.dis-dis}
  D_{\k \a p,\k' \a' p'}(t t')
  = -\frac{i}{\hbar} \< \hT\, \Delta \hat{\tau}_{\k \a p}(t) \,\Delta\hat{\tau}_{\k' \a' p'}(t') \>.
  \end{equation}
If we insert the last two equations in Eq.~(\ref{eq.WeDWe}) we find:
  \begin{eqnarray}\label{eq.Wph-1}
  && W_{\rm ph}(12)= \sum_{\substack{\k \a p \\ \k' \a' p'}}
  \int \!d(34)\, \e_{\rm e}^{-1}(13)
  \nabla_{3,\a}  V_\k(\br_3-\bm\tau_{\k p}^0)  \nonumber \\ 
   &&\times\, D_{\k \a p,\k' \a' p'}(t_3 t_4) \e_{\rm e}^{-1}(24) 
  \nabla_{4,\a'} V_{\k'}(\br_4-\bm\tau_{\k'p'}^0).
  \end{eqnarray}
In this expression $V_\k$ is the bare Coulomb potential of a nucleus or its ionic pseudo-potential.

At this point of the derivation, \citeauthor{Hedin1969} introduce the approximation that 
the electronic dielectric matrix in Eq.~(\ref{eq.Wph-1}) can be replaced
by its {\it static} counterpart.
This choice implies the Born-Oppenheimer adiabatic approximation.
In view of maintaining the formalism
as general as possible, we prefer to keep retardation effects, following 
the earlier works by \textcite{Bardeen1955} and \textcite{Baym1961}. 
We will come back to this aspect in Secs.~\ref{sec.phon-self} and \ref{eq.explic-elec}.

We stress that the sole approximation used until this point is to truncate the
density operator for the nuclei to the second order in the atomic displacements.
This is the standard {\it harmonic} approximation. Apart from this approximation,
which is useful to express $W_{\rm ph}$ in a tractable form, no other assumptions are made.
\textcite{Gillis1970} proposed a generalization of the results by \textcite{Baym1961}
which does not use the harmonic approximation. However, the resulting formalism 
is exceedingly complex, and has not been followed up.

\subsection{Phonon Green's function}\label{sec.pgf} 

In order to complete the set of self-consistent many-body equations for the coupled electron-phonon 
system, it remains to specify a prescription for calculating the displacement-displacement correlation 
function, $D_{\k \a p,\k' \a' p'}(t t')$. This function is seldom referred to as the `phonon Green's 
function', even though stricly speaking this name should be reserved for the quantity 
$-(i/\hbar) \< \hT\,  \ha_{\bq\nu}(t)\ha^\dagger_{\bq'\nu'}(t') \>$ which will be discussed
in Sec.~\ref{sec.phonons-bo}. In the following we describe the procedure originally devised by
\textcite{Baym1961}, and subsequently analyzed by \textcite{Keating1968}, \textcite{Hedin1969},
 \textcite{Gillis1970}, and \textcite{Maksimov1976}.

The starting point is the equation of motion for the displacement operators $\Delta\hat{\tau}_{\kappa\a p}(t)$.
In analogy with Eq.~(\ref{eq.heisen2}) we have:
$i\hbar \D / \D t  \Delta\hat{\bm\tau}_{\kappa p}(t) = [\Delta\hat{\bm\tau}_{\kappa p}(t),\hH]$.
Since we are considering the harmonic approximation and we expect the nuclei to oscillate around their
equilibrium positions, it is convenient to aim for an expression resembling Newton's equation.
This can be done by taking the time-derivative of the equation of motion:
  \begin{equation}\label{eq.heisen-disp-2}
  M_\k \frac{\D^2}{\D t^2} \Delta\hat{\bm\tau}_{\kappa p}
   = -\frac{M_\k}{\hbar^2}[[\Delta\hat{\bm\tau}_{\kappa p},\hH],\hH].
  \end{equation}
After evaluating the commutators using Eqs.~(\ref{eq.H-mb})-(\ref{eq.H-manybody-en}) and
performing the derivatives with respect to the nuclear displacements by means of Eq.~(\ref{eq.delta-taylor}),
we obtain (the steps are laborious but straightforward):
  \begin{eqnarray}\label{eq.newt-tmp2}
  &&M_\k \frac{\D^2}{\D t^2} \Delta\hat{\bm\tau}_{\kappa p}(t) =
    Z_\k \int\!\!d\br\,d\br' \, \hn^{(\k p)}(\br t) v( \br, \br') \nonumber\\
   &&\,\,\,\times \left\{ -\nabla^\prime \d(\br'\!-\!\btau_{\kappa p}^0) 
   + \nabla^\prime \!\left[\nabla^\prime \d(\br'\!-\!\btau_{\kappa p}^0)
   \cdot \Delta\hat{\bm\tau}_{\kappa p}(t)\right] \!\right\}\!\!.  \qquad
  \end{eqnarray}
Here $\hn^{(\k p)}(\br)$ is the total charge density of electrons and nuclei, {\it except} for the
contribution of the nucleus $\k$ in the unit cell $p$. In the second line 
$\nabla^\prime$ indicates that the derivatives are taken with respect to the variable $\br'$.
At this point, we can use the functional derivative technique as in Sec.~\ref{sec.GWph}
in order to determine an expression involving the displacement-displacement correlation function
from Eq.~(\ref{eq.dis-dis}). Here, instead of using $J(\br)$ as in Sec.~\ref{sec.sci-J} for the
nuclear density, it is convenient to work with the individual displacements, and introduce
a third perturbation $\hH_3(t) = \sum_{\k p} \bJ_{\k p}(t) \cdot \Delta\hat{\bm\tau}_{\kappa p}(t)$.
The extra terms $\bJ_{\k p}(t)$ have the meaning of external forces acting on the nuclei.
Using this perturbation, it is possible to write the displacement-displacement correlation
function in a manner similar to Eq.~(\ref{eq.den-den0}):
  \begin{equation}
  \frac{\d \< \Delta\hat{\tau}_{\kappa \a p}(t) \>}{\d F_{\k' \a' p'}(t')} = D_{\k \a p, \k' \a' p'}(t t').
  \end{equation}
This result was derived by \textcite{Baym1961} using a finite-temperature formalism. 
As in the case of the electron Green's function in Sec.~\ref{sec.GWph}, it can only be obtained 
by working in the interaction picture, and by taking into account the explicit time-dependence of the Hamiltonian
$\hH+\hH_3(t)$.  Also in the present case, we omit these details for the sake of conciseness.

If we take the expectation value of Eq.~(\ref{eq.newt-tmp2}) in the ground state,
after having added the new force term $-\bJ_{\k p}(t)$, and carry out the functional derivative 
with respect to $\bJ_{\k' p'}(t')$, we obtain:
   \begin{eqnarray}\label{eq.pure-pain}
   &&\hspace{-0.7cm}M_\k \frac{\D^2}{\D t^2} D_{\k \a p, \k' \a' p'} (t t')= -\delta_{\k \a p,\k' \a' p'}
     \delta(t t')  \nonumber \\
   && \hspace{-0.7cm}+Z_\k \int\!\!d\br\,d\br' 
   \left[ -\frac{\d \< \hn^{(\k p)}(\br t) \> }{\d F_{\k' \a' p'}(t')} 
    v( \br, \br') \nabla^\prime_{\!\a}\, \d(\br'-\btau_{\kappa p}^0) \right. \nonumber \\
   && \hspace{-0.7cm}\left. + \< \hn^{(\k p)}\!(\br t)\> v( \br, \br')  \nabla^\prime_{\!\a}\!
     \nabla^\prime_{\!\gamma}\, \d(\br'\!-\!\btau_{\kappa p}^0)
   D_{\k \gamma p, \k' \a' p'}(t t') \vphantom{\int}\right]\!\!,
   \end{eqnarray}
where the sum over the Cartesian directions $\gamma$ is implied.
The derivation of this result is rather cumbersome and involves the following considerations:
(i) the Dirac deltas in the first line come from the force terms $-\bJ_{\k p}$ added to
the Hamiltonian;
(ii) within the Harmonic approximation 
the expectation value $\< \hn^{(\k p)} \Delta\hat{\bm\tau}_{\kappa p} \>$ can
be replaced by $\< \hn^{(\k p)}\>\< \Delta\hat{\bm\tau}_{\kappa p} \>$ \cite{Baym1961};
(iii) the expectation value $\< \Delta\hat{\bm\tau}_{\kappa p} \>$ can be set to zero, because
at the end of the derivation one sets $|\bJ_{\k p}| = 0$ hence the expectation values of the
displacements vanish. We note that \textcite{Hedin1969} omitted the last line
of Eq.~(\ref{eq.pure-pain}) in their derivation, but this term was correctly included
by \textcite{Baym1961} and \textcite{Maksimov1976}.

The remaining functional derivative in Eq.~(\ref{eq.pure-pain}) can be expressed in terms of the nuclear charge
density using the same strategy which led to Eq.~(\ref{eq.W-D-tmp3}). The result is:
  \begin{eqnarray}\label{eq.hedin-b24}
  \hspace{-0.5cm}\frac{\d \< \hn^{(\k p)}(\br t) \> }{\d F_{\k' \a' p'}(t')} &=& 
  \!\!\int\! d\br'' dt'' \e_{\rm e}^{-1}(\br t, \br'' t'') 
  \frac{\d \< \hn_{\rm n}(\br'' t'') \> }{\d F_{\k' \a' p'}(t')} \nonumber \\
  &-& \!\!{\sum}_\gamma Z_\k D_{\k \gamma p, \k' \a' p'}(t t') \nabla_{\!\gamma} \d(\br\!-\!\bm\tau_{\k p}^0).
  \end{eqnarray} 
By inserting this result inside Eq.~(\ref{eq.pure-pain}) and using the expansion in 
Eq.~(\ref{eq.density-nuc-disp}), we finally obtain the equation of motion for the
displacement-displacement correlation function:
   \begin{eqnarray}
   && \hspace{-0.8cm}M_\k\frac{\D^2}{\D t^2} 
   D_{\k \a p, \k' \a' p'} (t t')= -\delta_{\k \a p,\k' \a' p'}\delta(tt')  \nonumber \\
   &&\hspace{-0.7cm} -\!\!\!\!\sum_{\k'' \a'' p''}\int \!\!dt'' 
   \Pi_{\k \a p, \k'' \a'' p''}(t t'') D_{\k'' \a'' p'',\k' \a' p'}(t''t'). \label{eq.phonon-eom}
   \end{eqnarray}
The quantity $\Pi_{\k p \a, \k' p' \a'}(t t')$ in this expression is called the `phonon self-energy' 
and is given by:
   \begin{eqnarray}
   &&\Pi_{\k \a p, \k' \a' p'}(t t') =\nonumber \\
   && \,\,\,\,\int\!\!d\br d\br'\! \left[
   Z_\k \nabla_{\!\a}\, \d(\br\!-\!\btau_{\kappa p}^0) W_{\rm e}(\br t, \br' t')\,
   Z_{\k'}\nabla^{\prime}_{\a'} \d(\br'\!-\!\bm\tau_{\k' p'}^0) \right.
    \nonumber \\
   && \left. \hspace{0.73cm}+\,\delta_{\k p, \k' p'}\delta(t t')\, \nabla_\a \< \hn (\br)\>\, v(\br,\br') \,
   Z_{\k'} \nabla^\prime_{\!\a'} \d(\br'\!-\!\btau_{\kappa' p'}^0) \right]\!\!.  \nonumber \\\label{eq.D-tmp-27}
   \end{eqnarray}
The derivation of Eq.~(\ref{eq.D-tmp-27}) 
is nontrivial and is not found consistently in the literature; it requires converting the
derivatives with respect to the position variables $\br$, $\br'$ into derivatives with respect
to the nuclear coordinates; integrating by parts in order to re-arrange the derivatives with 
respect to $\br$ and $\br'$; invoking the harmonic approximation; and considering that,
after setting the forces $\bJ_{kp}(t)=0$ and the field $\varphi(\br t)=0$ at the end,
the expectation value $\< \hn (\br t)\>$ does not depend on time. 
The term in the third line of Eq.~(\ref{eq.D-tmp-27}) is what \textcite{Baym1961} called
the `static force', since it arises from the forces experienced by the nuclei in their
equilibrium configuration. 

In order to simplify Eq.~(\ref{eq.D-tmp-27}) it is convenient to move from the time to the
frequency domain. We use the following convention for the Fourier 
transform of a function $f(t)$: $f(\w) \!= \!\int_{-\infty}^\infty dt f(t)e^{i\w t}$.
Since we~are considering equilibrium properties in absence of time-dependent external
potentials, the time variables enter in the above quantities only as differences \cite{Abrikosov1975}; 
for example $W_{\rm e}(\br t,\br' t') = W_{\rm e}(\br,\br',t-t')$. As a consequence,
Eq.~(\ref{eq.phonon-eom}) is rewritten as:
  \begin{eqnarray}
  && \hspace{-0.8cm}{\sum}_{\k'' \a'' p''} 
    \left[M_\k\w^2 \delta_{\k \a p,\k'' \a'' p''} - \Pi_{\k \a p, \k'' \a'' p''}(\w) \right]
    \nonumber \\ &&\hspace{0.8cm} \times\, D_{\k'' \a'' p'', \k' \a' p'} (\w) = \delta_{\k \a p,\k' \a' p'},
  \label{eq.D-freq-1}
  \end{eqnarray}
  \begin{table*}
  \begin{tabular}{c@{\hskip 0.4cm}l@{\hskip 0.5cm}l}
  \hline
  \\[-11pt]\hline\\[-11pt]
  Eq.\phantom{x} & Description & Expression \\
  \\[-11pt]\hline\\[-11pt]
  (\ref{eq.elec-density}) & Electronic charge density & $\< \hn_{\rm e}(1) \> 
  = -i\hbar {\sum}_{\sigma_1} G(11^+)$\\
  \\[-11pt]\\[-11pt]
  (\ref{eq.nuc-den-D}) & Nuclear charge density &
  $\<\hnn(\br t)\>  = n_{\rm n}^0(\br) -(i\hbar/2)\sum_{\k p, \a \a'} Z_\k
  \D^2 \d(\br-\bm\tau_{\k p}^0)/\D r_\a \D r_{\a'}
  D_{\k \a p,\k \a' p}(t^+ t)$ \\
  \\[-11pt]\\[-11pt]
  (\ref{eq.vtot-den}) & Total electrostatic potential 
   & $V_{\rm tot}(1) = \int\!\!d2\, v(12) \left[\< \hne(2) \> + \< \hnn(2) \> \right]$ \\ 
  \\[-11pt]\\[-11pt]
  (\ref{eq.all-mb}) &  Equation of motion, electrons & 
  $ \left[ i\hbar \D/\D t_1 + (\hbar^2/2m_{\rm e})\nabla^2(1)
  -V_{\rm tot}(1)\! \right]G(12) - \int\!\!d3\,\Sigma(13) G(32)  = \d(12)$ \\
  \\[-11pt]\\[-11pt]
  (\ref{eq.D-freq-1}) & Equation of motion, nuclei &
  $\sum_{\k''\a'' p''}\left[M_\k\w^2 \delta_{\k \a p,\k'' \a'' p''} - \Pi_{\k \a p, \k'' \a'' p''}(\w) \right]
    D_{\k'' \a'' p'', \k' \a' p'} (\w) = \delta_{\k \a p,\k' \a' p'}$ \\
  \\[-11pt]\\[-11pt]
  (\ref{eq.selfen-numbers}) & Electron self-energy 
   & $\Sigma(12) = i\hbar\int\! d(34) \,G(13) \,\Gamma(324) \left[ W_{\rm e}(41^+)+W_{\rm ph}(41^+) \right] $ \\
  \\[-11pt]\\[-11pt]
  (\ref{eq.W-irrpol-e}) & Screened Coulomb, electrons & $W_{\rm e}(12) = v(12) + \!\int \!d(34)\, 
  v(13) P_{\rm e}(34) W_{\rm e}(42)$ \\
  \\[-11pt]\\[-11pt]
  (\ref{eq.P-el}) & Electronic polarization & 
  $P_{\rm e}(12) = 
    -i\hbar \sum_{\sigma_1} \!\int \!d(34) \,G(13)\, G(41^+)\, \Gamma(342)$ \\
  \\[-11pt]\\[-11pt]
  (\ref{eq.eps-hedin-el}) & Electronic dielectric matrix &
  $\e_{\rm e}(12) = \d(12) - \int d(3) v(13) P_{\rm e}(32)$ \\
  \\[-11pt]\\[-11pt]
  (\ref{eq.vertex3}) & Vertex & $\Gamma(123) \!=\! 
  \d(12)\d(13) \!+\int \!\!d(4567) \left[\d\Sigma(12)/\d G(45)\right]
       G(46) G(75) \Gamma(673)$ \\
  \\[-11pt]\\[-11pt]
  (\ref{eq.Wph-1}) & Screened Coulomb, nuclei &
   $W_{\rm ph}(12)= \!\sum_{\k \a p, \k' \a' p'} \!\int \!d(34)\, \e_{\rm e}^{-1}(13)
  \nabla_{3,\a}  V_{\k}(\br_3\!-\!\bm\tau_{\k p}^0)$  \\
  \\[-11pt]\\[-11pt]
  &  & \hspace{1.3cm}$\times D_{\k \a p,\k' \a' p'}(t_3 t_4) \e_{\rm e}^{-1}(24)
  \nabla_{4,\a'} V_{\k'}(\br_4\!-\!\bm\tau_{\k' p'}^0)$ \\
  \\[-11pt]\\[-11pt]
  (\ref{eq.pi-all}) & Phonon self-energy & 
  $\Pi_{\k \a p, \k' \a' p'}(\w) =\!
   \sum_{\k'' p''}\! Z_\k Z_{\k''} (\D^2/\D r_\a \D r'_{\a'})$
   \\
  \\[-11pt]\\[-11pt]
   &  & \hspace{1.3cm}$\times\left[  
   \delta_{\k'p',\k''p''} W_{\rm e}(\br, \br',\w)
  -\!\delta_{\k p, \k' p'} W_{\rm e}(\br, \br',0)
  \right]_{\br = \btau_{\k p}^0, \br' = \btau_{\k'' p''}^0}\vspace{0cm}$ \\
  \\[-11pt]\hline\\[-11pt]
  \hline
  \end{tabular}
  \caption{Self-consistent Hedin-Baym equations for the coupled electron-phonon system in the
  harmonic approximation. 
  }
  \label{tab.hedin-baym}
  \end{table*}
\noindent
$\!\!$whereas the phonon self-energy in the frequency-domain at equilibrium reads:
  \begin{eqnarray}
  && \Pi_{\k \a p, \k' \a' p'}(\w) = \int\!\!d\br d\br'\! \left[
   Z_\k \nabla_{\!\a}\, \d(\br\!-\!\btau_{\kappa p}^0)\, W_{\rm e}(\br, \br',\w) \right. \nonumber \\
  &&\hspace{0.3cm}\left.
    +\,\delta_{\k p, \k' p'} \nabla_\a \< \hn (\br)\>\, v(\br,\br') \, \right]
   Z_{\k'} \nabla^\prime_{\!\a'} \d(\br'\!-\!\btau_{\kappa' p'}^0). \label{eq.pi-frequency}
  \end{eqnarray}
\noindent
The second line in this expression is conveniently rewritten by making use of the acoustic sum rule.
This sum rule is well-known in the theory of lattice dynamics of crystals \cite{Born1954}, and can be generalized
to the case of many-body Green's function approaches as follows \cite{Hedin1969}: 
  \begin{equation}\label{eq.asr}
  {\sum}_{\k'p'}\Pi_{\k \a p, \k' \a' p'}(\w=0) = 0 \,\,\mbox{ for any }\, \a,\a'.
  \end{equation}
This relation was first derived by \textcite{Baym1961} by imposing the condition that the nuclei in
the crystal must remain {\it near} their equilibrium positions due to fictitious restoring forces.
Physically this condition corresponds to considering a crystal which is held fixed in the laboratory
reference frame. In this approach, the crystal cannot translate or rotate 
as a whole. Similar relations were derived by 
\citeauthor{Sjolander1965} (\citeyear{Sjolander1965}) and 
\citeauthor{Gillis1970} (\citeyear{Gillis1970}).

If we combine Eqs.~(\ref{eq.pi-frequency}) and (\ref{eq.asr}), perform integrations by parts,
and carry out the integrations in $\br$ and $\br'$ we obtain:
  \begin{eqnarray}
  && \hspace{-0.7cm}\Pi_{\k \a p, \k' \a' p'}(\w) = 
   \sum_{\k''p''} Z_\k Z_{\k''} \left. \frac{\D^2}{\D r_\a \D r'_{\a'}} 
   \right|_{{\br \,\,= \btau_{\kappa p}^0, \br' = \btau_{\kappa'' p''}^0}}\!\!\!\!\!
  \nonumber \\
  && \hspace{0.0cm} \Big[ \delta_{\k' p',\k'' p''} W_{\rm e}(\br, \br',\w)
  -\!\delta_{\k p, \k' p'} W_{\rm e}(\br, \br',0) \Big],
    \label{eq.pi-all}
  \end{eqnarray}
which fulfils the sum rule in Eq.~(\ref{eq.asr}).

Eqs.~(\ref{eq.D-freq-1}) and (\ref{eq.pi-all}) completely define the nuclear dynamics in the
harmonic approximation. 
After obtaining the displacement-displacement correlation function $D_{\k \a p,\k' \a' p'}(t t')$
by solving this set of equations, 
it is possible to construct the expectation value of the nuclear density using Eqs.~(\ref{eq.density-nuc-disp})
and (\ref{eq.dis-dis}):
  \begin{equation}
  \<\hnn(\br t)\>  = n_{\rm n}^0(\br) \!-\!\frac{i\hbar}{2}\!\!\sum_{\k p, \a \a'}\!\! Z_\k 
  \frac{\D^2 \d(\br-\bm\tau_{\k p}^0)}{\D r_\a \D r_{\a'}}
  D_{\k \a p,\k \a' p}(t^+ t). 
  \label{eq.nuc-den-D}
  \end{equation}
We should emphasize that, according to Eq.~(\ref{eq.pi-all}), the coupling of the nuclear
displacements to the electrons is completely defined by the electronic
dielectric matrix through $W_{\rm e}$. Similarly, the nuclei affect the
electronic structure via the dielectric matrix which enters $W_{\rm ph}$ in Eq.~(\ref{eq.Wph-1})
and via the nuclear density inside $V_{\rm tot}$ in Eq.~(\ref{eq.vtot-den}).
From these considerations it should be clear that the {\it electronic} dielectric matrix
$\e_{\rm e}(\br,\br',\w)$ 
plays an absolutely central role in the the field-theoretic approach to the electron-phonon problem.

\subsection{Hedin-Baym equations}\label{sec.hedin-baym}

Apart from making use of the harmonic approximation, the set of equations given by
Eqs.~(\ref{eq.vtot-den}), (\ref{eq.all-mb}), (\ref{eq.selfen-numbers}), (\ref{eq.vertex3}), 
(\ref{eq.P-el}), (\ref{eq.elec-density}), 
(\ref{eq.W-irrpol-e}), (\ref{eq.eps-hedin-el}), (\ref{eq.Wph-1}), 
(\ref{eq.D-freq-1}), (\ref{eq.pi-all}), and (\ref{eq.nuc-den-D}) describe the coupled electron-phonon
system entirely from {\it first principles}.
This set of equations can be regarded as the most sophisticated description of interacting electrons
and phonons available today. Since the self-consistent equations for the electrons were
originally derived by \textcite{Hedin1965}, and those for the nuclei were derived first
by \textcite{Baym1961}, we will refer to the complete set as the {\it Hedin-Baym equations}.
Given the importance of these relations, we summarize them schematically in Table~\ref{tab.hedin-baym}.
The standard Hedin's equations for interacting electrons in the potential 
of clamped nuclei \cite{Hedin1965} are immediately recovered from the Hedin-Baym equations
by setting to zero the displacement-displacement correlation function of the nuclei, $D_{\k p \a, \k' p' \a'}=0$.

Table~\ref{tab.hedin-baym} provides a closed set of self-consistent equations whose solution yields
the Green's functions of a fully-interacting electron-phonon system, within the {\it harmonic} approximation.
We stress that these relations are fundamentally different from diagrammatic approaches. In fact, here
the coupled electron-phonon system is not addressed using Feynman-Dyson perturbation theory 
as it was done for example by \textcite{Keating1968}. Instead, in Table~\ref{tab.hedin-baym},
electrons and phonons are described non-perturbatively by means of a coupled set of nonlinear equations for the
exact propagators. In particular we emphasize that this approach does {\it not} require the
Born-Oppenheimer adiabatic approximation, and therefore it encompasses 
insulators, intrinsic as well as doped semiconductors, metals, and superconductors.

Almost every property related to electron-phonon interactions in solids that can be 
calculated today from first principles can be derived from these equations. Examples to
be discussed in Secs.~\ref{sec.green-recsp}-\ref{sec.transport} include the renormalization 
of the Fermi velocity, the band gap renormalization in semiconductors and insulators, 
the non-adiabatic corrections to vibrational frequencies, the Fr\"olich interaction,
and the lifetimes of electrons and phonons.
The generalization of these
results to the case of finite temperature should also be able to describe phonon-mediated superconductivity,
although this phenomenon is best addressed by studying directly the propagation of Cooper pairs
(see Sec.~\ref{sec.supercond}).

Baym's theory can in principle be extended to go beyond the harmonic approximation \cite{Gillis1970}.
However, the mathematical complexity of the resulting formalism is formidable, due to the appearance
of many additional terms which are neglected in the harmonic approximation.

\section{From a many-body formalism to practical calculations}\label{sec.green-recsp} 

The Hedin-Baym equations summarized in Table~\ref{tab.hedin-baym} define a rigorous
formalism for studying interacting electrons and phonons in metals, semiconductors,
and insulators entirely from first principles. However, a direct numerical solution
of these equations for real materials is currently out of reach, and
approximations are needed for practical calculations. The following sections establish the 
connection between the Hedin-Baym equations and standard expressions which are currently 
in use in {\it ab~initio} calculations of electron-phonon interactions.

\subsection{Effects of the electron-phonon interaction on phonons}\label{sec.phon-all}

\subsubsection{Phonons in the Born-Oppenheimer adiabatic approximation}\label{sec.phonons-bo}

The vibrational eigenmodes of the nuclei can be identified with the resonances of 
the displacement-displacement correlation function $D_{\k p \a, \k' p' \a'}(t t')$ in the
frequency domain. If we denote by ${\bf M}$ the diagonal matrix having
the nuclear masses $M_\k$ along its diagonal, then the formal solution of Eq.~(\ref{eq.D-freq-1}) 
can be written as:
  \begin{equation}\label{eq.phon-prop-mat}
  {\bf D}(\w) = \left[ \,{\bf M}\,\w^2 - {\bf \Pi}(\w) \,\right]^{-1},
  \end{equation}
where ${\bf D}$ is the matrix with elements $D_{\k p \a, \k' p' \a'}$.
The resonant frequencies of the system correspond to the solutions of the nonlinear
equations:
  \begin{equation}\label{eq.pi=mw2}
  \Omega_\nu(\w) - \w =0, \,\,\,\,\, \mbox {with } \nu=1,\dots 3M,\!\!\!\!\!\!
  \end{equation} 
where $\Omega^2_\nu(\w)$ is an eigenvalue of 
${\bf M}^{-1/2}\,{\bf \Pi}(\w)\,{\bf M}^{-1/2}$, parametric in the variable $\w$.

As expected, the study of lattice vibrations within a field-theoretic framework 
resembles the standard eigenvalue problem reviewed in Sec.~\ref{sec.vibr-theory}.
In particular, the matrix ${\bf \Pi}(\w)$ 
represents the many-body counterpart of the matrix of interatomic force
constants $C_{\kappa\a p,\kappa'\a' p'}$ introduced in Eq.~(\ref{eq.ifc}).
However, despite its formal simplicity, Eq.~(\ref{eq.pi=mw2}) conceals 
the full wealth of information associated with the many-body electronic screening
$\e_{\rm e}(\br,\br',\w)$ via Eq.~(\ref{eq.pi-all}).
In fact,  the phonon self-energy is generally {\it complex} 
and {\it frequency-dependent}. Therefore we can expect to find roots of Eq.~(\ref{eq.pi=mw2}) 
outside of the real frequency axis, as well as multiple roots for the same `eigenmode'.

The link between Eq.~(\ref{eq.pi=mw2}) and phonon calculations by means of DFT
is established by noting that DFT relies on the Born-Oppenheimer adiabatic approximation.
In the adiabatic approximation the nuclei are considered immobile during characteristic 
electronic timescales. Formally, this approximation 
is introduced by setting $\w=0$ in Eq.~(\ref{eq.pi-all}) \cite{Keating1968}. 
In practice, this assumption corresponds to stating that $\e_{\rm e}(\br,\br',\w)$ can be replaced
by $\e_{\rm e}(\br,\br',0)$ in the frequency range of the vibrational excitations. Obviously this is
not always the case, and important exceptions will be discussed in Sec.~\ref{sec.nonadiabatic}.

In order to see more clearly the connection with the formalism discussed in Sec.~\ref{sec.vibr-theory},
we partition the phonon self-energy into `adiabatic' and `non-adiabatic' contributions:
   \begin{equation} \label{eq.ad-nonad}
   {\bf \Pi}(\w) = {\bf \Pi}^{\rm A} + {\bf \Pi}^{\rm NA}(\w),
   \end{equation}
with ${\bf \Pi}^{\rm A}={\bf \Pi}(\w\!=\!0)$. 
As we will see below, the adiabatic term ${\bf \Pi}^{\rm A}$ will be taken to describe
`non-interacting' phonons, and the non-adiabatic self-energy $\Pi^{\rm NA}$ will be used
to describe the effects of electron-phonon interactions. 

In the early literature it is common to find
a different partitioning, whereby the non-interacting system is defined by the bare
interatomic force constants, corresponding to nuclei in the absence of electrons \cite{Grimvall1981}.
This alternative choice is not useful in modern calculations,
because the resulting non-interacting phonon dispersions are very different from the
fully-interacting dispersions. The present choice of using instead adiabatic phonons as the
non-interacting system, is more convenient in the context of modern {\it ab~initio} techniques,
since calculations of adiabatic phonon spectra are routinely performed within~DFPT.

In the remainder of this section we concentrate on the
adiabatic term, and we defer the discussion of the non-adiabatic self-energy to Sec.~\ref{sec.nonadiabatic}.
Using Eq.~(\ref{eq.pi-all}), we can rewrite the adiabatic self-energy as follows:
   \begin{eqnarray}
  &&\hspace{-0.7cm}\Pi_{\k \a p,\k' \a' p'}^{\rm A} =
    \sum_{\k'' p''} (\delta_{\k' p', \k'' p''}-\delta_{\k p, \k' p'}) \nonumber \\
  &&\hspace{0.0cm}  \times \left[
    \int \!d\br\, \frac{\D\, \<\hne(\br)\>}{\D \tau_{\k'' \a' p''}} \frac{\D V^{\rm en}(\br)}{\D \tau_{\k \a p}}
   + \frac{\D^2 U_{\rm nn}}{\D \tau_{\k \a p}\D \tau_{\k'' \a' p''}} \right]\!\!. \label{eq.bo-tmp1}
  \end{eqnarray}
In this expression $U_{\rm nn}$ is the nucleus-nucleus interaction energy from Eq.~(\ref{eq.V-manybody-nucl}),
$V^{\rm en}$ is the electron-nuclei interaction from Eq.~(\ref{eq.pot-en}), 
and all the derivatives are taken at the equilibrium coordinates. The derivation
of Eq.~(\ref{eq.bo-tmp1}) requires the use of the identity:
   \begin{equation} \label{eq.density-deriv}
    \frac{\D\, \<\hne(\br)\>}{\D \tau_{\k \a p}^0} \!= -{Z_\k}\!\int\!\! d\br' 
    [\e_{\rm e}^{-1}(\br,\br';0) \!-\! \delta(\br,\br')] \nabla'_{\a} \d(\br'\!-\!\btau^0_{\k p}). 
  \end{equation}
This identity follows from the same reasoning leading to Eq.~(\ref{eq.W-D-tmp3}), after considering
an external potential which modifies the 
position of the nucleus $\k$ in the cell $p$.
Equation~(\ref{eq.bo-tmp1}) can be recast in a familiar form by exploiting the acoustic sum rule
in Eq.~(\ref{eq.asr}). Indeed after a few tedious but straightforward manipulations we obtain:
  \begin{eqnarray}
  && \hspace{-0.8cm}\Pi_{\k \a p, \k' \a' p'}^{\rm A}=
  \int \!d\br\, \frac{\D \<\hne(\br)\>}{\D \tau_{\k' \a' p'}} 
  \frac{\D V^{\rm en}(\br)}{\D \tau_{\k \a p}} \nonumber \\
  &&\hspace{-0.1cm}+  
  \int \!d\br\, \< \hne(\br)\> \frac{\D^2 V^{\rm en}(\br)}{\D \tau_{\k \a p}\D \tau_{\k' \a' p'}} 
  + \frac{\D^2 U_{\rm nn}}{\D \tau_{\k \a p}\D \tau_{\k' \a' p'}}. \label{eq.hellman}
  \end{eqnarray}
In this form, one can see that the adiabatic self-energy 
gives precisely the interatomic force constants that we would obtain using the
Born-Oppenheimer approximation and the Hellman-Feynman theorem, compare Eq.~(\ref{eq.hellman}) for example with
\citeauthor{Baroni2001} (\citeyear{Baroni2001}, p.~517).

The difference between the $\Pi_{\k \a p, \k' \a' p'}^{\rm A}$ in Eq.~(\ref{eq.hellman}) and 
the $C_{\k \a p, \k' \a' p'}$ in  Eq.~(\ref{eq.ifc}) is that, in the former case,
the electron density response to atomic
displacements is governed by the exact many-body dielectric matrix $\e_{\rm e}(\br,\br',0)$
and electron density $\< \hne(\br)\>$, as shown by Eqs.~(\ref{eq.density-deriv}) and (\ref{eq.hellman}). 
As a result, $\Pi_{\k \a p, \k' \a' p'}^{\rm A}$ corresponds to force constants and electron density
{\it dressed} by all many-body interactions of the system (both electron-electron 
and electron-phonon interactions). In contrast, when the force constants in Eq.~(\ref{eq.ifc}) 
are calculated using DFT, the electron density response to an atomic displacement 
is evaluated using the RPA+$xc$ screening, that is $\e^{{\rm H}xc}(\br,\br')$
from Sec.~\ref{sec.dielec-mbody}, 
and the ground-state electron density is calculated at clamped nuclei.

The use of the adiabatic approximation in the study of phonons carries the important
advantage that the many-body force constants ${\bf \Pi}^{\rm A}$ form a real and
symmetric matrix. This can be seen by rewriting Eq.~(\ref{eq.pi-all}) for $\w=0$,
and using the relation $W_{\rm e}(\br,\br',\w) = W_{\rm e}(\br',\br,-\w)$ which follows
from the property $W(12)=W(21)$ (see Sec.~\ref{sec.GWph}).
Since ${\bf \Pi}^{\rm A}$ is real and symmetric, all its eigenvalues are guaranteed to be real. 
In this approximation, the excitations of the lattice correspond to sharp resonances in the 
displacement-displacement correlation function $\bD(\w)$, and it is meaningful to talk about 
phonons as long-lived excitations of the system. In fact these excitations are
infinitely long-lived in the harmonic approximation.
In practical calculations, the many-body ${\bf \Pi}^{\rm A}$ is invariably replaced
by the DFT interatomic force constants, and in this case the agreement of the calculated phonon frequencies 
with experiment is excellent in most cases. Illustrative examples can be found among others in
(\citeauthor{Yin1982}, \citeyear{Yin1982};
\citeauthor{Giannozzi1991}, \citeyear{Giannozzi1991};
\citeauthor{DalCorso1993}, \citeyear{DalCorso1993};
\citeauthor{Lee1994}, \citeyear{Lee1994};
\citeauthor{Kresse1995}, \citeyear{Kresse1995};
\citeauthor{DalCorso2000}, \citeyear{DalCorso2000};
\citeauthor{Bungaro2000}, \citeyear{Bungaro2000};
\citeauthor{Karki2000}, \citeyear{Karki2000};
\citeauthor{Baroni2001}, \citeyear{Baroni2001};
\citeauthor{Sanchez2007}, \citeyear{Sanchez2007};
\citeauthor{DalCorso2013}, \citeyear{DalCorso2013}). 

The most obvious criticism to the adiabatic approximation 
is that, in the case of {\it metals}, the assumption $\e_{\rm e}(\br,\br',\w) \!\simeq \!\e_{\rm e}(\br,\br',0)$ 
is inadequate. 
This can intuitively be understood by recalling that the dielectric function of the
homogeneous electron gas diverges when $\w,\bq\!\rightarrow\! 0$ \cite{Mahan1993}. In practical calculations,
this divergence is connected with vanishing denominators in Eq.~(\ref{eq.chi0-q}) for $\bq\!\rightarrow\! 0$. 
An approximate, yet very successful strategy for overcoming this problem, is to replace the occupation numbers 
in Eq.~(\ref{eq.chi0-q}) by smoothing functions such as the Fermi-Dirac distribution, and to describe 
the singular terms analytically \cite{deGironcoli1995}. Most first-principles calculations of phonon dispersion
relations in metals have been carried out using this strategy. Improvements upon this strategy
will be discussed in Sec.~\ref{sec.nonadiabatic}.

The adiabatic approximation leads naturally to the definition of an `adiabatic' propagator ${\bf D}^{\rm A}(\w)$,
which can be obtained from Eq.~(\ref{eq.phon-prop-mat}) after replacing the 
phonon self-energy by its static limit:
  \begin{equation}\label{eq.adiab-propag}
  {\bf D}^{\rm A}(\w) = \left[ \,{\bf M}\,\w^2 - {\bf \Pi}^{\rm A} \,\right]^{-1}.
  \end{equation}
Now, if we identify $\Pi_{\k \a p, \k' \a' p'}^{\rm A}$ with the interatomic force constant
$C_{\k \a p, \k' \a' p'}$ in  Eq.~(\ref{eq.ifc}), we can obtain an explicit expression
for the adiabatic phonon propagator in terms of the eigenmodes $e_{\k\a,\nu}(\bq)$
and eigenfrequencies $\w_{\bq\nu}$ introduced in Sec.~\ref{sec.vibr-theory}.
To this end we invert Eq.~(\ref{eq.adiab-propag}) using Eqs.~(\ref{eq.dynmat})-(\ref{eq.ortho2}),
and recall that the dynamical matrix is Hermitian and obeys the relation
$D^{{\rm dm},*}_{\k\a,\k'\a'}(\bq) = D_{\k\a,\k'\a'}^{\rm dm}(-\bq)$ \cite{Maradudin1968}. After tedious
but straightforward steps we find:
  \begin{equation}
  D^{\rm A}_{\k \a p,\k' \a' p'}(\w) ={\sum}_\nu \!\!\int\!\! \frac{d\bq}{\Omega_{\rm BZ}} \,
  S^*_{\bq\nu,\k\a p} S_{\bq\nu,\k'\a' p'}\, \frac{2\w_{\bq\nu}}{\w^2-\w_{\bq\nu}^2}, \label{eq.adiab-prop-explicit}
  \end{equation}
with the definition:
  \begin{equation}\label{eq.Ucrazy-def}
   S_{\bq\nu,\k\a p} = e^{i\bq\cdot\bR_{p}} (2 M_{\k} \w_{\bq\nu})^{-1/2}\, e_{\k\a,\nu}(\bq).
  \end{equation}
This result suggests that, as expected, the propagator should take a simple form in the eigenmodes
representation. In fact, by using the inverse transform of Eq.~(\ref{eq.Ucrazy-def}) we have:
$D^{\rm A}_{\bq\nu,\bq'\nu'}(\w) = \Omega_{\rm BZ}\,\delta(\bq\!-\!\bq') D^{\rm A}_{\bq\nu\nu'}(\w),$
with
  \begin{eqnarray}
  D^{\rm A}_{\bq\nu\nu'}(\w) = 2\w_{\bq\nu}/ (\w^2-\w_{\bq\nu}^2) \d_{\nu\nu'}.  \label{eq.adiab-prop-explicit-2}
  \end{eqnarray}
This result can alternatively be obtained starting from the ladder operators of
Appendix~\ref{sec.normalcoord}.
In fact, after using Eqs.~(\ref{eq.tau-from-x}), (\ref{eq.dis-dis}), and (\ref{eq.Ucrazy-def}) we find:
  \begin{equation}
  D^{\rm A}_{\bq\nu\nu'}(tt') = -i\< \hT\,  [\ha^\dagger_{\bq\nu}(t) \ha_{\bq\nu}(t')+
  \ha_{-\bq\nu}(t)\ha^\dagger_{-\bq\nu}(t') ]\> \d_{\nu\nu'}. \label{eq.schafer}
  \end{equation}
An explicit evaluation of the right-hand side
using the Heisenberg time evolution generated 
by the phonon 
Hamiltonian in Eq.~(\ref{eq.herm-compl}) yields precisely Eq.~(\ref{eq.adiab-prop-explicit-2}),
with the added advantage that it is easier to keep track of the time-ordering. The result is:
  \begin{equation}
  D^{\rm A}_{\bq\nu\nu}(\w) =  
   \frac{1}{\w-\w_{\bq\nu}+i\eta} - \frac{1}{\w+\w_{\bq\nu}-i\eta}, \label{eq.Schrieffer1}
  \end{equation}
with $\eta$ a positive real infinitesimal.
This alternative approach is very common in textbooks, see for example \textcite{Schrieffer1983} and
\textcite{Schafer2002}. However, it does not carry general validity in a field-theoretic
framework since it rests on the {\it adiabatic} approximation.

\subsubsection{Phonons beyond the adiabatic approximation}\label{sec.nonadiabatic}

In order to go beyond the adiabatic approximation, it is necessary to determine the complete
propagator ${\bf D}(\w)$ in Eq.~(\ref{eq.phon-prop-mat}). Formally this can be done by
combining Eqs.~(\ref{eq.phon-prop-mat}) and (\ref{eq.ad-nonad}) to obtain the following
Dyson-like equation:
  \begin{equation}\label{eq.dyson-ad-nonad}
  {\bf D}(\w) = {\bf D}^{\rm A}(\w) + {\bf D}^{\rm A}(\w){\bf \Pi}^{\rm NA}(\w){\bf D}(\w).
  \end{equation}
In this form it is apparent that the non-adiabatic phonon self-energy ${\bf \Pi}^{\rm NA}(\w)$
`dresses' the non-interacting phonons obtained within the adiabatic approximation,
as shown schematically in Fig.~\ref{phonon-selfenergy}(a).
It is convenient to rewrite the Dyson equation in such a way as to show more clearly the poles 
of the propagator. Using Eqs.~(\ref{eq.Ucrazy-def})-(\ref{eq.adiab-prop-explicit-2}) we find:
  \begin{figure}
  \includegraphics[width=\columnwidth]{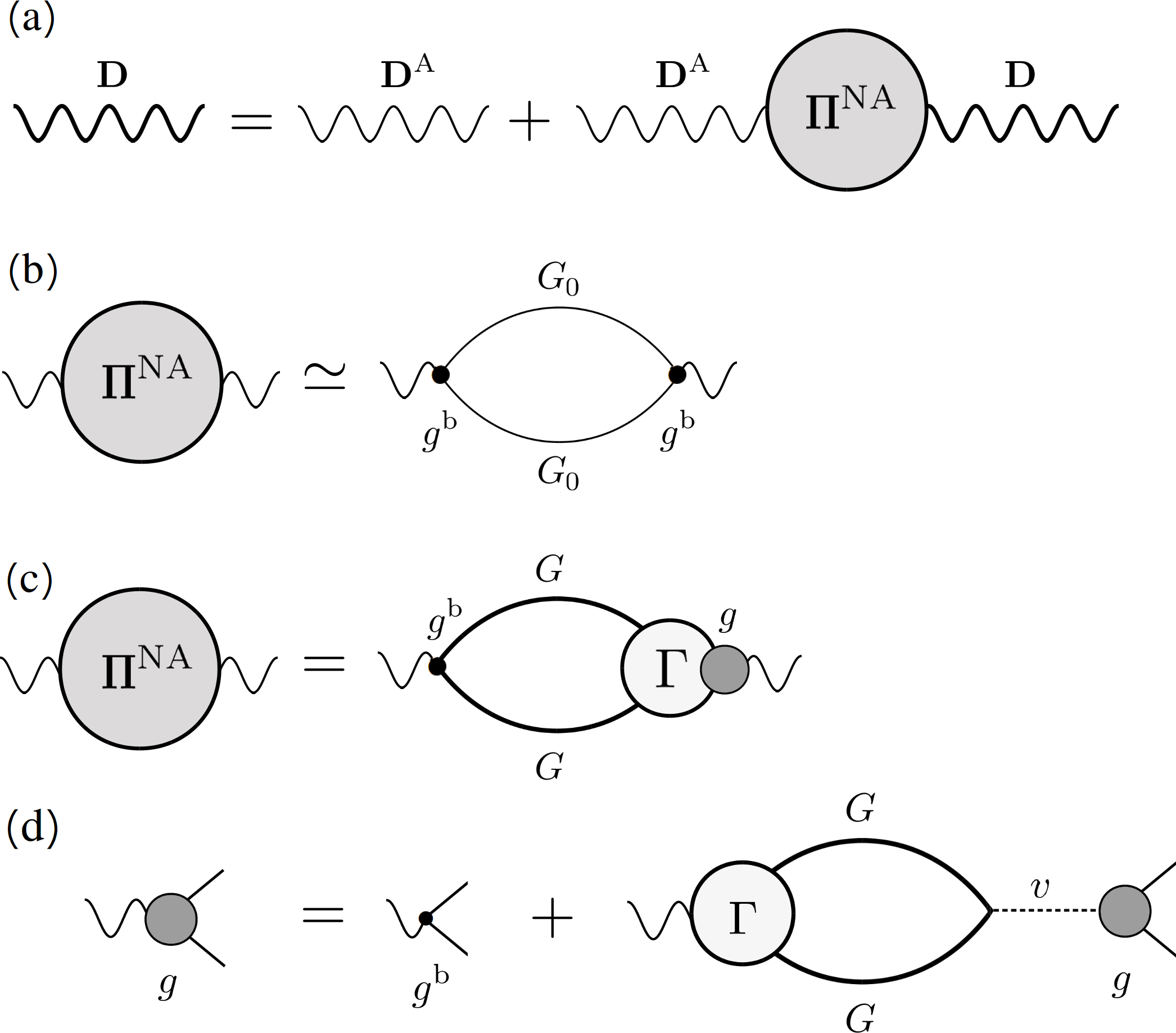}
  \caption[fig]{\label{phonon-selfenergy} 
  Diagrammatic representation of the phonon Green's function and self-energy.
  (a)~Dyson equation for the phonon propagator, Eq.~(\protect\ref{eq.dyson-ad-nonad}). 
  The thick wavy line represents the fully-interacting, non-adiabatic propagator; 
  the thin wavy line is the adiabatic propagator; the disc is the non-adiabatic self-energy.
  (b)~Lowest-order diagrammatic expansion of the phonon self-energy in terms
  of the bare electron-phonon vertices and the RPA electronic polarization.
  The small dots are the bare electron-phonon coupling functions, and
  the thin lines are the non-interacting (for example Kohn-Sham) electron Green's functions.
  This diagram is the simplest possible term which begins and ends with a phonon line.
  (c)~Non-perturbative representation of the phonon self-energy in terms of the
  bare coupling, the dressed coupling (large gray disc), the fully-interacting electron's
  Green's functions (thick lines), and the vertex $\Gamma$ from Eq.~(\protect\ref{eq.vertex}).
  This diagram was proposed by \textcite{Keating1968} and describes the first line of 
  Eq.~(\protect\ref{eq.gbarePg}).
  (d)~Schematic representation of the relation between the dressed electron-phonon coupling 
  $g$ and the bare coupling $g^{\rm b}$, from Eq.~(\protect\ref{eq.gscreened}). \textcite{Vogl1976}
  reports a similar diagram, although with the bare coupling function on the far right; the difference
  stems from the present choice of using the irreducible polarization $P=GG\Gamma$ instead
  of the reducible polarization employed by \citeauthor{Vogl1976}.
  }
  \end{figure}
  \begin{equation}\label{eq.dyson-pi}
  D^{-1}_{\bq\nu\nu'}(\w) =
  \frac{1}{2\w_{\bq\nu}} \left[ \delta_{\nu\nu'}(\w^2-\w_{\bq\nu}^2)
   -2\w_{\bq\nu}\Pi_{\bq\nu\nu'}^{\rm NA}(\w)\right],
  \end{equation}
where $\Pi_{\bq\nu\nu'}^{\rm NA}$ and $D^{-1}_{\bq\nu\nu'}$ are obtained using the transform
of Eq.~(\ref{eq.Ucrazy-def}) and its inverse, respectively.

From Eq.~(\ref{eq.dyson-pi}) we see that the non-adiabatic self-energy $\Pi^{\rm NA}$ 
modifies the adiabatic phonon spectrum in four distinct ways: (i) the real part of the
diagonal elements $\Pi_{\bq\nu\nu}^{\rm NA}$ shifts the adiabatic frequencies; (ii) the
imaginary part introduces spectral broadening; (iii)~the off-diagonal elements
of $\Pi_{\bq\nu\nu'}^{\rm NA}$ introduce a coupling between the adiabatic 
vibrational eigenmodes; (iv)~the frequency-dependence
of $\Pi^{\rm NA}_{\bq\nu\nu}(\w)$ might lead to multiple poles for the same mode $\nu$,
thereby introducing new structures in the phonon spectrum.

Today it is relatively common to calculate phonon linewidths arising from electron-phonon 
interactions 
\cite{Allen1972c,Bauer1998}.
Recently it has also become possible to study the frequency renormalization 
due to non-adiabatic effects \cite{Saitta2008,Calandra2010}.

The possibility of observing new features in vibrational spectra arising from the EPI
has not been studied from first principles, but the underlying 
phenomenology should be similar to that of plasmon satellites in photoelectron 
spectra (see Sec.~\ref{sec.polarons}).
Generally speaking we expect satellites whenever $\e_{\rm e}(\br,\br',\w)$ 
exhibits dynamical structure close to vibrational frequencies. This can happen,
for example, in the case of degenerate polar semicondutors, when phonon and plasmon
energies are in resonance.
In these cases, phonons and plasmons can combine into 
`coupled plasmon-phonon modes' \cite{Richter1984}, which are the electronic analogue of photon polaritons
(\citeauthor{Yu2010}, \citeyear{Yu2010}, p.~295). This phenomenon was 
predicted theoretically \cite{Varga1965}, and subsequently confirmed by Raman measurements 
on GaAs \cite{Mooradian1966}. We speculate that it should be possible to obtain 
coupled plasmon-phonon modes from the frequency-dependence of the phonon self-energy in Eq.~(\ref{eq.dyson-pi});
it would be interesting to perform first-principles calculations in order to
shed light on these aspects.

In practical calculations the non-adiabatic corrections to the adiabatic phonon spectrum 
are evaluated from Eq.~(\ref{eq.dyson-pi}) using first-order perturbation theory, by retaining only
the diagonal elements of $\Pi^{\rm NA}$. If we denote the complex zeros of $D^{-1}_{\bq\nu\nu}(\w)$
by $\tilde{\Omega}_{\bq\nu}=\Omega_{\bq\nu}-i\gamma_{\bq\nu}$,
in the case of non-degenerate eigenmodes Eq.~(\ref{eq.dyson-pi}) gives:
 \begin{equation}\label{eq.Omega-tilde}
 \tilde{\Omega}_{\bq\nu}^2=\w_{\bq\nu}^2 +2\w_{\bq\nu} \Pi_{\bq\nu\nu}^{\rm NA}(\tilde{\Omega}_{\bq\nu}),
 \end{equation}
therefore:
  \begin{eqnarray}
  \gamma_{\bq\nu} &=&  -\frac{\w_{\bq\nu}}{\Omega_{\bq\nu}}
    {\rm Im}\,\Pi_{\bq\nu\nu}^{\rm NA}(\Omega_{\bq\nu}-i\gamma_{\bq\nu}), \label{eq.pi-gamma}\\ 
  \Omega_{\bq\nu}^2 &=& \w_{\bq\nu}^2+\gamma_{\bq\nu}^2 + 2\w_{\bq\nu} 
    {\rm Re}\,\Pi_{\bq\nu\nu}^{\rm NA}(\Omega_{\bq\nu}-i\gamma_{\bq\nu}).\,\,\,\label{eq.pi-omega}
  \end{eqnarray}
Apart from the 
small $\gamma_{\bq\nu}^2$ term in the second line, these expressions
are identical to those provided by \textcite{Allen1972c} and \textcite{Grimvall1981}.
Since non-adiabatic corrections are usually small as compared to the adiabatic phonon
frequencies, the above expressions are often 
simplified further by using the additional approximations
$|\Omega_{\bq\nu}-\w_{\bq\nu}|\ll \w_{\bq\nu}$ and $|\gamma_{\bq\nu}|\ll \w_{\bq\nu}$,
leading to 
$\gamma_{\bq\nu} \simeq -{\rm Im}\,\Pi_{\bq\nu\nu}^{\rm NA}(\w_{\bq\nu})$ and
$\Omega_{\bq\nu} \simeq \w_{\bq\nu}+{\rm Re}\,\Pi_{\bq\nu\nu}^{\rm NA}(\w_{\bq\nu})$.
In these forms it becomes evident that the real part of the self-energy 
shifts the adiabatic phonon frequencies, and the imaginary part is responsible
for the spectral broadening of the resonances. Using these expressions in Eq.~(\ref{eq.dyson-pi})
and going back to the time domain, it is seen that, as a result of the EPI,
phonons acquire a finite lifetime given by $\tau^{\rm ph}_{\bq\nu} = (2\gamma_{\bq\nu})^{-1}$.

\subsubsection{Expressions for the phonon self-energy used in {\it ab~initio} 
calculations}\label{sec.phon-self}

In the literature on electron-phonon interactions, the 
phonon self-energy $\Pi$ is almost invariably expressed in terms of an electron-phonon vertex $g$
and the electron Green's function $G$ as $\Pi = |g|^2 G G$ in symbolic notation, see for example 
\citeauthor{Grimvall1981} (\citeyear{Grimvall1981}, p.~195).
While this has become common 
practice also in {\it ab~initio} calculations,
the origin of this choice is not entirely transparent. 
One could derive the above expression directly from Eq.~(\ref{eq.epi-hamilt}), using
standard Green's function techniques. 
However, this procedure does not answer
the key question on {\it how} to calculate the electron-phonon matrix elements~$g$.

Closer inspection of the theory reveals that this is a rather nontrivial point.
In fact, on the one hand, a straightforward expansion of the second-quantized Hamiltonian
of Eq.~(\ref{eq.H-mb-3}) in terms of the nuclear coordinates leads to `bare' electron-phonon matrix elements,
$g^{\rm b}$, which contain the bare Coulomb interaction between
electrons and nuclei. 
On the other hand, if we go back
to Sec.~\ref{sec.general-matel}, we see that the electron-phonon matrix elements in DFT are
`dressed' by the self-consistent response of the electrons. The difference between bare and dressed
vertex is not only quantitative, but also qualitative: for example in metals the bare vertex is 
long-ranged, while the screened vertex is short-ranged.

The relation between bare and dressed electron-phonon vertices and the derivation
of explicit expressions for the phonon self-energy have been discussed by many authors, see for 
example \textcite{Scalapino1969,Rickayzen1980}. In short the argument is that the lowest-order 
Feynman diagram starting and ending with a phonon line must contain precisely two {\it bare}
electron-phonon vertices, as shown in Fig.~\ref{phonon-selfenergy}(b). By construction this diagram corresponds
to having $\Pi = |g^{\rm b}|^2GG$. In order to make the transition from 
the bare vertex to the dressed vertex it is necessary to collect together all the proper electronic 
polarization diagrams around the vertex. However, these steps have been carried out only
for the homogeneous electron gas \cite{Scalapino1969,Rickayzen1980}. In the following we show
how the dressed electron-phonon vertex emerges from a non-perturbative analysis based on the
Hedin-Baym equations. 

The non-adiabatic phonon self-energy $\Pi^{\rm NA}$ introduced in Sec.~\ref{sec.nonadiabatic}
can be written explicitly by combining Eqs.~(\ref{eq.pi-frequency}) and (\ref{eq.ad-nonad}):
  \begin{eqnarray}
  && \hspace{-0.5cm}\Pi^{\rm NA}_{\k \a p, \k' \a' p'}(\w) = \int\!\!d\br d\br'
   Z_\k \nabla_{\!\a}\, \d(\br\!-\!\btau_{\kappa p}^0) \nonumber \\ 
  && \hspace{-0.3cm} \times \left[W_{\rm e}(\br, \br',\w)
    -W_{\rm e}(\br, \br',0) \right]
   Z_{\k'} \nabla^\prime_{\!\a'} \d(\br'\!-\!\btau_{\kappa' p'}^0). \label{eq.pi-na-expl}
  \end{eqnarray}
Using the Dyson equation for the screened Coulomb interaction,
it can be seen that this expression does indeed contain electron-phonon matrix elements.
In fact, by inserting Eq.~(\ref{eq.W-irrpol-e}) into Eq.~(\ref{eq.pi-na-expl}) we find
terms like $vP_{\rm e}W_{\rm e}$,
and the electron-phonon matrix elements will arise from taking the gradients of $v$ 
and $W_{\rm e}$, respectively. By working in the eigenmodes
representation via Eq.~(\ref{eq.Ucrazy-def}), after lengthy manipulations
this procedure yields:
  \begin{eqnarray}
  \hspace{-.6cm}\hbar\, \Pi^{\rm NA}_{\bq\nu,\bq'\nu'}(\w) &=& \!\!\int \!\!d\br d\br'\,
   g^{\rm b}_{\bq\nu}(\br) P_{\rm e}(\br,\br',\w)\, g^{\rm cc}_{\bq'\nu'}(\br',\w)\nonumber \\
  &-& \!\!\int \!\!d\br d\br'\, g^{\rm b}_{\bq\nu}(\br) P_{\rm e}(\br,\br',\,0)\,
  g^*_{\bq'\nu'}(\br',\hspace{0.9pt}0),\,\,\,\, \label{eq.gbarePg}
  \end{eqnarray}
where we introduced electron-phonon `coupling functions' as follows.
The {\it bare} coupling $g^{\rm b}$ is defined~as:
  \begin{equation}
  g^{\rm b}_{\bq\nu}(\br) = \Delta_{\bq\nu} V^{\rm en}(\br), \label{eq.gbare}
  \end{equation}
where $V^{\rm en}$ is the potential of the nuclei from Eq.~(\ref{eq.pot-en});
in practical calculations this quantity is replaced by the usual ionic pseudo-potentials.
The meaning of the variation $\Delta_{\bq\nu}$ is the same as in Eqs.~(\ref{eq.dV-all-1})-(\ref{eq.dV-all-3}).
The {\it dressed} couplings $g$ and $g^{\rm cc}$ are defined as \cite{Hedin1969}:
  \begin{eqnarray}
  g_{\bq\nu}(\br,\w) &=& \int\!d\br'\, \e_{\rm e}^{-1}(\br,\br',\w) \,g^{\rm b}_{\bq\nu}(\br'),   \label{eq.gscreened1}\\
  g^{\rm cc}_{\bq\nu}(\br,\w) &=& \int\!d\br'\, \e_{\rm e}^{-1}(\br,\br',\w) \,g^{\rm b,*}_{\bq\nu}(\br').
  \label{eq.gscreened}
  \end{eqnarray}
Since the dielectric matrix is real at $\w\!=\!0$, we have
the simple relation $g^{\rm cc}_{\bq\nu}(\br,0)\! =\! g^*_{\bq\nu}(\br,0)$.
In order to derive Eq.~(\ref{eq.gbarePg}) it is best to carry out the algebra in the time domain.
We emphasize that the result expressed by
Eq.~(\ref{eq.gbarePg}) is {\it non-perturbative}, and relies solely on the harmonic approximation.

Equation~(\ref{eq.gbarePg})
is in agreement with the standard result for the homogeneous electron gas \cite{Scalapino1969}.
The same expression was also obtained by \textcite{Keating1968} using a detailed diagrammatic
analysis. Keating's diagrammatic representation of the self-energy 
is shown in Fig.~\ref{phonon-selfenergy}(c), and can be obtained from Eq.~(\ref{eq.gbarePg}) 
by noting that, in symbolic notation, $P_{\rm e} = GG \Gamma$ from Eq.~(\ref{eq.P-el}), therefore
$\Pi = g^{\rm b} GG \Gamma g$.
For completeness we also show in Fig.~\ref{phonon-selfenergy}(d) a diagrammatic representation
of the dressed electron-phonon coupling function $g$ as given by Eq.~(\ref{eq.gscreened}). This representation
is obtained by observing that $g= \e^{-1}g^{\rm b}$, $\e = 1-vP$, and
$P = GG \Gamma$, therefore $g = g^{\rm b} + vGG\Gamma g$.

In view of practical first-principles calculations it is useful to have a simplified expression
for the non-adiabatic phonon self-energy in Eq.~(\ref{eq.gbarePg}).
To this aim we make the following approximations:
\begin{itemize}
\item[(i)] The vertex function in Eq.~(\ref{eq.P-el})
  is set to $\Gamma(123) = \d(12)\d(13)$. This is the same approximation at the heart of the $GW$
  method \cite{Hedin1965,Hybertsen1986,Onida2002};
\item[(ii)] The fully-interacting electron Green's function $G$ is replaced by its non-interacting
  counterpart, using the Kohn-Sham eigenstates/eigenvalues evaluated with the nuclei
  held in their equilibrium positions;
\item[(iii)] The fully-interacting dielectric matrix in Eq.~(\ref{eq.gscreened})
  is replaced by the RPA+$xc$ response obtained from a DFT calculation, as discussed 
  in Sec.~\ref{sec.general-matel}; 
\item[(iv)] The frequency-dependence of the screened electron-phonon coupling defined 
  in  Eq.~(\ref{eq.gscreened}) is neglected:
  $g^{\rm cc}_{\bq\nu}(\br,\w)\!\simeq \!g^{\rm cc}_{\bq\nu}(\br,0) \!=\! g^*_{\bq\nu}(\br,0)$. This approximation is
  ubiquitous in the literature but it is never mentioned explicitly;
\item[(v)] For notational simplicity, we consider a spin-degenerate system with time-reversal symmetry;
  this simplification is easily removed.
\end{itemize}
Using these assumptions we can rewrite the component of Eq.~(\ref{eq.gbarePg}) for $\bq=\bq'$ as: 
  \begin{eqnarray}
  &&\hspace{-0.7cm}\hbar\,\Pi^{\rm NA}_{\bq\nu\nu'}(\w) = 
  2 \sum_{mn} \!\int\! \!\frac{d\bk}{\Omega_{\rm BZ}} g^{\rm b}_{mn\nu}(\bk,\bq) g^*_{mn\nu'}(\bk,\bq) 
   \nonumber \\
  &&\hspace{-0.3cm}\times \left[ \frac{f_{m\bk+\bq}-f_{n\bk}}{\ve_{m\bk+\bq}-\ve_{n\bk}-\hbar(\w+i\eta)} - 
     \frac{f_{m\bk+\bq}-f_{n\bk}}{\ve_{m\bk+\bq}-\ve_{n\bk}} \right]\!\!. \label{eq.phon-self-dft}
  \end{eqnarray}
We note that the components of the phonon self-energy for $\bq\!\ne\!\bq'$ vanish due to the periodicity
of the crystalline lattice.
In Eq.~(\ref{eq.phon-self-dft}) the sums run over all the Kohn-Sham states, with occupations $f_{n\bk}$, and $\eta$ is
a real positive infinitesimal. In this case we indicate explicitly the factor of 2 arising
from the spin degeneracy.
The matrix element $g_{mn\nu}(\bk,\bq)$ is the same as in Eq.~(\ref{eq.matel}), and
it is precisely the quantity calculated by most linear-response codes.
The matrix element $g^{\rm b}_{mn\nu}(\bk,\bq)$ 
is obtained from $g_{mn\nu}(\bk,\bq)$ by replacing the variation of the Kohn-Sham potential
by the corresponding variation of the ionic (pseudo)potentials.
The field-theoretic phonon self-energy given by Eq.~(\ref{eq.phon-self-dft}) is in agreement with the
expression derived by \textcite{Calandra2010} starting from time-dependent density-functional
perturbation theory. 

The presence of both the bare electron-phonon
matrix element and the screened matrix element in Eq.~(\ref{eq.phon-self-dft}) 
has not been fully appreciated in the literature, and most {\it ab~initio} calculations 
employ an approximate self-energy whereby $g^{\rm b}$ is replaced by $g$.
The replacement of the bare matrix elements by their screened counterparts in the phonon self-energy 
goes a long way back, and can be found already in the seminal work by \textcite{Allen1972c}.
As a result many investigators (including the author) calculated phonon lifetimes using the 
following expression \cite{Grimvall1981}:
  \begin{eqnarray}
  \frac{1}{\tau^{\rm ph}_{\bq\nu}} &=& 
  \frac{2\pi}{\hbar} 2 \sum_{mn} \!\int\! \!\frac{d\bk}{\Omega_{\rm BZ}} |g_{mn\nu}(\bk,\bq)|^2
  (f_{n\bk}-f_{m\bk+\bq})\nonumber \\ &\times& \d(\ve_{m\bk+\bq}-\ve_{n\bk}-\hbar\w_{\bq\nu}).
      \label{eq.gamma-allen}
  \end{eqnarray}
This is obtained from Eq.~(\ref{eq.phon-self-dft}) by taking the imaginary part and by 
making the replacement $g^{\rm b}\! \rightarrow \!g$.
While Eq.~(\ref{eq.gamma-allen}) can be derived from the Fermi golden rule in a independent-particle
approximation (see \citeauthor{Albers1976}, \citeyear{Albers1976}, Appendix~B), 
the choice of the electron-phonon matrix elements is somewhat arbitrary.
In future calculations of the phonon self-energy it will be important to assess
the importance of using the correct vertex structure, that is replacing 
$|g_{mn\nu}(\bk,\bq)|^2$ by $g^{\rm b}_{mn\nu}(\bk,\bq) g^*_{mn\nu}(\bk,\bq)$
in Eq.~(\ref{eq.gamma-allen}). 

\label{pag.nonadiab}
In general, the effects of the non-adiabatic self-energy
on the phonon spectrum are expected to be significant only in the case of metals and
small-gap semiconductors. In fact, by combining Eqs.~(\ref{eq.pi-gamma}), (\ref{eq.pi-omega}),
and (\ref{eq.phon-self-dft}) it is seen that $\Pi^{\rm NA}$ can be large only when
occupied and empty single-particle states are separated by an energy of the order
of the characteristic phonon energy. In such a case, we can expect a shift of the adiabatic
phonon frequencies, and a concomitant broadening of the lines. A clear illustration of
these effects was provided by \textcite{Maksimov1996}, who analyzed a simplified model
of a metal with linear bands near the Fermi level.

Calculations of phonon linewidths based on Eq.~(\ref{eq.gamma-allen})
have been reported by several authors\footnote{See for example 
\citeauthor{Butler1979}, \citeyear{Butler1979};
\citeauthor{Bauer1998}, \citeyear{Bauer1998};
\citeauthor{Shukla2003}, \citeyear{Shukla2003};
\citeauthor{Lazzeri2006b}, \citeyear{Lazzeri2006b};
\citeauthor{Park2008b}, \citeyear{Park2008b};
\citeauthor{Giustino2007b}, \citeyear{Giustino2007b};
\citeauthor{Heid2010}, \citeyear{Heid2010}.} and have become commonplace
in first-principles studies of electron-phonon physics. On the other hand, calculations
of the non-adiabatic phonon frequencies using Eq.~(\ref{eq.phon-self-dft}) have only
been reported in 
(\citeauthor{Lazzeri2006}, \citeyear{Lazzeri2006};
\citeauthor{Piscanec2007}, \citeyear{Piscanec2007};
\citeauthor{Caudal2007}, \citeyear{Caudal2007};
\citeauthor{Saitta2008}, \citeyear{Saitta2008};
\citeauthor{Calandra2010}, \citeyear{Calandra2010}), using the approximation that the
bare vertex $g^{\rm b}$ can be replaced by the screened vertex $g$. Examples of such
calculations will be reviewed in Sec.~\ref{sec.nonadiab}.

Equation~(\ref{eq.phon-self-dft}) suggests several avenues worth exploring
in the future: firstly, the use of the bare vertex should not pose a challenge in
practical calculations, since this quantity is already being calculated in linear-response
DFT codes. Testing the impact of the bare vertex on phonon linewidths and frequency
renormalizations will be important. Secondly, Eq.~(\ref{eq.phon-self-dft}) contains
off-diagonal couplings, which are usually ignored. It will be interesting to check the
effect of using the complete matrix self-energy. Thirdly, the dynamical structure
of the self-energy may contain interesting information, such as for example spectral
satellites and coupled phonon-plasmon modes.
Lastly, the move from Eq.~(\ref{eq.gbarePg})
to Eq.~(\ref{eq.phon-self-dft}) involves the approximation that the frequency-dependence
of the electron-phonon matrix elements can be neglected. 
The validity of this approximation is uncertain, and there are no reference
{\it ab~initio} calculations on this. However, we note that
frequency-dependent electron-phonon matrix elements have been employed 
systematically in theoretical models of doped semiconductors 
(\citeauthor{Mahan1993}, \citeyear{Mahan1993}, Sec. 6.3).

Before closing this section we note that the formalism discussed here is based
on zero-temperature Green's functions. In order to extend the present results
to finite temperature it is necessary to repeat all derivations using the
Matsubara representation, and then perform the analytic continuation of the self-energy
to the real frequency axis. Detailed derivations can be found in \cite{Baym1961}
and \cite{Gillis1970}, and will not be repeated here. Fortunately it turns out
that Eq.~(\ref{eq.phon-self-dft}) can be extended to
finite temperature by simply replacing the occupation factors $f_{n\bk}$ and 
$f_{m\bk+\bq}$ by the corresponding Fermi-Dirac distributions.\footnote{\label{note.FDBE}Throughout
this article, when $f_{n\bk}$ and $n_{\bq\nu}$ have the meaning of Fermi-Dirac and Bose-Einstein
distributions, respectively, they are defined as follows:
$f_{n\bk} = f[(\ve_{n\bk}\!-\!\ve_{\rm F})/\kt]$ with $f(x)=1/(e^x+1)$ and
$\ve_{\rm F}$ being the Fermi energy;
$n_{\bq\nu} = n(\hbar\w_{\bq\nu}/\kt)$ with $n(x)= 1/(e^x-1)$.}

\subsection{Effects of the electron-phonon interaction on electrons}

\subsubsection{Electron self-energy: Fan-Migdal and Debye-Waller terms}\label{sec.elec-SE}

In Sec.~\ref{sec.phon-all} we discussed the link between 
the Hedin-Baym equations summarized in Table~\ref{tab.hedin-baym} and {\it ab~initio} 
calculations of phonons. We first identified
a Hermitian eigenvalue problem for the vibrational frequencies via the adiabatic approximation, 
and then we improved upon this description by means of a non-adiabatic self-energy. 
In this section, we adopt a similar strategy in order to discuss 
electronic excitation energies: first we identify an approximation to the Hedin-Baym 
equations which does {\it not} include any electron-phonon interactions, and then we introduce 
an electron self-energy to incorporate such interactions.

The single most common approximation in first-principles electronic structure calculations is
to describe nuclei as classical particles clamped in their equilibrium positions.
Within this approximation the expectation value of the nuclear charge density operator
in Eq.~(\ref{eq.n-density-op}), $\< \hnn(\br) \>$, is replaced by the first term in
Eq.~(\ref{eq.density-nuc-disp}), $n_{\rm n}^0(\br)$. From Eq.~(\ref{eq.nuc-den-D}) we see
that this approximation formally corresponds to setting to zero the 
displacement-displacement correlation function of the nuclei. This observation
suggests that, in order to unambiguously single out electron-phonon interactions in the Hedin-Baym equations,
we need to define a non-interacting problem by setting $D_{\k \a p, \k' \a' p'}=0$,
and identify the electron-phonon interaction with the remainder.
In the following, we write an equation of motion for the electrons analogous to Eq.~(\ref{eq.all-mb}),
except with the nuclei clamped in their equilibrium positions;
then we use a Dyson-like equation to recover the fully-interacting electron Green's function.

The equation of motion for the electron Green's function
at {\it clamped nuclei}, which we denote as $G^{\rm cn}$, reads:
  \begin{eqnarray}\label{eq.all-mb-clamped}
  && \hspace{-0.7cm}\left[ i\hbar \frac{\D}{\D t_1} + \frac{\hbar^2}{2m_{\rm e}}\nabla^2(1)
  -V_{\rm tot}^{\rm cn}(1)\! \right]G^{\rm cn}(12) \nonumber \\
  &&\hspace{1cm} - \int\!\!d3\,\Sigma_{\rm e}^{\rm cn}(13) G^{\rm cn}(32)  = \d(12).
  \end{eqnarray}
Here the potential $V_{\rm tot}^{\rm cn}$ differs from its counterpart $V_{\rm tot}$ of Eq.~(\ref{eq.vtot-den})
in that the total density of electrons and nuclei, $\<\hn\>$, is replaced by the density calculated at clamped
nuclei, $\<\hat{n}^{\rm cn}\>$:
  \begin{equation}
  V_{\rm tot}^{\rm cn}(1) = \int\!\!d 2 \,v(1,2)\, \< \hn^{\rm cn}(2)\>, 
  \end{equation}
where we defined:
  \begin{equation}
  \< \hn^{\rm cn}(1)\> = -i\hbar {\sum}_{\sigma_1} G^{\rm cn}(11^+) + n_{\rm n}^0(\br_1).  \label{eq.density-cl}
  \end{equation}
The term $\Sigma_{\rm e}^{\rm cn}$ in Eq.~(\ref{eq.all-mb-clamped}) represents the electronic part
of Hedin's self-energy in Eq.~(\ref{eq.selfen-numbers}), evaluated at clamped nuclei:  
  \begin{equation}\label{eq.sigma-e-cl}
  \Sigma_{\rm e}^{\rm cn}(12) = i\hbar\!\! \int\!\! d(34) \,G^{\rm cn}(13) \,\Gamma^{\rm cn}(324) 
  \,W_{\rm e}^{\rm cn}(41^+).
  \end{equation}
In this expression, the vertex $\Gamma^{\rm cn}$ and the screened Coulomb interaction $W_{\rm e}^{\rm cn}$
are both evaluated via the Hedin-Baym equations at clamped nuclei.
Equations~(\ref{eq.all-mb-clamped})-(\ref{eq.sigma-e-cl}) lead directly to
the well-known Hedin's equations~\cite{Hedin1965}. 
Hedin's equations and the associated $GW$ method at clamped nuclei
are addressed in a number of excellent reviews 
(\citeauthor{Hedin1969}, \citeyear{Hedin1969};
\citeauthor{Hybertsen1986}, \citeyear{Hybertsen1986};
\citeauthor{Aryasetiawan1998}, \citeyear{Aryasetiawan1998};
\citeauthor{Onida2002}, \citeyear{Onida2002})
hence they will not be discussed here.

In order to recover the complete Hedin-Baym equation of motion, Eq.~(\ref{eq.all-mb}), starting
from Eqs.~(\ref{eq.all-mb-clamped})-(\ref{eq.sigma-e-cl}), it is sufficient to introduce the
Dyson equation:
  \begin{equation}\label{eq.dyson-ep}
  G(12) = G^{\rm cn}(12) + \!\int\!d(34)\, G^{\rm cn}(13)\, \Sigma^{\rm ep}(34)\, G(42),
  \end{equation}
together with the electron self-energy $\Sigma^{\rm ep}$ arising from electron-phonon interactions:
  \begin{equation}
  \Sigma^{\rm ep} = \Sigma^{\rm FM}+\Sigma^{\rm DW}+\Sigma^{\rm dGW}, \label{eq.selfen-ep}
  \end{equation}
where we have defined:
  \begin{eqnarray}
  \hspace{-0.5cm} \Sigma^{\rm FM}(12) &=& i\hbar\!\!\int\!\! d(34)\, G(13)\, \Gamma(324)\, W_{\rm ph}(41^+), \label{eq.FM} \\
   \hspace{-0.5cm} \Sigma^{\rm DW}(12) &=&
   \int\!\!d3\, v(13) \left[\< \hn(3) \> -\< \hn^{\rm cn}(3)\> \right]\d(12), \label{eq.DW}\\
   \hspace{-0.5cm} \Sigma^{\rm dGW}(12) &=& \Sigma_{\rm e}(12) - \Sigma_{\rm e}^{\rm cn}(12). \label{eq.selfen-ep-3rdterm}
  \end{eqnarray}
We emphasize that Eqs.~(\ref{eq.all-mb-clamped})-(\ref{eq.selfen-ep-3rdterm}) are just an alternative
formulation of the Hedin-Baym equations in Table~\ref{tab.hedin-baym}. 
The advantage of this formulation
is that it better reflects standard {\it practice}, whereby the DFT equations and the $GW$ quasiparticle
corrections are evaluated at clamped nuclei. Equations~(\ref{eq.all-mb-clamped})-(\ref{eq.selfen-ep-3rdterm}) 
are formally exact within the harmonic approximation. 

\begin{figure}[t!]
  \includegraphics[width=\columnwidth]{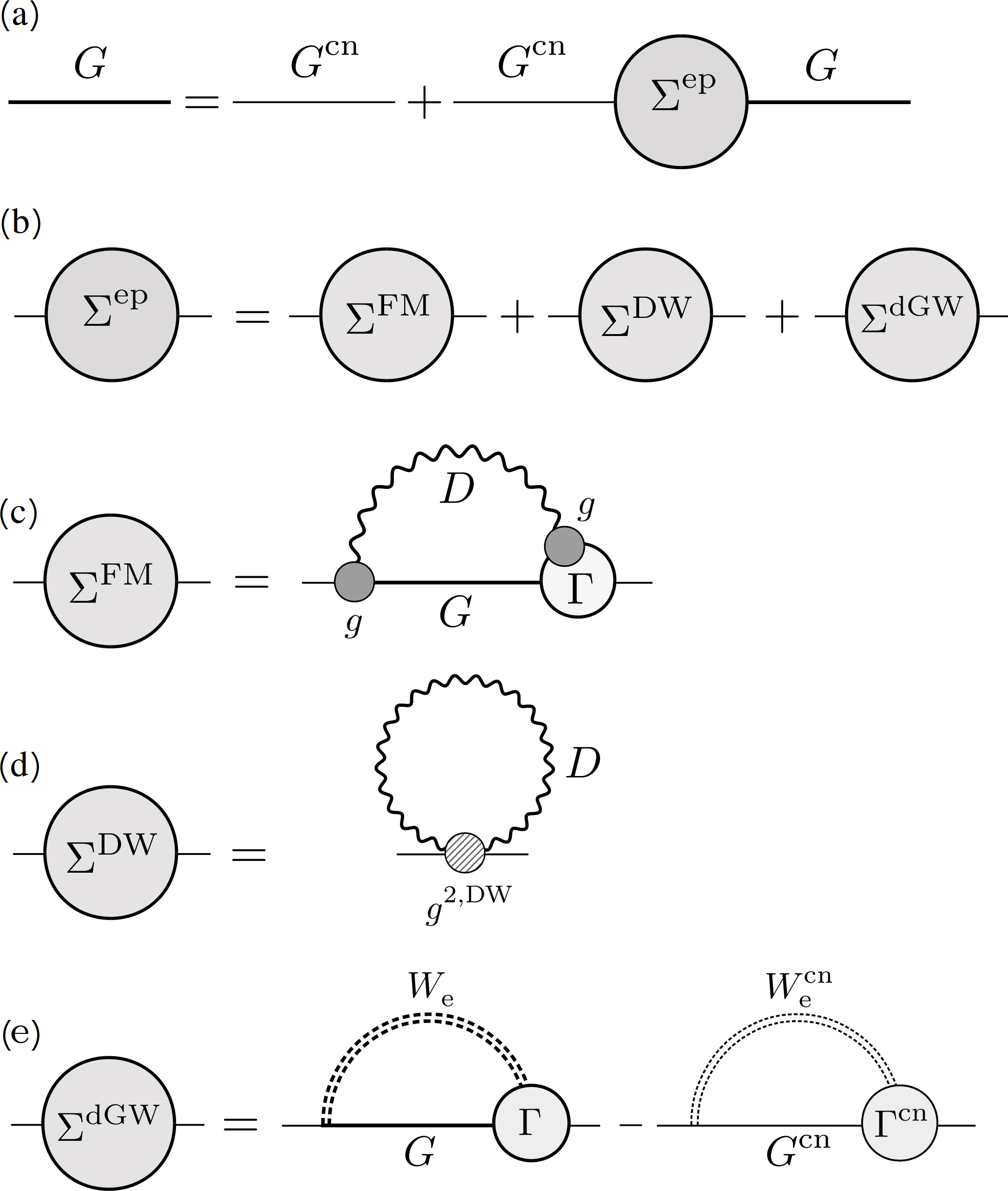}
  \caption{\label{fig.elec-selfenergy}
  Diagrammatic representation of the electron Green's function and electron-phonon self-energy.
  (a)~Dyson equation for the electron Green's function, Eq.~(\protect\ref{eq.dyson-ep}).
  The thick straight line represents the fully-dressed electron propagator, the thin straight line 
  is the propagator calculated at clamped nuclei, and the disc is the electron-phonon self-energy.
  (b)~Decomposition of the electron-phonon self-energy into Fan-Migdal self-energy, Eq.~(\protect\ref{eq.FM}),
  Debye-Waller contribution, Eq.~(\protect\ref{eq.DW}), and the remainder given by
  Eq.~(\protect\ref{eq.selfen-ep-3rdterm}).
  (c)~Fan-Migdal electron-phonon self-energy expressed in terms of the dressed electron-phonon coupling
  function 
  (dark grey disc as in Fig.~\protect\ref{phonon-selfenergy}), 
  the fully-interacting electron's Green's functions (thick straight line),  
  the fully interacting phonon propagator (thick wavy line), and the  vertex $\Gamma$ from 
  Eq.~(\protect\ref{eq.vertex}).
  (d)~Debye-Waller contribution resulting from the fully interacting phonon propagator (thick wavy line)
  and the matrix element in Eq.~(\protect\ref{eq.matel-dw-dft}) (hatched disc).
  (e)~Correction to Hedin's $GW$ self-energy arising from the modification of the electronic
  structure induced by the electron-phonon interaction. $W_{\rm e}$ is the screened
  Coulomb interaction of Eq.~(\protect\ref{eq.W-irrpol-e}) (bold dashed double line). $W_{\rm e}^{\rm cn}$ 
  is the screened Coulomb interaction evaluated at clamped nuclei (thin dashed double line).
  $\Gamma^{\rm cn}$ is the vertex of Eq.~(\protect\ref{eq.vertex}), but evaluated at clamped nuclei. 
  }
  \end{figure}

A schematic representation of the Dyson equation for the electron Green's function and the 
decomposition of the electron self-energy are given in Fig.~\ref{fig.elec-selfenergy}.
The self-energy contribution $\Sigma^{\rm FM}$ in Eq.~(\ref{eq.FM}) is a dynamic correction to the electronic excitation
energies, and is analogous to the $GW$ self-energy at clamped nuclei. Indeed, in the same way as the correlation 
part of the standard $GW$ self-energy describes the effect of the dynamic electronic polarization upon the 
addition of electrons or holes to the system, the term $G W_{\rm ph}$ in Eq.~(\ref{eq.FM})
describes the effect of the dynamic polarization of the lattice. 

In the semiconductors community,
the self-energy obtained from Eq.~(\ref{eq.FM}) by setting $\Gamma(123)=\d(13)\d(23)$ 
is commonly referred to as the {\it Fan} self-energy 
(\citeauthor{Fan1951}, \citeyear{Fan1951}; \citeauthor{Allen1976}, \citeyear{Allen1976};
\citeauthor{Allen1981}, \citeyear{Allen1981}; \citeauthor{Cardona2001}, \citeyear{Cardona2001}).
In the metals and superconductors communities, the same term 
is traditionally referred to as the self-energy in the {\it Migdal} approximation
(\citeauthor{Migdal1958}, \citeyear{Migdal1958};
\citeauthor{Engelsberg1963}, \citeyear{Engelsberg1963};
\citeauthor{Scalapino1969}, \citeyear{Scalapino1969};
\citeauthor{Schrieffer1983}, \citeyear{Schrieffer1983}).
By extension we refer to the self-energy $\Sigma^{\rm FM}$ in Eq.~(\ref{eq.FM})
as the `Fan-Migdal' (FM) self-energy.

The static term $\Sigma^{\rm DW}$ in Eq.~(\ref{eq.DW}) corresponds to the difference between 
the self-consistent potential $V_{\rm tot}$ calculated for the fully-interacting system, and the
same potential evaluated with the nuclei clamped in their equilibrium positions, $V_{\rm tot}^{\rm cn}$.
Intuitively this term corresponds to a time-independent correction to the `crystal potential' that
arises from the
fuzziness of the nuclear charge density around the equilibrium nuclear positions.
This term is similar to the one appearing in the study of the temperature dependence of 
X-ray diffraction and neutron diffraction spectra 
(\citeauthor{Mermin1966}, \citeyear{Mermin1966}; \citeauthor{Ashcroft1976}, \citeyear{Mermin1966}),
and is commonly referred to as the {\it Debye-Waller} (DW) term 
(\citeauthor{Antoncik1955}, \citeyear{Antoncik1955};
\citeauthor{Walter1970}, \citeyear{Walter1970};
\citeauthor{Cardona2005}, \citeyear{Cardona2005}).
\textcite{Hedin1969} did not include this term in their classic work,
however this contribution was discussed by \citeauthor{Allen1976} (\citeyear{Allen1976}; see also 
\citeauthor{Allen1978}, \citeyear{Allen1978}).

The last term Eq.~(\ref{eq.selfen-ep}), $\Sigma^{\rm dGW}$, is the correction to the standard Hedin 
self-energy arising from the fact that the fully-interacting electron Green's function 
and density are slightly different from those evaluated at clamped nuclei, owing to the
electron-phonon interaction. 
The magnitude of this term corresponds to a fraction of the $GW\Gamma$ quasiparticle corrections at clamped nuclei. 
Since $\Sigma^{\rm dGW}$ has never been investigated so far, we will not
discuss this term further.

\subsubsection{Expressions for the electron self-energy used in {\it ab~initio} 
calculations}\label{eq.explic-elec}

A complete self-consistent solution of Eqs.~(\ref{eq.all-mb-clamped})-(\ref{eq.selfen-ep-3rdterm})
from first principles 
is not possible at present, and one has to replace the various entries of
Eq.~(\ref{eq.selfen-ep}) by the best approximations available. In practice,
one resorts to either DFT or to $GW$ calculations; recent progress will be
reviewed in Secs.~\ref{sec.kinks} and \ref{sec.semicond}.

Using Eqs.~(\ref{eq.Wph-1}), (\ref{eq.adiab-prop-explicit})-(\ref{eq.Ucrazy-def}), and 
(\ref{eq.gscreened}) we can rewrite the Fan-Migdal self-energy as follows:
  \begin{eqnarray}
  && \hspace{-0.7cm}
  \Sigma^{\rm FM}(12) = i\sum_{\nu\nu'} \int \frac{d\w}{2\pi}\frac{d\bq}{\Omega_{\rm BZ}}d(34)  
      e^{-i\w(t_4-t_1^+)} \nonumber \\
  &&\hspace{-0.2cm} 
  \times G(13)\, \Gamma(324)\, g_{\bq\nu}^{\rm cc}(\br_4,\w)\, D_{\bq\nu\nu'}(\w)\, g_{\bq\nu'}(\br_1,\w).
 \label{eq.fan-all}
  \end{eqnarray}
This shows that the Fan-Migdal self-energy is, in symbolic notation, of the type $\Sigma = g^2 D G \Gamma$;
a graphical representation of this term is given in Fig.~\ref{fig.elec-selfenergy}(c).
In order to make the last expression amenable to {\it ab~initio} calculations, it is common
to make the following approximations, which are similar to those introduced earlier for the
phonon self-energy:
\begin{itemize} \label{pag.approx-SE}
\item[(i)] The vertex $\Gamma(123)$ is set to $\d(13)\d(23)$;
\item[(ii)] The fully-interacting electron Green's function is replaced by the Kohn-Sham 
 Green's function evaluated at clamped nuclei;
\item[(iii)] The fully-interacting phonon propagator $D_{\bq\nu\nu'}(\w)$ is replaced by the adiabatic propagator
$D^{\rm A}_{\bq\nu\nu'}(\w)$ from Eq.~(\ref{eq.Schrieffer1});
\item[(iv)] The screened electron-phonon vertex is evaluated using the RPA+$xc$ electronic screening
from a DFT calculation;
\item[(v)] The frequency dependence of the electron-phonon coupling is neglected,
  $g_{\bq\nu}(\br,\w)\simeq g_{\bq\nu}(\br,0)$.
\end{itemize}
After using these approximations in Eq.~(\ref{eq.fan-all}), we obtain the following result 
for the $\bk\!=\!\bk'$ matrix elements of the Fan-Migdal self-energy in the basis of Kohn-Sham states:
  \begin{eqnarray}
  && \Sigma^{\rm FM}_{n n'\bk}(\w) = \frac{1}{\hbar}
    \sum_{m\nu} \!\int\!\! \frac{d\bq}{\Omega_{\rm BZ}} g_{mn\nu}^*(\bk,\bq) g_{mn'\nu}(\bk,\bq) \nonumber \\
   && \times\! \left[ \frac{1-f_{m\bk+\bq}}{\w\!-\!\ve_{m\bk+\bq}/\hbar-\w_{\bq\nu}+i\eta} + 
     \frac{f_{m\bk+\bq}}{\w\!-\!\ve_{m\bk+\bq}/\hbar+\w_{\bq\nu}-i\eta} \right]\!. \nonumber \\
   \label{eq.fan-final}
  \end{eqnarray}
Here $f_{m\bk+\bq}=1$ for occupied Kohn-Sham states and 0 otherwise, and
the matrix element $g_{mn\nu}(\bk,\bq)$ is obtained from Eq.~(\ref{eq.matel}).
As for the phonon self-energy in Eq.~(\ref{eq.phon-self-dft}), also in this case the components of the electron 
self-energy for $\bk\!\ne\!\bk'$ vanish due to the periodicity of the lattice.
The result in Eq.~(\ref{eq.fan-final}) is obtained by closing the contour of the frequency integration in the
upper complex plane, owing to the $t_1^+$ in the exponential of Eq.~(\ref{eq.fan-all}).
The infinitesimals inside the electron and phonon propagators, which reflect the
time-ordering, are crucial to obtain the correct result \cite{Schrieffer1983}.
The spin label is omitted in Eq.~(\ref{eq.fan-final}) since this contribution to the self-energy
is diagonal in the spin indices. 

The result above is only valid at zero temperature.
The extension to finite temperature requires going through the Matsubara representation,
and then continuing the self-energy from the imaginary axis to the real axis. 
The procedure is described in many textbooks, see for example Sec.~3.5 of \cite{Mahan1993}.
The result is that at finite temperature the square brackets of Eq.~(\ref{eq.fan-final}) are 
to be modified as follows:
  \begin{equation}\label{eq.fan-final-T}
  \hspace{-0.05cm}\left[\! \frac{1\!-\!f_{m\bk+\bq}}{\cdots + i\eta} \!+\! \frac{f_{m\bk+\bq}}{\cdots -i\eta} 
   \!\right] 
   \!\! \rightarrow \!\!
   \left[\! \frac{1\!-\!f_{m\bk+\bq}\!+\!n_{\bq\nu}}{\cdots+i\eta} \!+\! \frac{f_{m\bk+\bq}\!+\!n_{\bq\nu}}
    {\cdots + i\eta}\! \right]\!\!, 
  \end{equation}
where $f_{m\bk+\bq}$ and $n_{\bq\nu}$ are now Fermi-Dirac and Bose-Einstein distribution functions, respectively
(see footnote~\ref{note.FDBE}).
The change of sign in the imaginary infinitesimal on the second fraction has to do with the fact
that in the Matsubara formalism the analytic continuation from the imaginary frequency 
axis to the real axis through the upper complex plane leads to the so-called {\it retarded} self-energy, 
that is a self-energy with all poles below the real axis \cite{Abrikosov1975,Mahan1993}. 

The Debye-Waller term in Eq.~(\ref{eq.DW}) can also be written in a form which is convenient
for practical calculations, by expanding the total density operator $\hn(3)$ to second order in the
atomic displacements. Using Eqs.~(\ref{eq.vtot-den}) and (\ref{eq.dis-dis}) we find:
  \begin{equation}
  \Sigma^{\rm DW}(12) = \d(12)\, \frac{i\hbar}{2}\!\!\!\sum_{\substack{\k \a p\\\k' \a' p'}} \!\!\frac{\D^2  V_{\rm tot}(1)}
  {\D\tau_{\k\a p}^0\D\tau_{\k'\a' p'}^0} D_{\k \a p,\k' \a' p'}(t_1^+,t_1). \label{eq.dw-D}
  \end{equation}
In order to arrive at this result, it is necessary to make the additional approximation that the
electronic field operators and the operators for the nuclear displacements are uncorrelated, 
that is $\< \hn \,\Delta \hat{\tau}_{\k \a p} \> = 
\< \hn \> \< \Delta \hat{\tau}_{\k \a p} \>$, and similarly for the second power of the displacements.
This requirement was noticed by \textcite{Gillis1970}, and is trivially satisfied if we describe
phonons within the adiabatic approximation of Sec.~\ref{sec.phonons-bo}.
Equation~(\ref{eq.dw-D}) motivates the diagrammatic representation of the Debye-Waller self-energy
shown in Fig.~\ref{fig.elec-selfenergy}(d), whereby the phonon line begins and ends 
at the same time point.
We note that Eq.~(\ref{eq.dw-D}) involves the variation of
the screened potential $V_{\rm tot}$;
this result, which we here derived starting from Schwinger's functional derivative
technique, is also obtained when starting from a perturbative diagrammatic analysis
\cite{Marini2015}.
The Debye-Waller self-energy can be simplified further if we make use of the following approximations,
in the same spirit as for the Fan-Migdal self-energy:
\begin{itemize}
\item[(vi)] The fully-interacting phonon propagator is replaced by the adiabatic propagator
$D^{\rm A}_{\bq\nu\nu'}(\w)$ from Eq.~(\ref{eq.Schrieffer1});
\item[(vii)] The total many-body potential $V_{\rm tot}$ of Eq.~(\ref{eq.vtot-den}) 
  is replaced by the Kohn-Sham potential $V^{\rm KS}(\br)$ evaluated at clamped nuclei.
  Strictly speaking, the Kohn-Sham effective potential includes also contributions from exchange
  and correlation, which in the Hedin-Baym equations are all contained in the electron self-energy.
  However, the present discussion holds unchanged if we add any local and frequency-independent
  potential to $V_{\rm tot}$ in Eq.~(\ref{eq.all-mb}), while removing the same potential from the self-energy 
  \cite{Keating1968}.
\end{itemize}
Using these simplifications together with Eqs.~(\ref{eq.adiab-prop-explicit}) and (\ref{eq.Ucrazy-def}),
we can write the $\Sigma^{\rm DW}$ in the basis of Kohn-Sham eigenstates as follows:
  \begin{equation}
  \Sigma^{\rm DW}_{nn'\bk} = {\sum}_\nu 
   \!\int\!\! \frac{d\bq}{\Omega_{\rm BZ}} \, g^{\rm DW}_{nn'\nu\nu}(\bk,\bq,-\bq),
   \label{eq.dw-DFT}
  \end{equation}
where the Debye-Waller matrix element $g^{\rm DW}$ is obtained 
from Eq.~(\ref{eq.matel-dw-dft}), and the presence of only the diagonal terms $\nu\!=\!\nu'$
is a result of the Kronecker delta in Eq.~(\ref{eq.adiab-prop-explicit-2}).
In going from Eq.~(\ref{eq.dw-D}) to Eq.~(\ref{eq.dw-DFT}) the frequency integration is performed
by using Eq.~(\ref{eq.Schrieffer1}), after closing the contour in the lower half plane.
The resulting expression is diagonal in the spin indices.

The expression for the Debye-Waller term in Eq.~(\ref{eq.dw-DFT}) is only valid 
at zero temperature. In this case the extension to finite temperature is immediate since the self-energy
does not involve the electron Green's function, hence we only 
need to evaluate the canonical average of Eq.~(\ref{eq.schafer}) at equal times. The result is that
Eq.~(\ref{eq.dw-DFT}) is simply to be modified as follows:
  \begin{equation}   \label{eq.dw-DFT-T}
  g^{\rm DW}_{nn'\nu\nu}(\bk,\bq,-\bq) \rightarrow g^{\rm DW}_{nn'\nu\nu}(\bk,\bq,-\bq) (2n_{\bq\nu}+1),
  \end{equation}
with $n_{\bq\nu}$ being the Bose-Einstein occupations (see footnote~\ref{note.FDBE}).

The Debye-Waller contribution to the electron self-energy is almost invariably ignored
in the literature on metals and superconductors, but it is well-known in the theory
of temperature-dependent band structures of semiconductors 
(\citeauthor{Allen1976}, \citeyear{Allen1976};
\citeauthor{Allen1981}, \citeyear{Allen1981};
\citeauthor{Marini2008}, \citeyear{Marini2008};
\citeauthor{Giustino2010}, \citeyear{Giustino2010};
\citeauthor{Ponce2015}, \citeyear{Ponce2015}).
Neglecting $\Sigma^{\rm DW}$ in metals is partly justified by the fact that 
this term is frequency-independent,
therefore it is expected to be a slowly-varying function over each Fermi surface sheet.
A detailed first-principles analysis of this aspect is currently lacking.

\subsubsection{Temperature-dependence of electronic band structures}\label{sec.tempdep-theory}

Once determined the electron self-energy as in Sec.~\ref{eq.explic-elec} it is possible
to study the modification of the electronic structure induced by the EPI.
To this aim it is convenient to rewrite Eq.~(\ref{eq.dyson-ep}) in the basis of Kohn-Sham
eigenstates:
  \begin{equation}\label{eq.eq.Gm1-ep-ks0}
  G^{-1}_{nn'\bk}(\w) = G^{{\rm cn},-1}_{nn'\bk}(\w) - \Sigma^{\rm ep}_{nn'\bk}(\w).
  \end{equation}
Assuming that the electronic structure problem at clamped nuclei has been solved using
DFT or DFT+$GW$ calculations, the  Green's function $G^{\rm cn}$ can be written in terms of simple
poles at the Kohn-Sham or quasiparticle eigenvalues $\ve_{n\bk}$ \cite{Hedin1969}. In this case
Eq.~(\ref{eq.eq.Gm1-ep-ks0}) reduces to:
  \begin{equation}\label{eq.Gm1-ep-ks}
  G^{-1}_{nn'\bk}(\w) = (\hbar\w - \tilde{\ve}_{n\bk})\d_{nn'} - \Sigma^{\rm ep}_{nn'\bk}(\w),
  \end{equation}
where $\tilde{\ve}_{n\bk}=\ve_{n\bk}\pm i\hbar\eta$ with the upper/lower sign corresponding 
to occupied/unoccupied states. The spin indices are omitted since these self-energy contributions 
do not mix states with opposite spin.

In order to gain insight into the effects of the electron-phonon interaction,
we start from the drastic approximation
that $\Sigma^{\rm ep}$ only leads to a small shift of the quasiparticle poles, from the `non-interacting'
energies $\ve_{n\bk}$ to the renormalized energies $\tilde{E}_{n\bk}=E_{n\bk}+i\Gamma_{n\bk}$.
In this approximation, the fully interacting Green's function is expressed as a sum of simple poles,
given by the zeros of Eq.~(\ref{eq.Gm1-ep-ks}):
  \begin{eqnarray}\label{eq.qp-shift}
  E_{n\bk} &=& \ve_{n\bk} + {\rm Re}\,\Sigma^{\rm ep}_{nn\bk}(\tilde{E}_{n\bk}/\hbar), \\
  \Gamma_{n\bk} &=& {\rm Im}\, \Sigma^{\rm ep}_{nn\bk}(\tilde{E}_{n\bk}/\hbar). \label{eq.qp-width} 
  \end{eqnarray}
As in the case of vibrational frequencies in Eq.~(\ref{eq.Omega-tilde}), 
we are considering for simplicity non-degenerate electronic states, and making the assumption
that the off-diagonal elements of the self-energy $\Sigma^{\rm ep}_{nn'\bk}$ with $n\!\ne\! n'$
can be neglected. In more general situations the right-hand side of Eq.~(\ref{eq.Gm1-ep-ks}) needs 
to be to diagonalized, or alternatively the off-diagonal terms $\Sigma^{\rm ep}_{nn'\bk}$ need to be treated
perturbatively. The energies $E_{n\bk}$ obtained from Eq.~(\ref{eq.qp-shift}) yield the
band structure renormalized by the EPIs, to be discussed below. The imaginary part
$\Gamma_{n\bk}$ in Eq.~(\ref{eq.qp-width}) is connected with the quasiparticle lifetimes
and will be discussed in Sec.~\ref{sec.el-lifetime}.

Equation~(\ref{eq.qp-shift}) is to be solved self-consistently for $E_{n\bk}$ and $\Gamma_{n\bk}$.
When Eq.~(\ref{eq.qp-shift}) is used in combination with the standard approximations to the Fan-Migdal
and Debye-Waller self-energies given by Eqs.~(\ref{eq.fan-final}) and (\ref{eq.dw-DFT}), the
result that one obtains is equivalent to describing electron-phonon couplings to second order in Brillouin-Wigner 
perturbation theory \cite{Mahan1993}. Similarly one recovers the more basic Rayleigh-Schr\"odinger 
perturbation theory by making the replacements $E_{n\bk}\!\rightarrow \!\ve_{n\bk}$ and 
$\Gamma_{n\bk} \!\rightarrow \!0$ in Eq.~(\ref{eq.qp-shift}).

By combining Eqs.~(\ref{eq.fan-final})-(\ref{eq.fan-final-T}),
(\ref{eq.dw-DFT})-(\ref{eq.dw-DFT-T}), and (\ref{eq.qp-shift}), we obtain the temperature-dependent 
`band structure renormalization' arising from the EPI:
  \begin{eqnarray} 
  E_{n\bk} &=& \ve_{n\bk}+{\sum}_\nu\!\int\!\! \frac{d\bq}{\Omega_{\rm BZ}} 
    \sum_{m} |g_{mn\nu}(\bk,\bq)|^2  \nonumber \\
   & \times &
    \left.  {\rm Re} \left[
        \,\frac{1-f_{m\bk+\bq}+n_{\bq\nu}}{E_{n\bk}-\ve_{m\bk+\bq}-\hbar\w_{\bq\nu}
       +i\Gamma_{n\bk}} \right. \right. \nonumber \\ && \left.
     \,\,\,\,\,\,\,+ 
     \,\frac{\qquad f_{m\bk+\bq}+n_{\bq\nu}}{E_{n\bk}-\ve_{m\bk+\bq}+\hbar\w_{\bq\nu}+i\Gamma_{n\bk}} 
      \,\right]  \nonumber \\
   &  +& {\sum}_\nu\!\int\!\! \frac{d\bq}{\Omega_{\rm BZ}}
        g^{\rm DW}_{nn\nu\nu}(\bk,\bq,-\bq) (2n_{\bq\nu}+1). 
   \label{eq.AH-BW}
  \end{eqnarray}
For practical calculations it is important to bear in mind that this result rests on 
the approximations (i)-(vii) introduced at p.~\pageref{pag.approx-SE}, as well as the
harmonic approximation.

The theory of temperature-dependent band structures developed by \textcite{Allen1976} makes two
additional approximations on top of Eq.~(\ref{eq.AH-BW}): Brillouin-Wigner perturbation theory
is replaced by Rayleigh-Schr\"odinger perturbation theory; and the phonon energies in the denominators
are neglected. Using these additional approximations Eq.~(\ref{eq.AH-BW}) becomes:
  \begin{eqnarray}
  E_{n\bk} =  \ve_{n\bk}&+&\sum_\nu\!\!\int\!\!\! \frac{d\bq}{\Omega_{\rm BZ}}
    \!\left[ \sum_{m} \!\frac{|g_{mn\nu}(\bk,\bq)|^2\!\!}{\ve_{n\bk}\!-\!\ve_{m\bk+\bq}} \right. 
        \nonumber \\ &+& \left. 
      g^{\rm DW}_{nn\nu\nu}(\bk,\bq,\!-\bq) \vphantom{\sum_{m}} \right] (2n_{\bq\nu}+1),
   \label{eq.AH-RS}
  \end{eqnarray}
which is referred to as the `adiabatic Allen-Heine formula'.
By setting $T\!=\!0$ the Bose-Einstein factors $n_{\bq\nu}$ vanish and we have the so-called
`zero-point renormalization' of the energy bands, $\Delta E_{n\bk}^{\rm ZP}=E_{n\bk}(T\!=\!0)-\ve_{n\bk}$.
This is the modification of the electronic energies evaluated at clamped nuclei,
which arises from the zero-point fluctuations of the atoms around their equilibrium sites.

An expression that is essentially identical to Eq.~(\ref{eq.AH-RS}) can also be obtained directly
from Eq.~(\ref{eq.epi-hamilt}) using second-order Raleigh-Schr\"odinger 
perturbation theory in Fock space, following the same
lines as in (\citeauthor{Kittel1963}, \citeyear{Kittel1963}, p.~134). A detailed derivation
of the formalism starting from Eq.~(\ref{eq.epi-hamilt}) was given by \textcite{Chakraborty1978}.

Historically, the Allen-Heine theory \cite{Allen1976} was developed by starting 
from a straightforward
perturbation theory expansion of the electron energies in terms of the atomic displacements
within the adiabatic approximation, followed by a canonical average of the displacements using
Bose-Einstein statistics. It is
reassuring that, after making a few well-defined approximations,
a field-theoretic method leads to the same result.

Equation~(\ref{eq.AH-RS}) was employed in many semi-empirical calculations.\footnote{
See for example the works of 
\citeauthor{Allen1981} (\citeyear{Allen1981}, \citeyear{Allen1983}),
\citeauthor{Lautenschlager1985} (\citeyear{Lautenschlager1985}),
\citeauthor{Gopalan1987} (\citeyear{Gopalan1987}),
\citeauthor{Zollner1992} (\citeyear{Zollner1992}),
\citeauthor{Olguin2002} (\citeyear{Olguin2002}).
Detailed reviews of early calculations and comparison to experiments can be found 
in \cite{Cardona2001,Cardona2005,Cardona2005b}.} \label{note.allen-heine-early}
More recently, this expression was used in 
the context of first-principles DFT calculations by \textcite{Marini2008} and \textcite{Giustino2010}.
DFT calculations of band structure renormalization based on Eqs.~(\ref{eq.AH-BW}) or (\ref{eq.AH-RS}) 
are becoming increasingly popular, and the latest developments will be reviewed in Sec.~\ref{sec.temper}. 

\begin{figure}[t]
  \includegraphics[width=\columnwidth]{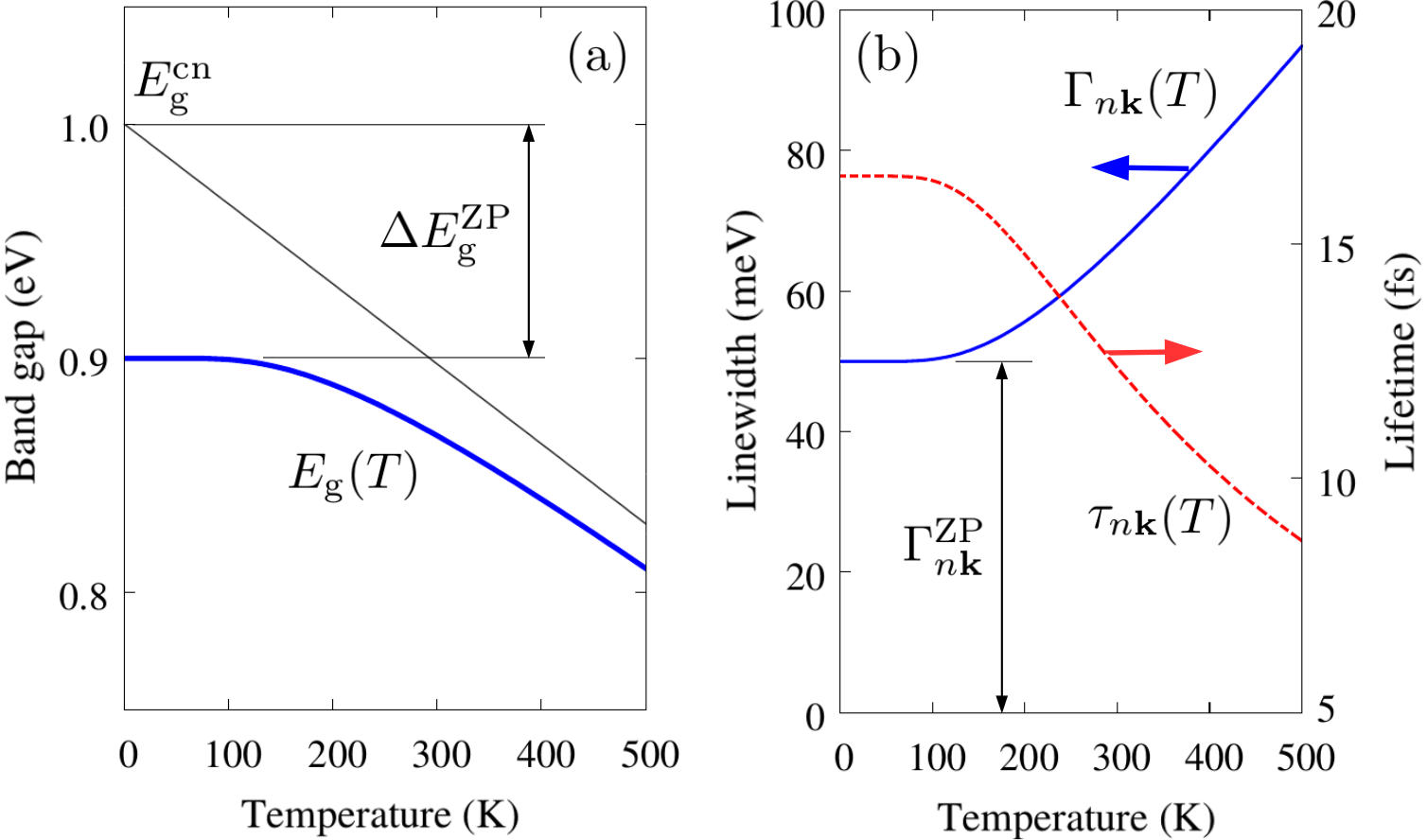}
  \caption{\label{fig.varshni} (Color online)
  Temperature-dependent band gap and lifetimes in an idealized 
  semiconductor or insulator. (a)~Temperature dependence of the band gap according to Eq.~(\ref{eq.varshni})
  (thick solid blue line). The straight thin black line is the asymptotic expansion at high temperature; 
  this line intercepts the vertical axis at the band gap calculated with clamped nuclei, $E_{\rm g}^{\rm cn}$.
  The difference between the latter value and the band gap at $T\!=\!0$ including the EPI gives the
  zero-point renormalization, $\Delta E_{\rm g}^{\rm ZP}$. (b)~Temperature dependence of the electron
  linewidth (solid blue line) and lifetimes (dashed red line) using the same model as in (a). The
  zero-point broadening is $\Gamma_{n\bk}^{\rm ZP}$. This simplified trend is only
  valid when the electron energy is at least one phonon energy away from a band extremum,
  so that both phonon emission and phonon absorption processes are allowed.
  The parameters of the model are: $E_{\rm g}^{\rm cn}=1$~eV, $\Delta E_{\rm g}^{\rm ZP}=100$~meV, 
  $\hbar\w_0=100$~meV, $\Gamma_{n\bk}^{\rm ZP}=50$~meV; these values are representative of common
  semiconductors.
  }
  \end{figure}

The nature of the band structure renormalization by electron-phonon interactions
can be understood at a qualitative
level by considering a drastically simplified model, consisting of a semiconductor
with parabolic and nondegenerate valence and conduction bands, with the band extrema coupled
to all other states by a dispersionless phonon mode of frequency $\w_0$. If the
Debye-Waller matrix elements are much smaller than the Fan-Migdal matrix elements,
then the dominant contributions to Eq.~(\ref{eq.AH-RS}) arise
from denominators such as $\ve_{n\bk}-\ve_{n\bk+\bq} \simeq \pm\hbar^2 |\bq|^2/2 m^*_n$, where
the upper/lower sign is for the valence/conduction band, and $m^*_n$ are the effective masses. 
As a result the temperature
dependence of the band gap will take the form:
  \begin{equation}
  E_{\rm g}(T) = E_{\rm g}^{\rm cn} - \left|\Delta E_{\rm g}^{\rm ZP}\right|
  [1+ 2\,n(\hbar\w_0/ k_{\rm B}T)], \label{eq.varshni}
  \end{equation} 
where $E_{\rm g}^{\rm cn}$ is the gap at clamped nuclei,
and $|\Delta E_{\rm g}^{\rm ZP}|$ is the zero-point correction. The negative sign in the last
expression arises from
the opposite curvatures of the valence and conduction bands. In this example the
band gap decreases with temperature: this is a well known effect in semiconductor physics,
and is often referred to as the `Varshni effect' \cite{Varshni1967}. The
first measurements of such effects were performed by \textcite{Fan1949}, and 
stimulated the development of the first theory of temperature-dependent band gaps \cite{Fan1951}.
A schematic illustration of this qualitative model is provided in Fig.~\ref{fig.varshni}(a).
The redshift of the gap as a function of temperature is seen in many albeit not all semiconductors.
For example, copper halides \cite{Gobel1998} and lead halide perovskites \cite{Dinnocenzo2014} 
exhibit an `inverse Varshni' effect, that is a blueshift of the gap with temperature;
in addition some chalcopyrites exhibit a non-monotonic temperature dependence 
of the band gap \cite{Bhosale2012}.
We also point out that the qualitative model shown in Fig.~\ref{fig.varshni}(a) 
does not take into account the subtle temperature-dependence of the band gap renormalization
at very low temperature. These effects were recently investigated by \textcite{Allen2016}.

\subsubsection{Carrier lifetimes}\label{sec.el-lifetime}

While the real part of the poles in Eq.~(\ref{eq.qp-shift}) describes the energy level renormalization
induced by the electron-phonon coupling, the imaginary part $\Gamma_{n\bk}$ in Eq.~(\ref{eq.qp-width}) 
carries information on the spectral broadening, which will be discussed in Sec.~(\ref{sec.kinks-sat}), 
and on quasiparticle lifetimes, which we discuss below.

After transforming $G_{nn'\bk}(\w)$ from Eq.~(\ref{eq.Gm1-ep-ks})
into the time domain it is seen that,
for an electron or hole added to the system at 
time $t$ in the state $|n\bk \>$,
 the probability amplitude to persist in the same state
 decreases as $\exp[\Gamma_{n\bk}(t'-t)/\hbar]$.
Using Eqs.~(\ref{eq.fan-final}) and (\ref{eq.qp-width}) it can be seen that
$\Gamma_{n\bk}<0$ for an electron added to the
system and $\Gamma_{n\bk}>0$ for a hole.
Therefore the average time spent by the particle in the state $|n\bk\>$ 
is $\tau_{n\bk} = \hbar/(2|\Gamma_{n\bk}|)$.

A popular expression for the electron and hole lifetimes is obtained by making
the replacement $\tilde{E}_{n\bk}\!\rightarrow\! \ve_{n\bk}$ in Eq.~(\ref{eq.fan-final}),
and by taking the absolute value of the imaginary part. We find:
  \begin{eqnarray}
  && \hspace{-0.65cm} \frac{1}{\tau_{n\bk}} = \frac{2\pi}{\hbar}
  \sum_{m\nu} \!\int\!\! \frac{d\bq}{\Omega_{\rm BZ}} |g_{nm\nu}(\bk,\bq)|^2\nonumber \\
  && \times
  \left| (1-f_{m\bk+\bq})\d(\ve_{n\bk} -\hbar\w_{\bq\nu}-\!\ve_{m\bk+\bq}) \right.\nonumber \\
  && \left. \hspace{0.9cm} -f_{m\bk+\bq}\,\,\d(\ve_{n\bk}+\hbar\w_{\bq\nu} -\!\ve_{m\bk+\bq})\right|. 
  \label{eq.fermirule-gamma}
  \end{eqnarray}
A more accurate expression is discussed after Eq.~(\ref{eq.Zfac}) in the next section.
The extension of the above result to finite temperature is obtained by taking the absolute value of the
imaginary part of Eq.~(\ref{eq.fan-final-T}):
  \begin{eqnarray}
  && \hspace{-0.5cm} \frac{1}{\tau_{n\bk}} = \frac{2\pi}{\hbar} 
  \sum_{m\nu} \!\int\!\! \frac{d\bq}{\Omega_{\rm BZ}} |g_{nm\nu}(\bk,\bq)|^2\nonumber \\
  && \times
  \left[ (1-f_{m\bk+\bq}+n_{\bq\nu})\d(\ve_{n\bk} -\hbar\w_{\bq\nu}-\!\ve_{m\bk+\bq}) \right.+\nonumber \\
  && \left. \hspace{1cm} (f_{m\bk+\bq}+n_{\bq\nu})\d(\ve_{n\bk}+\hbar\w_{\bq\nu} -\!\ve_{m\bk+\bq})\right].
  \label{eq.fermirule}
  \end{eqnarray}
We emphasize the change of sign in the third line, resulting from the analytic continuation to the
retarded self-energy. Equation~(\ref{eq.fermirule}) coincides with the expression that
one would obtain by using the standard Fermi golden rule \cite{Grimvall1981}. 
The intuitive interpretation of this result is that the quasiparticle lifetime is reduced
by processes of phonon emission and absorption, corresponding to the second and third lines 
of Eq.~(\ref{eq.fermirule}), respectively. We note that in deriving 
Eq.~(\ref{eq.fermirule}) we did not consider the Debye-Waller self-energy; 
this is because the diagonal matrix elements of $\Sigma^{\rm DW}$ are purely real, 
hence they do not contribute to the quasiparticle widths \cite{Lautenschlager1986}.
{\it Ab initio} calculations of carrier lifetimes using
Eq.~(\ref{eq.fermirule}) were first reported by \textcite{Eiguren2002,Eiguren2003}.
These applications and more recent developments will be reviewed in Sec.~\ref{sec.transport-lifet}.

If we evaluate Eq.~(\ref{eq.fermirule}) for the same simplified model introduced for the temperature renormalization,
and we neglect the phonon energy in the Dirac delta functions, we obtain $\Gamma_{n\bk}(T) = 
\Gamma_{n\bk}^{\rm ZP} \, [1+ 2\,n(\hbar\w_0/ k_{\rm B}T)]$ where $\Gamma_{n\bk}^{\rm ZP}$
is the linewidth at $T=0$. The dependence of the linewidth and the corresponding lifetime
on temperature for this model are shown in Fig.~\ref{fig.varshni}(b). This trend is typical
in semiconductors \cite{Lautenschlager1987,Lautenschlager1987b}.

\subsubsection{Kinks and satellites}\label{sec.kinks-sat}

In many cases of interest, the use of Brillouin-Wigner perturbation theory as given by 
Eqs.~(\ref{eq.qp-shift})-(\ref{eq.qp-width}) is not sufficient to provide an adequate
description of EPIs, and it becomes necessary to go back to the complete Dyson equation,
Eq.~(\ref{eq.Gm1-ep-ks}). Generally speaking a direct solution of the Dyson equation is important
in all those cases where the electronic energy scales are comparable to phonon energies,
namely in metals (including superconductors), narrow-gap semiconductors, and doped semiconductors.
In order to study these systems, it is convenient to introduce an auxiliary function called
the `spectral density function', or simply spectral function.

In its simplest version the spectral function is defined as \cite{Abrikosov1975,Mahan1993}:
  \begin{equation}
  A(\bk,\w) = -\frac{1}{\pi}\, {\sum}_n{\rm Im}\, G^{\,\rm ret}_{nn\bk}(\w),
  \end{equation}
where the superscript `ret' stands for `retarded', and simply indicates that all poles
of the Green's function $G_{nn'\bk}(\w)$ in the upper complex plane must be replaced by their complex conjugate.
The spectral function is positive definite and carries the meaning of a `many-body momentum-resolved 
density of states' \cite{Abrikosov1975}.
This is precisely the function that is probed by angle-resolved photoelectron spectroscopy
experiments or ARPES \cite{Damascelli2003}.
Using Eq.~(\ref{eq.Gm1-ep-ks}) the spectral function can be rewritten as:
  \begin{equation}\label{eq.specfun2}
  A(\bk,\w) = \sum_n \frac{-(1/\pi)\,{\rm Im }\,\Sigma^{\rm ep}_{nn\bk}(\w)}{\left[\hbar\w\!-\!\ve_{n\bk}\!-\!
   {\rm Re}\,\Sigma^{\rm ep}_{nn\bk}(\w)\right]^2\!+\!\left[{\rm Im }\,\Sigma^{\rm ep}_{nn\bk}(\w)\right]^2}.
  \end{equation}
In order to obtain the correct spectral function, it is important to use the
{\it retarded} self-energy. This is done by using Eq.~(\ref{eq.fan-final-T})
for the Fan-Migdal term, while the static Debye-Waller term remains unchanged.

It is often convenient to approximate the spectral function as a sum of quasiparticle peaks.
To this aim, one performs a linear expansion of Eq.~(\ref{eq.specfun2}) around each quasiparticle 
energy $E_{n\bk}$, to obtain:
  \begin{equation}\label{eq.specfun-laurent}
  A(\bk,\w) = \!\sum_n Z_{n\bk} \frac{-(1/\pi)\,Z_{n\bk} \,{\rm Im }\,\Sigma^{\rm ep}_{nn\bk}(E_{n\bk}/\hbar)}
     {\left[\hbar\w\!-\!E_{n\bk}\right]^2\!+\!\left[Z_{n\bk} \,{\rm Im }\,\Sigma^{\rm ep}_{nn\bk}(E_{n\bk}/\hbar)\right]^2}.
  \end{equation}
This is a sum of Lorentzians with strength $Z_{n\bk}$ and width $Z_{n\bk}\, {\rm Im }\,\Sigma^{\rm ep}_{nn\bk}
(E_{n\bk}/\hbar)$. Here the `quasiparticle strength' is defined as the homonymous quantity appearing 
in $GW$ calculations \cite{Hedin1969}:
  \begin{equation}\label{eq.Zfac}
  Z_{n\bk}= 
  \left[1 - \hbar^{-1}\D{\rm Re}\Sigma^{\rm ep}_{nn\bk}(\w)/\D\w\big|_{\w=E_{n\bk}/\hbar}\right]^{-1}\!\!\!\!.
  \end{equation}
The result expressed by Eq.~(\ref{eq.specfun-laurent}) shows that, in a rigorous field-theoretic approach,
the quasiparticle broadening and lifetime given by Eqs.~(\ref{eq.fermirule-gamma}) 
and (\ref{eq.fermirule}) should be renormalized by $Z_{n\bk}$ and $Z_{n\bk}^{-1}$, respectively,
and should be evaluated using the quasiparticle energy $E_{n\bk}$ instead of $\ve_{n\bk}$.
This result can also be derived from Eq.~(\ref{eq.qp-width}) by performing
a Taylor expansion of the self-energy along the imaginary axis and using the Cauchy-Riemann conditions.

\begin{figure}[t]
  \includegraphics[width=\columnwidth]{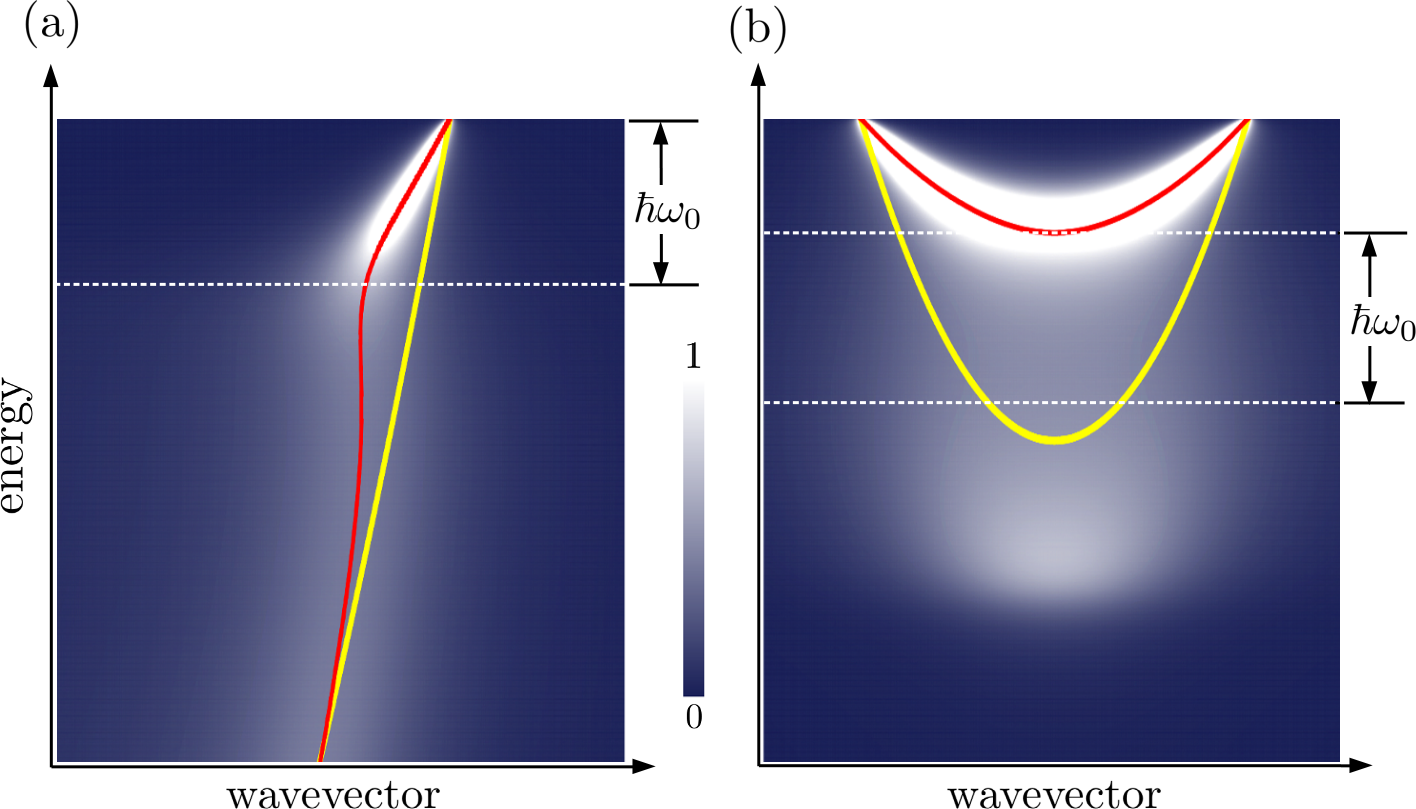}
  \caption{\label{fig.kink-sat} (Color online)
  Two-dimensional maps of the electron spectral function $A(\bk,\w)$ for electrons
  coupled to a dispersionless phonon of frequency $\w_0$. The non-interacting bands are given by
  $\ve(\bk) = -\ef + \hbar^2 |\bk|^2/2 m^*$,
  and the Fermi level coincides with the top of the energy window.
  The matrix element is $|g|^2 = \hbar\w_0 / N_{\rm F}$
  when the electron energies differ by less than the cutoff $\ve_{\rm max}$, and zero otherwise
  ($N_{\rm F}$ is the density of states at the Fermi level).
  (a) Spectral function for the case $\ef=10\,\hbar\w_0$ (white on blue/black), 
  non-interacting band structure (solid line, yellow/light gray),
  and fully-interacting band structure within Brillouin-Wigner perturbation theory (solid line,
  red/dark gray).
  (b) Spectral function for the case $\ef=2\, \hbar\w_0$. 
  The model parameters are: $m^*=0.1\, m_{\rm e}$, 
  $\hbar\w_0=100$~meV, $\eta = 20$~meV, $\ve_c=5$~eV,
  For clarity the calculated spectral functions are cut off at the value
  3~eV$^{-1}$ and normalized.
  The self-energy is shifted by a constant so as to have $\Sigma^{\rm ep}(0)=0$;
  this correction guarantees the fulfillment of Luttinger's theorem about the volume enclosed
  by the Fermi surface \cite{Luttinger1960}.
  }
  \end{figure}

In order to illustrate the typical features of the spectral function, we consider
a model system characterized by one parabolic conduction band. The
occupied electronic states couple to all states within an energy cutoff via a 
dispersionless phonon mode and a constant electron-phonon matrix element.
A simplified version of this model was discussed by \textcite{Engelsberg1963}
by considering a constant density of electronic states. By evaluating the spectral
function in Eq.~(\ref{eq.specfun2}) using the Fan-Migdal self-energy and neglecting
the Debye-Waller term, we obtain the results shown in Fig.~\ref{fig.kink-sat}
for two sets of parameters.

In Fig.~\ref{fig.kink-sat}(a) the Fermi energy is much larger than the characteristic phonon energy.
This case is representative of a metallic system with electron bands nearly linear 
around the Fermi level. Here the electron-phonon interaction leads to (i) a reduction of the band 
velocity in proximity of the Fermi level, and (ii) a broadening of the spectral function beyond the 
phonon energy~$\hbar\w_0$. 
A detailed analysis of these features for a slightly simpler model system, including 
a discussion of the analytic properties of the Green's function, can be found 
in the work by \citeauthor{Engelsberg1963}. 

The solid line (red/dark gray) in Fig.~\ref{fig.kink-sat}(a) shows the renormalized band structure obtained 
from Brillouin-Wigner perturbation theory, Eq.~(\ref{eq.qp-shift}). We see that these 
solutions track the maxima of the spectral function $A(\bk,\w)$. 
The renormalized bands exhibit a characteristic `S-shape' near the Fermi level, 
corresponding to multiple solutions 
of Eq.~(\ref{eq.qp-shift}) for the same wavevector $\bk$. Starting from the late 1990s 
such S-shaped energy-momentum dispersion curves have been observed in a number of ARPES experiments,
and have become known in the literature as the `photoemission kink' \cite{Valla1999}.
First-principles calculations of kinks were first reported by \textcite{Eiguren2003,Giustino2008}, 
and will be reviewed in Sec.~\ref{sec.kinks}.

In Fig.~\ref{fig.kink-sat}(b) the Fermi energy is comparable to the characteristic phonon energy.
This case is representative of a degenerately doped semiconductor close to a conduction
band minimum. Here the electron-phonon interaction leads to two distinct spectral features: (i) a parabolic band 
with a heavier mass, which is well described by the Brillouin-Wigner solutions (solid line, red/dark gray),
and (ii) a polaron satellite that is visible further down. In this example, it is clear that
Eq.~(\ref{eq.qp-shift}) is unable to describe the satellite, and that the spectral function
carries qualitatively new information about the system. Polaron satellites resembling 
Fig.~\ref{fig.kink-sat}(b) have been observed in ARPES experiments on doped oxides 
(\citeauthor{Moser2013}, \citeyear{Moser2013}; 
\citeauthor{Chen2015}, \citeyear{Chen2015};
\citeauthor{Cancellieri2016}, \citeyear{Cancellieri2016};
\citeauthor{Wang2016}, \citeyear{Wang2016})
and recently calculated from first principles \cite{Verdi2016}.

\subsubsection{Model Hamiltonians, polarons, and the cumulant expansion}\label{sec.polarons}

At the end of this section it is worth mentioning complementary {\it non first-principles}
approaches for studying the effects of EPIs on the electronic properties of solids. 
Model EPI Hamiltonians can be derived from Eq.~(\ref{eq.epi-hamilt}) by choosing {\it a priori} 
explicit expressions for the electron band energies, the vibrational frequencies,
and the coupling matrix elements. Examples of model Hamiltonians are those
of \textcite{Froehlich1954}, \textcite{Holstein1959a}, Su, Schrieffer, and Heeger (\citeyear{Su1979}),
the Hubbard-Holstein model \cite{Berger1995}, the Peierls-Hubbard model \cite{Campbell1984},
the `$t$-$J$' Holstein model \cite{Rosch2004}, and the Su-Schrieffer-Heeger-Holstein model
\cite{Perroni2004}.
These models involve the tight-binding approximation, the Einstein
phonon spectrum, and electron-phonon couplings to first order in the atomic displacements.
Using these model Hamiltonians it is possible to go beyond
the approximations introduced in Sec.~\ref{eq.explic-elec}, and
obtain non-perturbative solutions by means of canonical Lang-Firsov transformations, path-integral methods,
exact diagonalization, variational or quantum Monte Carlo techniques \cite{Alexandrov2008,
Alexandrov2010}. These models have been used extensively to explore many aspects of polaron physics,
for example the ground-state energy of polarons (weak or strong coupling), their spatial extent
(large or small polarons), and transport properties (band-like or hopping-like).

Given the considerable body of literature on model EPI Hamiltonians, it is natural to ask
whether one could bring {\it ab~initio} calculations of EPIs to a similar level of sophistication.
The main limitation of current first-principles approaches is that, given the complexity of the
calculations, the electron self-energies are evaluated using the bare propagators, as 
in Eq.~(\ref{eq.fan-final}). As a consequence, higher-order interaction diagrams beyond the
Migdal approximation \cite{Migdal1958} are omitted altogether.

A promising avenue for going beyond the Migdal approximation consists of introducing 
higher-order diagrams via the `cumulant expansion' approach 
(\citeauthor{Hedin1969}, \citeyear{Hedin1969};
\citeauthor{Langreth1970}, \citeyear{Langreth1970};
\citeauthor{Aryasetiawan1996}, \citeyear{Aryasetiawan1996}).
In the cumulant expansion method, instead of calculating the electron Green's function
via a Dyson equation, one evaluates the time evolution of the Green's 
function by formulating the problem in the interaction picture, in symbols: 
$G_{nn\bk}(t) = (i/\hbar) \exp[-i(\ve_{n\bk}/\hbar)t+C_{nn\bk}(t)]$ 
\cite{Aryasetiawan1996}. The distinctive advantage of this approach is that
the `cumulant' $C_{nn\bk}(t)$ can be obtained from a low-order self-energy, for example 
the Fan-Migdal self-energy in Eq.~(\ref{eq.fan-final}), and the  exponential `resummation'
automatically generates higher-order diagrams (\citeauthor{Mahan1993}, \citeyear{Mahan1993}, pag.~523).
Detailed discussions of the cumulant expansion formalism
are given by \textcite{Sky2015} and \textcite{Gumhalter2016}.

The cumulant method provides an interesting point of contact between {\it ab~initio} and 
model Hamiltonian approaches. In fact, the cumulant expansion is closely related
to the `momentum average approximation' introduced by \textcite{Berciu2006} for
studying the Green's function of the Holstein polaron.

The cumulant expansion has proven successful in {\it ab~initio}
calculations of electron-electron interactions, in particular 
valence band satellites in semiconductors
(\citeauthor{Kheifets2003}, \citeyear{Kheifets2003};
\citeauthor{Guzzo2011}, \citeyear{Guzzo2011}, \citeyear{Guzzo2012}, \citeyear{Guzzo2014};
\citeauthor{Lischner2013}, \citeyear{Lischner2013};
\citeauthor{Kas2014}, \citeyear{Kas2014};
\citeauthor{Caruso2015}, \citeyear{Caruso2015}, \citeyear{Caruso2015b}b).
In the context of EPIs, the {\it ab~initio} cumulant expansion method has been applied to
elemental metals by \textcite{Story2014}, and to the ARPES spectra of $n$-doped TiO$_2$ 
by \textcite{Verdi2016}. In the latter work the cumulant method correctly reproduced 
the polaron satellites observed in the experiments of \textcite{Moser2013}.

The study of polarons using {\it ab~initio} many-body techniques is yet to begin, however a first
calculation of the spectral function of Fr\"ohlich polarons and an approximate polaron 
wavefunction have recently been reported by \textcite{Verdi2016}.

\section{Efficient calculations of matrix elements and their integrals}\label{sec.wannier}

The study of EPIs from first principles requires
evaluating Brillouin-zone integrals of functions that exhibit strong fluctuations.
This requirement can be appreciated by inspecting Eqs.~(\ref{eq.phon-self-dft})
and (\ref{eq.fan-final}): there the denominators become large whenever the difference
between two electronic eigenvalues approaches a phonon energy. 
As a result, while in DFT total energy calculations the Brillouin zone is typically
discretized using meshes of the order of 10$\times$10$\times$10 points,
the numerical convergence of EPI calculations requires much finer grids, 
sometimes with as many as $10^6$ wavevectors \cite{Giustino2007b,Ponce2015}. 
Determining vibrational frequencies $\w_{\bq\nu}$ and perturbations $\Delta_{\bq\nu} v^{\rm KS}(\br)$
for such a large number of wavevectors is a prohibitive task, since every calculation
is roughly as expensive as one total energy minimization.

These difficulties stimulated the development of specialized numerical techniques for
making calculations of EPIs affordable. In the following sections two such techniques 
are reviewed: electron-phonon Wannier interpolation and Fermi-surface harmonics.

\subsection{Wannier interpolation}

\subsubsection{Maximally-localized Wannier functions}

In addition to the standard description of electrons in solids in terms of Bloch waves, as in Eq.~(\ref{eq.utilde}),
it is possible to adopt an alternative point of view whereby electrons are described as linear combinations 
of localized orbitals called `Wannier functions' \cite{Wannier1937}. 
The most general relation between Wannier functions and Bloch waves can be written 
as follows.
One considers electron bands $\ve_{n\bk}$ with eigenfunctions $\psi_{n\bk}$, where the index $n$ is restricted to
a set of bands that are separated from all other bands by finite energy gaps above and below. 
These bands are referred to as `composite energy bands' \cite{Marzari1997}. 
Wannier functions are defined as:
  \begin{equation}\label{eq.wannier-def}
  \wf_{m p} (\br) = N_p^{-1} {\sum}_{n\bk} e^{i\bk\cdot (\br-\bR_p)}\, U_{nm\bk}\, u_{n\bk}(\br),
  \end{equation}
where $U_{nm\bk}$ is a unitary matrix in the indices $m$ and $n$.
From this definition and Eq.~(\ref{eq.psum}) it follows that Wannier functions are normalized
in the supercell, $\< \wf_{m p} | \wf_{m' p'} \>_{\rm sc} = \d_{mp, m'p'}$.
Furthermore, since $u_{n\bk}$ is lattice-periodic, Wannier functions have the property
$\wf_{m p} (\br) = \wf_{m 0} (\br-\bR_p)$.
The inverse transformation of Eq.~(\ref{eq.wannier-def}) is obtained
by using the unitary character of $U_{nm\bk}$ together with Eq.~(\ref{eq.psum}):
  \begin{equation}\label{eq.wannier-def-inv}
  u_{n\bk}(\br) = {\sum}_{mp} e^{-i\bk\cdot (\br- \bR_{p})} U^\dagger_{mn\bk} \wf_{m p} (\br).
  \end{equation}
The unitary matrix $U_{nm\bk}$ is completely arbitrary,
therefore there exists considerable freedom in the construction of Wannier functions. For example
by requiring that $U_{nm,-\bk} = U_{nm\bk}^*$ one can make Wannier functions real-valued.
\citeauthor{Marzari1997} exploited this degree of freedom 
to construct Wannier functions that are {\it maximally localized}.

A comprehensive and up-to-date review of the theory and applications of maximally-localized
Wannier functions (MLWFs) is given by 
\textcite{Marzari2012}. Here we only recall that, in order to minimize the spatial extent 
of a function in a periodic solid, one needs to use a modified definition of the position operator,
since the standard position operator is unbounded in an infinite crystal. This procedure is now 
well-established and it is linked to the development of the modern theory of dielectric 
polarization \cite{King-Smith1993,Resta1994}. Nowadays it is possible to determine MLWFs routinely 
\cite{Mostofi2008}.
The original algorithm of \textcite{Marzari1997} was also extended to deal with 
situations where a composite set of bands cannot be identified. This happens notably in metals
for electronic states near the Fermi energy. For these cases, \textcite{Souza2001} 
developed a band `disentanglement' procedure, which extracts a subset of composite bands 
out of a larger set of states.

For the purposes of the present article, the most important property of MLWFs is that they
are exponentially localized in insulators, in the sense that $|\wf_{m0}(\br)| \sim |\br|^{-\a}
\exp(-h|\br|)$ for large $|\br|$, with $\a, h>0$ real parameters. This property was demonstrated
in one spatial dimension by \textcite{Kohn1959} and \textcite{He2001}, and in two and three
dimensions by \textcite{Brouder2007}, under the condition that the system exhibits time-reversal symmetry.
In the case of metallic systems, no exponential localization is expected. However, the 
Wannier functions obtained in metals using the disentanglement procedure of \textcite{Souza2001}
are typically highly localized. 

MLWFs are usually comparable in size to atomic orbitals, and this makes them ideally suited 
for Slater-Koster interpolation of band structures, as shown by \textcite{Souza2001}.
This concept was successfully employed in a number of applications requiring accurate
calculations of band velocities, effective masses, density of states, Brillouin-zone integrals,
and transport coefficients 
(\citeauthor{Wang2006}, \citeyear{Wang2006};
\citeauthor{Yates2007}, \citeyear{Yates2007};
\citeauthor{Wang2007}, \citeyear{Wang2007};
\citeauthor{Pizzi2014}, \citeyear{Pizzi2014}).

\subsubsection{Interpolation of electron-phonon matrix elements}\label{sec.wannier-nonpolar}

Wannier functions were introduced in the study of EPIs by \textcite{Giustino2007,Giustino2007b}. The starting
point is the definition of the electron-phonon matrix element in the Wannier representation:\footnote{
We note that $g_{mn\k\a}(\bR_p,\bR_{p'})$ 
has dimensions of energy by length, at variance with Eq.~(\ref{eq.matel}).
For consistency here we use a definition that differs from that given in \cite{Giustino2007b}
by a factor $N_p$; this factor is inconsequential.
}
  \begin{equation}\label{eq.g-wann-def}
  g_{mn\k\a}(\bR_p,\bR_{p'})\! =\!
  \< \wf_{m0} (\br) | \frac{\D V^{\rm KS}\!} {\D \tau_{\k\a}} (\br-\bR_{p'})|
  \wf_{n 0}(\br-\bR_p)\>_{\rm sc},
  \end{equation}
where the subscript `sc' indicates that the integral is over the BvK supercell.
The relation between these quantities 
and the standard EPI matrix elements $g_{mn\nu}(\bk,\bq)$ is found by replacing
Eq.~(\ref{eq.wannier-def-inv}) inside
Eq.~(\ref{eq.matel}), and using Eqs.~(\ref{eq.deltavq})-(\ref{eq.dV-all-3})
\cite{Giustino2007b}:
  \begin{eqnarray}
  && g_{mn\nu}(\bk,\bq) = {\sum}_{pp'}
  e^{i(\bk\cdot \bR_p+\bq\cdot \bR_{p'})}  \nonumber \\
    && \hspace{0.2cm} \times\!\!\!\!\sum_{\,\,\,m'n'\k\a}\!\!\!\!
  U_{m m'\bk+\bq}\, g_{m'n'\k\a}(\bR_p,\bR_{p'})\, U^\dagger_{n'n\bk} {\rm u}_{\k\a,\bq\nu},
  \label{eq.wan2blo}
   \hspace{0.6cm}
  \end{eqnarray}
where we defined
$  {\rm u}_{\k\a,\bq\nu} = (\hbar / 2 M_\k \w_{\bq\nu})^\frac{1}{2}\, e_{\kappa\a,\nu}(\bq)$
and $e_{\kappa\a,\nu}(\bq)$ are the vibrational eigenmodes of Eq.~(\ref{eq.dynmat2}).
The inverse relation is:
  \begin{eqnarray}
  &&g_{mn\k\a}(\bR_p,\bR_{p'}) = \frac{1}{N_p N_{p'}} \sum_{\bk,\bq} 
  e^{-i(\bk\cdot \bR_p+\bq\cdot\bR_{p'})} \nonumber \\
  && \hspace{0.4cm}\times \sum_{m'n'\nu} {\rm u}_{\k\a,\bq\nu}^{-1} \, 
   U^\dagger_{m m'\bk+\bq} \, g_{m'n'\nu}(\bk,\bq) \, U_{n'n\bk},  \hspace{0.6cm}
  \label{eq.blo2wan}
  \end{eqnarray}
with ${\rm u}_{\k\a,\bq\nu}^{-1} = (\hbar/2M_\k\w_{\bq\nu})^{-\frac{1}{2}}\, e^*_{\kappa\a,\nu}(\bq)$.
The last two equations define a generalized Fourier transform of the electron-phonon matrix elements
between reciprocal space and real space. In Eq.~(\ref{eq.blo2wan}) we have $N_p$ and $N_{p'}$
to indicate that the BvK supercells for electronic band structures and phonon dispersions
may not coincide.

\begin{figure}[t!]
  \includegraphics[width=0.75\columnwidth]{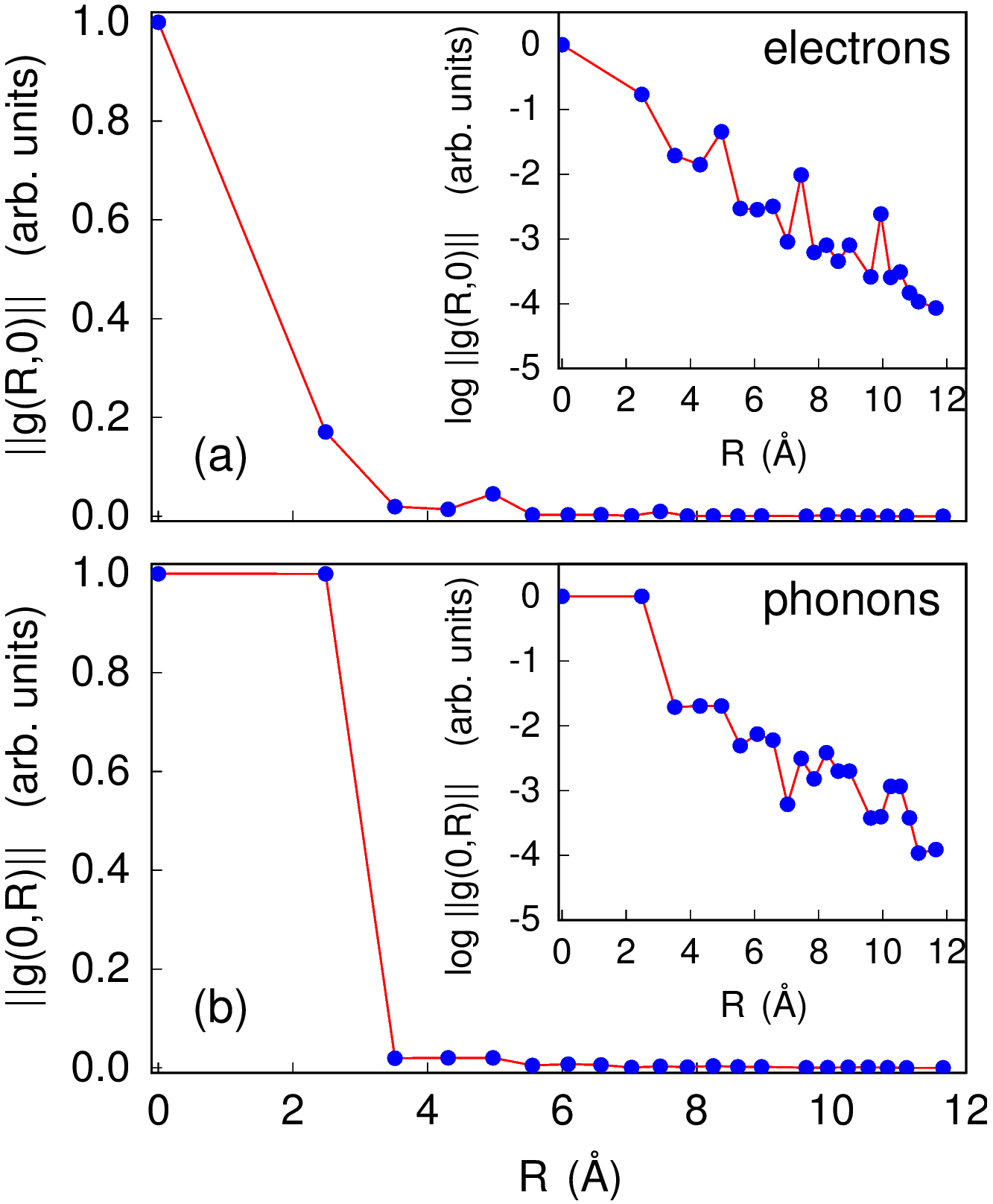}
  \caption[fig]{\label{fig.wannier} (Color online)
  Spatial decay of the electron-phonon matrix elements of diamond in the
  Wannier representation: (a) $\max |g_{mn\k\a}(\bR_p,0)|$ vs. $|\bR_p|$,
  and (b) $\max |g_{mn\k\a}(0,\bR_{p'})|$ vs. $|\bR_{p'}|$. The maximum values
  are taken over all subscript indices, and the data are
  normalized to the largest value. The insets show the same quantities
  in logarithmic scale. The calculations were performed using the local-density approximation
  to DFT. 
  Reproduced with permission from \cite{Giustino2007b}, copyright
  (2007) by the American Physical Society.
  }
  \end{figure}

If the quantity $g_{mn\k\a}(\bR_p,\bR_{p'})$
decays rapidly as a function of $|\bR_p|$ and $|\bR_{p'}|$, then only a small number of
matrix elements in the Wannier representation will be sufficient to generate
$g_{mn\nu}(\bk,\bq)$ anywhere in the Brillouin zone by means of Eq.~(\ref{eq.wan2blo}).
The dependence of the matrix elements on $\bR_p$ and $\bR_{p'}$ can be analyzed by considering
the following bound:
$|g_{mn\k\a}(\bR_p,\bR_{p'})| \le \int_{\rm sc} d\br \,|\wf_{m0}^* (\br)\wf_{n 0}(\br-\bR_p)|
\times \int_{\rm sc} d\br \,|\D V^{\rm KS}/\D \tau_{\k\a} (\br-\bR_{p'})|$.
The first term guarantees that the matrix element decays in the variable $\bR_p$
at least as fast as MLWFs. As a result the worst case scenario corresponds to the choice $\bR_p=0$.
In this case, the matrix element $|g_{mn\k\a}(0,\bR_{p'})|$ decays with the variable $\bR_{p'}$ 
at the same rate as the {\it screened} electric dipole potential generated by the atomic displacement 
$\Delta \tau_{\k\a}$. In non-polar semiconductors and insulators,
owing to the analytical properties of the dielectric matrix \cite{Pick1970}, this
potential decays at least as fast as a quadrupole, that is $|\bR_{p'}|^{-3}$.
As a result, all matrix elements in reciprocal space are finite
for $\bq\rightarrow 0$ \cite{Vogl1976} and hence amenable to interpolation. 
In the case of metals the asymptotic trend of $\D V^{\rm KS}/\D \tau_{\k\a}$ 
is dictated by Fermi-surface nesting, 
leading to Friedel oscillations that decay as $|\bR_{p'}|^{-4}$ 
(\citeauthor{Fetter2003}, \citeyear{Fetter2003}, pp.~175--180). These oscillations are connected
to the Kohn anomalies in the phonon dispersion relations \cite{Kohn1959b}.
In practical calculations, Friedel oscillations are usually not an issue since they are suppressed
by the numerical smearing of the Fermi-Dirac occupations, and a Yukawa-type exponential
decay is recovered. 
The case of polar materials is more subtle and will be discussed in Sec.~\ref{sec.polar}.
Figure~\ref{fig.wannier} illustrates the spatial decay of $|g_{mn\k\a}(\bR_p,\bR_{p'})|$
as a function of $\bR_p$ and $\bR_{p'}$ for the prototypical case of diamond.

\begin{figure}[t!]
  \includegraphics[width=0.65\columnwidth]{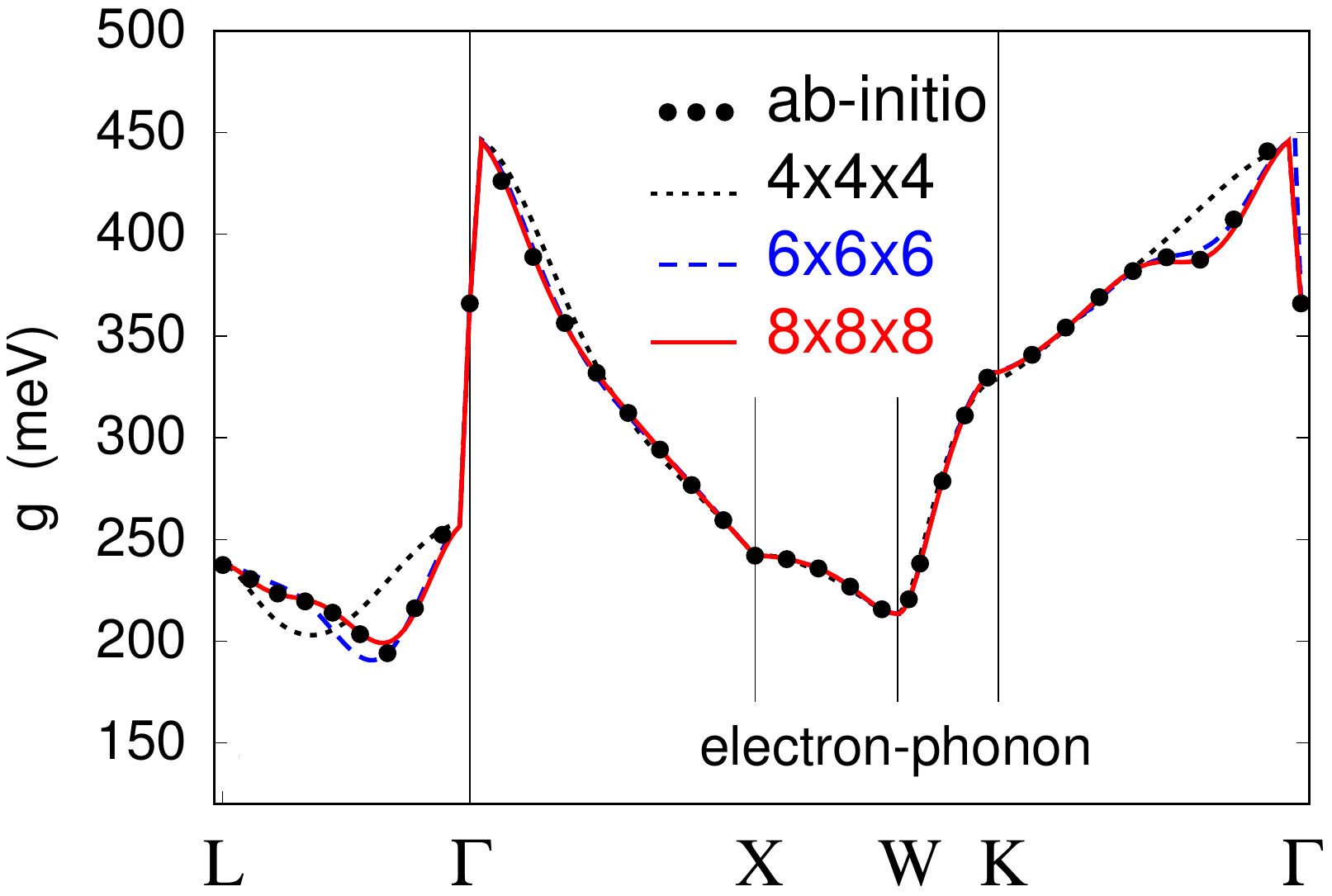}
  \caption[fig]{\label{fig.diam-interp} (Color online)
  Comparison between Wannier-interpolated electron-phonon matrix elements and
  explicit DFPT calculations, for diamond.
  The interpolated matrix elements were calculated starting from a coarse $4^3$ Brillouin-zone 
  grid (dotted line, black), a $6^3$ grid (dashed line, blue), and a $8^3$ grid (solid line, red).
  The dots indicate explicit DFPT calculations. In this example $|n{\bf k}\>$ is set to the
  valence band top at $\Gamma$; $|m{\bk}+{\bf q}\>$ spans $\Lambda_3$, $\Delta_5$, and $\Sigma_2$ bands,
  and the phonon is set to the highest optical branch. 
  Reproduced with permission from \cite{Giustino2007b}, copyright
  (2007) by the American Physical Society.
  }
  \end{figure}

The interpolation strategy is entirely analogous to standard techniques for generating
phonon dispersion relations using the interatomic force constants \cite{Gonze1997b}: one first
determines matrix elements in the Bloch representation using DFPT on a corse grid in the 
Brillouin zone, as in Sec.~\ref{sec.matel-dft}. Then MLWFs are determined using the procedures
of \textcite{Marzari1997,Souza2001}. This yields the rotation matrices $U_{mn\bk}$ to be used
in Eq.~(\ref{eq.blo2wan}). The Fourier transform to real space is performed via Eq.~(\ref{eq.blo2wan}).
At this point, one assumes that matrix elements outside of the Wigner-Seitz supercell
defined by the coarse Brillouin-zone grid can be neglected, and uses Eq.~(\ref{eq.wan2blo})
in order to obtain the matrix elements $g_{mn\nu}(\bk,\bq)$ on very fine grids.
The last step requires the knowledge of the rotation matrices $U_{mn\bk}$ also on the
fine grids; these matrices are obtained from the Wannier interpolation of the band structures,
as described by \textcite{Souza2001}. The operation is computationally inexpensive and
enables the calculation of millions of electron-phonon matrix elements. The procedure
can now be applied routinely \cite{Noffsinger2010,Ponce2016}. Figure~\ref{fig.diam-interp} shows the matrix 
elements obtained using this method, as compared to explicit DFPT calculations.

Wannier interpolation of electron-phonon matrix elements was successfully employed
in a number of applications, ranging from metal and superconductors to semiconductors
and nanoscale systems.\footnote{
See for example 
(\citeauthor{Park2007}, \citeyear{Park2007};
\citeauthor{Giustino2008}, \citeyear{Giustino2008};
\citeauthor{Park2008}, \citeyear{Park2008};
\citeauthor{Noffsinger2009}, \citeyear{Noffsinger2009}, \citeyear{Noffsinger2010};
\citeauthor{Calandra2010}, \citeyear{Calandra2010};
\citeauthor{Giustino2010}, \citeyear{Giustino2010};
\citeauthor{Vukmirovic2012}, \citeyear{Vukmirovic2012};
\citeauthor{Noffsinger2012}, \citeyear{Noffsinger2012};
\citeauthor{Margine2013}, \citeyear{Margine2013}, \citeyear{Margine2014};
\citeauthor{Park2014}, \citeyear{Park2014};
\citeauthor{Bernardi2014}, \citeyear{Bernardi2014};
\citeauthor{Verdi2015}, \citeyear{Verdi2015};
\citeauthor{Sjakste2015}, \citeyear{Sjakste2015};
\citeauthor{Bernardi2015}, \citeyear{Bernardi2015}).
}

\subsubsection{Electron-phonon matrix elements in polar materials}\label{sec.polar}

In the case of polar materials, that is systems exhibiting nonzero Born effective charges \cite{Pick1970}, 
the interpolation scheme discussed in Sec.~\ref{sec.wannier-nonpolar} breaks down. 
In fact, in these systems the dominant contribution to the potential $\D V^{\rm KS}/\D \tau_{\k\a}$ 
in Eq.~(\ref{eq.g-wann-def}) is a dipole, which decays as $|\bR_{p'}|^{-2}$. As a consequence some of
the matrix elements in reciprocal space diverge as $|\bq|^{-1}$ for $\bq\rightarrow 0$, and 
cannot be interpolated straightforwardly from a coarse grid to a fine grid.
Physically this singularity corresponds to the `Fr\"ohlich
electron-phonon coupling' \cite{Froehlich1954}.

The adaptation of the Wannier interpolation method to the case of polar materials was recently
given by \textcite{Sjakste2015} and by \textcite{Verdi2015}. In both works the basic idea is to
separate the matrix elements into a short-range contribution, $g_{mn\nu}^{\mathcal S}({\bf k},{\bf q})$, 
which is amenable to standard Wannier interpolation, and a long-range contribution, 
$g_{mn\nu}^{\mathcal L}({\bf k},{\bf q})$, which is singular and is dealt with analytically.
The strategy is analogous to that in use for calculating LO-TO splittings in polar materials
\cite{Gonze1997b}.
The starting point is to define the long-range component of the matrix elements by considering
the potential generated by the Born charges of all the atoms, when displaced according to
a given vibrational eigenmode. The derivation relies on standard electrostatics and
can be found in \cite{Verdi2015}:
  \begin{eqnarray} \label{gL}
  && g_{mn\nu}^{\mathcal L}({\bf k},{\bf q}) =
  i\frac{4\pi}{\Omega} \frac{e^2 }{4\pi\varepsilon_0}
  \sum_{\kappa}
  \left(\frac{\hbar}{2 N_p M_\kappa \omega_{{\bf q}\nu}}\right)^{\!\!\frac{1}{2}} 
   \sum_{{\bf G}\ne -{\bf q}} \nonumber \\
   &&\,\,\,\,\frac{ ({\bf q}+{\bf G})\cdot{\bf Z}^*_\kappa 
  \cdot {\bf e}_{\kappa\nu}({\bf q}) }
  {\,\,\,\,({\bf q}+{\bf G})\cdot\bm\epsilon^\infty\!\cdot({\bf q}+{\bf G})}
  \langle \psi_{m{\bf k+q}} |e^{i({\bf q}+{\bf G})\cdot({\bf r}-\bm\tau_{\kappa})}| \psi_{n{\bf k}} 
  \rangle_{\rm sc}. \,\,\,\nonumber \\
  \end{eqnarray}
In this expression ${\bf Z}^*_\kappa$ and $\bm\epsilon^\infty$ denote the Born effective charge tensors
and the electronic permittivity tensor (that is, the permittivity evaluated at clamped nuclei). This expression
is the generalization of Fr\"ohlich's model to the case of anisotropic crystalline
lattices and multiple phonon modes \cite{Froehlich1954}. The result can be derived alternatively using the analytical
properties of the dielectric matrix \cite{Pick1970} as discussed by \textcite{Vogl1976}.

\begin{figure}[t!]
  \includegraphics[width=0.65\columnwidth]{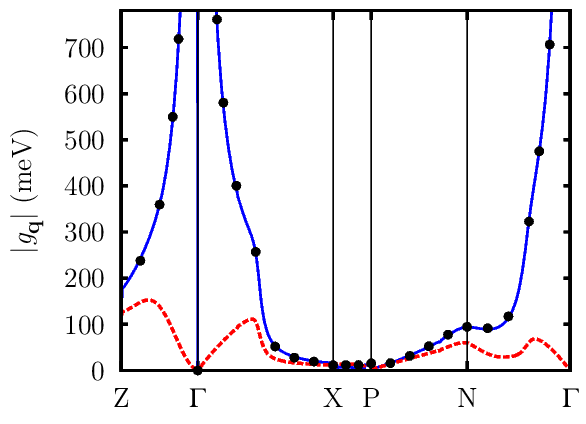}
  \caption[fig]{\label{fig.verdi} (Color online)
  Wannier interpolation of electron-phonon matrix elements for anatase TiO$_2$.
  The initial state $|n\bk\>$ is set to the bottom of the conduction band at $\Gamma$, 
  the final state $|m\bk+\bq\>$ spans the bottom of the conduction band along 
  high-symmetry lines, and the phonon is the highest LO mode. 
  The dots correspond to explicit DFPT calculations. The red dashed line 
  is the short-range component of the matrix elements, $g^{\mathcal S}$. The solid 
  curve in blue represents the matrix elements $g^{\mathcal S}+g^{\mathcal L}$, as obtained from 
  the modified Wannier interpolation of Sec.~\ref{sec.polar}. The interpolation
  was performed starting from a coarse 4$\times$4$\times$4 unshifted grid.
  Reproduced with permission from \cite{Verdi2015}, copyright
  (2015) by the American Physical Society.
  }
  \end{figure}

In order to perform Wannier interpolation, one subtracts Eq.~(\ref{gL})
from the matrix elements computed on a coarse grid, interpolates the remainig short-range part,
and then adds back Eq.~(\ref{gL}) on the fine grid. This process requires the interpolation
of the brakets $\langle \cdots \>_{\rm sc}$ in the second line of Eq.~(\ref{gL}). 
\citeauthor{Verdi2015} showed that, for small ${\bf q}+{\bf G}$, these brakets can be interpolated via the relation
$\langle \psi_{m{\bf k+q}} |e^{i({\bf q}+{\bf G})\cdot{\bf r}}| \psi_{n{\bf k}}\>_{\rm sc}
=\big[U_{\bk+\bq}U_{\bk}^\dagger\big]_{mn}$, where the rotation matrices
$U_{mn\bk}$ are obtained as usual from the procedure of \textcite{Marzari1997,Souza2001}.\footnote{
Eq.~(4) of \textcite{Verdi2015} misses a factor $e^{-i({\bf q}+{\bf G})\cdot\bm\tau_{\kappa}}$;
this factor needs to be retained in order to correctly describe the acoustic modes near $\bq=0$.
In practical calculations the $\bG$-vector sum in Eq.~(\ref{gL}) is 
restricted to small $|\bq+\bG|$ via the cutoff function $e^{-a|\bq+\bG|^2}$;
the results are independent of the choice of the cutoff parameter $a$.
} 
Figure~\ref{fig.verdi} shows an example of Wannier interpolation for the prototypical
polar semiconductor TiO$_2$:
it is seen that the singularity is correctly captured by the modified interpolation
method.

At the end of this section, we mention that other interpolation schemes are equally
possible
(\citeauthor{Eiguren2008b}, \citeyear{Eiguren2008b};
\citeauthor{Prendergast2009}, \citeyear{Prendergast2009};
\citeauthor{Agapito2013}, \citeyear{Agapito2013};
\citeauthor{Gunst2016}, \citeyear{Gunst2016}). 
For example \textcite{Eiguren2008b} proposed to interpolate only the local component
of $\Delta V^{\rm KS}_{\bq\nu}$, while calculating explicitly the nonlocal part of 
the perturbation as well as the Kohn-Sham wavefunctions in the Bloch representation.
Furthermore, Eq.~(\ref{eq.wan2blo}) remains unchanged if MLWFs are replaced
by a basis of localized atomic orbitals, and all the concepts discussed in this section
remain valid. An interpolation scheme using local orbitals was recently demonstrated
by \textcite{Gunst2016}. 

\subsection{Fermi surface harmonics}

In the study of {\it metallic} systems, one is often interested in describing EPIs only for
electronic states in the vicinity of the Fermi surface. 
In these cases, besides the Wannier interpolation discussed in
Secs.~\ref{sec.wannier-nonpolar}-\ref{sec.polar}, it is possible to perform efficient
calculations using `Fermi-surface harmonics' (FSH). FSHs were introduced by \textcite{Allen1976b} 
and recently revisited by \textcite{Eiguren2014}.

The basic idea underlying FSHs is to replace expensive three-dimensional Brillouin-zone 
integrals by inexpensive one-dimensional integrals in the energy variable. To this aim,
\textcite{Allen1976b} proposed to expand
functions of the band index $n$ and wavevector $\bk$, say $A_{n\bk}$, in products of pairs
of functions, one depending on the energy, $A_L(\ve)$, and one depending on 
the wavevector, $\Phi_L(\bk)$:
  \begin{equation}\label{eq.fsh1}
  A_{n\bk} = {\sum}_L\, A_L(\ve_{n\bk}) \,\Phi_L(\bk).
  \end{equation}
In this expression, the Fermi-surface harmonics $\Phi_L(\bk)$ (to be defined below)
are constructed so as to obey the following orthogonality condition:
  \begin{equation}\label{eq.fsh2}
  N_p^{-1}{\sum}_{n\bk} \d(\ve_{n\bk}-\ve) \,\Phi_L(\bk)\, \Phi_{L'}(\bk) = N(\ve) \d_{LL'},
  \end{equation}
where 
$N(\ve) = N_p^{-1}{\sum}_{n\bk} \d(\ve_{n\bk}-\ve)$ is the
density of states. Using Eqs.~(\ref{eq.fsh1})-(\ref{eq.fsh2}) one finds:
  \begin{equation}
  A_L(\ve) = N(\ve)^{-1} 
  N_p^{-1}{\sum}_{n\bk} 
  \d(\ve_{n\bk}-\ve) \,\Phi_L(\bk)\, A_{n\bk}.
  \end{equation}
\citeauthor{Allen1976b} showed that in the FSH representation a linear system of the kind
$A_{n\bk} = N_p^{-1}\sum_{n'\bk'} M_{n\bk,n'\bk'} B_{n'\bk'}$ transforms into 
$A_L(\ve) = \sum_{L'} \int d\ve' N(\ve') M_{LL'}(\ve,\ve')B_{L'}(\ve')$. 
Linear systems of this kind are common in the solution
of the Boltzmann transport equation (Sec.~\ref{sec.transport}) 
and the Eliashberg equations for the superconducting gap (Sec.~\ref{sec.eliashberg}).
If one could perform the expansion using only a few harmonics,
then the transformation would be advantageous, since the integrals over the wavevectors 
would have been absorbed in the expansion coefficients.

In the original proposal of \citeauthor{Allen1976b}, the harmonics $\Phi_L(\bk)$ were defined as polynomials 
in the band velocities, however the completeness of the basis set was not established.
In a recent work, \textcite{Eiguren2014} proposed to construct these functions as eigenstates
of a modified Helmholtz equation:
  \begin{equation}\label{eq.hfsh}
  |{\bf v}_{\bk}| \nabla^2_{\bk} \Phi_L(\bk) = \w_L  \Phi_L(\bk),
  \end{equation}
where ${\bf v}_{\bk}=\hbar^{-1}\nabla_\bk \ve_{n\bk}$ is the band velocity for states {\it at} the Fermi surface, 
and $\w_L$ is the eigenvalue for the harmonic $\Phi_L$.
The new definition in Eq.~(\ref{eq.hfsh}) 
maintains the properties of the original FSHs, and carries the added advantage
that the basis set is complete. In this case the subscript $L$ in $\Phi_L(\bk)$ labels the eigenstates
of the Helmholtz equation. \citeauthor{Eiguren2014} demonstrated the construction of `Helmholtz FSHs' for
prototypical metals such as Cu, Li, and MgB$_2$. 

Recent examples of the application of Fermi surface harmonics to
first-principles calculations of EPIs include work on the photoemission kink 
of YBa$_2$Cu$_3$O$_7$ \cite{Heid2008}, and on the Seebeck coefficient of Li \cite{Xu2014}.

\section{Non-adiabatic vibrational frequencies and linewidths}\label{sec.nonadiab}

As discussed in Secs.~\ref{sec.nonadiabatic}-\ref{sec.phon-self}, 
the electron-phonon interaction can lead to a renormalization of the {\it adiabatic}
vibrational frequencies and to a broadening of the spectral lines. 

\begin{figure}[t!]
  \includegraphics[width=0.55\columnwidth]{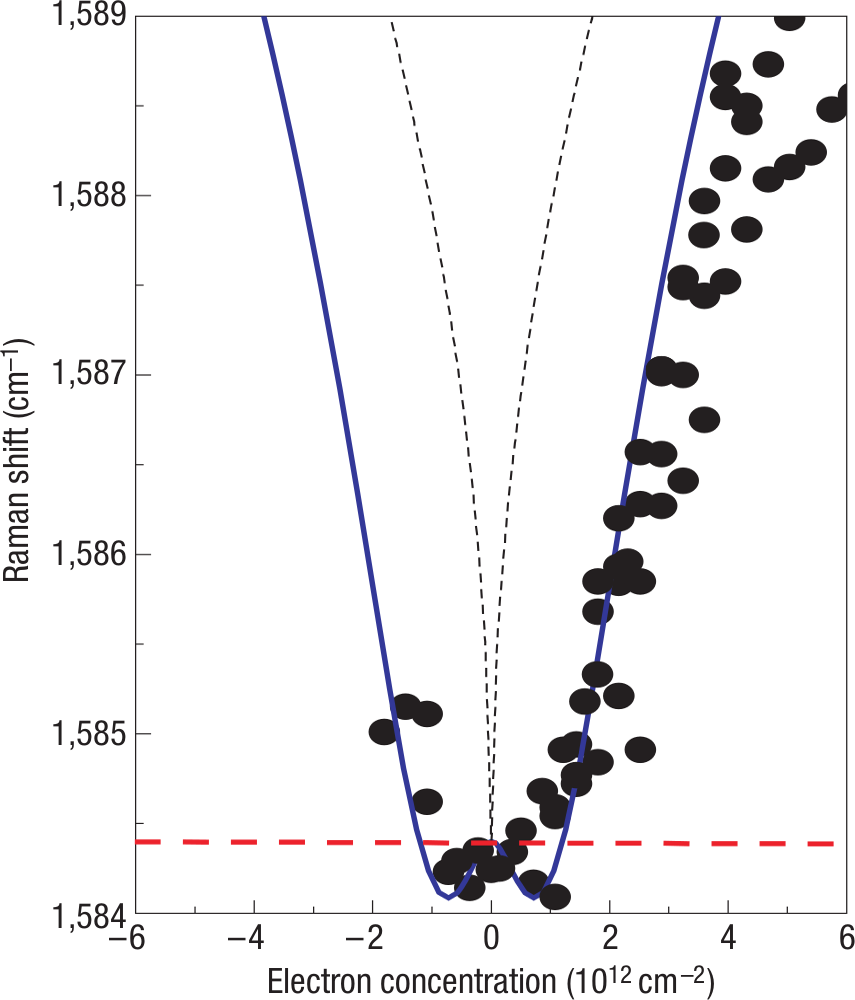}
  \caption[fig]{\label{fig.pisana2007} (Color online)
  Frequency of the Raman $G$ band of graphene vs.~carrier concentration.
  The black filled disks are from Raman measurements of gated
  graphene on a silicon substrate at 295~K. The thick horizontal dashed line (red) shows the variation
  of the $E_{2g}$ mode frequency with doping, within the adiabatic 
  approximation. The solid blue line shows the variation of the frequency 
  calculated by including non-adiabatic frequency renormalization.
  Reproduced with permission from \cite{Pisana2007}, copyright
  (2007) by Macmillan Publishers Ltd.
  }
  \end{figure}

The first {\it ab~initio} investigations of the effects of the non-adiabatic
renormalization of phonon frequencies were reported by \textcite{Lazzeri2006}
and \textcite{Pisana2007}. In these works the authors concentrated on the 
$E_{2g}$ phonon of graphene, which is found at the wavenumber 
$\w/2\pi c = 1585$~cm$^{-1}$ at room temperature
($c$ is the speed of light).
This phonon corresponds to an in-plane C--C stretching vibration with $\bq=0$, and
has been studied extensively via Raman spectroscopy.
In the graphene literature this mode is referred to as the `Raman $G$ band'. 
Figure~\ref{fig.pisana2007} shows a comparison between calculated 
and measured $E_{2g}$ phonon frequencies, as a function of doping, from \cite{Pisana2007}. 
The calculations were performed (i) within the adiabatic approximation 
and (ii) by including the non-adiabatic frequency renormalization using Eq.~(\ref{eq.phon-self-dft}).\footnote{
In the works reviewed in this section the authors used Eq.~(\ref{eq.phon-self-dft})
with the bare matrix elements $g^{\rm b}_{mn\nu}(\bk,\bq)$ replaced by the screened matrix elements
$g_{mn\nu}(\bk,\bq)$. All calculations were performed within DFT, using
either the LDA or gradient-corrected DFT functionals.}
From Fig.~\ref{fig.pisana2007} we see that the adiabatic theory is unable to reproduce
the experimental data. On the contrary, the calculations including non-adiabatic effects
nicely follow the measured Raman shift. This is a clear example of the limits of the
adiabatic Born-Oppenheimer approximation and a demonstration of the importance
of the phonon self-energy in Eq.~(\ref{eq.phon-self-dft}).

The fact that the adiabatic approximation is inadequate for the $E_{2g}$ phonon of 
doped graphene should have been expected from the discussion on p.~\pageref{pag.nonadiab}.
In fact, graphene is a zero-gap semiconductor, therefore 
electrons residing
in the vicinity of the Dirac points can make `virtual' transitions with $|\bq|\!=\!0$ and
energies comparable to that of the $E_{2g}$ mode. As a result, the condition underlying
the adiabatic approximation, $|\ve_{m\bk+\bq}-\ve_{n\bk}| \gg \hbar\w_{\bq\nu}$, does not hold
in this case. 

The importance of non-adiabatic effects was confirmed also in the case of metallic single-walled
carbon nanotubes 
(\citeauthor{Piscanec2007}, \citeyear{Piscanec2007};
\citeauthor{Caudal2007}, \citeyear{Caudal2007}). 
In these works, the authors studied the phonon dispersion
relations in the vicinity of $|\bq|\!=\!0$, and they found that the difference between adiabatic and
non-adiabatic dispersions is concentrated around the zone center.
This finding is consistent with earlier models of non-adiabatic effects. In fact \textcite{Maksimov1996}
showed that, for metals with linear electron bands crossing the Fermi level,
$\Pi^{\rm NA}$ is only significant for wavevectors $|\bq|\sim \w/v_{\rm F}$,
where $\w$ is the phonon energy and $v_{\rm F}$ the Fermi velocity.
This result can be derived from Eq.~(\ref{eq.phon-self-dft}).

\begin{figure}[t!]
  \includegraphics[width=0.75\columnwidth]{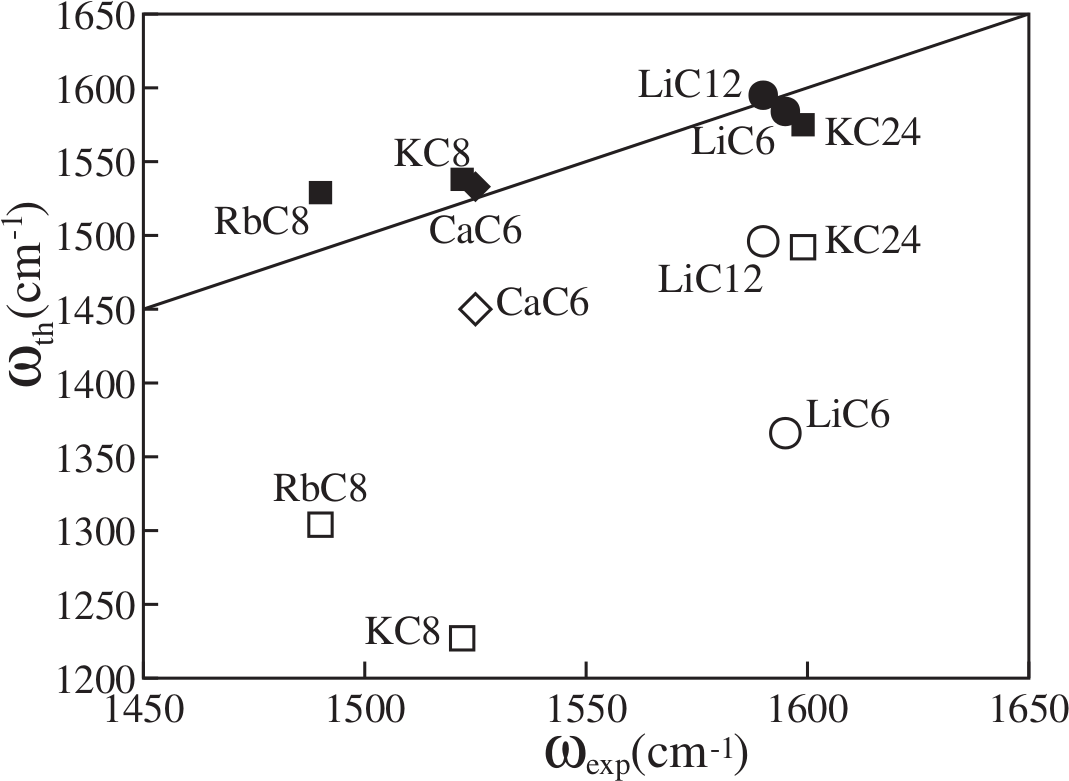}
  \caption[fig]{\label{fig.saitta2008} 
  Comparison between measured ($\w_{\rm exp}$) and calculated ($\w_{\rm th}$)
  vibrational frequencies of the $E_{2g}$ mode of graphite intercalation
  compounds. Open symbols are adiabatic DFPT calculations, filled symbols are
  calculations including the non-adiabatic corrections.
  The line corresponds to $\w_{\rm th} = \w_{\rm exp}$.
  Reproduced with permission from \cite{Saitta2008}, copyright
  (2008) by the American Physical Society.
  }
  \end{figure}

In the previous examples, the non-adiabatic renormalization of the vibrational frequencies
is measurable but very small, typically of the order of 1\% of the corresponding adiabatic frequencies.
\textcite{Saitta2008} considered the question as to whether one could find materials exhibiting
large non-adiabatic renormalizations, and considered several graphite intercalation compounds,
namely LiC$_6$, LiC$_{12}$, KC$_8$, KC$_{24}$, RbC$_8$, CaC$_6$, SrC$_6$, BaC$_6$, 
as well as other metallic systems such as MgB$_2$, Mg, and Ti. In order to calculate the
non-adiabatic renormalization at a reduced computational cost, the real part of the phonon
self-energy was approximated as follows: 
  $\hbar\,{\rm Re}\,\Pi^{\rm NA}_{\bq=0,\nu\nu} \simeq N_{\rm F}\, \< |g_{nn\nu}(\bk,\bq=0)|^2 \>_{\rm BZ}$,
where $\< \cdots \>_{\rm BZ}$ stands for the average taken over the wavevectors $\bk$ in the
Brillouin zone, and $N_{\rm F}$ is the density of states at the Fermi level. 
This expression can be derived from Eq.~(\ref{eq.phon-self-dft}) 
by replacing the bare matrix elements by their screened counterparts,
and by neglecting the `interband' contributions $m\ne n$ in the sum. 
Figure~\ref{fig.saitta2008} shows a comparison between vibrational frequencies from experiment
and those calculated with or without including the non-adiabatic self-energy.
It is clear that the non-adiabatic frequencies are in much better agreement with experiment
than the corresponding adiabatic calculations. Furthermore, in these compounds the renormalization
can reach values as large as $\sim$300~cm$^{-1}$.
In contrast to this, the renormalization in Mg and Ti was found to be of only a few wavenumbers
in~cm$^{-1}$. 

The case of MgB$_2$ proved more puzzling: here the calculated non-adiabatic frequency
is 761~cm$^{-1}$, while experiments reported 600~cm$^{-1}$. In order to explain this discrepancy,
\citeauthor{Saitta2008} reasoned that a more accurate calculation would require taking into
account the relaxation time of the electrons, as pointed out by \textcite{Maksimov1996}.
This would act so as to partly wash out non-adiabatic effects. In the field-theoretic language of
Sec.~\ref{sec.green-recsp}, this observation corresponds to stating that when one approximates
the {\it dressed} electron propagator $G$ in Fig.~\ref{fig.elec-selfenergy}(c) using the
non-interacting Kohn-Sham Green's function, one should include at the very least 
the effects of finite electron 
lifetimes (due to electron-electron, electron-impurity, and electron-phonon scattering), 
for example via the imaginary part of Eq.~(\ref{eq.fan-final}).

\begin{figure}[t!]
  \includegraphics[width=0.7\columnwidth]{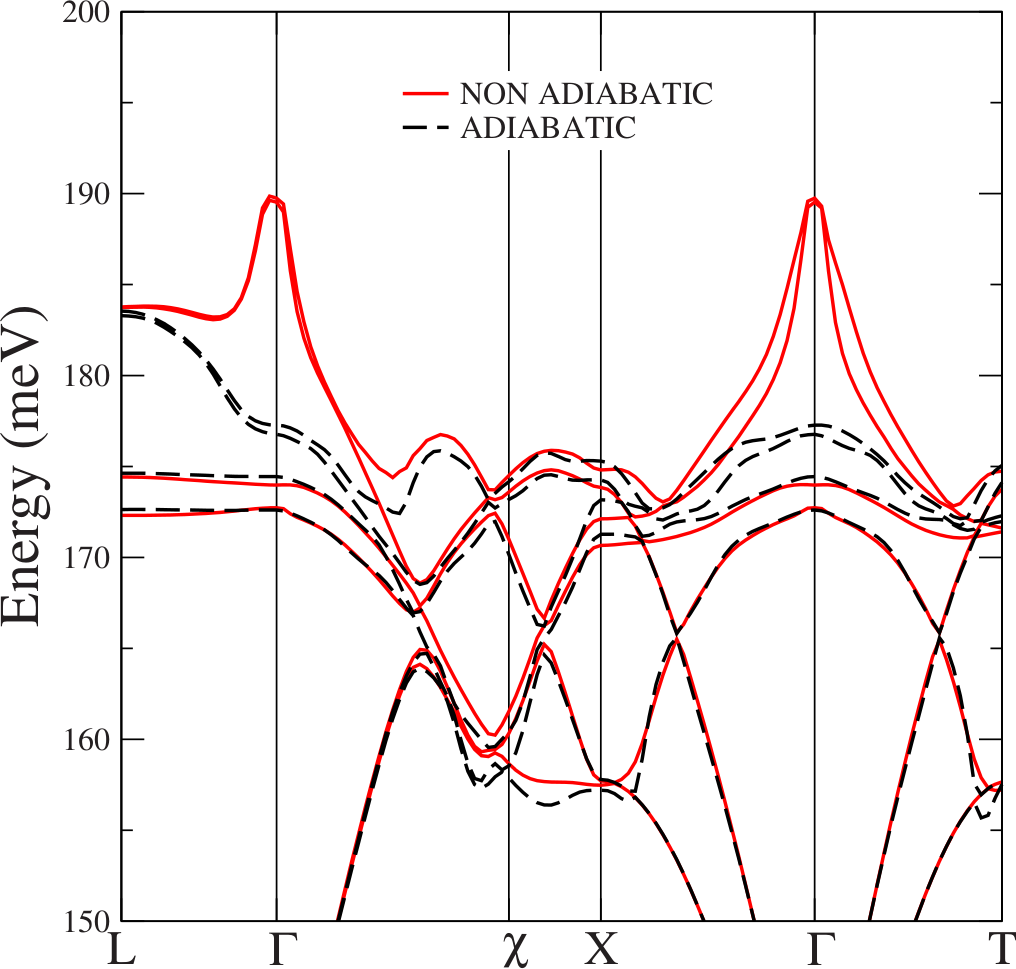}
  \caption[fig]{\label{fig.calandra2010} (Color online)
  Phonon dispersion relations of CaC$_6$ calculated using Wannier interpolation.
  The dashed lines (black) and the solid lines (red) represent the standard adiabatic calculation
  and the non-adiabatic phonon dispersions, respectively.
  Reproduced with permission from \cite{Calandra2010}, copyright
  (2010) by the American Physical Society.
  }
  \end{figure}

The calculations discussed so far in this section addressed the
non-adiabatic renormalization of zone-center phonons. The generalization to 
calculations of complete phonon dispersions was made by \textcite{Calandra2010}.
In their work, \citeauthor{Calandra2010} employed Wannier interpolation 
(see Sec.~\ref{sec.wannier}) in order to calculate the non-adiabatic phonon
self-energy of Eq.~(\ref{eq.phon-self-dft}) throughout the Brillouin zone.
Figure~\ref{fig.calandra2010} shows a comparison between the standard DFPT phonon 
dispersion relations of CaC$_6$ and the dispersions obtained after incorporating
the non-adiabatic self-energy. It is seen that also in this case non-adiabatic
effects are most pronounced at small $\bq$, and can be as large as 7\% of the
adiabatic frequency.

In their work \citeauthor{Calandra2010}
approximated the bare matrix element $g^{\rm b}_{mn\nu}(\bk,\bq)$ appearing 
in Eq.~(\ref{eq.phon-self-dft}) by the screened matrix element $g_{mn\nu}(\bk,\bq)$.
This replacement was justified by reasoning that the error is of second-order
in the induced electron density, hence it should be negligible.

The broadening of vibrational spectra arising from the electron-phonon interaction
is almost invariably calculated from first principles using Eq.~(\ref{eq.gamma-allen}).
Since the integration of the Dirac delta is computationally costly, it is common
to rewrite that equation by neglecting the phonon energy in the delta function and
by taking the low-temperature limit, as proposed by \textcite{Allen1972c}:
 \begin{equation}
  \frac{\gamma_{\bq\nu}}{\pi \w_{\bq\nu}} \!\simeq 2 \!\sum_{mn} \!\int\! \!\!\frac{d\bk}{\Omega_{\rm BZ}}
     |g_{mn\nu}(\bk,\bq)|^2
  \d(\ve_{n\bk}-\ve_{\rm F}) \d(\ve_{m\bk+\bq}-\ve_{\rm F}),\label{eq.doubledelta}
 \end{equation}
where $\ve_{\rm F}$ is the Fermi energy.
Oddly enough, this is a sort of adiabatic approximation to the non-adiabatic theory.
The main advantage is that this expression is positive definite, hence easier
to converge numerically as compared to the complete expression in Eq.~(\ref{eq.gamma-allen}). The disadvantages
are that the temperature dependence is lost, and that one cannot resolve fine features on the
scale of a phonon energy. There exists a vast literature on first-principles calculations
of phonon linewidths using Eq.~(\ref{eq.doubledelta}), mostly in connection with
electron-phonon superconductors.\footnote{
See for example early frozen-phonon calculations
(\citeauthor{Dacarogna1985b}, \citeyear{Dacarogna1985b};
\citeauthor{Chang1986}, \citeyear{Chang1986};
\citeauthor{Lam1986}, \citeyear{Lam1986}) and
more recent DFPT calculations 
(\citeauthor{Savrasov1994}, \citeyear{Savrasov1994};
\citeauthor{Bauer1998}, \citeyear{Bauer1998};
\citeauthor{Shukla2003}, \citeyear{Shukla2003};
\citeauthor{Heid2010}, \citeyear{Heid2010}).
Earlier calculations not based on DFT are reviewed by \textcite{Grimvall1981}.}
Equation~(\ref{eq.doubledelta}) is now implemented in several large software projects 
\cite{Giannozzi2009,Gonze2009}, and it is used routinely. 

Phonon linewidths range from very small values such as $\sim$1~meV in Nb \cite{Bauer1998}
to large values such as $\sim$30~meV in MgB$_2$ \cite{Shukla2003}. The agreement between
calculations and neutron scattering or Raman measurements is usually reasonable.

Calculations of phonon linewidths using the more accurate expression in Eq.~(\ref{eq.gamma-allen})
are computationally more demanding and have been reported less extensively in the literature.\footnote{
See for example the works of \textcite{Lazzeri2006b}; \textcite{Giustino2007b,Bonini2007,Park2008b}.}

So far we considered the effect of EPIs on the frequencies and lifetimes of vibrational
excitations in solids. Another important phenomenon which modifies frequencies and lifetimes 
is {\it anharmonicity}. Anharmonic effects result from third and higher-order terms in the Taylor
expansion of the total potential energy $U$ in the atomic displacements (Sec.~\ref{sec.vibr-theory}).
These effects can be interpreted as additional interactions that couple the 
oscillators of the harmonic lattice; for example, third-order anharmonic effects reduce 
phonon lifetimes via energy up- or down-conversion processes involving three phonons.

Anharmonic effects can be described using a many-body field-theoretic 
formalism \cite{Cowley1963}, in complete analogy with the theory of EPIs discussed in Sec.~\ref{sec.green}.
The calculation of anharmonic effects from first principles goes through the evaluation of 
third- and fourth-order derivatives of the total potential energy $U$ in the adiabatic approximation.
Third-order coefficients are routinely computed using DFPT
(\citeauthor{Debernardi1995}, \citeyear{Debernardi1995}; \citeauthor{Deinzer2003}, \citeyear{Deinzer2003};
\citeauthor{Lazzeri2003}, \citeyear{Lazzeri2003};
\citeauthor{Broido2007}, \citeyear{Broido2007}; \citeauthor{Bonini2007}, \citeyear{Bonini2007}).
In thoses cases where the harmonic approximation fails completely, `self-consistent phonon'
techniques can be employed \cite{Hooton1955,Koheler1996}; recent implementations and calculations
can be found in \citeauthor{Errea2011} (\citeyear{Errea2011}, \citeyear{Errea2014}), 
\textcite{Hellman2011,Hellman2013,Monserrat2013}.

\section{Electron-phonon interactions in photoelectron spectroscopy}\label{sec.kinks}

In Sec.~\ref{sec.kinks-sat} we have seen how the electron-phonon
interaction in metals can lead to band structure `kinks' near the Fermi energy.
This is illustrated by the model calculation in Fig.~\ref{fig.kink-sat}(a).
The experimental investigation of these features started in the late 1990s, following the
development of high-resolution angle-resolved photoelectron spectroscopy (ARPES). 
Since in ARPES only the component of the
photoelectron momentum parallel to the sample surface is conserved \cite{Damascelli2003}, 
complete energy vs.  wavevector dispersion relations can be measured directly only for 2D or quasi-2D 
materials. Accordingly, the first observations of kinks were reported for the surface states of elemental
metals\footnote{\textcite{Valla1999,Hengsberger1999}} and for the CuO$_2$ planes of copper oxide 
superconductors.\footnote{\citeauthor{Valla1999a} (\citeyear{Valla1999a}), \citeauthor{Lanzara2001}
(\citeyear{Lanzara2001}), \citeauthor{Johnson2001} (\citeyear{Johnson2001}).}
{\it Ab initio} calculations of ARPES kinks can be performed by using 
the diagonal part of the Fan-Migdal self-energy (the Debye-Waller self-energy will
be discussed at the end of this section). To this aim, it is common to rewrite
Eqs.~(\ref{eq.fan-final})-(\ref{eq.fan-final-T}) at finite temperature using a spectral
representation:
  \begin{eqnarray}
  && \Sigma^{\rm FM}_{n n\bk}(\w,T) =
   \int_{-\infty}^{+\infty}\!\! \!\!\!\!\!\!d\ve \!
    \int_0^\infty \!\!\!\!\!d\ve'  \,\,
   \alpha^2\!F_{n\bk}(\ve,\ve')
  \nonumber \\ && \,\,\times\!
  \left[ \frac{1\!-\!f(\ve/\kt)\!+\!n(\ve'/\kt)}{\hbar\w\!-\!\ve-\ve'+i\hbar \eta} +
  \frac{f(\ve/\kt)\!+\!n(\ve'/\kt)}{\hbar\w\!-\!\ve+\ve'+i\hbar \eta} \right]\!. \nonumber \\
  \label{eq.sigma-a2F}
  \end{eqnarray}
Here the function $\a^2F$ is the so-called the `Eliashberg function' and is defined as:
  \begin{eqnarray} \label{eq.eliashb-func}
  \alpha^{\!2}\!F_{\!n\bk}(\ve,\ve') & = &\sum_{m\nu} \int\!\! \frac{d\bq}{\Omega_{\rm BZ}}
  |g_{nm\nu}(\bk,\bq)|^{^2} \nonumber \\ & \times &\d(\ve-\ve_{m\bk+\bq})  \d(\ve'-\hbar\w_{\bq\nu}).
  \end{eqnarray}
This quantity is positive, temperature-independent, and contains all the materials-specific 
parameters. Physically
it is proportional
to the scattering rate of an electron in the state $|n\bk\>$ into any electronic state at the energy~$\ve$, 
by emitting or absorbing any one phonon of energy~$\ve'$. 
One complication related to the Eliashberg function is that in the literature many variants
of Eq.~(\ref{eq.eliashb-func}) can be found, each stemming from specific approximations; 
some of these expressions are summarized by \citeauthor{Grimvall1981}
(\citeyear{Grimvall1981}, pp.~107--109). 
 
The first {\it ab~initio} calculations of the phonon-induced electron self-energies and 
photoemission kinks were reported by \textcite{Eiguren2003} for the surface state of the 
Be(0001) surface.  Since the evaluation
of Eqs.~(\ref{eq.sigma-a2F}) and (\ref{eq.eliashb-func}) is computationally demanding, 
\citeauthor{Eiguren2003} employed simplified expressions which involve three 
approximations: (i) the Eliashberg function is replaced by its isotropic average,
 $\alpha^2\!F_n(\ve,\ve') =
\int\! d\bk \,\d(\ve_{n\bk}\!-\!\ve)$ $\alpha^2\!F_{n\bk}(\ve,\ve') /
\!\int\! d\bk\, \d(\ve_{n\bk}\!-\!\ve)$;
(ii) phonon energies are neglected next to electron energies (as in the adiabatic
approximation),
and (iii) particle-hole symmetry is assumed.
Using these approximations the imaginary part of the Fan-Migdal self-energy becomes:
  \begin{eqnarray}\label{eq.eiguren}
  && |{\rm Im}\,\Sigma^{\rm FM}_{n}(\w,T)| = \pi \!\int_0^\infty \!\!\!\!\!d\ve'\,
  \alpha^2\!F_n(\hbar\w,\ve') \{ 1 +2 n(\ve'/\kt)  \nonumber \\ && \hspace{1.3cm}
    +f[(\hbar\w+\ve')/\kt]- f[(\hbar\w-\ve')/\kt] \},\hspace{0.5cm}
  \end{eqnarray}
where the average of the self-energy is defined as for the Eliashberg function.
The real part of the self-energy can be found starting from the same approximations, and is given
by \textcite{Grimvall1981}.

  \begin{figure}[t!]
  \includegraphics[width=0.9\columnwidth]{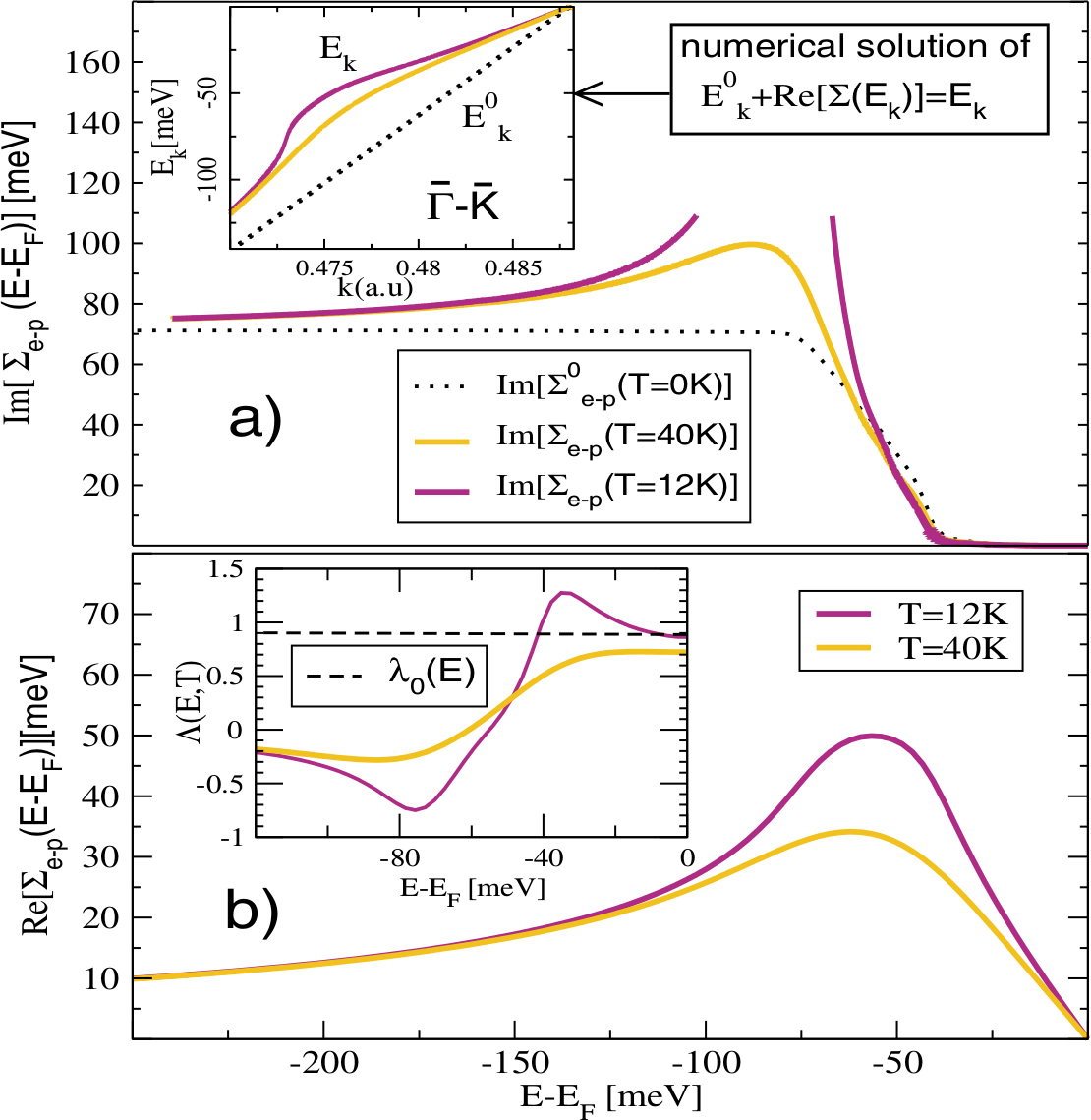}
  \caption[fig]{\label{fig.eiguren2003} (Color online)
  Calculated Fan-Migdal self-energy of the surface state at the Be(0001) surface.
  (a)~Imaginary part of the self-energy, obtained from Eq.~(\ref{eq.eiguren}).
  The dashed (black) line is the self-energy evaluated using the DFT/LDA bands; the solid
  lines (color/grayscale) 
  correspond to the self-energy calculated by taking into account the renormalization of the
  DFT band structure by the electron-phonon interaction.
  (b)~Real part of the self-energy, using the same color/grayscale code as in (a). 
  The inset in (a) compares the renormalized band structure (color) with the `bare'
  DFT band (black dashed line). The inset of (b) shows the renormalization of the
  band velocity induced by the electron-phonon interaction.
  Reproduced with permission from \cite{Eiguren2003}, copyright
  (2003) by the American Physical Society.
  }
  \end{figure}

Figure~\ref{fig.eiguren2003} shows the self-energy of the surface state at the Be(0001)
surface calculated by \textcite{Eiguren2003} using Eq.~(\ref{eq.eiguren}). The imaginary
part resembles a step-function, with an onset around the energy threshold
for phonon emission by a hole (40--80~meV in this case).
At a qualitative level, this trend can
be rationalized by replacing the Eliashberg function in Eq.~(\ref{eq.eiguren}) 
by a Dirac delta at a characteristic
phonon energy $\hbar\w_{\rm ph}$. In this case, the hole self-energy becomes proportional
to $f[(\hbar\w+\hbar\w_{\rm ph})/\kt]$. At $T\!=\!0$ this is precisely a step function with onset 
at $-\hbar\w_{\rm ph}$.
The real part of the self-energy
vanishes for $|\w|\gg \w_{\rm max}$, with $\w_{\rm max}$ being the largest phonon frequency.
This can be seen in Fig.~\ref{fig.eiguren2003}(b), and is a consequence of the
approximation of particle-hole symmetry.
\citeauthor{Eiguren2003} also determined the renormalization
of the surface state band structure arising from electron-phonon interactions, using
Eq.~(\ref{eq.qp-shift}); this is shown in the
inset of Fig.~\ref{fig.eiguren2003}(a). Overall the calculations of \textcite{Eiguren2003}
showed good agreement with photoelectron spectroscopy experiments, both in the shape
and magnitude of the self-energy \cite{LaShell2000}.

In addition to the above calculations, several studies of the electron-phonon self-energy at metal surfaces
were reported, namely for the Cu(111) and Ag(111) surfaces \cite{Eiguren2002},
the Al(100), Ag(111), Cu(111), and Au(111) surfaces \cite{Eiguren2003b},
and the W(110) surface \cite{Eiguren2008}. Building on 
these studies, \textcite{Eiguren2009} performed a detailed analysis of
the {\it self-consistent} solutions of the complex Dyson equation for the quasiparticle energies,
Eq.~(\ref{eq.qp-shift})-(\ref{eq.qp-width}), and illustrated the key concepts in the cases
of the W(110) surface and for the phonon-mediated superconductor MgB$_2$.

Equation~(\ref{eq.eiguren}) or closely-related approximations were also employed in the study 
of electron and hole lifetimes in bulk Be \cite{Sklyadneva2005}, Pb \cite{Sklyadneva2006}, 
and Mg \cite{Leonardo2007}; the photoemission kink in YBa$_2$Cu$_3$O$_7$ \cite{Heid2008}; 
and the spectral function of Ca-intercalated graphite \cite{Sanna2012}.

\begin{figure}[t!]
  \includegraphics[width=0.85\columnwidth]{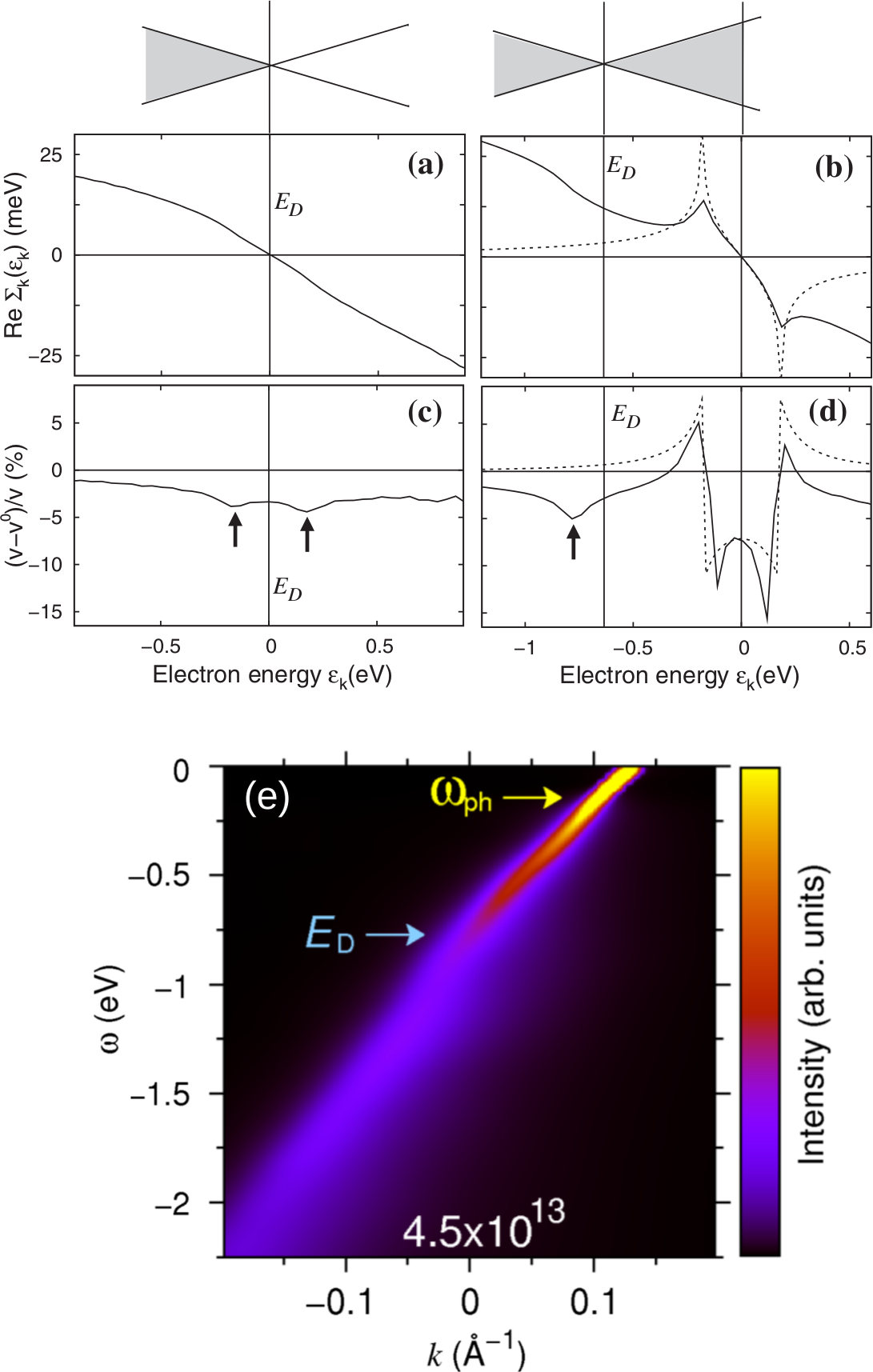}
  \caption[fig]{\label{fig.park2007} (Color online)
  (a), (b)~Calculated real part of the Fan-Migdal self-energy in pristine and $n$-doped graphene, respectively
  (solid black lines). The doping level is 4$\cdot10^{13}$~cm$^{-2}$. The dashed lines correspond
  to a simplified analytical model where particle-hole symmetry is assumed.
  (c),~(d)~Electron band velocity renormalization resulting from the self-energies in (a) and (b).
  All calculations in (a)-(d) were performed using DFT/LDA.
  Reproduced with permission from \cite{Park2007}, copyright
  (2007) by the American Physical Society.
  (e)~Calculated spectral function of $n$-doped graphene for one of the branches of the Dirac cone. 
  $\w_{\rm ph}$ indicates the characteristic phonon energy leading to the photoemission kink;
  $E_{\rm D}$ denotes the energy of the Dirac point. The calculations include $GW$ quasiparticle
  corrections. 
  Reproduced with permission from \cite{Park2009b}, copyright
  (2009) by the American Chemical Society.
  }
  \end{figure}

In the case of complex systems the validity of the approximations (i)--(iii) leading to Eq.~(\ref{eq.eiguren})
is not warranted, and a direct calculation of the Fan-Migdal self-energy using Eqs.~(\ref{eq.sigma-a2F})
and (\ref{eq.eliashb-func}) is necessary. Calculations of the complete 
self-energy were reported by \textcite{Park2007} for graphene, by \textcite{Giustino2008} 
for the high-temperature superconductor La$_{1-x}$Sr$_x$Cu$_2$O$_4$, and by \textcite{Margine2016}
for Ca-intercalated bilayer graphene.
Figure~\ref{fig.park2007} shows the calculated self-energy and spectral function of graphene
calculated by \textcite{Park2007,Park2009b}. The structure of the self-energy is similar to that
shown in Fig.~\ref{fig.eiguren2003}, with one important exception: ${\rm}\,\Sigma^{\rm FM}_{nn\bk}(\w)$
does not vanish a few phonon energies away from the Fermi level, but tends instead towards a 
linear asymptote. 
\textcite{Calandra2007} performed a combined {\it ab~initio}/analytical study of the effects 
of the electron-phonon interaction on the electron bands of graphene and obtained very similar results.
A linear asymptote in the real-part of the self-energy is a general feature of systems which
do {\it not} exhibit particle-hole symmetry. For another example see the work on copper oxides 
by \textcite{Giustino2008}.

Using the Fan-Migdal self-energy, it is possible to calculate the renormalization of the
band velocity induced by the electron-phonon interaction.
Let us denote by ${\bf v}_{n\bk} = \hbar^{-1}\nabla_\bk \ve_{n\bk}$ 
the DFT band velocity and ${\bf V}_{n\bk} = \hbar^{-1}\nabla_\bk E_{n\bk}$ the band velocity after including
electron-phonon interactions. Using Eq.~(\ref{eq.qp-shift}) with $\Gamma_{n\bk}\!=\!0$
we find that these two quantities are simply related by
${\bf V}_{n\bk} = Z_{n\bk} \,{\bf v}_{n\bk} = {\bf v}_{n\bk}/(1+\lambda_{n\bk})$,
where $Z_{n\bk}$ is the quasiparticle strength of Eq.~(\ref{eq.Zfac}), and
$\lambda_{n\bk}$ is the `mass enhancement parameter' or `electron-phonon coupling strength' \cite{Grimvall1981}:
  \begin{equation}\label{eq.lambdank}
  \lambda_{n\bk} = Z_{n\bk}^{-1}-1 =  - \hbar^{-1}\D\,{\rm Re}\Sigma_{nn\bk}(\w)/\D\w\big|_{\w=E_{n\bk}/\hbar}.
  \end{equation}
In the study of EPIs in metals, the coupling strength $\lambda_{n\bk}$ 
is of significant interest since it is related to the superconducting transition
temperature of phonon-mediated superconductors (see 
Secs.~VIII.1 and \ref{sec.supercond}).

The velocity renormalization in graphene calculated using Eq.~(\ref{eq.lambdank}) is shown 
in Fig.~\ref{fig.park2007}(c) and (d), while a calculation of the complete spectral function $A(\bk,\w)$
is shown in Fig.~\ref{fig.park2007}(e). Here the characteristic photoemission kink
is visible between 150--200~meV but it is not very pronounced, 
since in this case $\lambda_{n\bk}\sim 0.1$ \cite{Park2009b}. These results are 
in good agreement with measured
photoelectron spectra \cite{Bostwick2007,Zhou2007}.

Incidentally, we remark that in the analysis of ARPES data it is common to extract the coupling strength
$\lambda_{n\bk}$ directly from the ratio of the band velocities above and below the 
electron-phonon kink. 
However, this procedure is subject to a significant uncertainty, since the `bare' velocity is not
known and must be approximated by fitting the fully-interacting dispersions using {\it ad hoc}
models. For example, in the vicinity of Van Hove singularities this procedure leads to a significant 
overestimation of the electron-phonon coupling strength $\lambda_{n\bk}$ 
(\citeauthor{Park2008}, \citeyear{Park2008}; \citeauthor{Bianchi2010}, \citeyear{Bianchi2010}). 

In addition to photoemission kinks, recent ARPES measurements
revealed the existence of polaron satellites in doped oxides, namely TiO$_2$ \cite{Moser2013}
and SrTiO$_3$ 
\cite{Chen2015,Cancellieri2016,Wang2016}. The phenomenology is similar to what was discussed
in relation to Fig.~\ref{fig.kink-sat}(b). The first theoretical studies along this direction 
were reported by \textcite{Story2014}, who applied the cumulant expansion approach
to the case of the electron-phonon self-energy; by \textcite{Antonius2015}
who identified satellites in the spectral functions of LiF and MgO;
and by \textcite{Verdi2016}, who calculated the ARPES spectra of $n$-doped TiO$_2$.

At the end of this section, it is worth coming back to the Debye-Waller self-energy.\label{page.dw-constant}
So far we only discussed the Fan-Migdal self-energy, starting from 
Eq.~(\ref{eq.sigma-a2F}), and we ignored the Debye-Waller self-energy appearing in Eq.~(\ref{eq.selfen-ep}).
This omission reflects the fact that, in the literature on electron-phonon interactions
in metals, the DW term has always been disregarded.
In order to rationalize this approximation, we rewrite the DW self-energy
as follows, by combining Eqs.~(\ref{eq.dw-DFT}) and (\ref{eq.matel-dw-dft}):
\begin{equation}\label{eq.dw-pot}
  \Sigma^{\rm DW}_{nn\bk} = \< u_{n\bk}| V_{\rm DW}|u_{n\bk}\>_{\rm uc},
  \end{equation}
with $V_{\rm DW}(\br) = 
   \Omega_{\rm BZ}^{-1}{\sum}_\nu \int d\bq\, (n_{\bq\nu}+1/2)\Delta_{\bq\nu} \Delta_{-\bq\nu} v^{\rm KS}(\br)$.
The subscript `uc' indicates that the integral is over one unit cell.
From Eq.~(\ref{eq.dw-pot}) we see that $V_{\rm DW}$ acts like a static local {\it potential}; 
indeed the first calculations including
DW effects were performed by directly modifying the ionic pseudopotentials \cite{Antoncik1955,Walter1970}.
From Eq.~(\ref{eq.dw-pot}) we also see that the only variation in the DW self-energy comes from
the Bloch amplitudes~$u_{n\bk}$. Let us consider the limiting situation of the homogeneous electron gas.
In this case $|u_{n\bk}(\br)|^2 = 1/\Omega$
(Sec.~\ref{sec.matel-approx}), therefore $\Sigma^{\rm DW}_{nn\bk}$ is a constant, independent
of $\bk$.
This behavior should be contrasted with the Fan-Migdal self-energy, which
exhibits significant structure near the Fermi energy, as it can be seen in Fig.~\ref{fig.kink-sat}.

In more realistic situations, such as doped semiconductors, 
$\bk\cdot{\bf p}$ perturbation theory \cite{Kittel1963} can be used to show that $\Sigma^{\rm DW}_{nn\bk}$
varies smoothly as a function of $\bk$ within the {\it same} band.
In contrast with this scenario, $\Sigma^{\rm DW}_{nn\bk}$ exhibits 
large variations across {\it different} bands. This carries important
consequences for the calculation of temperature-dependent band gaps (Sec.~\ref{sec.temper}).

\subsubsection{Electron mass enhancement in metals}\label{sec.mass-enhancement}

We now come back to the mass enhancement parameter $\lambda_{n\bk}$
introduced in Eq.~(\ref{eq.lambdank}), since this quantity played a central role in the 
development of the theory of EPIs in metals.

The notion of `mass enhancement' becomes clear when we consider a parabolic band as
in the model calculations of Fig.~\ref{fig.kink-sat}. Near the Fermi surface
the non-interacting dispersions are given by $\ve_{n\bk} = \hbar \,\bk_{\rm F} \cdot \hbar 
\,(\bk-\bk_{\rm F})/ m^*$, where $\bk_{\rm F}$ is a wavevector on the Fermi surface,
and the electron velocity is ${\bf v}_{n\bk}=\hbar \,\bk_{\rm F}/m^*$. After taking into
account the EPI, the electron velocity is renormalized to
${\bf V}_{n\bk}={\bf v}_{n\bk}/(1+\lambda_{n\bk})$. Since the magnitude of the Fermi
momentum is unchanged (see caption of Fig.~\ref{fig.kink-sat}) this renormalization 
can be interpreted as an effective increase of the band mass from 
$m^*$ to $m_{\rm ep} = m^*(1+\lambda_{n\bk})$.
This reasoning holds for metals with parabolic bands and for
doped semiconductors near band extrema, and does not take into account the Debye-Waller
self-energy.

The electron mass enhancement is reflected into the increase of the heat capacity of
metals at low temperature. In fact, below the Debye temperature the electronic contribution
to the heat capacity dominates over the lattice contribution \cite{Kittel1976}.
Since the heat capacity is proportional to the density of states at the Fermi level, 
and the density of states is inversely proportional to the band velocity, it
follows that the heat capacity is directly proportional to the electron mass.
This property can be used as a means to determine the strength of the electron-phonon 
coupling in simple metals from specific heat measurements \cite{Grimvall1975}.

The general theory of the effects of electron-phonon interactions on the heat capacity
and other thermodynamic functions was developed by \textcite{Eliashberg1963}, \textcite{Prange1964},
and \textcite{Grimvall1969}. A field-theoretic analysis of the effect of EPIs on thermodynamic
functions was developed by \textcite{Eliashberg1963} starting from the
identities of \textcite{Luttinger1960b} in the zero-temperature limit. 
Eliashberg's analysis was subsequently extended to all temperatures by \textcite{Grimvall1969}.
Here we only quote Grimvall's result relating the electronic entropy to
the Fan-Migdal self-energy of Sec.~\ref{sec.elec-SE}:
  \begin{equation}\label{eq.entropy}
  S_{\rm e} = \frac{N_{\rm F} \kb \hbar^3}{(\kb T)^2}\int_0^\infty \frac{\w
         \left[ \w -\hbar^{-1}{\rm Re}\Sigma^{\rm FM}_{nn\bk}(\w,T)  \right]
       }{{\rm cosh}^2(\hbar\w/2\kb T)} d\w.
  \end{equation}
In order to derive this relation, Grimvall started by expressing the thermodynamic potential
of the coupled electron-phonon system in terms of the electron and phonon propagators
and self-energies, and identified the electronic contribution by neglecting terms
of order $(m_{\rm e}/M_0)^{1/2}$ as well as electron-electron interactions 
(\citeauthor{Grimvall1969}, \citeyear{Grimvall1969}, Appendix).

Below the Debye temperature, an explicit expression for the entropy in Eq.~(\ref{eq.entropy}) 
can be obtained by 
noting that the function ${\rm cosh}^{-2}(\hbar\w/2\kb T)$ is nonvanishing only for
$\hbar\w \lesssim 2\kb T$; in this range Eq.~(\ref{eq.lambdank}) yields
$\Sigma^{\rm FM}_{nn\bk}(\w,T) \simeq -\lambda_{n\bk}\, \hbar\w$, therefore the integration in
Eq.~(\ref{eq.entropy}) can be carried out explicitly. As a result, the low-temperature heat capacity 
can be written as:
  \begin{equation}\label{eq.specheat}
  C_{\rm e} = T \,\frac{\D S_{\rm e}}{\D T} = \frac{2}{3}\pi^2 \kb^2 N_{\rm F} 
      (1+\lambda_{n\bk}) T.
  \end{equation}
If we ignore the EPI by setting $\lambda_{n\bk}=0$,
this expression reduces to the standard textbook result for the free electron gas \cite{Kittel1976}.
At high temperature Eq.~(\ref{eq.specheat}) is no longer valid, and one has to evaluate 
the integral in Eq.~(\ref{eq.entropy}) using the complete frequency-dependent FM self-energy.
The main result of this procedure is that at high temperature the electronic heat capacity is no
longer renormalized by EPIs. A detailed discussion of this aspect is provided by
\citeauthor{Grimvall1969} (\citeyear{Grimvall1969, Grimvall1981}).

Early examples of DFT calculations of mass enhancement parameters and comparisons with 
specific heat measurements in simple metals can be found 
in \textcite{Dacarogna1985b}, \textcite{Savrasov1994}, and \textcite{Liu1996}.

\section{Electron-phonon effects in the optical properties of semiconductors}\label{sec.semicond}

\subsection{Temperature dependence of band gaps and band structures}\label{sec.temper}

\subsubsection{Perturbative calculations based on the Allen-Heine theory}\label{sec.temper}

In Sec.~\ref{sec.tempdep-theory} we discussed how the electron-phonon interaction induces
a `renormalization' of the electronic energy levels, and thereby gives rise to `temperature-dependent
band structures'. These effects have been studied in detail using the Fan-Midgal and the Debye-Waller 
self-energies, either via the Raleigh-Schr\"odinger approximation to Eq.~(\ref{eq.AH-BW}), or via
its adiabatic counterpart given by Eq.~(\ref{eq.AH-RS}). 

Equation~(\ref{eq.AH-RS}) was first employed in a number of calculations based on empirical
pseudopotentials, following the seminal work of \textcite{Allen1976}. The key references can be
found on p.~\pageref{note.allen-heine-early}, footnote. \textcite{Allen1981} offer
a clear introduction to the basic theory, a discussion of the computational methodology, as 
well as an historical perspective on earlier calculations. 

The evaluation of the Debye-Waller contribution to the self-energy requires the calculation of the
second-order variations of the Kohn-Sham potential with respect to the ionic displacements, Eq.~(\ref{eq.matel-dw-dft}).
From a computational standpoint, this is challenging because one would have to use second-order DFPT,
as discussed at the end of Sec.~\ref{sec.matel-dft}. 
In order to avoid this complication, it is common practice to
recast all second-order derivatives as products of first-order derivatives.
This strategy was introduced by \citeauthor{Allen1976} in the case of monoatomic crystals, 
and extended to polyatomic unit cells by \citeauthor{Allen1981}.
The key observation behind this approach is that one can impose translational
invariance of the theory to second order in the nuclear displacements. Specifically,
the variation of the Kohn-Sham eigenvalues ensuing an arbitrary displacement of the nuclei 
should not change if all nuclei are {\it further} displaced by the same amount. 
Using time-independent perturbation theory,
this condition yields the following two sum rules:
  \begin{eqnarray}
  &&{\sum}_{\k p} \< \psi_{n\bk} |\D_{\k\a p} V^{\rm KS}| \psi_{n\bk} \>_{\rm sc} = 0, \label{eq.asr-allen-1} \\
  && {\sum}_{\k' p'} \< \psi_{n\bk} | \D^2_{\k\a p,\k'\a' p'} V^{\rm KS}| \psi_{n\bk} \>_{\rm sc}
     = - 2\, {\rm Re} {\sum}_{\k' p'}{\sum}_{m\bq}{\!\!\!\!\!\!\!\vphantom{\sum}}^{\prime} \nonumber \\
    &&  
   \frac{ \< \psi_{n\bk} | \D_{\k\a p}\! V^{\rm KS}| \psi_{m\bk+\bq} \>_{\rm sc} 
        \< \psi_{m\bk+\bq} | \D_{\k'\a' p'} \!V^{\rm KS} | \psi_{n\bk} \>_{\rm sc\!}
       }{\ve_{n\bk}-\ve_{m\bk+\bq}}\!. \,\,\,\qquad \label{eq.asr-allen-2}
  \end{eqnarray}
Here $\D_{\k\a p} V^{\rm KS}$ is a short-hand notation for $\D V^{\rm KS}/\D \tau_{\k\a p}$ and similarly for the
second derivative; the primed summation indicates that eigenstates $\psi_{m\bk+\bq}$ such that
$\ve_{n\bk}=\ve_{m\bk+\bq}$ are skipped.
The first sum rule is equivalent to stating that the electron-phonon
matrix elements $g_{mn\nu}(\bk,\bq)$ associated with the three translational modes at $|\bq|\!=\!0$ must vanish;
this is an alternative formulation of the `acoustic sum rule'.
The second sum rule,  Eq.~(\ref{eq.asr-allen-2}), suggests to 
express the matrix elements of the second derivatives of
the potential in terms of first-order derivatives. However, Eq.~(\ref{eq.asr-allen-2})
cannot be used as it stands, since it involves a {\it sum} of matrix elements on the left-hand
side. In order to proceed, \citeauthor{Allen1976} employed the `rigid-ion' approximation,
whereby $V^{\rm KS}$ is written as a sum of atom-centered contributions
(see Sec.~\ref{sec.metals}).
Under this approximation all the terms $\k p \ne \k'p'$ 
on the left-hand side of Eq.~(\ref{eq.asr-allen-2}) are neglected, and an explicit expression for
$\< \psi_{n\bk} | \D^2_{\k\a p,\k\a p} V^{\rm KS}| \psi_{n\bk} \>_{\rm sc}$ is obtained.

In view of practical DFT implementations, \textcite{Giustino2010} used the sum rule in
Eq.~(\ref{eq.asr-allen-2}) in order to rewrite the Debye-Waller self-energy as follows:
  \begin{equation}
  \Sigma^{\rm DW}_{nn\bk} = 
  -{\sum}_{\nu m}{\!\!\!\!\!\!\!\vphantom{\sum}}^{\prime}\,\,\,\,\,
   \int\!\!\frac{d\bq}{\Omega_{\rm BZ}} \frac{g_{mn\nu}^{2,{\rm DW}}(\bk,\bq)}
  {\ve_{n\bk}-\ve_{m\bk}} (2n_{\bq\nu}+1).\label{eq.dw-giustino2010}
  \end{equation}
Here $g_{mn\nu}^{{\rm DW}}(\bk,\bq)$ is an `effective' matrix element, and it is obtained
from the standard DFPT matrix elements by means
of inexpensive matrix multiplications:
  \begin{eqnarray}\label{eq.2010-1}
  \hspace{-.5cm} g_{mn\nu}^{2,{\rm DW}\!}(\bk,\bq)&\! =\!&\sum_{\substack{\k\a\\\,\k'\a'}} \! 
   \frac{t_{\k\a,\k'\a'}^\nu(\bq)\!}{2\w_{\bq\nu}}
  h_{mn,\k\a\!}^*(\bk) h_{mn,\k'\a'\!}(\bk), \\
  t_{\k\a,\k'\a'}^\nu(\bq) &\!=\!& \frac{e_{\k\a\nu}(\bq)e^*_{\k\a'\nu}(\bq)}{ M_\k}+
  \frac{e_{\k'\a\nu}(\bq) e^*_{\k'\a'\nu}(\bq)}{ M_{\k'}}, \,\,\,\,\,\,\,\,\,\, \label{eq.dw-giustino2010-t} \\
  h_{mn,\k\a}(\bk) &\!=\!& {\sum}_\nu (M_\k \,\w_{0\nu})^\frac{1}{2}\, e_{\k\a\nu}(0) g_{mn\nu}(\bk,0).
  \label{eq.2010-3}
  \end{eqnarray}
In the case of the three translational modes at $|\bq|\!=\!0$, these definitions are replaced by
$g_{mn\nu}^{\rm DW}(\bk,\bq)=0$, see the discussion at the end of p.~\pageref{page.3modes}.
The derivation of Eqs.~(\ref{eq.dw-giustino2010})-(\ref{eq.2010-3}) 
requires using Eqs.~(\ref{eq.tau-from-x}), (\ref{eq.dV-all-1})-(\ref{eq.dV-all-3}), and (\ref{eq.matel}),
as well as taking the canonical average over the adiabatic nuclear quantum states.

Equation~(\ref{eq.dw-giustino2010}) involves a summation over unoccupied Kohn-Sham states,
and so does the Fan-Migdal self-energy in Eq.~(\ref{eq.AH-RS}). The numerical convergence of these
sums is challenging, since one needs to evaluate a very large number of unoccupied electronic states.
To address this issue, \textcite{Gonze2011} developed a procedure whereby only
a subset of unoccupied states is required, along the lines of the DFPT equations of
Sec.~\ref{sec.matel-dft}.

The first {\it ab~initio} calculations using the formalism of \citeauthor{Allen1976} were 
reported by \textcite{Marini2008}, who investigated the temperature dependence
of the optical absorption spectrum of silicon and boron nitride.
In this work \citeauthor{Marini2008} included excitonic effects by combining the Bethe-Salpeter 
formalism \cite{Onida2002} with the Allen-Heine theory, and obtained good agreement
with experiments by calculating the direct absorption peaks using DFT/LDA phonons and matrix elements
(indirect optical absorption will be discussed in Sec.~\ref{sec.phon-ass}).

The second application of the Allen-Heine theory using DFT/LDA was reported by \textcite{Giustino2010}
for the case of diamond. Here the authors investigated the temperature dependence of the direct band gap
of diamond, and found that the Fan-Migdal and the Debye-Waller self-energies are of comparable
magnitude.
The calculations captured the characteristic Varshni effect (Fig.~\ref{fig.varshni}), and were
able to reproduce the measured redshift of the band gap in the temperature range 80-800~K.
These calculations were based on the adiabatic version of the Allen-Heine theory, 
and employed Eqs.~(\ref{eq.2010-1})-(\ref{eq.2010-3}) for the Debye-Waller self-energy.
The calculations by \citeauthor{Giustino2010} confirmed the large ($>0.5$~eV) zero-point 
renormalization of the direct gap of diamond predicted by \textcite{Zollner1992}
using the empirical pseudopotential method. 

The unusually large zero-point correction to the electronic structure of diamond stimulated
further work on this system:
\textcite{Cannuccia2011,Cannuccia2012} calculated the gap renormalization in 
diamond by employing both the adiabatic version of the Allen-Heine theory, 
as well as the non-adiabatic 
Green's function approach, as described in Sec.~\ref{sec.kinks-sat}. Their calculations confirmed the
large zero-point renormalization, and showed that the adiabatic
theory underestimates the effect to some extent. \citeauthor{Cannuccia2012} 
(\citeyear{Cannuccia2012,Cannuccia2013}) also analyzed the quasiparticle renormalization and the 
spectral function.

\begin{figure}[t!]
  \includegraphics[width=0.8\columnwidth]{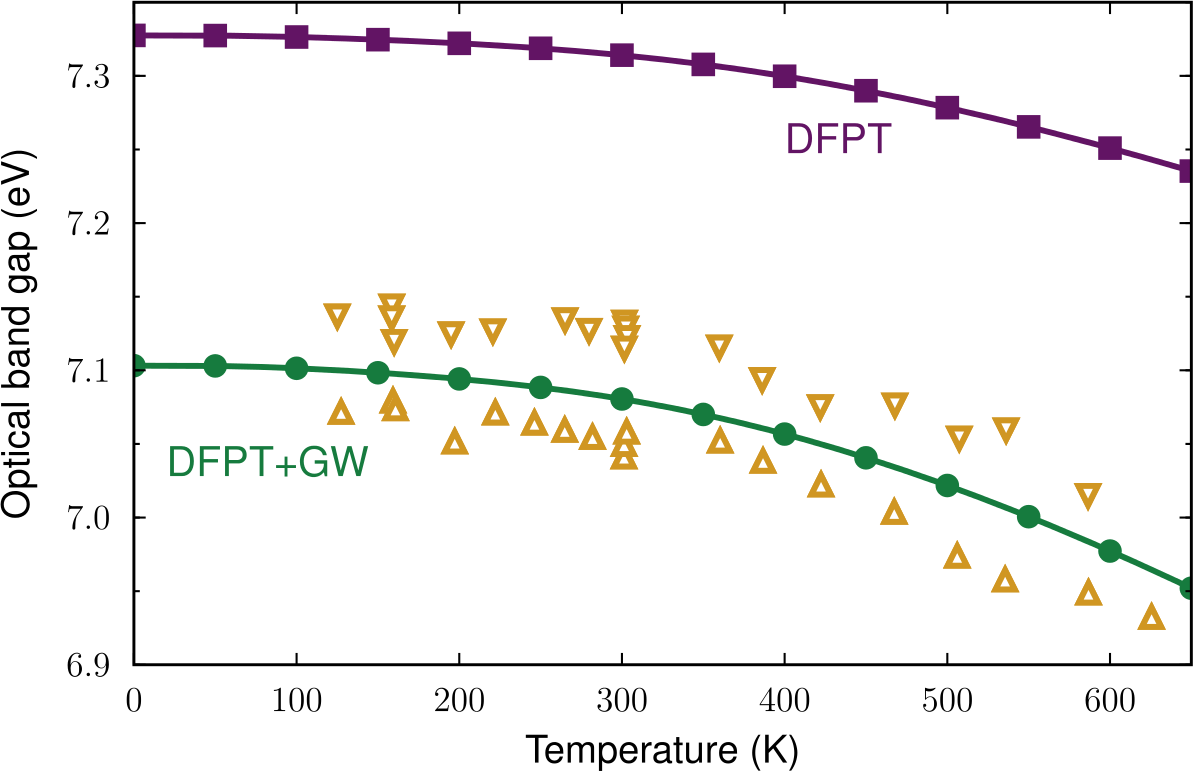}
  \caption[fig]{\label{fig.antonius2014} (Color online)
  Temperature dependence of the direct band gap of diamond calculated using
  the Allen-Heine theory. The upper curve shows the results obtained
  within DFPT at the LDA level. The lower
  curve was obtained via $GW$ calculations in the frozen-phonon approach.
  Triangles are experimental data. The zero-point renormalization calculated by including $GW$ quasiparticle
  corrections is 628~meV.
  Reproduced with permission from \cite{Antonius2014}, copyright
  (2014) by the American Physical Society. 
  }
  \end{figure}

\textcite{Antonius2014} revisited the electron-phonon interaction in diamond by assessing
the reliability of the rigid-ion approximation
and the importance of many-body $GW$ quasiparticle 
corrections to the DFT/LDA band structure. The main findings were that
the rigid-ion approximation introduces a very small error in diamond, of the order of $\sim$10~meV, 
while $GW$ quasiparticle corrections can increase the electron-phonon renormalization of 
the band gap by as much as $\sim$200~meV. The temperature dependence of the band gap of diamond calculated 
by \citeauthor{Antonius2014} is shown in Fig.~\ref{fig.antonius2014}.

Further work on diamond was reported by \textcite{Ponce2014b}, who compared {\it ab~initio}
calculations based on the Allen-Heine theory with explicit frozen-phonon calculations
(Sec.~\ref{sec.nonpert}). The corrections to the rigid-ion approximation were
found to be smaller than 4~meV in all cases considered. \textcite{Ponce2014b,Ponce2014} also
reported a detailed assessment of the accuracy of the various levels of approximation in the
calculation of the zero-point renormalization of energy levels, as well as a thorough comparison
between the results of different first-principles implementations.

The electron-phonon renormalization of band structures was also investigated
in a number of other systems. For example \textcite{Kaway2014} studied zinc-blend GaN by
combining the Allen-Heine theory  with the Bethe-Salpeter approach. 
\textcite{Ponce2015} investigated silicon, diamond, BN, $\alpha$-AlN, and $\beta$-AlN
using both the adiabatic version of the Allen-Heine theory and the non-adiabatic
Green's function method of Eqs.~(\ref{eq.fan-final})-(\ref{eq.fan-final-T}). 
\textcite{Friedrich2015} investigated the zero-point renormalization
in LiNbO$_3$ using the adiabatic Allen-Heine theory.
\textcite{Villegas2016} studied the anomalous temperature dependence of the band gap
of black phosphorous.
\textcite{Antonius2015}
investigated diamond, BN, LiF, and MgO, focusing on the dynamical aspects and the spectral
function (see Sec.~\ref{sec.kinks}). The works by \textcite{Ponce2015,Antonius2015} 
were the first to report complete band structures at finite temperature.

In \cite{Ponce2015} the authors paid special attention to the numerical convergence of 
the self-energy integrals with respect to the limit $\eta\rightarrow 0$ of the broadening 
parameter; they noted that in the case of polar crystals the adiabatic correction to the
electron energies of band extrema, as given by Eq.~(\ref{eq.AH-RS}), diverges in the limit 
of dense Brillouin-zone sampling. This behavior stems from the polar singularity in the
electron-phonon matrix elements, Eq.~(\ref{gL}). In fact, near band extrema the
bands are approximately parabolic, and the integrand in the adiabatic Fan-Migdal self-energy
goes as $q^{-4}$ for $q\rightarrow 0$, while the volume element goes only as
$d\bq = 4\pi q^2 dq$. 
This problem can be avoided by first performing the integration over $\bq$ in principal value, without
neglecting phonon frequencies, and then taking the limit $\w_{\bq\nu}\rightarrow 0$
so as to recover the adiabatic approximation (\citeauthor{Fan1951}, \citeyear{Fan1951}, Sec.~IV); 
in this way the adiabatic
approximation can still be employed without incurring into a singularity in the calculations. 
A practical strategy to correctly perform the principal value integration
in first-principles calculations
was recently proposed by \textcite{Nery2016}; here the authors treat the singularity
via an explicity analytic integration near $\bq=0$.
The complications arising in polar materials 
can also be avoided at once by using directly the more accurate expression
in Eq.~(\ref{eq.AH-BW}) based on Brillouin-Wigner perturbation theory, or even better by
calculating the spectral functions as in \cite{Kaway2014,Antonius2015}. In particular,
Eq.~(\ref{eq.AH-BW}) shows that in more accurate approaches the infinitesimal 
$\eta$ should be replaced by the finite {\it physical} linewidth $\Gamma_{n\bk}$.

Although temperature-dependent band structures of polar materials were recently reported 
\cite{Kaway2014,Ponce2015,Antonius2015}, the specific role
of the Fr\"ohlich coupling discussed in Sec.~\ref{sec.polar} received only little attention
so far. The only {\it ab~initio} investigations which specifically addressed the role 
of polar couplings in this context are from \textcite{Botti2013} and \textcite{Nery2016}.
In order to understand the strategy of \citeauthor{Botti2013}, 
we refer to the Hedin-Baym equations in Sec.~\ref{sec.GW-1}.
\citeauthor{Botti2013} proposed that, instead of splitting the screened Coulomb
interaction $W$ into electronic and nuclear contributions as in Eq.~(\ref{eq.W=We+Wp}), 
one could try to directly calculate the screened Coulomb interaction $W$ {\it including}
the lattice screening, as in Eqs.~(\ref{eq.selfen-numbers}) and (\ref{eq.W-numbers}).
In order to make the calculations tractable, \citeauthor{Botti2013} evaluated
the total dielectric matrices using a simplified model based on the
Lyddane-Sachs-Teller relations. The resulting formalism combines $GW$ calculations
and experimentally measured LO-TO splittings. 
The zero-point renormalization of the band gaps calculated by \citeauthor{Botti2013} for
LiF, LiCl, NaCl, and MgO were all $>1$~eV. This is an interesting result and deserves
further investigation. We note incidentally that the Allen-Heine theory
and that of \citeauthor{Botti2013} can both be derived from the Hedin-Baym equations. Therefore 
the approach of \citeauthor{Botti2013} should effectively correspond to calculating the Fan-Migdal 
self-energy by retaining only the long-range part of the polar electron-phonon matrix elements.
In the case of the work by \textcite{Nery2016}, the authors reported a
Fr\"ohlich contribution to the zero-point renormalization of the band gap of GaN of 45~meV,
to be compared with the total renormalization arising from all modes of 150~meV.

\subsubsection{Non-perturbative adiabatic calculations}\label{sec.nonpert}

An alternative approach to the calculation of temperature-dependent band structures
consists of avoiding perturbation theory and electron-phonon matrix elements altogether,
and replacing the entire methodology discussed in Sec.~\ref{sec.temper} by straightforward 
finite-differences calculations. To see how this alternative strategy works we perform
a Taylor expansion of the Kohn-Sham eigenvalues $\ve_{n\bk}$ to second order
in the atomic displacements $\Delta \tau_{\k\a p}$, and then average the result on a nuclear
wavefunction identified by the quantum numbers $\{n_{\bq\nu}\}$. After using Eq.~(\ref{eq.tau-from-x})
one obtains:
  \begin{equation}\label{eq.AH-findiff}
  \<\ve_{n\bk}\>_{\{n_{\bq\nu}\}} = \ve_{n\bk} +
  {\sum}_\nu\!\!\int\!\!\frac{d\bq}{\Omega_{\rm BZ}} (n_{\bq\nu}+1/2) \frac{\D \ve_{n\bk}}{\D n_{\bq\nu}},
  \end{equation}
where we used the formal definition $\D/\D n_{\bq\nu} = \Delta_{\bq\nu}\Delta_{-\bq\nu}$,
and the variations $\Delta_{\bq\nu}$ are the same as 
in Eqs.~(\ref{eq.dV-all-1})-(\ref{eq.dV-all-3}). The nuclear wavefunctions are obtained
from the ground-state in Eq.~(\ref{eq.phonon-gs}) by applying the ladder operators,
as discussed in Appendix~\ref{sec.normalcoord}. The above expression can be generalized to finite temperature
by considering a canonical average over all possible nuclear states. The result 
maintains the same form as in Eq.~(\ref{eq.AH-findiff}), except that we now have 
the Bose-Einstein occupations $n_{\bq\nu}(T)$ (see footnote~\ref{note.FDBE}).
Equation~(\ref{eq.AH-findiff}) is precisely the conceptual starting point of the 
Allen-Heine theory of temperature-dependent band structures, and appeared for the first
time in \cite{Allen1981}. If the variations
$\Delta_{\bq\nu}\Delta_{-\bq\nu} \ve_{n\bk}$ are calculated in second-order perturbation
theory, one obtains precisely the formalism of \citeauthor{Allen1976}.

It has been proposed that the coefficients $\D \ve_{n\bk}/\D n_{\bq\nu}$ could alternatively
be obtained from the derivatives of the vibrational frequencies with respect to the electronic 
occupations, $\hbar \D \w_{\bq\nu}/\D f_{n\bk}$ \cite{Allen1979,KingSmith1989,Ponce2014b}. A formal
derivation of the link between these alternative approaches can be found 
in (\citeauthor{Allen1979}, \citeyear{Allen1979}, Appendix; the authors refer to this as Brooks' theorem). 
Incidentally, the first {\it ab~initio} calculation of temperature-dependent
band gaps relied on this approach \cite{KingSmith1989}.

Most commonly, the coefficients $\D \ve_{n\bk}/\D n_{\bq\nu}$ in Eq.~(\ref{eq.AH-findiff}) are evaluated 
using frozen-phonon supercell calculations, via the second derivative
of the eigenvalue $\ve_{n\bk}$ with respect to collective atomic displacements along
the vibrational eigenmodes $e_{\k\a\nu}(\bq)$.
This approach was employed by \textcite{Capaz2005} to study the temperature dependence
of the band gaps in carbon nanotubes (within a tight-binding model), and by \textcite{Bester2013} 
to obtain the zero-point renormalization and temperature dependence of the gaps of silicon 
and diamond quantum dots.
Recent examples include works on diamond, silicon, SiC \cite{Monserrat2014b}, 
as well as CsSnI$_3$ \cite{Patrick2015}.

Frozen-phonon supercell calculations based on Eq.~(\ref{eq.AH-findiff}) carry the advantage
that the rigid-ion approximation, which is necessary to obtain 
Eqs.~(\ref{eq.dw-giustino2010})-(\ref{eq.2010-3}), is no longer required. Therefore this
approach is more accurate in principle. In practice, however, the calculations are challenging as they 
require large supercells, and the derivatives must be evaluated for every vibrational mode of the supercell.
Several computational strategies were developed to tackle this 
challenge. \textcite{Patrick2013} proposed to perform the averages leading 
to Eq.~(\ref{eq.AH-findiff}) via importance-sampling Monte Carlo integration.
\textcite{Monserrat2016} described a constrained Monte Carlo scheme which improves 
the variance of the Monte Carlo estimator. 
Recently \textcite{Zacharias2016} showed that it is possible to 
perform these calculations more efficiently by replacing the stochastic sampling
by a suitable choice of an `optimum' configuration; the result becomes exact in the
thermodynamic limit of large supercells. In order to reduce the computational cost
associated with the use of large supercells,
\textcite{Lloyd2015} introduced `non-diagonal' supercells,
which allow one to access phonon wavevectors belonging to a uniform grid of $N_p$ points
using supercells containing only $N_p^{1/3}$ unit~cells.

The merit of these non-perturbative approaches is that they treat explicitly the nuclear wavefunctions,
and enable exploring effects which go beyond the Allen-Heine theory.
For example \textcite{Monserrat2013,Monserrat2014,Monserrat2015b} and \textcite{Engel2015}
were able to investigate
effects beyond the harmonic approximations in several systems, such as LiH, LiD, high-pressure He, 
molecular crystals of CH$_4$, NH$_3$, H$_2$O, HF, as well as Ice. 
In all these cases the authors found large zero-point effects on the band gaps.

Finally, we mention that the calculation of electronic properties at finite temperature 
via the Allen-Heine theory and its variants is closely related to what one would obtain 
using path-integral molecular dynamics simulations \cite{DellaSala2004,Ramirez2006,Ramirez2008}, 
or even classical molecular dynamics simulations at high enough temperatures \cite{Franceschetti2007}. 

\subsection{Phonon-assisted optical absorption}\label{sec.phon-ass}

In addition to modifying the electron energy levels in solids, the electron-phonon interaction
plays an important role in the optical properties of semiconductors and insulators, as it is
responsible for phonon-assisted optical transitions. Phonon-assisted
processes could be analyzed by considering the many-body electronic screening function $\epsilon_{\rm e}(12)$
in Eq.~(\ref{eq.eps-hedin-el}), by using the electron Green's function $G$ dressed by the electron-phonon
self-energy $\Sigma^{\rm ep}$ as in Eq.~(\ref{eq.dyson-ep}). Since this would require us a lengthy
detour, here we simply reproduce the standard result of second-order
time-dependent perturbation theory \cite{Bassani1975,Ridley1993}:
\begin{eqnarray}
  &&\a(\omega) = 
  \frac{\pi e^2}{\epsilon_0 \,c \,\Omega}\frac{1}{\w \,n_{\rm r}(\w)}
   \int \!\!\frac{d \bk\, d \bq}{\Omega_{\rm BZ}^2} \sum_{mn\nu}\sum_{s=\pm 1} 
  (f_{n{\bf k}}-f_{m{\bf k+q}}) \nonumber \\ 
  &&\times \Bigg| {\bf e}\!\cdot \!{\sum}_p\!\left[ 
  \frac{{\bf v}_{np}({\bf k})g_{pm\nu}({\bf k},{\bf q})}{\varepsilon_{p{\bf k}}
  -\varepsilon_{n{\bf k}}-\hbar\omega} + \frac{g_{np\nu}({\bf k},{\bf q}){\bf v}_{pm}
  ({\bf k}+{\bf q})}{\varepsilon_{p{\bf k+q}}-\varepsilon_{n{\bf k}} +s\hbar\omega_{{\bf q}\nu}}
  \right]\!\!\Bigg|^{2} \nonumber \\
  &&\times \left( n_{{\bf q}\nu}+1/2+s/2\right)
    \delta(\varepsilon_{m{\bf k+q}}\!-\!\varepsilon_{n{\bf k}}
  \!-\!\hbar\omega \!+\!s\hbar\omega_{{\bf q}\nu}).  \label{eq.hbb}
  \end{eqnarray}
In this expression $\a(\w)$ is the absorption coefficient for visible light, {\bf e} is the photon 
polarization, ${\bf v}_{mn}$ are the matrix elements of the electron velocity operator, and $n_{\rm r}(\w)$ is 
the real part of the refractive index. The two denominators in the second line corresponds to indirect
processes whereby a photon is absorbed and a phonon is absorbed or emitted (left), and processes
whereby a phonon is absorbed or emitted, and subsequently a photon is absorbed (right). The above expression
relies on the electric dipole approximation and is therefore valid for photon energies up to a few 
electronvolts. The theory leading to Eq.~(\ref{eq.hbb}) was originally 
developed by Hall, Bardeen, and Blatt (\citeyear{Hall1954}).

  \begin{figure}[t!]
  \includegraphics[width=0.7\columnwidth]{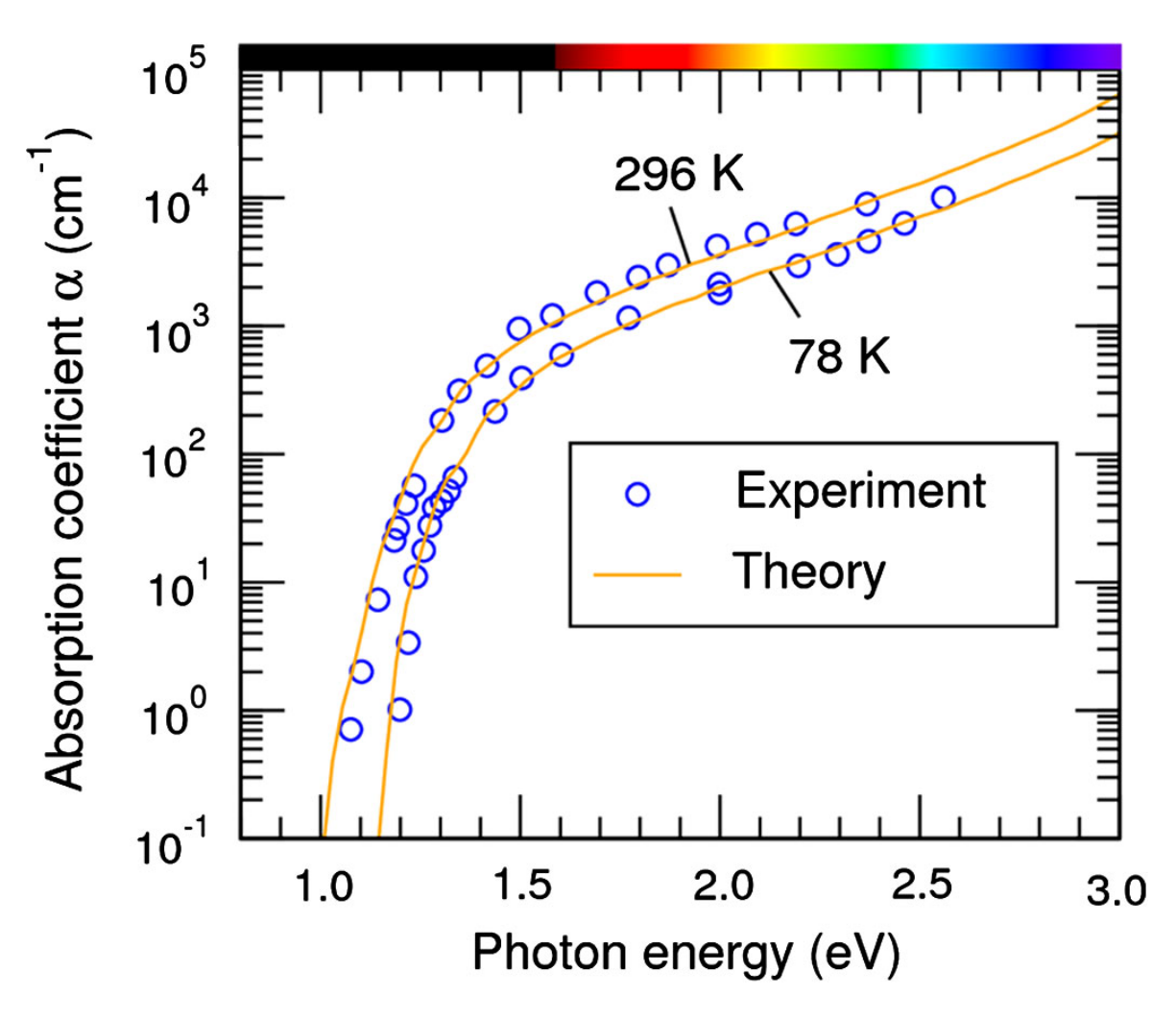}
  \caption[fig]{\label{fig.indirect} (Color online)
  Phonon-assisted optical absorption in silicon: comparison between first-principles calculations
  (solid lines, orange) and experiment (circles, blue). The calculations were performed using the theory of
  \textcite{Hall1954}, as given by Eq.~(\ref{eq.hbb}). Spectra calculated at different temperatures were
  shifted horizontally so as to match the experimental onsets.
  Reproduced with permission from \cite{Noffsinger2012}, copyright
  (2012) by the American Physical Society.
  }
  \end{figure}

The first {\it ab~initio} 
calculation employing Eq.~(\ref{eq.hbb}) was reported by \textcite{Noffsinger2012} for the prototypical
case of silicon. The authors employed DFT for computing phonons and electron-phonon matrix
elements, and the $GW$ method for the quasiparticle band structures. The sampling of the Brillouin
zone was achieved by means of the interpolation strategy described in Sec.~\ref{sec.wannier}.
Figure~\ref{fig.indirect} shows that the calculations by \citeauthor{Noffsinger2012}
are in very good agreement with experiment throughout the energy range of indirect absorption.

Further work along similar lines was reported by \textcite{Kioupakis2010}, who calculated the
indirect optical absorption by free carriers in GaN; and \textcite{Kioupakis2015b}, who studied
the indirect absorption by free carriers in the transparent conducting oxide SnO$_2$. Recently,
the {\it ab~initio} theory of phonon-assisted absorption was also extended to the case of indirect Auger
recombination by \textcite{Kioupakis2015a}.

One limitation of the theory by \citeauthor{Hall1954} is that
the indirect absorption onset is {\it independent} of temperature. This is seen by noting
that the Dirac delta functions in Eq.~(\ref{eq.hbb}) contain the band structure energies at {\it clamped} nuclei.
The generalization to incorporate temperature-dependent band structures as discussed
in Sec.~\ref{sec.temper} is nontrivial. \textcite{Patrick2014,Zacharias2016} 
showed that the electron-phonon renormalization of the band structure
modifies the energies of {\it real} transitions but leaves unchanged the energies
of {\it virtual} transitions; in other words the Allen-Heine renormalization
should be incorporated only in the Dirac delta functions and
in the first denominator in Eq.~(\ref{eq.hbb}).
In order to avoid these complications at once, \textcite{Zacharias2015} developed an alternative approach
which relies on the `semiclassical' approximation of \textcite{Williams1951} and \textcite{Lax1952}.
In this approximation, the initial states in the optical transitions are described quantum-mechanically,
and the final states are replaced by a quasiclassical continuum. In the formulation
of \citeauthor{Zacharias2015} the imaginary part of the temperature-dependent dielectric function takes the form:
  \begin{equation}\label{eq.marios}
  \e_2(\w;T)  =\frac{1}{Z}{\sum}_{\{n_{\bq\nu}\}} e^{-E_{\{n_{\bq\nu}\}} /k_{\rm B}T}
  \< \e_2(\w) \>_{\{n_{\bq\nu}\}},
  \end{equation}
where $\e_2(\w)$ denotes the imaginary part of the dielectric function at clamped nuclei,
and the expectation values have the same meaning as in Eq.~(\ref{eq.AH-findiff}).
$E_{\{n_{\bq\nu}\}}$ is the energy of the quantum nuclear state specified by the
quantum number $\{n_{\bq\nu}\}$ and $Z$ is the canonical partition function.
\citeauthor{Zacharias2015} demonstrated that this approach provides an {\it adiabatic} approximation
to Eq.~(\ref{eq.hbb}), and seamlessly includes the temperature dependence of the electronic structure
within the Allen-Heine theory. Using techniques similar to those of Sec.~\ref{sec.nonpert}, 
the authors calculated the indirect optical absorption lineshape of silicon
at various temperatures and obtained very good agreement with experiment. 
These results were recently extended to the temperature-dependent optical
spectra of diamond and gallium arsenide by \textcite{Zacharias2016}.

\section{Carrier dynamics and transport}\label{sec.transport}

\subsection{Electron linewidths and lifetimes}\label{sec.transport-lifet}

In Sec.~\ref{sec.el-lifetime} we have seen how the Fan-Migdal self-energy can be used in order
to evaluate the quasiparticle lifetimes (or equivalently linewidths) resulting from the 
electron-phonon interaction. The first {\it ab~initio} calculations of this kind were reported by 
\textcite{Eiguren2002,Eiguren2003} in the study of the decay of metal surface states, and 
by \textcite{Sklyadneva2005,Sklyadneva2006} and \textcite{Leonardo2007} in the study of electron 
lifetimes of elemental metals. Some of these calculations and the underlying approximations 
were reviewed in Sec.~\ref{sec.kinks}. Calculations of quasiparticle linewidths
were also employed to study the temperature-dependent broadening of the optical
spectra in semiconductors. For example \textcite{Giustino2010} and \textcite{Ponce2015}
investigated the broadening of the direct absorption edge of diamond and silicon, respectively.
In both cases good agreement with experiment was obtained. More recently, the same approach was
employed to study the broadening of photoluminescence peaks in lead-iodide perovskites \cite{Herz2016}.
In this case, it was found that the standard Fermi golden rule expression, Eq.~(\ref{eq.fermirule}), 
significantly overestimates the experimental data. The agreement with experiment is restored
by taking into account the quasiparticle renormalization $Z_{n\bk}$; see discussion following
Eq.~(\ref{eq.Zfac}). 

\begin{figure}[t!]
  \includegraphics[width=\columnwidth]{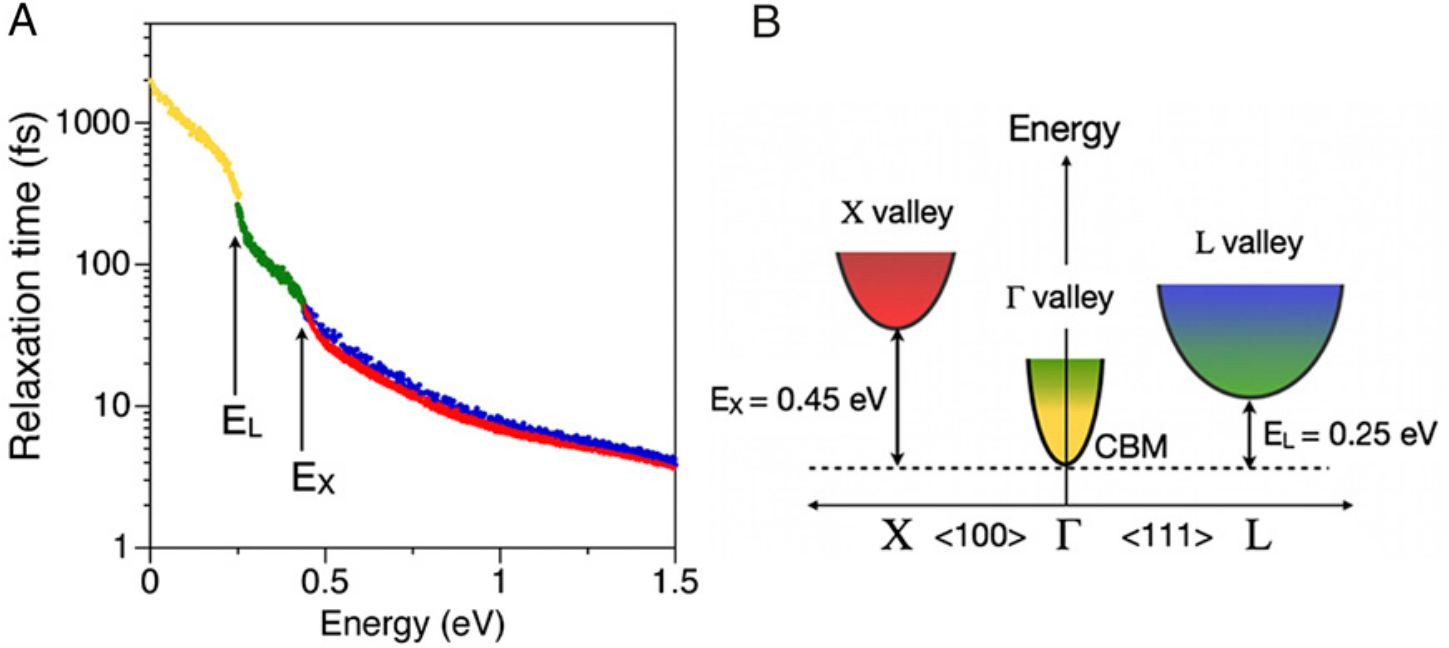}
  \caption[fig]{\label{fig.bernardi2015} (Color online)
  Electron relaxation times in GaAs resulting from electron-phonon scattering.
  (a)~Calculated relaxation times as a function of electron energy with respect to the conduction
  band bottom. The color code (gray shades) of the data points identifies the valley where each electronic
  state belongs. (b)~Schematic representation of the conduction band valleys in GaAs.
  Reproduced with permission from \cite{Bernardi2015}, copyright
  (2015) by the National Academy of Sciences.
  }
  \end{figure}

While these works were primarily concerned with the broadening of the spectral lines
in photoemission or optical experiments, Eq.~(\ref{eq.fermirule}) can also be used 
to study carrier lifetimes in time-resolved experiments. 
The first {\it ab~initio} study in this direction was reported from \textcite{Sjakste2007},
who investigated the thermalization of hot electrons in GaAs and the exciton lifetimes in GaP.
In the case of GaAs, \citeauthor{Sjakste2007} found thermalization rates
in quantitative agreement with time-resolved luminescence and transient optical absorption measurements.
Work along similar lines was also reported for the intervalley scattering times in Ge \cite{Tyuterev2011}.
Recently, the thermalization rates of hot electrons in GaAs were revisited by \textcite{Bernardi2015}.
The authors employed Eq.~(\ref{eq.fermirule}) and the Wannier interpolation technique described 
in Sec.~\ref{sec.wannier} in order to finely sample the electron-phonon scattering processes near the
bottom of the conduction band, see Fig.~\ref{fig.bernardi2015}. Based on these calculations 
they were able to interpret transient absorption measurements in terms of the carrier lifetimes
within each valley. Another interesting application of Eq.~(\ref{eq.fermirule}) was reported
by \textcite{Bernardi2014}, who investigated the rate of hot carrier thermalization in silicon,
within the context of photovoltaics applications. 

Very recently \textcite{Sangalli2015,Sangalli2015b} employed the lifetimes calculated using
Eq.~(\ref{eq.fermirule}) in order to study carrier dynamics in silicon in real time.
Strictly speaking, these developments lie outside of the scope of equilibrium Green's 
functions discussed in Sec.~\ref{sec.green}, and require concepts based on 
non-equilibrium Green's functions \cite{Kadanoff1962}. However, the basic ingredients of the 
electron-phonon calculations remain unchanged.

In all calculations discussed in this section, the electron-phonon matrix elements were obtained
within DFT. However, in order to accurately describe electron-phonon scattering near band extrema 
\textcite{Bernardi2014,Bernardi2015} and \textcite{Herz2016} employed $GW$ quasiparticle band structures.
This is important in order to obtain accurate band effective masses, which affect
the carrier lifetimes via the density of states.

\subsection{Phonon-limited mobility}

The carrier lifetimes $\tau_{n\bk}$ of Eq.~(\ref{eq.fermirule}) are also useful in the calculation
of electrical mobility, conductivity, and resistivity, within the context of the semiclassical
model of electron dynamics in solids \cite{Ashcroft1976}. In the semiclassical model, one describes
the electronic response to an external perturbation by taking the fermionic occupations $f_{n\bk}$ 
to represent the probability density function in the phase space defined by the unperturbed band 
structure. The probabilities $f_{n\bk}$ are then determined using a standard Boltzmann equation.
A comprehensive discussion of these methods can be found in the classic book of \textcite{Ziman1960}.

Here we only touch upon the key result which is needed in {\it ab~initio} calculations
of electrical conductivity. In the semiclassical model, the electrical current is calculated as
${\bf J} = -2 e \,\Omega_{\rm BZ}^{-1}\sum_n\int \!d\bk\, \bv_{n\bk} f_{n\bk}$. In the absence of
external fields and thermal gradients, the occupations $f_{n\bk}$ reduce to the standard Fermi-Dirac 
occupations at equilibrium, $f_{n\bk}^0$, and the current vanishes identically. Upon introducing an electric
field $\bE$, the electrons respond by adjusting their occupations. In this model it is assumed that
the variation $f_{n\bk}-f_{n\bk}^0$ is so small that the electronic density is essentially
the same as in the unperturbed system. The
modified occupations can be calculated using the linearized Boltzmann transport equation \cite{Ziman1960}:
  \begin{eqnarray}
  && \frac{\D f_{n\bk}^0}{\D\ve_{n\bk}} \,\bv_{n\bk} \cdot (-e) \bE
  = -{\sum}_\nu\!\int\!\!\frac{d\bq}{\Omega_{\rm BZ}} 
  \,\Gamma_{mn\nu}(\bk,\bq) \nonumber \\
  && \qquad\qquad \times \left[ (f_{n\bk}-f_{n\bk}^0)-(f_{m\bk+\bq}-f_{m\bk+\bq}^0)\right]\!,\,\,\,\,\,
   \label{eq.boltz.1}
  \end{eqnarray}
where the kernel $\Gamma_{mn\nu}(\bk,\bq)$ is defined as:
  \begin{eqnarray}
  &&\Gamma_{mn\nu}(\bk,\bq) = {\sum}_{s=\pm 1}
      \frac{2\pi}{\hbar}|g_{mn\nu}(\bk,\bq)|^2 f^0_{n\bk}(1-f^0_{m\bk+\bq})\nonumber \\
      &&\qquad\times( n_{\bq\nu}\!+\!1/2\! -\!s/2) \,\d(\ve_{n\bk}+s\hbar\w_{\bq\nu}-\ve_{m\bk+\bq}).
      \qquad\,\, \label{eq.boltz.2}
  \end{eqnarray}
The left-hand side of Eq.~(\ref{eq.boltz.1}) represents the collisionless term of the Boltzmann
equation, that is the change in occupations due to the particle drift under the electric field.
The right-hand side
represents the change of occupations resulting from electrons scattered in or out of
the state $|n\bk\>$ by phonon emission or absorption. The rates in Eq.~(\ref{eq.boltz.2}) are
simply derived from Fermi's golden rule \cite{Grimvall1981}. 
By solving Eq.~(\ref{eq.boltz.1}) self-consistently for all $f_{n\bk}$ 
it is possible to calculate the current, and from there the conductivity. 
The connection with the carrier lifetimes $\tau_{n\bk}$ of Eq.~(\ref{eq.fermirule}) is obtained
within the so-called `energy relaxation time approximation'. In this approximation the incoming
electrons are neglected in Eq.~(\ref{eq.boltz.1}), that is the last term $(f_{m\bk+\bq}-f_{m\bk+\bq}^0)$
in the second line is ignored. 
As a result the entire right-hand side of the equation simplifies to $-(f_{n\bk}-f_{n\bk}^0)/\tau_{n\bk}$.

\begin{figure}[t!]
  \includegraphics[width=0.8\columnwidth]{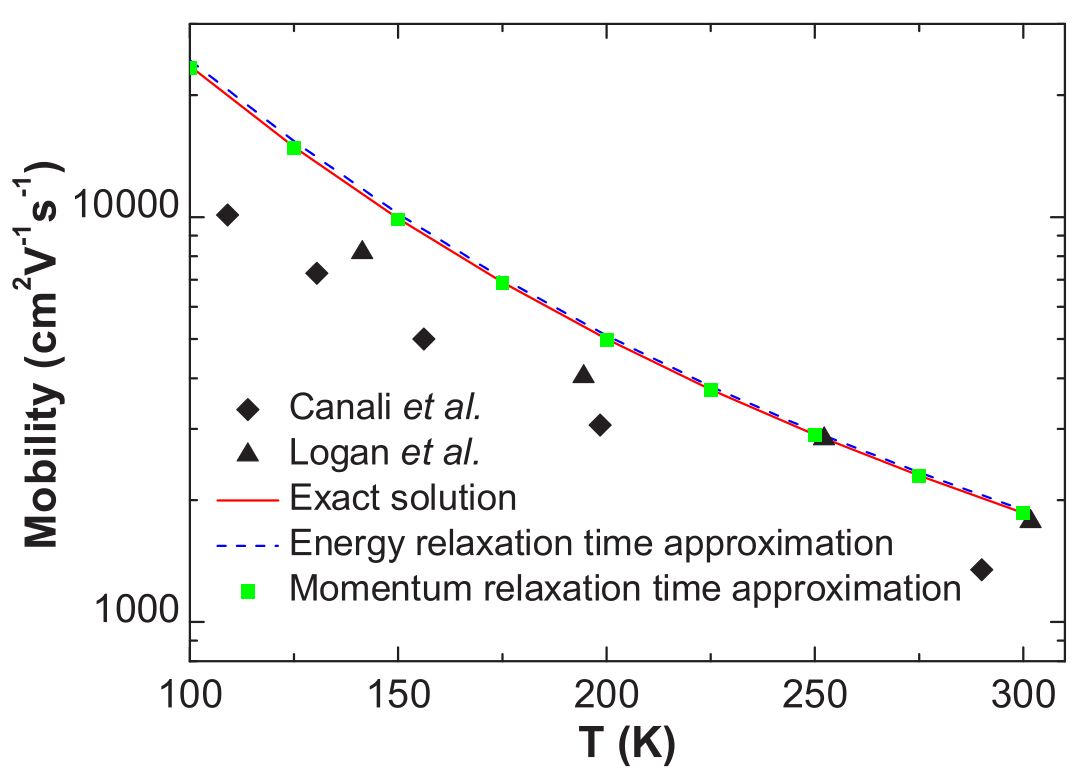}
  \caption[fig]{\label{fig.li2015} (Color online)
  Temperature-dependent mobility of $n$-type silicon. The solid line (red) indicates the mobility
  calculated using the linearized Boltzmann transport equation, Eq.~(\ref{eq.boltz.1});
  the dashed line (blue) corresponds to the energy relaxation time approximation.
  The triangles and diamonds are experimental data points. 
  Reproduced with permission from \cite{Li2015}, copyright
  (2015) by the American Physical Society.
  }
  \end{figure}

The direct solution of Eq.~(\ref{eq.boltz.1}) is computationally challenging, and fully {\it ab~initio} 
calculations were reported only very recently by \textcite{Li2015} for Si, MoS$_2$, and Al,
and by \textcite{Fiorentini2016} for $n$-doped Si.
Figure~\ref{fig.li2015} shows that the mobility of $n$-type silicon calculated by \citeauthor{Li2015} 
is in good agreement with experiment. The theory overestimates the measured 
values to some extent, and this might have to do with the limitations of the DFT matrix elements 
(see Sec.~\ref{sec.beyonddft}).
In addition to the carrier mobility, \citeauthor{Fiorentini2016} employed the
{\it ab~initio} Boltzmann formalism to calculate thermoelectric properties,  
such as the Lorenz number and the Seebeck coefficient.

The first {\it ab~initio} calculation of mobility was reported by \textcite{Restrepo2009} for 
the case of silicon, within the energy relaxation time approximation. Other recent calculations
using various approximations to Eq.~(\ref{eq.boltz.1})
focused on silicon 
(\citeauthor{Wang2011}, \citeyear{Wang2011}; \citeauthor{Liao2015}, \citeyear{Liao2015}), 
graphene 
(\citeauthor{Borysenko2010}, \citeyear{Borysenko2010};
\citeauthor{Park2014}, \citeyear{Park2014};
\citeauthor{Sohier2014}, \citeyear{Sohier2014};
\citeauthor{Restrepo2014}, \citeyear{Restrepo2014};
\citeauthor{Gunst2016}, \citeyear{Gunst2016};
\citeauthor{Kim2016}, \citeyear{Kim2016}),
MoS$_2$ 
(\citeauthor{Kaasbjerg2012}, \citeyear{Kaasbjerg2012};
\citeauthor{Li2013}, \citeyear{Li2013};
\citeauthor{Restrepo2014}, \citeyear{Restrepo2014};
\citeauthor{Gunst2016}, \citeyear{Gunst2016}), 
silicene 
(\citeauthor{Li2013}, \citeyear{Li2013}; \citeauthor{Gunst2016}, \citeyear{Gunst2016}),
SrTiO$_3$ and KTiO$_3$ (\citeauthor{Himmetoglu2014}, \citeyear{Himmetoglu2014}; 
\citeauthor{Himmetoglu2016}, \citeyear{Himmetoglu2016}).

{\it Ab initio} calculations of the resistivity of metals are less challenging than for semiconductors, 
and started appearing already with the work of \textcite{Bauer1998}. Most calculations on metals
are based on Ziman's resistivity formula, see \citeauthor{Grimvall1981} (\citeyear{Grimvall1981}, p.~210).
An interesting recent example can be found in the work by \textcite{Xu2014} on the transport coefficients of
lithium. We also highlight related work on phonon-limited transport
in organic crystals (\citeauthor{Hannewald2004}, \citeyear{Hannewald2004}, \citeyear{Hannewald2004b};
\citeauthor{Ortmann2009}, \citeyear{Ortmann2009};
\citeauthor{Vukmirovic2012}, \citeyear{Vukmirovic2012}).

\section{Phonon-mediated superconductors}\label{sec.supercond}

The last application of {\it ab~initio} calculations of EPIs
that we will consider is the study of phonon-mediated superconductivity \cite{Schrieffer1983}. 
This research field is so vast
that any attempt at covering it in a few pages would not make justice to the subject.
For this reason, it was decided to limit the discussion to those novel theoretical
and methodological developments which are aiming at fully {\it predictive} calculations,
namely the `anisotropic Migdal-Eliashberg theory' (Sec.~\ref{sec.eliashberg}),
and the `density functional theory for superconductors' (Sec.~\ref{sec.scdft}). 
For completeness, in Sec.~\ref{sec.sc-standard} we also summarize the most popular 
equations employed in the study of phonon-mediated superconductors.
All calculations described in this section were performed at the DFT level.

\subsection{McMillan/Allen-Dynes formula} \label{sec.sc-standard}

Most {\it ab~initio} calculations on phonon-mediated superconductors
rely on a semi-empirical expression for the critical temperature, first introduced
by \textcite{McMillan1968} and then refined by \textcite{Allen1975}:
  \begin{equation}\label{eq.mcmillan}
  \kt_{\rm c} = \frac{\hbar\w_{\rm log}}{1.2}\exp\left[-\frac{1.04(1+\lambda)}{\lambda 
  - \mu^* (1+0.62\,\lambda)} \right].
  \end{equation}
Here $T_{\rm c}$ is the superconducting critical temperature, $\w_{\rm log}$ is a
`logarithmic' average of the phonon frequencies \cite{Allen1975}, $\lambda$ is the
electron-phonon `coupling strength', and $\mu^*$ is a parameter describing the Coulomb repulsion.
The functional form of Eq.~(\ref{eq.mcmillan}) was derived by \citeauthor{McMillan1968}
by determining an approximate solution of the Eliashberg gap equations (see Sec.~\ref{sec.eliashberg}).
$\lambda$ and $\w_{\rm log}$ are calculated from the isotropic
version of the Eliashberg function in Eq.~(\ref{eq.eliashb-func}) as follows
(\citeauthor{McMillan1968}, \citeyear{McMillan1968};
\citeauthor{Allen1975}, \citeyear{Allen1975}; \citeauthor{Grimvall1981}, \citeyear{Grimvall1981};
\citeauthor{AllenMitrovic}, \citeyear{AllenMitrovic}):
  \begin{eqnarray}
  \a^2F(\w) &=&
  \frac{1}{N_{\rm F}}\!
 \int \!\frac{d\bk\,d\bq}{\Omega_{\rm BZ}^2} \sum_{mn\nu}
    |g_{mn\nu}(\bk,\bq)|^2\nonumber \\ 
    &\times&\d(\ve_{n\bk}\!-\!\ve_{\rm F})\d(\ve_{m\bk+\bq}\!-\!\ve_{\rm F}) \d(\hbar\w\!-\!\hbar\w_{\bq\nu}),
   \qquad  \\
   \lambda &=& 2\int_0^\infty \frac{\a^2F(\w)}{\w}d\w, \label{eq.lambda-sc}\\
  \w_{\rm log} &=& \exp\left[\frac{2}{\lambda}\int_0^\infty\!\! d\w \,\frac{\a^2F(\w)}{\w}\log \w\right],
   \label{eq.wlog}
  \end{eqnarray}
where $N_{\rm F}$ is the density of states at the Fermi level and the matrix elements $g_{mn\nu}(\bk,\bq)$
are the same as in Eq.~(\ref{eq.matel}).
The remaining parameter $\mu^*$ \cite{Morel1962}
is obtained as $1/\mu^* = 1/\mu + \log(\w_{\rm p}/\w_{\rm ph})$,
where $\hbar\w_{\rm p}$ is the characteristic plasma energy of the system, $\hbar\w_{\rm ph}$
the largest phonon energy, and $\mu$ is the average electron-electron Coulomb repulsion across
the Fermi surface. More specifically: 
$\mu = N_{\rm F}\<\< V_{n\bk,n'\bk'} \>\>_{\rm FS}$, where $\<\<\cdots\>\>_{\rm FS}$
denotes a double average over the Fermi surface, and
$V_{n\bk,n'\bk'}=\< \bk' n',-\bk'n'|W|\bk n,-\bk n\>$, with  
$W$ being the screened Coulomb interaction of Sec.~\ref{sec.sci-J} 
(\citeauthor{Lee1995}, \citeyear{Lee1995}; \citeauthor{Lee1996}, \citeyear{Lee1996}).\label{pag.coul}

The coupling strength $\lambda$ is related to the mass enhancement parameter $\lambda_{n\bk}$ 
discussed in Sec.~\ref{sec.mass-enhancement}.
The main difference between $\lambda$ and $\lambda_{n\bk}$ is that
the former represents an average over the Fermi surface, while the latter refers
to the Fermi velocity renormalization of a specific electron band. While these quantities
are related, they do not coincide and hence cannot be used interchangeably.

Equations~(\ref{eq.mcmillan})-(\ref{eq.wlog}) involve a number of approximations. For example,
it is assumed that the superconductor is isotropic and exhibits a single superconducting gap.
Furthermore, almost invariably the effective Coulomb potential $\mu^*$ is treated as an adjustable
parameter, on the grounds that it should be in the range $\mu^* = 0.1\text{-}0.2$.
This procedure introduces a large uncertainty in the determination 
of $T_{\rm c}$, especially at moderate coupling strengths. 

\subsection{Anisotropic Migdal-Eliashberg theory}\label{sec.eliashberg} 

A first-principles approach to the calculation of the superconducting critical temperature
is provided by the anisotropic Migdal-Eliashberg theory 
(\citeauthor{Migdal1958}, \citeyear{Migdal1958}; \citeauthor{Eliashberg1960}, \citeyear{Eliashberg1960}). 
This is a field-theoretic approach to the superconducting pairing, formulated in the language
of finite-temperature Green's functions. At variance with 
the Hedin-Baym equations of Table~\ref{tab.hedin-baym}, the Migdal-Eliashberg theory is
best developed within the Nambu-Gor'kov formalism (\citeauthor{Gorkov1958}, \citeyear{Gorkov1958};
\citeauthor{Nambu1960}, \citeyear{Nambu1960}), which enables describing the propagation
of electron quasiparticles and of superconducting Cooper pairs on the same footing
\cite{Scalapino1969,Schrieffer1983}.
A comprehensive presentation of the Migdal-Eliashberg theory is provided by \textcite{AllenMitrovic}.
Their article served as the starting point of current first-principles implementations of the theory.

In the Migdal-Eliashberg theory, one solves the two coupled equations:
 \begin{eqnarray}
 &&Z_{n\bk}(i\omega_j) =1+
 \frac{\pi \kt}{N_{\rm F}} \!\sum_{n'{\bf k}' j'}
 \frac{ \omega_{j'}/\w_j }{\sqrt{\hbar^2\omega_{j'}^2+\Delta_{n'{\bf k}'}^2(i\omega_{j'})} }
 \nonumber \\
  &&\hspace{0.3cm}\times \, \lambda_{n\bk,n'\bk'}(i\omega_{j}\!-\!i\omega_{j'}) 
  \delta(\ve_{n'{\bf k}'}\!-\!\ve_{\rm F}), \label{eq.me1}\\
 &&Z_{n\bk}(i\omega_j) \Delta_{n\bk}(i\omega_j) =
 \frac{\pi \kt}{N_{\rm F}} \!\sum_{n'{\bf k}' j'} 
 \frac{ \Delta_{n'\bk'}(i\omega_{j'}) }{ \sqrt{\hbar^2\omega_{j'}^2+\Delta_{n'\bk'}^2(i\omega_{j'})} } \nonumber\\
 &&\hspace{0.3cm}\times\left[ \lambda_{n\bk,n'\bk'}(i\omega_{j}\!-\!i\omega_{j'})\!-\!N_{\rm F}
 V_{n\bk,n'\bk'}\right]\delta(\ve_{n'{\bf k}'}\!-\!\ve_{\rm F}),\label{eq.me2}
 \end{eqnarray}
where $\sum_{\bk'}$ stands for $\Omega_{\rm BZ}^{-1}\int d\bk'$.
In these equations, $T$ is the absolute temperature, $Z_{n\bk}(i\omega_j)$ 
is the quasiparticle renormalization function, and is analogous to $Z_{n\bk}$ 
in Eq.~(\ref{eq.lambdank}). $\Delta_{n\bk}(i\omega_j)$ the superconducting gap function.
The functions $Z_{n\bk}(i\omega_j)$
and $\Delta_{n\bk}(i\omega_j)$ are determined along the imaginary frequency axis,
at the fermion Matsubara frequencies $i\w_j=i(2j+1)\pi \kt/\hbar$ with $j$ an integer.
The anisotropic and frequency-dependent generalization of Eq.~(\ref{eq.lambda-sc}) to be used in the Migdal-Eliashberg
equations is:
 \begin{equation}
 \lambda_{n\bk,n'\bk'}(i\w) = 
  \frac{N_{\rm F}}{\hbar} {\sum}_{\nu} 
 \frac{2\omega_{\bq\nu} }{\omega_{\bq\nu}^2+\w^2} | g_{nn'\nu}(\bk,\bq)|^2,
 \end{equation}
with $\bq=\bk'\!-\!\bk$. Equations~(\ref{eq.me1})-(\ref{eq.me2}) are to be solved self-consistently
for each temperature $T$. The superconducting critical temperature is then obtained as the
highest temperature for which a nontrivial solution is obtained, that is a solution with 
$\Delta_{n\bk}(i\omega_j)\ne 0$. From the superconducting gap along the imaginary axis
it is then possible to obtain the gap at real frequencies by analytic continuation \cite{Marsiglio1988},
and from there one can compute various thermodynamic functions.

  \begin{figure}[t!]
  \includegraphics[width=0.8\columnwidth]{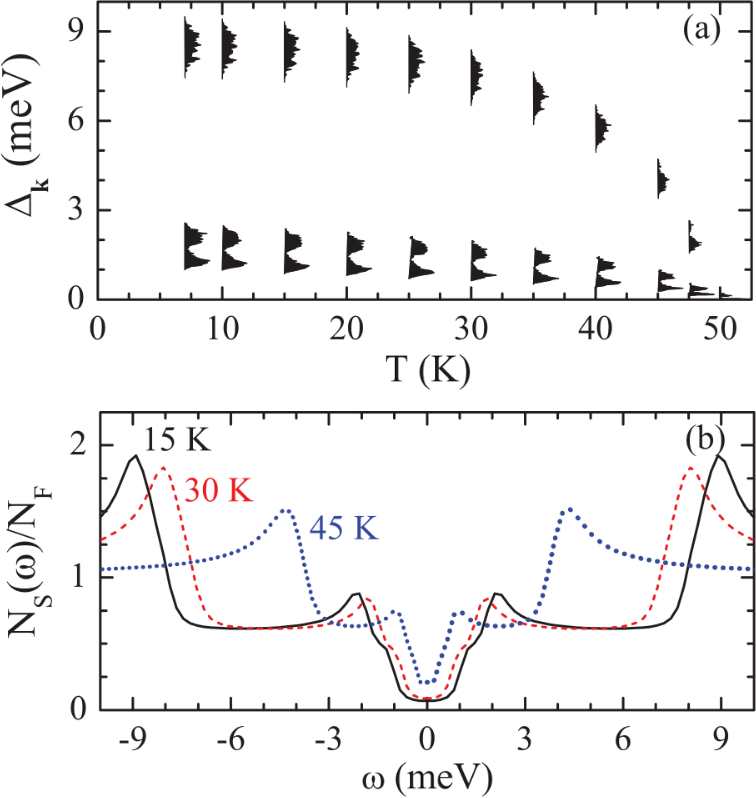}
  \caption[fig]{\label{fig.margine} (Color online)
  (a)~Energy distribution of the superconducting gap function of MgB$_2$ as a function
  of temperature, calculated using the anisotropic Migdal-Eliashberg theory. The gap vanishes
  at the critical temperature (in this calculation $T_{\rm c}=50$~K). Two distinct superconducting
  gaps can be seen at each temperature. (b)~Density of electronic states in the superconducting
  state of MgB$_2$ at various temperatures calculated within the Migdal-Eliashberg theory.
  Reproduced with permission from \cite{Margine2013}, copyright
  (2013) by the American Physical Society.
  }
  \end{figure}

The first {\it ab~initio} implementation of the anisotropic Migdal-Eliashberg theory was
reported by \textcite{Choi2002,Choi2002b} and \textcite{Choi2003} in a study of the superconducting properties
of MgB$_2$. The authors succeeded to explain the anomalous heat capacity of MgB$_2$ in terms
of two distinct superconducting gaps, and obtained a $T_{\rm c}$ in good agreement with experiment.
These calculations were later extended to MgB$_2$ under pressure \cite{Choi2009b} 
and other hypothetical borides \cite{Choi2009}. \textcite{Margine2013} demonstrated an implementation
of the Migdal-Eliashberg theory based on the Wannier interpolation scheme of Sec.~\ref{sec.wannier},
and reported applications to Pb and MgB$_2$. The superconducting gap and superconducting
density of states of MgB$_2$ calculated by \citeauthor{Margine2013} are shown in Fig.~\ref{fig.margine}.
In all these calculations, the Coulomb repulsion was described empirically via $\mu^*$, and this
partly accounts for the slight discrepancy between the calculated $T_{\rm c}$ of 50~K and the experimental
$T_{\rm c}$ of 39~K \cite{Nagamatsu2001}. 
Additional calculations based on the anisotropic Migdal-Eliashberg theory include
a study of doped graphene \cite{Margine2014}, as well as investigations of 
Li-decorated monolayer graphene \cite{Zheng2016} and Ca-intercalated
bilayer graphene \cite{Margine2016}. In this latter work the authors incorporated
Coulomb interactions from first principles, after calculating $\mu^*$ via the screened 
Coulomb interaction in the random-phase approximation. The calculated $T_{\rm c}=7$-8~K was in reasonable
agreement with the experimental value of 4~K \cite{Ichinokura2016}.
The Migdal-Eliashberg theory has also been extended to describe the
superconducting state as a function of applied magnetic field;
a complete {\it ab~initio} implementation was successfully demonstrated with an
application to MgB$_2$ \cite{Aperis2015}. Very recently \textcite{Sano2016} performed
{\it ab~initio} Migdal-Eliashberg calculations including retardation effects on high-pressure sulfur hydrides,
obtaining good agreement with experiment.
Interestingly in this work the authors also checked the effect of the zero-point renormalization
of the electron bands within the Allen-Heine theory, and found that it accounts for a change
in $T_{\rm c}$ of up to 20~K.

\subsection{Density functional theory for superconductors} \label{sec.scdft}

Another promising {\it ab~initio} approach to the calculation of the superconducting critical temperature
is the density functional theory for superconductors \cite{Luders2005,Marques2005}.
The starting point of this approach is a generalization of the Hohenberg-Kohn theorem \cite{Hohenberg1964} 
to a system described by three densities: the electron density in the normal state,
the density of superconducting pairs, and the nuclear density. Based on this premise,
\citeauthor{Luders2005} mapped the fully-interacting system into an equivalent
Kohn-Sham system \cite{Kohn1965} of non-interacting nuclei and non-interacting, yet superconducting, 
electrons.
The resulting Kohn-Sham equations for the electrons take the form of Bogoliubov-de Gennes equations
\cite{Bogoliubov1958},
whereby electrons are paired by an effective gap function $\Delta(\br,\br')$. 

In its simplest formulation, the density functional theory for superconductors determines the
expectation value of the pairing field over Kohn-Sham eigenstates,
$\Delta_{n\bk} = \< u_{n\bk}(\br) | \Delta(\br,\br') | u_{n\bk}(\br')\>$,
using the following gap equation:
  \begin{eqnarray}\label{eq.scdft1}
  \Delta_{n\bk} = -\mathcal{Z}_{n\bk}\Delta_{n\bk}-\sum_{n'\bk'}
  \frac{\mathcal{K}_{n\bk,n'\bk'}\Delta_{n'\bk'}}{2 E_{n'\bk'}}
      \tanh\left(\frac{E_{n'\bk'}}{2\kt}\right),\nonumber\\
  \end{eqnarray}
where $E_{n\bk}^2 = \ve_{n\bk}^2 + |\Delta_{n\bk}|^2$. 
In this expression, the kernel $\mathcal{K}$ contains information about the phonon-mediated
pairing interaction and the Coulomb repulsion between electrons, $\mathcal{K}=\mathcal{K}^{\rm ep}+
\mathcal{K}^{\rm ee}$, and $\mathcal{Z}$ contains information about the electron-phonon
interaction. More specifically, $\mathcal{K}^{\rm ep}$ and $\mathcal{Z}$ are evaluated starting
from the electron-phonon matrix elements $g_{mn\nu}(\bk,\bq)$ and the DFT electron band structure
and phonon dispersions, as in the Migdal-Eliashberg theory. $\mathcal{K}^{\rm ee}$
is approximated using the screened Coulomb interaction $V_{n\bk,n'\bk'}$ introduced below
Eq.~(\ref{eq.wlog}).
Complete expressions for $\mathcal{Z}$ and $\mathcal{K}$ can be found in \cite{Marques2005}.

  \begin{figure}[t!]
  \includegraphics[width=\columnwidth]{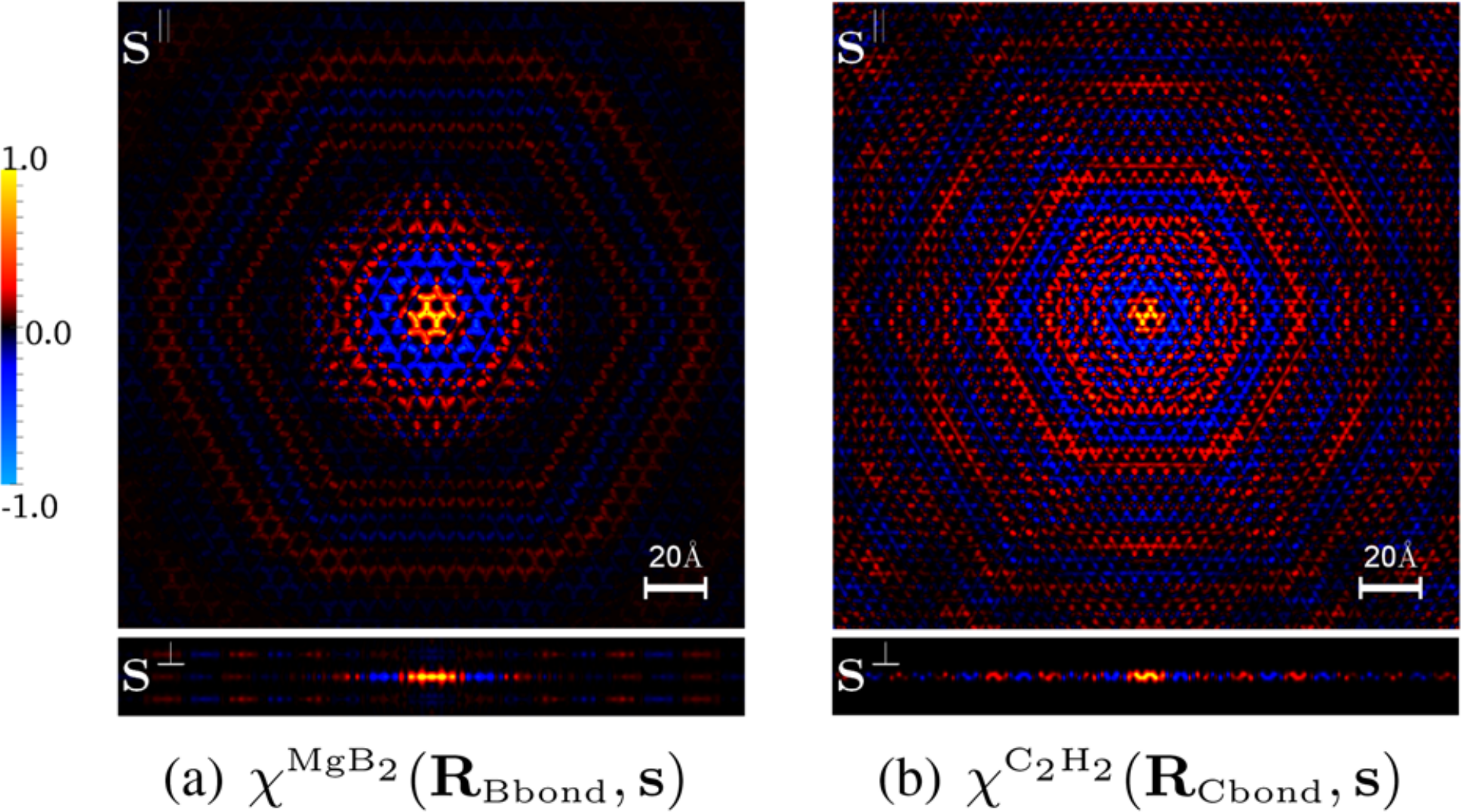}
  \caption[fig]{\label{fig.linscheid} (Color online)
  Superconducting order parameter $\chi(\br,\br')$ 
  in real space, calculated for (a)~MgB$_2$ and (b)~hole-doped 
  graphane. The plots show a top view (top) and a side view (bottom) of the hexagonal layers
  in each case. The variable ${\bf s}\!=\!\br-\br'$ is the relative coordinate in the order parameter,
  while the center-of-mass coordinate is placed in the middle of a B-B bond or a C-C bond. 
  Reproduced with permission from \cite{Linscheid2015c}, copyright
  (2015) by the American Physical Society.
  }
  \end{figure}

Equation~(\ref{eq.scdft1}) is reminiscent of the gap equation in the Bardeen-Cooper-Schrieffer (BCS) theory
\cite{Schrieffer1983}, with the difference that the {\it ab~initio} kernel $\mathcal{K}$ replaces
the model interaction of the BCS theory, and the function $\mathcal{Z}$ introduces quasiparticle
renormalization as in the Migdal-Eliashberg theory, see Eq.~(\ref{eq.me1}).
At variance with the Migdal-Eliashberg theory, the gap function in the density functional theory 
for superconductors does not carry an explicity frequency dependence. Nevertheless, retardation
effects are fully included through the dependence of the kernels $\mathcal{Z}$ and $\mathcal{K}$
on the electron bands and the phonon dispersions. An important advantage of this theory is that
the Coulomb potential $\mu^*$ is not required, since the electron-electron repulsion is 
seamlessly taken into account by means of the kernel $\mathcal{K}^{\rm ee}$.

The density functional theory for superconductors was successfully employed to study the superconducting
properties of MgB$_2$ \cite{Floris2005}, Li, K, and Al under pressure \cite{Profeta2006,Sanna2006}, 
Pb \cite{Floris2007}, Ca-intercalated graphite \cite{Sanna2007}, high-pressure hydrogen
\cite{Cudazzo2008,Cudazzo2010}, CaBeSi \cite{Bersier2009}, layered nitrides \cite{Akashi2012},
alkali-doped fullerides \cite{Akashi2013}, compressed sulfur hydrides \cite{Akashi2015},
and intercalated layered carbides, silicides, and germanides \cite{Flores2015}.

An interesting recent development of the theory was the determination of the superconducting
order parameter in real space, $\chi(\br,\br') = \< \hat{\psi}_\uparrow(\br) \hat{\psi}_\downarrow(\br')\>$ 
\cite{Linscheid2015c}. In the density functional
theory for superconductors, the order parameter
is obtained from the superconducting gap using the relation $\chi_{n\bk} = \Delta_{n\bk}/(2|E_{n\bk}|) 
\tanh\left[E_{n\bk}/(2\kt)\right]$. 
Figure~\ref{fig.linscheid} shows the order parameter
calculated by \citeauthor{Linscheid2015c} for both MgB$_2$ and 
hole-doped graphane \cite{Savini2010}. The plots show Friedel-like oscillations of the superconducting
density as a function of the relative coordinates between two paired electrons.

Further developments of the superconducting density functional theory include the
study of non-phononic pairing mechanisms, such as plasmon-assisted superconductivity \cite{Akashi2013c},
and the extension to magnetic systems \cite{Linscheid2015a,Linscheid2015b}.

\section{Electron-phonon interactions beyond the local density approximation to DFT}\label{sec.beyonddft}

The calculations of electron-phonon interactions reviewed in Sec.~\ref{sec.nonadiab}-\ref{sec.supercond}
have in common the fact that most investigations used the local density approximation to DFT or
a generalized gradient approximation (GGA) such as the PBE functional \cite{Perdew1996}. Although the
LDA and the GGA
do represent the workhorse of electron-phonon calculations from first principles, there is growing evidence 
that these choices can lead to an underestimation of the electron-phonon coupling strength.
At a conceptual level we can understand this point by rewriting the electron-phonon matrix element
after combining Eqs.~(\ref{eq.matel}) and (\ref{eq.gbare}), (\ref{eq.gscreened1}):
  \begin{equation}
  g_{mn\nu}(\bk,\bq)= \< u_{m\bk+\bq} |\!\! \int\!\!d\br'\, \e_{\rm e}^{-1}(\br,\br',\w) \Delta_{\bq\nu} v^{\rm en}(\br') 
    | u_{n\bk} \>_{\rm uc}.
  \end{equation}
In DFT the many-body dielectric matrix $\e_{\rm e}$ appearing in this expression 
is replaced by the RPA$+xc$ screening $\e^{{\rm H}xc}$
from Eq.~(\ref{eq.eps-rpa+xc}). Given the DFT band gap problem, we expect 
$\e^{{\rm H}xc}$ to overestimate the screening, thereby leading to matrix elements $g_{mn\nu}(\bk,\bq)$
which are underestimated to some extent.

Several groups investigated this point on quantitative grounds. \textcite{Zhang2007} studied the
electron-phonon coupling in a model copper oxide superconductor, CaCuO$_2$. By calculating
the vibrational frequencies of the half-breathing Cu-O stretching mode, the authors established
that the local spin-density approximation (LSDA) yields phonons which are
too soft (65.3~meV) as compared to experiment (80.1~meV). In contrast, the introduction of
Hubbard corrections in a LSDA$+U$ scheme restored agreement with experiment (80.9~meV). 
Since the electron-phonon matrix elements are connected to the phonon frequencies via the
phonon self-energy, Eq.~(\ref{eq.phon-self-dft}), a corresponding underestimation of the
matrix elements can be expected. These results were supported by the work of \textcite{Floris2011},
who developed DFPT within LSDA$+U$, and applied their
formalism to the phonon dispersions of antiferromagnetic MnO and NiO.
Here the authors found that the DFT underestimates measured LO energies by as much as 15~meV
in MnO, while the use of LSDA$+U$ leads to good agreement with experiment. 
Related work was reported by \textcite{Hong2012}, who investigated 
the multiferroic perovskites CaMnO$_3$, SrMnO$_3$, BaMnO$_3$, LaCrO$_3$, LaFeO$_3$, and the
double perovskite La$_2$CrFeO$_6$.
Here the authors calculated the variation of the vibrational frequencies between the ferromagnetic
and the antiferromagnetic phases of these compounds as a function of the Hubbard $U$ parameter,
and compared DFT$+U$ calculations with hybrid-functional calculations.

\textcite{Lazzeri2008} investigated the effect of quasiparticle $GW$ corrections on the
electron-phonon coupling of graphene and graphite, for the $A_1'$ phonon at $K$ and the $E_{2g}$
phonon at $\Gamma$. They evaluated the intraband electron-phonon matrix elements using a
frozen-phonon approach, noting that $g_{nn\nu}(\bk,\bq\!=\!0)$ represents precisely the shift
of the Kohn-Sham energy $\ve_{n\bk}$ upon displacing the atoms according to the $\nu$-th phonon eigenmode
at $\bq\!=\!0$.
\citeauthor{Lazzeri2008} found that the matrix elements increase by almost 40\% from DFT to $GW$. 
The $GW$ values led to slopes in the phonon dispersions near $K$ in very good agreement with 
inelastic X-ray scattering data \cite{Gruneis2009b}. Similar results, albeit less dramatic,
were obtained by \textcite{Gruneis2009} for the potassium-intercalated graphite KC$_8$. 

\textcite{Janssen2010} studied the electron-phonon coupling in the C$_{60}$ molecule as a model 
for superconducting alkali-doped fullerides. They employed the PBE0 hybrid functional  \cite{PBE0}
with a fraction of exact exchange $\a\!=\!30$\%, 
and obtained an enhancement of the total coupling strength $\lambda$ of 42\% as compared
to PBE. This work was followed up by \textcite{Faber2011}, who used
the $GW$ approximation and obtained a similar enhancement of 48\%. We also point out an earlier
work by \textcite{Saito2002} based on the B3LYP functional, reporting similar results.

\textcite{Yin2013} investigated the effects of using the $GW$ approximation and the HSE hybrid
functional \cite{HSE} on the electron-phonon coupling in the superconducting 
bismuthates Ba$_{1-x}$K$_x$BiO$_3$ and chloronitrides $\beta$-ZrNCl, as well as MgB$_2$.
In the case of Ba$_{1-x}$K$_x$BiO$_3$ the authors obtained a three-fold increase in the coupling
strength $\lambda$ from PBE to HSE. This enhancement brought the critical temperature calculated
using Eq.~(\ref{eq.mcmillan}) to 31~K, very close to the experimental value of 32~K.
Similarly, in the case of $\beta$-ZrNCl, \citeauthor{Yin2013} obtained a 50\% increase of $\lambda$, 
bringing the calculated critical temperature, 18~K, close to the experimental value of 16~K.
Instead, in the case of MgB$_2$, they noticed only a slight increase of the electron-phonon
coupling as compared to the standard LDA. 

Another application of hybrid functionals to the study of EPIs was reported
by \textcite{Komelj2015}. Here the authors investigated the sensitivity of the superconducting
critical temperature of the H$_3$S phase of sulfur hydride to the exchange and correlation functional.
They found that the PBE0 functional enhances the critical temperature by up to 25\%
as compared to PBE, bringing $T_{\rm c}$ from 201-217~K to 253-270~K (the spread in values is related
to the choice of the parameter $\mu^*$).

\textcite{Mandal2014} reported work on the superconductor FeSe based on dynamical mean-field
theory (DMFT). In this case DMFT yielded a three-fold enhancement of the coupling strength for
selected modes.

As already mentioned in Sec.~\ref{sec.temper}, \textcite{Antonius2014} performed $GW$
calculations of the electron-phonon coupling in diamond using a frozen-phonon approach. They
found that quasiparticle corrections lead to a uniform enhancement of the electron-phonon
matrix elements. The net effect is an increase of the zero-point renormalization of the band gap
by 40\% as compared to standard LDA calculations.
\textcite{Monserrat2016b} confirmed this result and found a $GW$ correction 
of comparable magnitude in the case of silicon. However, \citeauthor{Monserrat2016b} also found
that the $GW$ corrections to the zero-point band gap renormalization of LiF, MgO, and TiO$_2$
are very small ($\sim$5\% of the PBE value), therefore at present it is not possible to draw general
conclusions.

Finally, we mention that \textcite{Faber2015} examined possible strategies for systematically 
incorporating $GW$ corrections in electron-phonon calculations. By using diamond, graphene,
and C$_{60}$ as test cases, the authors showed that a `constant screening' approximation is
able to reproduce complete $GW$ results with an error below 10\% at reduced computational cost. 
This approximation amounts
to evaluating the variation of the Green's function $G$ in a frozen-phonon calculation,
while retaining the screened Coulomb interaction $W$ of  the unperturbed ground state.

All these recent developments point to the need of moving beyond local exchange and correlation
functionals in the study of
electron-phonon interactions from first principles. In the future, it will be important to
devise accurate computational methods for calculating not only the intraband electron-phonon
matrix elements (as in the frozen-phonon method) but also matrix elements between all states and
for scattering across the entire Brillouin zone. 

For the sake of completeness we emphasize that the underestimation of the EPI matrix elements
by semilocal DFT functionals does not propagate in the same way into different materials properties.
This is readily understood by examining two fundamental quantities, 
the Allen-Heine renormalization of electron bands, Eq.~(\ref{eq.AH-RS}), and
the adiabatic phonon frequencies, as obtained from Eqs.~(\ref{eq.bo-tmp1}) and (\ref{eq.density-deriv}).
In the former case the electronic screening enters as $\epsilon_\infty^{-2}$;
in the latter case the screening contributes through a term which scales with $\epsilon_\infty^{-1/2}$.
As a result, in the hypothetical case of a semiconductor for which
DFT underestimated the electronic permittivity by 20\%, we would have an error
of $\sim$40\% in the energy renormalization, and of $\sim$10\% in the phonon frequencies.
This example is an oversimplification of the problem, but it shows 
that different properties relating to the EPI could be affected to a very different degree
by the inherent limitations of DFT functionals.

\section{Conclusions}\label{sec.conclusions}

The study of electron-phonon interactions has a long and distinguished history, 
but it is only during the past two decades that {\it quantitative} and {\it predictive}
calculations have become possible. First-principles calculations of electron-phonon 
couplings are finding an unprecedented variety of applications in many areas of condensed 
matter and materials physics, from spectroscopy to transport, from metals to semiconductors 
and superconductors. 
In this article we discussed the standard DFT formalism for performing calculations of
electron-phonon interactions, we showed how most equations can be derived from a
field-theoretic framework using a few well-defined approximations, and we reviewed recent 
applications of the theory to many materials of current interest.

As calculation methods improve relentlessly and quantitative comparisons 
between theory and experiment become increasingly refined, 
new and more complex questions arise.
Much is still left to do, both in the fundamental theory of electron-phonon interactions,
and in the development of more accurate and more efficient computational methods.

For one, we are still using theories where the coupling matrix elements are
calculated using the adiabatic local density approximation to DFT.
The need for moving beyond standard DFT and beyond the adiabatic approximation
can hardly be overemphasized. Progress is being made on the incorporation of nonlocal
corrections into electron-phonon matrix elements, for example using hybrid functionals 
or $GW$ techniques, but very little is known
about retardation effects. It is expected that such effects may be important
in the study of heavily doped oxides and semiconductors, both in their normal and 
superconducting states (\citeauthor{Mahan1993}, \citeyear{Mahan1993}, Sec.~6.3.A), 
but {\it ab~initio} investigations are currently missing.
 This is truly uncharted territory and a systematic
effort in this direction is warranted.

In this article we emphasized that it is possible to formulate 
a compact, unified theory of electron-phonon interactions starting from 
a fully {\it ab~initio} field-theoretic approach. The only assumption which
is absolutely crucial to the theory is the {\it harmonic} approximation.
Abandoning the harmonic approximation leads to the appearance of several
new terms in the equations, and the resulting formalism becomes considerably more complex
than in Table~\ref{tab.hedin-baym}. 
 Despite these difficulties, given the importance of anharmonic effects in
many systems of current interest, extending the theory to the case of anharmonic phonons
and multi-phonon interactions constitutes a pressing challenge.
{\it Ab initio} investigations of anharmonic effects on the temperature
dependence of band gaps have recently 
been reported (\citeauthor{Monserrat2013}, \citeyear{Monserrat2013};
\citeauthor{Antonius2015}, \citeyear{Antonius2015}).
Since these studies rely on non-perturbative adiabatic calculations in supercells, it would
be highly desirable to establish a clear formal connection of these methods with
the rigorous field-theoretic approach of Sec.~\ref{sec.green}. Along the same line,
it would be important to clarify the relation between many-body approaches,
adiabatic supercells calculations, and more traditional classical or path-integral 
molecular dynamics simulations.

The study of electron-phonon interactions has long been dominated by Fr\"ohlich-like
Hamiltonians, whereby the electron-phonon coupling is retained only to linear order in
the atomic displacements. This is the case for all the model 
Hamiltonians mentioned in Sec.~\ref{sec.polarons}.
It is now clear that quadratic couplings, leading
to the Debye-Waller contributions in the optical spectra of semiconductors, are by no
means negligible and should be investigated more systematically. For
example, in the current literature it is invariably assumed that Debye-Waller contributions are 
negligible in metals near the Fermi surface; while this is probably the case for the simplest
elemental metals, what happens in the case of multiple Fermi-surface 
sheets is far from clear, and should be tested by direct calculations.

The identification of the correct matrix elements to be calculated is not always 
a trivial task, as it was discussed for the case of the non-adiabatic phonon self-energy. 
In the future it will be important to pay attention to these aspects, especially
in view of detailed comparison with experiment. For now, the issue on whether
the phonon self-energy arising from EPIs should be calculated using bare or screened EPI
matrix elements (Sec.~\ref{sec.nonadiab}) is to be considered an open question, and calls 
for further investigation.

The theory and applications reviewed in this article focused on non-magnetic systems.
The rationale for this choice is that a complete many-body theory 
of electron-phonon interactions for magnetic systems is not available yet. Recent investigations of 
spin-phonon couplings were conducted by assuming that the spin and the vibrational degrees
of freedom can be decoupled, as in the Born-Oppenheimer approximation. Under this assumption
it is possible to investigate how the spin configuration responds to a frozen phonon,
or alternatively how the vibrational frequencies depend on the spin configuration
(see for example \citeauthor{Chan2007}, \citeyear{Chan2007};
\citeauthor{Lazewski2010}, \citeyear{Lazewski2010};
\citeauthor{Lee2011}, \citeyear{Lee2011};
\citeauthor{Cao2015}, \citeyear{Cao2015}). 
In all these cases it would be desirable to employ a more rigorous many-body theory of spin-phonon
interactions.
The Hedin-Baym equations discussed in Sec.~\ref{sec.green}
maintain their validity in the case of spin-polarized systems,
provided collinear spins are assumed. 
In more general situations, where it is important to consider
noncollinear spins, external magnetic perturbations, or spin-dependent interactions
such as spin-orbit and Rashba-Dresselhaus couplings, it becomes necessary to generalize
the equations in Table~\ref{tab.hedin-baym}. 
Although such a generalization has not been reported yet, the
work of \citeauthor{Aryasetiawan2008} (\citeyear{Aryasetiawan2008}) constitutes
a promising starting point. In that work the Schwinger functional derivative technique
(see Sec.~\ref{sec.GW-1}) was used to extend Hedin's equations at clamped nuclei
to systems containing spin-dependent interactions.
Generalizing \citeauthor{Aryasetiawan2008}'s work to incorporate nuclear vibrations
will be important for the study of electron-phonon
interactions in many systems of current interest, from multifunctional
materials to topological quantum matter.

At this time it is not possible to predict how this fast-moving field will evolve over the
years to come. However, the impressive progress made during the past decade gives us
confidence that this interesting research area 
will continue to thrive, and will keep surprising us with fascinating challenges and 
exciting new opportunities.

\begin{acknowledgments}
I am indebted with Samuel Ponc\'e, Carla Verdi, and Marios Zacharias for a critical
reading of the manuscript, and with many colleagues who have kindly provided
feedback on the arXiv preprint version of this article.
This article is the result of many stimulating discussions with
friends and colleagues over the last few years. I would like to thank in particular 
Philip Allen,
Stefano Baroni,
Lilia Boeri,
Nicola Bonini,
Fabio Caruso,
Hyoung Joon Choi,
Marvin Cohen,
Michel C\^ot\'e,
Andrea Dal Corso,
Claudia Draxl,
Asier Eiguren,
Giulia Galli,
Paolo Giannozzi,
Stefano de Gironcoli,
Xavier Gonze,
G\"oran Grimvall,
Eberhard Gross,
Branko Gumhalter,
Sohrab Ismail-Beigi,
Emmanuil Kioupakis,
Georg Kresse,
Steven Louie,
Roxana Margine,
Andrea Marini,
Miguel Marques,
Nicola Marzari,
Takashi Miyake,
Bartomeu Monserrat,
Jeffrey Neaton,
Richard Needs,
Jesse Noffsinger,
Cheol-Hwan Park,
Christopher Patrick,
Warren Pickett,
Samuel Ponc\'e,
Paolo Radaelli,
Lucia Reining,
John Rehr,
Patrick Rinke,
Angel Rubio,
Matthias Scheffler,
Young-Woo Son,
Alexander Tkatchenko,
Chris Van de Walle,
Carla Verdi,
Matthieu Verstraete,
and Marios Zacharias.
I also would like to thank
Marco Bernardi,
Asier Eiguren,
Xavier Gonze,
Eberhard Gross,
Wu Li,
Steven Louie, and 
Francesco Mauri
for kindly allowing me to reproduce their figures.
This work was supported by the Leverhulme Trust (Grant RL-2012-001), the Graphene Flagship 
(Horizon 2020 grant no. 696656-GrapheneCore1), and the UK Engineering and Physical Sciences Research Council
(Grants No. EP/J009857/1 and No. EP/M020517/1).
\end{acknowledgments}

\appendix

\section{Born-von~K\'arm\'an boundary conditions}\label{sec.bvk}

In this Appendix we provide more details on the notation related to the Born-von~K\'arm\'an boundary conditions
used throughout this article.
The crystalline unit cell is defined by the primitive lattice vectors $\ba_i$ with $i=1,2,3$,
and the $p$-th unit cell is identified by the vector $\bRp = {\sum}_i \, n_i \ba_i $ with $n_i$ 
integers between 0 and $N_i-1$. The BvK supercell contains $N_p = N_1\!\times\! N_2\!\times\! N_3$ unit cells.
The primitive vectors of the reciprocal lattice are denoted by $\bb_j$, and fulfil the duality condition
$\ba_i \cdot \bb_j = 2\pi \delta_{ij}$.
We consider Bloch wavevectors $\bq$ belonging to a uniform grid in one unit cell 
of the reciprocal lattice: $\bq = {\sum}_j (m_j/N_j)\bb_j$ with $m_j$ being integers between 0 and $N_j\!-\!1$. 
This grid contains the same number of $\bq$-vectors as the number of unit cells in the BvK 
supercell. From these definitions the standard sum rules follow:
  \begin{eqnarray}
  {\sum}_\bq \exp(i\bq\cdot\bRp) = N_p \delta_{p 0}, \,\,\,\,
  {\sum}_p \exp(i\bq\cdot\bRp) = N_p \delta_{\bq 0}. \nonumber \\ \label{eq.psum}
  \end{eqnarray}
If $\bG$ is a reciprocal-lattice vector, the replacement of any of the 
$\bq$-vectors by $\bq+\bG$ in these expressions and in all expressions presented in this article
is inconsequential, since $\exp(i\bG\cdot\bRp)=1$. 
Similarly any replacement of $\bRp$ by $\bRp+\bT$ where $\bT$ is a lattice vectors of the
BvK supercell is inconsequential. 
Owing to these properties we are at liberty to replace the $\bq$-grid defined above with 
a Wigner-Seitz grid, i.e.~the first Brillouin zone, and the supercell with a Wigner-Seitz supercell.
These choices are useful for practical calculations in order to exploit the symmetry operations of the crystal,
and to truncate the interatomic force constants, given by Eq.~(\ref{eq.ifc}), outside a Wigner-Seitz supercell.

\section{Ladder operators in extended systems}\label{sec.normalcoord}

In this Appendix we describe the 
construction of the phonon ladder operators $\ha_{\bq\nu}/\ha^\dagger_{\bq\nu}$,
and derive the phonon Hamiltonian given by Eq.~(\ref{eq.herm-compl}).
We show how the definition of the ladder operators depends on the behavior of 
the wavevector $\bq$ under inversion.

The normal modes introduced in Eq.~(\ref{eq.dynmat2}) can be used to define a linear 
coordinate transformation of the ionic displacements as follows:
  \begin{equation}\label{eq.x-from-tau}
  z_{\bq\nu} = N_p^{-\frac{1}{2}}\sum_{\kappa\a p} e^{-i\bq\cdot\bRp} 
  (M_\kappa/M_0)^{\frac{1}{2}}\,
  e^*_{\kappa\a,\nu}(\bq) \,\Delta\tau_{\kappa\a p}. 
  \end{equation}
Here $z_{\bq\nu}$ is referred to as `complex normal coordinate' \cite{Bruesch1982}.
The exponential and the masses in Eq.~(\ref{eq.x-from-tau}) are chosen so as
to obtain Eq.~(\ref{eq.herm-compl}) starting from Eq.~(\ref{eq.harm}). 
Since there are $3MN_p$ degrees of freedom, and since the complex normal coordinates correspond to
$2\times 3MN_p$ real variables, this coordinate transformation carries some redundancy.
Indeed by combining Eqs.~(\ref{eq.e-cc}) and (\ref{eq.x-from-tau}) it is seen that:
  \begin{equation}\label{eq.xminus}
  z_{-\bq\nu} = z^*_{\bq\nu}.
  \end{equation}
The inverse relation of Eq.~(\ref{eq.x-from-tau}) is: 
  \begin{equation}\label{eq.tau-from-x-2}
  \Delta\tau_{\kappa\a p}  = N_p^{-\frac{1}{2}} (M_0/M_\k)^\frac{1}{2}
  {\sum}_{\bq\nu} e^{i\bq\cdot\bRp} e_{\kappa\a,\nu}(\bq)\, z_{\bq\nu}.
  \end{equation}
The right-hand side is real-valued after Eqs.~(\ref{eq.e-cc}) and~(\ref{eq.xminus}).
In preparation for the transition to a quantum description of lattice vibrations, 
it is useful to identify $3MN_p$ independent normal coordinates. This can be done by partitioning the grid 
of $\bq$-vectors in three sets. We call $\mathcal{A}$ the set of vectors which are invariant under inversion, 
that is $-\bq + \bG = \bq$ for some reciprocal lattice vector $\bG$ (including $|\bG|\!=\!0$). The 
center of the Brillouin zone and the centers of its faces 
belong to this set. The remaining vectors can be separated further in $\mathcal{B}$ and $\mathcal{C}$, 
in such a way that all the vectors in $\mathcal{C}$ are obtained from those in $\mathcal{B}$ by inversion
(modulo a reciprocal lattice vector).
After defining $z_{\bq\nu}\!=\!x_{\bq\nu}\!+iy_{\bq\nu}$,
Eq.~(\ref{eq.tau-from-x-2}) can be rewritten as:
  \begin{eqnarray}
  \Delta\tau_{\kappa\a p} &=&  N_p^{-\frac{1}{2}}(M_0/M_\kappa)^\frac{1}{2}
  \left[
  {\sum}_{\bq \in \mathcal{A},\nu} e_{\kappa\a,\nu}(\bq) 
  \,x_{\bq\nu} \right. \nonumber \\
  & +& \left.
  2 {\rm Re} \,{\sum}_{\bq \in \mathcal{B},\nu}  \,e^{i\bq\cdot\bRp} e_{\kappa\a,\nu}(\bq)
  (x_{\bq\nu}+iy_{\bq\nu})\right]\!\!. \hspace{0.5cm}\label{eq.realvar}
  \end{eqnarray}
The $\bq$-vectors of the set $\mathcal{C}$ have been grouped together with those in $\mathcal{B}$
by taking the real part in the second line. 
It can be verified that in this expression there are exactly $3M N_p$ real coordinates,
therefore we can choose the $x_{\bq\nu}$ for $\bq$ in $\mathcal{A}$  and the pairs $x_{\bq\nu}$, $y_{\bq\nu}$ 
for $\bq$ in $\mathcal{B}$ as the independent variables. These variables are referred to as 
`real normal coordinates' \cite{Bruesch1982}.

Using Eqs.~(\ref{eq.harm})-(\ref{eq.e-cc}), (\ref{eq.psum}), and (\ref{eq.realvar})
the nuclear Hamiltonian can be written in terms of $3MN_p$ independent harmonic oscillators
in the real normal coordinates:
  \begin{eqnarray}\label{eq.hamilt-real-var}
  \hH_{\rm p} &= &
  \!\frac{1}{2}{\sum}_{\bq\in \mathcal{B},\nu} \hbar\w_{\bq\nu}( -\D^2/\D \tilde{x}_{\bq\nu}^2\!
    -\D^2/\D \tilde{y}_{\bq\nu}^2\! +  \tilde{x}_{\bq\nu}^2 +  \tilde{y}_{\bq\nu}^2  ) \nonumber \\
  &+& \!\frac{1}{2}{\sum}_{\bq\in \mathcal{A},\nu} \hbar\w_{\bq\nu} (-\D^2/\D \tilde{x}_{\bq\nu}^2 +   
   \tilde{x}_{\bq\nu}^2 ),
  \end{eqnarray}
where for ease of notation we performed the scaling:
  \begin{eqnarray}
   & \tilde{x}_{\bq\nu} = x_{\bq\nu}/2\, l_{\bq\nu} \hspace{2.4cm} & \mbox{ for $\bq$ in }\mathcal{A},
         \label{eq.xA} \\
   & \tilde{x}_{\bq\nu} = x_{\bq\nu}/l_{\bq\nu}, \,\, \tilde{y}_{\bq\nu} = y_{\bq\nu}/l_{\bq\nu} \,\,\,
    & \mbox{ for $\bq$ in }\mathcal{B}, \label{eq.xB}
  \end{eqnarray}
with $l_{\bq\nu}$ being the zero-point displacement amplitude of Eq.~(\ref{eq.zeropdisp2}).\label{page.3modes}
In the case of $|\bq|\!=\!0$ there are three normal modes for which $\w_{\bq\nu}\!=\!0$,
and the corresponding potential terms $\tilde{x}_{\bq\nu}^2$ must be removed from Eq.~(\ref{eq.hamilt-real-var}).

The eigenstates of Eq.~(\ref{eq.hamilt-real-var}) are found by introducing
the real ladder operators for each normal coordinate \cite{Tannoudji1977}:
  \begin{equation}\label{eq.tann1}
  \ha_{\bq\nu,x} = (\tilde{x}_{\bq\nu}+ \D/\D \tilde{x}_{\bq\nu})/\sqrt{2},
  \end{equation}
and similarly for $\ha_{\bq\nu,y}$. 
With these definitions Eq.~(\ref{eq.hamilt-real-var}) becomes:
  \begin{eqnarray}
  \hH_{\rm p} &= &
  {\sum}_{\bq\in \mathcal{B},\nu} \hbar\w_{\bq\nu} \left(
     \ha^\dagger_{\bq\nu,x} \ha_{\bq\nu,x} 
   + \ha^\dagger_{\bq\nu,y} \ha_{\bq\nu,y} + 1
    \right)  \nonumber \\
  &+&{\sum}_{\bq\in \mathcal{A},\nu} \hbar\w_{\bq\nu} \left( \ha^\dagger_{\bq\nu,x} \ha_{\bq\nu,x} + 1/2 \right).
  \end{eqnarray}
The eigenstates of this Hamiltonian are products of simple harmonic oscillators \cite{Merzbacher1998},
and the ground state is:
  \begin{equation}\label{eq.phonon-gs} 
  \chi_0(\{\btau_{\k p} \}) = A\, e^{-\frac{1}{2}\left({\sum}_{\bq\in\mathcal{A},\nu} \tilde{x}_{\bq\nu}^2
  +{\sum}_{\bq\in\mathcal{B},\nu} \tilde{x}_{\bq\nu}^2 + \tilde{y}_{\bq\nu}^2
  \right)},
  \end{equation}
with $A$ a normalization constant. The relations between the positions $\btau_{\k p}$
and the normal coordinates $\tilde{x}_{\bq\nu}$, $\tilde{y}_{\bq\nu}$ are given by
Eqs.~(\ref{eq.x-from-tau}), (\ref{eq.xA})-(\ref{eq.xB}), and~(\ref{eq.zeropdisp2}).

The eigenstates of $\hH_{\rm p}$ can be generated by applying 
$\ha^\dagger_{\bq\nu,x}$ and $\ha^\dagger_{\bq\nu,y}$ to the ground state $\chi_0$.
However this approach is not entirely satisfactory,
since we cannot assign separate quantum numbers to modes with wavevectors $\bq$ or~$-\bq$.
In order to avoid this inconvenience we observe that, for each normal mode,
the first line of Eq.~(\ref{eq.hamilt-real-var}) defines an effective
isotropic two-dimensional harmonic oscillator.
The degenerate eigenstates of this oscillator can
be combined to form eigenstates of the angular momentum; this leads to right and
left circular quanta with the same energy and definite angular momentum \cite{Tannoudji1977}.
This analogy motivates the consideration of the following linear combinations, for $\bq$ in~$\mathcal{B}$:
  \begin{eqnarray}
  \ha^+_{\bq\nu} & = & (\ha_{\bq\nu,x}+i \ha_{\bq\nu,y})/\sqrt{2}, \label{eq.apq-tmp} \\
  \ha^-_{\bq\nu} & = & (\ha_{\bq\nu,x}-i \ha_{\bq\nu,y})/\sqrt{2}. \label{eq.amq-tmp}
  \end{eqnarray}
Since both $\ha_{\bq\nu,x}$ and $\ha_{\bq\nu,y}$ lower the energy of an eigenstate
by the same quantum of energy $\hbar\w_{\bq\nu}$, the resulting states are degenerate
and their linear combinations are also eigenstates for the same eigenvalue.
As a consequence we can generate all the eigenstates of the Hamiltonian $\hH_{\rm p}$
by acting on the ground state $\chi_0$ with the creation operators $\ha^{+,\dagger}_{\bq\nu}$ 
and $\ha^{-,\dagger}_{\bq\nu}$. 
In this reasoning the wavevectors $\bq$ belong to $\mathcal{B}$;
if we now consider
Eqs.~(\ref{eq.xminus}), (\ref{eq.apq-tmp}), and (\ref{eq.amq-tmp}) we see that formally we also have
$\ha^-_{\bq\nu} = \ha^+_{-\bq\nu}$.
Therefore it is natural to associate $\ha^-_{\bq\nu}$ to 
phonons propagating along the direction $-\bq$.

These observations suggest replacing the real ladder operators of Eq.~(\ref{eq.tann1}) by the
complex ladder operators $\ha^+_{\bq\nu}$ and $\ha^-_{-\bq\nu}$ for $\bq$ in $\mathcal{B}$ 
and $\mathcal{C}$, respectively. In the case of $\bq$ in $\mathcal{A}$ we keep the real 
operators $\ha_{\bq\nu,x}$.
These definitions can be turned into the compact expressions:
  \begin{eqnarray}
   \ha_{\bq\nu} =&  \ha_{\bq\nu,x} \hspace{2cm}&\qquad\!\!\mbox{ for $\bq$ in }\mathcal{A}, \label{eq.complex-ladder2}\\
   \ha_{\bq\nu} =&\,\, 
    (\ha_{\bq\nu,x}+i \ha_{\bq\nu,y})/\sqrt{2}  &\qquad\!\!\mbox{ for $\bq$ in }\mathcal{B}, \mathcal{C}.
     \label{eq.complex-ladder}
  \end{eqnarray}
Using these operators the nuclear Hamiltonian of Eq.~(\ref{eq.hamilt-real-var}) takes the well-known form
given by Eq.~(\ref{eq.herm-compl}).
Any eigenstate of $\hH_{\rm p}$ can now be generated as
$\prod_{\bq\nu} (n_{\bq\nu}!)^{-\frac{1}{2}} (\ha^{\dagger}_{\bq\nu})^{n_{\bq\nu}} \chi_0$.
In this form we see that it is possible to assign independently a number of phonons $n_{\bq\nu}$ to 
each wavevector $\bq$ and each mode~$\nu$.
Using Eqs.~(\ref{eq.xA})-(\ref{eq.tann1}) and (\ref{eq.complex-ladder2})-(\ref{eq.complex-ladder})
we also have the basic identity: 
  \begin{equation}\label{eq.z-from-a}
  z_{\bq\nu} = l_{\bq\nu} \,(\ha_{\bq\nu}+\ha^\dagger_{-\bq\nu}).
  \end{equation}
By combining this last expression with Eq.~(\ref{eq.tau-from-x-2}) we obtain Eq.~(\ref{eq.tau-from-x}).

\end{document}